\documentclass[11pt,a4paper,twoside]{report} 
\usepackage{feynmf,german,amsmath,amssymb,latexsym,epsfig,floatflt,subfigure}
\renewcommand{\slash}[1]{#1\!\!\!/}
\newcommand{\scr}{\scriptscriptstyle}
\newcommand{\decay}{\;\rule[0.5ex]{0.4pt}{1ex}\hspace*{-4pt}\rightarrow}

\newcommand{\muk}{\mu^{\scr +} K^0}
\newcommand{\epi}{e^{\scr +} \pi^0}
\newcommand{\et}{e^{\scr +} \eta}
\newcommand{\ek}{e^{\scr +} K^0}
\newcommand{\mpi}{\mu^{\scr +} \pi^0}
\newcommand{\mt}{\mu^{\scr +} \eta}
\newcommand{\ero}{e^{\scr +} \rho^0}
\newcommand{\eo}{e^{\scr +} \omega}
\newcommand{\eks}{e^{\scr +} K^{*0}}
\newcommand{\mro}{\mu^{\scr +} \rho^0}
\newcommand{\mo}{\mu^{\scr +} \omega}
\newcommand{\nep}{\nu^{\scr C}_{\scr e} \pi^+}
\newcommand{\nek}{\nu^{\scr C}_{\scr e} K^+}
\newcommand{\nmpi}{\nu^{\scr C}_{\scr \mu} \pi^+}
\newcommand{\nmk}{\nu^{\scr C}_{\scr \mu} K^+}
\newcommand{\nero}{\nu^{\scr C}_{\scr e} \rho^+}
\newcommand{\neks}{\nu^{\scr C}_{\scr e} K^{*+}}
\newcommand{\nmro}{\nu^{\scr C}_{\scr \mu} \rho^+}
\newcommand{\nmks}{\nu^{\scr C}_{\scr \mu} K^{*+}}
\newcommand{\ntp}{\nu^{\scr C}_{\scr \tau} \pi^+}
\newcommand{\ntk}{\nu^{\scr C}_{\scr \tau} K^+}
\newcommand{\ntro}{\nu^{\scr C}_{\scr \tau} \rho^+}
\newcommand{\ntks}{\nu^{\scr C}_{\scr \tau} K^{*+}}
\newcommand{\epin}{e^{\scr +} \pi^-}
\newcommand{\mpin}{\mu^{\scr +} \pi^-}
\newcommand{\eron}{e^{\scr +} \rho^-}
\newcommand{\mron}{\mu^{\scr +} \rho^-}
\newcommand{\nepn}{\nu^{\scr C}_{\scr e} \pi^0}
\newcommand{\nekn}{\nu^{\scr C}_{\scr e} K^0}
\newcommand{\nmpin}{\nu^{\scr C}_{\scr \mu} \pi^0}
\newcommand{\nmkn}{\nu^{\scr C}_{\scr \mu} K^0}
\newcommand{\neron}{\nu^{\scr C}_{\scr e} \rho^0}
\newcommand{\neksn}{\nu^{\scr C}_{\scr e} K^{*0}}
\newcommand{\nmron}{\nu^{\scr C}_{\scr \mu} \rho^0}
\newcommand{\nmksn}{\nu^{\scr C}_{\scr \mu} K^{*0}}
\newcommand{\neon}{\nu^{\scr C}_{\scr e} \omega}
\newcommand{\nmon}{\nu^{\scr C}_{\scr \mu} \omega}
\newcommand{\netn}{\nu^{\scr C}_{\scr e} \eta}
\newcommand{\nmtn}{\nu^{\scr C}_{\scr \mu} \eta}
\newcommand{\ntpn}{\nu^{\scr C}_{\scr \tau} \pi^0}
\newcommand{\ntkn}{\nu^{\scr C}_{\scr \tau} K^0}
\newcommand{\nttn}{\nu^{\scr C}_{\scr \tau} \eta}
\newcommand{\ntron}{\nu^{\scr C}_{\scr \tau} \rho^0}
\newcommand{\nton}{\nu^{\scr C}_{\scr \tau} \omega}
\newcommand{\ntksn}{\nu^{\scr C}_{\scr \tau} K^{*0}}
\begin{document}

\thispagestyle{empty}

\begin{titlepage}

\vspace*{-2cm}

\begin{tabular}{llr}
 & \hspace{1.5cm} FACHBEREICH PHYSIK & \\
 & \hspace{1.5cm} BERGISCHE UNIVERSIT"AT & \\
 & \hspace{1.5cm} GESAMTHOCHSCHULE WUPPERTAL & \\
\begin{picture}(0.,0.)
\put(-0.2,0.35){\includegraphics[scale=0.1]{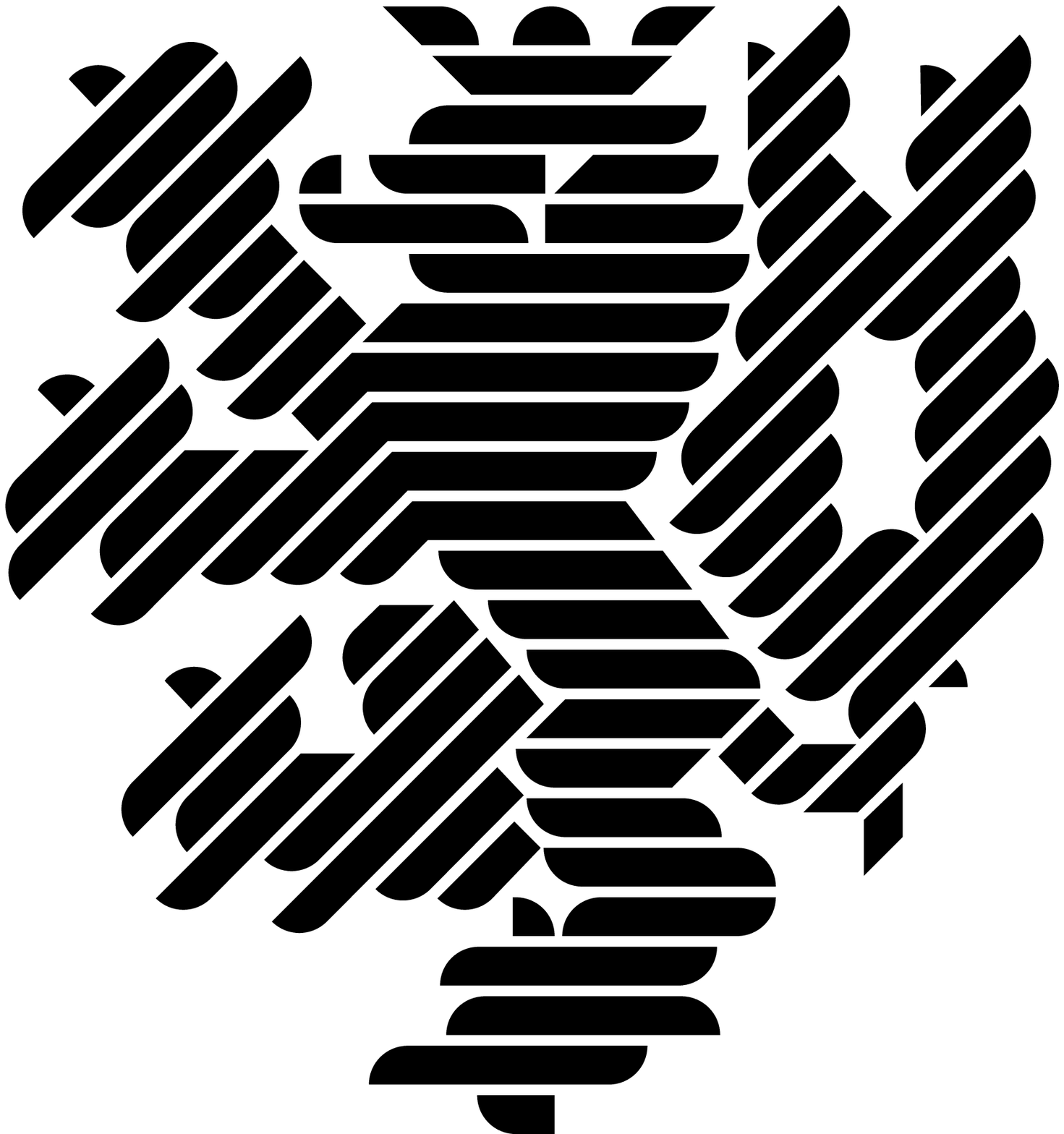}}
\end{picture}
\end{tabular}

\vspace{2.5cm}

\begin{center}
 
{\Huge \textbf{Nukleonenzerfall und \\
Neutrinoeigenschaften in einem \\
Massenmodell auf der \\
Grundlage einer \\
$SO(10)$-Grand Unif\/ied-Theorie \\ \mbox{}}}

\vspace{4cm}

{\large 
Dissertation zur Erlangung des Doktorgrades der Naturwissenschaften \\
am Fachbereich Physik der Bergischen Universit\"at \\
Gesamthochschule Wuppertal \\
\vspace{1cm}
vorgelegt von \\ 
{\bf Carsten Merten}}

\vspace{4cm}

{\large {\bf WUB-DIS 99-14} \\
Dezember 1999}

\end{center}

\end{titlepage}

\thispagestyle{empty}

\mbox{ }

\newpage

\thispagestyle{empty}

\mbox{ }

\newpage

\thispagestyle{empty}

\mbox{ }

\begin{titlepage}

\vspace*{-2cm}

\begin{tabular}{llr}
 & \hspace{1.5cm} FACHBEREICH PHYSIK & \\
 & \hspace{1.5cm} BERGISCHE UNIVERSIT"AT & \\
 & \hspace{1.5cm} GESAMTHOCHSCHULE WUPPERTAL & \\
\begin{picture}(0.,0.)
\put(-0.2,0.35){\includegraphics[scale=0.1]{Pics/loewe.eps}}
\end{picture}
\end{tabular}

\vspace{2.5cm}

\begin{center}
 
{\Huge \textbf{Nukleonenzerfall und \\
Neutrinoeigenschaften in einem \\
Massenmodell auf der \\
Grundlage einer \\
$SO(10)$-Grand Unif\/ied-Theorie \\ \mbox{}}}

\vspace{4cm}

{\large 
Dissertation zur Erlangung des Doktorgrades der Naturwissenschaften \\
am Fachbereich Physik der Bergischen Universit\"at \\
Gesamthochschule Wuppertal \\
\vspace{1cm}
vorgelegt von \\ 
{\bf Carsten Merten}}

\vspace{4cm}

{\large {\bf WUB-DIS 99-14} \\
Dezember 1999}

\end{center}

\end{titlepage}

\thispagestyle{empty}

\vspace*{3cm}

\begin{verse}

{\it \hspace*{1cm} Wo Du das Nichts erblickst, ist eine Kraft, \\
     \hspace*{1cm} Verborgen, unerreichbar allem B"osen, \\
     \hspace*{1cm} Die aus sich selbst -- sich und die Welt erschafft. \\
     \hspace*{1cm} Und die vermag's, das R"atsel aufzul"osen. \\
     \hspace*{1cm} \\
     \hspace*{1cm} \rm{Michael Ende}, \it{Das Gauklerm"archen} }

\end{verse}

\newpage

\thispagestyle{empty}

\vspace*{1.5cm}

\begin{center}
  \LARGE{ {\bf Abstract}}
\end{center}

\noindent The Standard Model (SM) of elementary particle physics is a 
very successful theory. Its predictions have been tested experimentally 
to a high level of accuracy. However, the SM is not considered to be a 
fundamental theory of nature. It contains a lot of arbitrary parameters
especially in the fermionic sector and it cannot give small neutrino 
masses which are indicated by recent experiments like Super-Kamiokande.

\noindent Grand Unif\/ied Theories (GUTs) can solve several weaknesses of the
SM. They unify the SM interactions and lead to relations between the
quark and lepton mass matrices, thus reducing the arbitrariness in the
fermionic sector. Furthermore, GUT models with an intermediate
symmetry breaking scale are able to produce small neutrino masses by 
means of the see-saw mechanism. All GUTs include baryon and lepton 
number violating interactions which mediate proton and bound neutron decay.

\noindent In this work a mass model based on a $SO(10)$ GUT with a global
$U(1)$ family symmetry is discussed which leads to an asymmetric 
Nearest Neighbour Interaction structure for the fermionic mass
matrices. As a result of the analysis one gets three solutions of the model
which include several large left- and right-handed fermion mixings. 
Those mixings are not observable in the SM where only the CKM quark
mixing matrix can be measured, but they have testable ef\/fects on the 
branching ratios of nucleon decays in theories beyond the SM. One
f\/inds that decay channels with $e^+$ in the f\/inal state are suppressed 
while channels with $\mu^+$ and $\nu^{\scr C}$ are enhanced compared to 
models with small mixings. The total nucleon lifetimes obtained should be
observable by future experiments.
The $SO(10)$ model also predicts the masses and mixings of the light 
neutrinos. They are in the right range to explain the
anomalies of solar and atmospheric neutrinos by means of oscillations,
preferring the small angle MSW solution for the solar neutrino def\/icit.

\newpage

\thispagestyle{empty}

\setcounter{page}{0}

\mbox{ }

\newpage

\tableofcontents

\listoftables

\listoffigures

\chapter*{Einleitung}
\addcontentsline{toc}{chapter}{Einleitung}

\pagestyle{myheadings}\markboth{EINLEITUNG}{EINLEITUNG}

\thispagestyle{plain}

\noindent Das Standardmodell der Elementarteilchenphysik ist eine
"au"serst erfolgreiche Theorie. Es beschreibt auf konsistente Weise 
die starke, schwache und elektromagnetische Wechselwirkung, und seine
Vorhersagen sind experimentell mit gro"ser Genauigkeit best"atigt worden.
Nach der Entdeckung des top-Quarks steht lediglich der Nachweis
des Higgs-Bosons, dessen Existenz von der Theorie gefordert wird, noch aus.

Es gibt jedoch verschiedene Anhaltspunkte, die darauf hindeuten, da"s
das Standardmodell keine wirklich fundamentale Theorie darstellt,
sondern vielmehr als ef\/fektive Nieder\-ener\-gie-N"aherung einer solchen
den heute experimentell zug"anglichen Energiebereich beschreibt. So
sind die Neutrinos im Rahmen des Standardmodells masselose Teilchen,
w"ahrend die Evidenz f"ur nichtverschwindende Neutrinomassen durch
Experimente wie Super-Kamiokande in den letzten Jahren deutlich
zugenommen hat. Auch die beobachtete Baryonasymmetrie des Universums
l"a"st sich nicht befriedigend erkl"aren.

Eine weitere formale Schw"ache des Standardmodells besteht in der 
gro"sen Zahl freier Parameter, welche durch die Theorie nicht festgelegt
sind, sondern an experimentelle Ergebnisse angepa"st werden
m"ussen. Es gibt 18 solcher Parameter, von denen allein 13 im Fermionsektor
liegen; dies sind die Fermionmassen und die Quarkmischungen. Abgesehen
davon liefert das Standardmodell keine wirkliche Vereinheitlichung der
drei Wechselwirkungen, da die zugeh"origen Kopplungskonstanten v"ollig
unabh"angig voneinander sind.

Um diese Schwachpunkte zumindest teilweise beseitigen zu k"onnen, ist
eine Reihe von Versuchen gemacht worden, das Standardmodell in eine
umfassendere Theorie einzubetten. Viele von ihnen liefern jedoch kaum
Vorhersagen, welche in naher Zukunft experimentell verif\/iziert werden
k"onnen. Einen in dieser Hinsicht vielversprechenden Ansatz
liefern die sogenannten Grand Unif\/ied-Theorien, deren Grundidee darin
besteht, die drei Kr"afte des Standardmodells in einer einzigen
Wechselwirkung zu vereinheitlichen. Formal geschieht das durch die
Konstruktion einer Eichtheorie mit einfacher Symmetriegruppe, welche
die Standardmodell-Symmetriegruppe als Untergruppe
enth"alt. Motivation hierf"ur ist in erster Linie die Tatsache, da"s
die skalenabh"angigen Kopplungen des Standardmodells sich bei sehr
gro"sen Energien von $10^{14}-10^{15}$ GeV n"aherungsweise in einem 
Punkt tref\/fen. Mittels spontaner Symmetriebrechung erh"alt man aus 
der Grand Unif\/ied-Theorie als ef\/fektive N"aherung bei niedrigen
Energien wieder das Standardmodell.

Die wohl bemerkenswerteste Vorhersage von Grand Unif\/ied-Theorien
besteht in der Instabilit"at des Protons und des gebundenen Neutrons 
aufgrund von baryon- und leptonzahlverletzenden Wechselwirkungen. 
Einerseits sind Zerf"alle von Nukleonen bisher nicht beobachtet
worden, andererseits ist die relevante Wechselwirkung in Grand 
Unif\/ied-Theorien wegen der gro"sen Masse der zugeh"origen Eichbosonen 
bei niedrigen Energien auch "uberaus schwach. Die einfachste der 
Grand Unif\/ied-Theorien, welche auf der Gruppe $SU(5)$ beruht, ist 
experimentell ausgeschlossen worden, da sie unter anderem
zu kurze Lebensdauern f"ur die Nukleonen liefert. Modelle auf der 
Grundlage der $SO(10)$ wie das hier untersuchte besitzen diesen 
Mangel nicht.

Da sich die laufenden Standardmodell-Kopplungen nicht exakt in einem
Punkt tref\/fen, ist eine direkte Vereinheitlichung der drei
Wechselwirkungen nicht m"oglich; es mu"s eine zus"atzliche
intermedi"are Symmetrie vorhanden sein. Die Skala, bei welcher diese
Symmetrie in das Standardmodell gebrochen wird, liegt "ublicherweise
im Bereich $10^{10}-10^{12}$ GeV und hat somit die richtige
Gr"o"senordnung, damit "uber den See-Saw-Mechanismus sehr kleine, aber 
nichtverschwindende Neutrinomassen erzeugt werden k"onnen.
Weitere Vorz"uge von Grand Unif\/ied-Theorien sind Beziehungen 
zwischen den Massenmatrizen der Quarks und der Leptonen. 
Infolgedessen k"onnen durch Kenntnis der Massenmatrizen der geladenen 
Fermionen Vorhersagen f"ur die Neutrinomassen und -mischungen gewonnen 
werden. Weiterhin kann im Rahmen von $SO(10)$-Modellen mit
intermedi"arer Massenskala und schweren Majorana-Neutrinos die 
Baryonasymmetrie des Universums erkl"art werden.

Ein anderer Versuch, den Fermionsektor des Standardmodells besser zu
verstehen, besteht in der Untersuchung von ph"anomenologisch
motivierten Ans"atzen f"ur die Massenmatrizen der Fermionen. Diese 
Ans"atze zeichnen sich durch Symmetrieanforderungen oder
sogenannte Texturen, das hei"st Nullen als Matrixeintr"age an
bestimmten Stellen, aus. Auf die zugrundeliegende Theorie jenseits
des Standardmodells, welche den Ansatz realisiert, wird im allgemeinen
nicht weiter eingegangen. Damit will man die Zahl der freien
Parameter des Standardmodells reduzieren und Beziehungen zwischen 
Massen und Mischungen erhalten.

Schlie"slich kann man beide Zug"ange kombinieren und Massenmodelle auf
der Grundlage von Grand Unif\/ied-Theorien konstruieren. Das er"of\/fnet
die M"oglichkeit, die Vorz"uge dieser Theorien gegen"uber dem
Standardmodell auszunutzen und die Ans"atze wegen der gegebenen Beziehungen
zwischen den Fermionmassenmatrizen weniger willk"urlich zu machen.
Ferner haben in solchen Modellen alle Fermionmischungen Einf\/lu"s auf
zumindest prinzipiell observable Gr"o"sen wie Nukleonzerfallsraten,
w"ahrend im Standardmodell lediglich eine bestimmte Kombination der 
linksh"andigen Quarkmischungen, die CKM-Matrix, beobachtbar
ist. Daraus folgen "uberpr"ufbare Konsequenzen, an denen man den Erfolg 
des Ansatzes messen kann.

Gegenstand dieser Arbeit wird ein Massenmodell im Rahmen einer 
$SO(10)$-Theorie sein. Es wird ein asymmetrischer 
"`Nearest Neighbour Interaction"'-Ansatz f"ur die 
Massenmatrizen der Fermionen benutzt, welcher durch eine globale 
$U(1)$-Familiensymmetrie realisiert wird.
Dieser Ansatz f"uhrt auf voneinander unabh"angige rechts- und 
linksh"andige Mischungen und bietet ausdr"ucklich die M"oglichkeit, 
da"s diese betragsm"a"sig gro"s sind. 
In vielen Massenmodellen werden gro"se Mischungen im Bereich 
der geladenen Fermionen mit dem Hinweis auf die relativ kleinen 
CKM-Mischungen der Quarks au"ser Acht gelassen.
In der Tat besitzen alle gefundenen L"osungen des untersuchten 
Modells mehrere gro"se Mischungen, was zu Verzweigungsraten der 
Nukleonen f"uhrt, die sich von denen im Fall verschwindender
Mischungen deutlich unterscheiden. 
Zus"atzlich sind die erhaltenen Neutrinoeigenschaften, welche ebenfalls
Modellvorhersagen darstellen, in der Lage, die Anomalien der Sonnen-
und atmosph"arischen Neutrinos durch Oszillationsl"osungen zu erkl"aren. 

In den letzten Jahren hat sich die Forschungst"atigkeit haupts"achlich
auf supersymmetrische Grand Unif\/ied-Theorien, welche spontan in das
minimale supersymmetrische Standardmodell gebrochen werden,
beschr"ankt. Wird die Supersymmetrie, eine Symmetrie zwischen
Fermionen und Bosonen, bei vergleichsweise niedrigen
Energien von $\lesssim 1$ TeV ef\/fektiv gebrochen, so tref\/fen sich die
Kopplungskonstanten des supersymmetrische Standardmodells bei
etwa $10^{16}$ GeV genau in einem Punkt. Da die Massen der
Superpartner dann nicht sehr viel gr"o"ser als die der gew"ohnlichen 
Teilchen sind, bietet sich die M"oglichkeit, das Divergenzverhalten
der Theorie zu verbessern und das Hierarchieproblem zu l"osen. 
Allerdings l"a"st die Abwesenheit von experimentellen
Hinweisen auf eine bei kleinen Energien gebrochene Supersymmetrie in 
der Natur ein solches Szenario zunehmend unwahrscheinlicher erscheinen.
Auch eine f"ur die Erzeugung von Neutrinomassen "uber den
See-Saw-Mechanismus erforderliche intermedi"are Massenskala
l"a"st sich in supersymmetrischen Grand Unif\/ied-Theorien nicht auf
nat"urliche Weise realisieren.
Desweiteren sind supersymmetrische Modelle in ihrer Vorhersagekraft 
durch den Einf\/lu"s zahlreicher unbekannter Gr"o"sen, wie zum Beispiel 
die Massen der Superpartner, stark eingeschr"ankt. Deshalb wird hier
der Standpunkt vertreten, da"s Modelle ohne Supersymmetrie weiter
untersucht werden sollten. Dies schlie"st eine Brechung der
Supersymmetrie bei sehr hohen Energien keineswegs aus. 

Im ersten Kapitel wird zun"achst ein "Uberblick "uber die
Grundkonzepte und wichtigsten Eigenschaften des Standardmodells 
gegeben, auch dessen Grenzen werden genauer betrachtet. Kapitel\,2 
behandelt den Themenkomplex der Grand Unif\/ied-Theorien. Nach
einer Schilderung der grundlegenden Ideen und der allgemeinen
Vorgehensweise bei der Konstruktion solcher Modelle werden 
die $SU(5)$ als Prototyp der Grand Unif\/ied-Theorien und die $SO(10)$
als das dieser Arbeit zugrundeliegende Beispiel ausf"uhrlich
diskutiert; desweiteren wird auf Vor- und Nachteile der Modelle
eingegangen. Gegenstand des dritten Kapitels sind die theoretischen 
Grundlagen und der experimentelle Status von Neutrino-Oszillationen. 
Die drei bis heute beobachteten Neutrino-Anomalien und ihre m"oglichen 
Oszillationsl"osungen durch massive Neutrinos werden
vorgestellt. Kapitel\,4 beginnt mit vorbereitenden Arbeiten wie der 
Bestimmung der Symmetriebrechungsskalen und Kopplungen.
Mittelpunkt des Kapitels ist der Ansatz f"ur das betrachtete 
$SO(10)$-Massenmodell und dessen numerische L"osung. Als ein wichtiges
Resultat erh"alt man daraus Voraussagen "uber die Eigenschaften im
Neutrinosektor der Theorie, welche sich als ph"anomenologisch sinnvoll
erweisen. In Kapitel\,5 schlie"slich werden f"ur die analysierten 
L"osungen die partiellen und totalen Zerfallsraten der Nukleonen 
berechnet. Diese stellen eine wesentliche und in absehbarer Zeit 
experimentell "uberpr"ufbare Vorhersage des untersuchten Modells dar.

Ziel und Motivation dieser Untersuchung lassen sich abschlie"send mit 
den einleitenden Worten aus \cite{dhr} sehr tref\/fend zusammenfassen:

{\it "`The Standard Model is unlikely to be a fundamental
theory; it contains 18 arbitrary parameters, 13 of which are the
fermion masses and mixing angles. In a fundamental theory, these
should be calculable from a few inputs just as the hydrogen spectral
lines follow from Quantum Mechanics. We are very far from such a
theory of fermion masses. We would be fortunate to have an analogue
of Balmer's formula since it might lead us to the fundamental
theory. The framework described here is, at best, an attempt to
obtain such a formula."'}

\pagestyle{headings}

\chapter{Das Standardmodell}
\section{Grundkonzepte des Standardmodells}
Das Standardmodell (SM) der Elementarteilchenphysik \cite{ism} ist
eine renormierbare Eichtheorie, welche die Theorie der starken 
Wechselwirkung, die 
Quantenchromodynamik (QCD), und das Glashow-Weinberg-Salam-Modell der 
elektroschwachen Wechselwirkung zusammenfa"st. Es basiert auf der Invarianz
unter lokalen $SU(3)_{\scr C} \otimes SU(2)_{\scr L} \otimes
U(1)_{\scr Y}$-Eich\-trans\-for\-ma\-tionen; die Symmetriegruppe 
$G_{\scr \textrm{SM}}$ ist ein direktes Produkt aus drei Faktoren.

Die QCD \cite{gwp} beruht auf der Eichgruppe $SU(3)_{\scr C}$ und 
beschreibt die Wechselwirkung von Teilchen mit Farbladung, den Quarks
und Gluonen. Die Quarks sind die fermionischen Grundbausteine der stark
wechselwirkenden Materie und treten in drei verschiedenen Farbzust"anden
auf. Unter $SU(3)_{\scr C}$-Transformationen verhalten sie sich wie
die Fundamentaldarstellung ${\bf 3}$.
In der Natur werden allerdings nur farbneutrale Kombinationen von
drei Quarks (Baryonen) bzw. Quark und Antiquark (Mesonen) beobachtet;
dieses Ph"anomen wird als Conf\/inement bezeichnet.
Die Gluonen sind die mit den acht Generatoren der $SU(3)_{\scr C}$ 
verbundenen Vektorbosonen und transformieren sich gem"a"s der adjungierten 
Darstellung. Durch den Austausch von Gluonen wird die starke
Wechselwirkung zwischen den Quarks vermittelt.

Die Theorie der elektroschwachen Wechselwirkung \cite{gws} besitzt eine 
$SU(2)_{\scr L} \otimes U(1)_{\scr Y}$-Eich\-sym\-me\-trie, die durch
den Higgs-Mechanismus spontan in die $U(1)_{\scr \textrm{em}}$ der
Quantenelektrodynamik (QED) gebrochen ist. Eine Grundeigenschaft dieses
Modells ist die Parit"atsverletzung, da das Transformationsverhalten 
der Fermionen von deren Chiralit"at abh"angt. Man zerlegt die durch 
Dirac-Spinoren $\Psi$ dargestellten Fermionfelder gem"a"s 
\begin{equation}
\Psi_{\scr L,R} = \dfrac{1}{2} (1 \mp \gamma_{\scr 5}) \Psi
\; , \quad \Psi = \Psi_{\scr L} + \Psi_{\scr R}
\end{equation}
in ihre links- und rechtsh"andigen Komponenten, sogenannte
Weyl-Spinoren. 
Die links\-h"an\-di\-gen Quarks und Leptonen sind in 
$SU(2)_{\scr  L}$-Doubletts angeordnet, w"ahrend die 
rechts\-h"an\-di\-gen Fermionen Singuletts bilden
und an der schwachen Wechselwirkung nicht teilnehmen.

\noindent Die elektrische Ladung der Teilchen ergibt sich aus ihrer
Hyperladung $Y$ und der Komponente $T^{}_{\scr 3}$ des schwachen
Isospins ($T^{}_a = \sigma^{}_a/2$ mit $a=1,2,3$ sind
die Generatoren der $SU(2)_{\scr  L}$-Lie-Algebra) aus der Beziehung
\begin{equation} \label{hyp}
Q = T^{}_{\scr 3} + Y
\end{equation}
In Erweiterungen des SM erweist es sich aus
gruppentheoretischen Gr"unden als sinnvoll, statt der
rechtsh"andigen Teilchen die linkh"andigen Komponenten der
Antiteilchen zu betrachten. Die Antiteilchen erh"alt man durch die
Anwendung der Ladungskonjugation $\mathcal{C}$
\begin{equation}
\mathcal{C} \, \Psi \, \mathcal{C}_{}^{-1} \equiv
\Psi^{\scr C}_{} = C \, \bar \Psi^{\scr T}_{} \quad \textrm{mit} \quad
C = i \, \gamma_{\scr 2} \, \gamma_{\scr 0} \; , \quad 
\Psi^{\scr C}_{\scr L,R} \equiv (\Psi_{\scr R,L})^{\scr C}_{}
\end{equation}
Weiterhin gelten die Identit"aten
\begin{equation} \label{ct}
\Psi_{\scr R,L} = C \, ( \, \bar \Psi^{\scr C}_{\scr L,R} \, )^T
 \quad \textrm{und} \quad
\bar \Psi_{\scr R,L} = ( \, \Psi^{\scr C}_{\scr L,R} \, )^T \, C
\end{equation}

\noindent Die Fermionen treten im SM in drei Familien mit jeweils
gleichen Quantenzahlen auf. Jede dieser Familien transformiert sich
nach der Darstellung 
\begin{equation} \label{smfrep}
({\bf 3,2})_{\scr \frac{1}{6}} \oplus ({\bf \bar 3,1})_{\scr -\frac{2}{3}}
\oplus ({\bf \bar 3,1})_{\scr \frac{1}{3}} 
\oplus ({\bf 1,2})_{\scr -\frac{1}{2}}
\oplus ({\bf 1,1})_{\scr 1}
\end{equation}
der SM-Symmetriegruppe $G_{\scr \textrm{SM}}$. Der erste Eintrag
bezeichnet hierbei die $SU(3)_{\scr C}$-Dar\-stel\-lung, der zweite die 
$SU(2)_{\scr L}$-Darstellung und der Index die $U(1)_{\scr Y}$
-Hyperladungs\-quan\-ten\-zahl (die komplex konjugierte Darstellung
wird durch einen Querstrich gekennzeichnet). 
In Tabelle \ref{SMFerm} ist der fermionische Teilcheninhalt des SM 
aufgef"uhrt. 
\begin{table}[h]
\begin{center}
\begin{tabular}{|c|l|r||c|l|r|}
\hline
Quarks & $G_{\scr \textrm{SM}}$ & $Q$ & Leptonen & $G_{\scr \textrm{SM}}$ & 
$Q$ \\
\hline \hline
$\begin{pmatrix} u^{}_{\scr L} \\ d^{}_{\scr L} \end{pmatrix}$,
$\begin{pmatrix} c^{}_{\scr L} \\ s^{}_{\scr L} \end{pmatrix}$,
$\begin{pmatrix} t^{}_{\scr L} \\ b^{}_{\scr L} \end{pmatrix}$ &
$({\bf 3,2})_{\scr \frac{1}{6}}$ &
$\begin{matrix} +\frac{2}{3} \\ -\frac{1}{3} \end{matrix}$ &
$\begin{pmatrix} \nu^{}_{\scr eL} \\ e^{\scr -}_{\scr L} \end{pmatrix}$,
$\begin{pmatrix} \nu^{}_{\scr \mu L} \\ \mu^{\scr -}_{\scr L} \end{pmatrix}$,
$\begin{pmatrix} \nu^{}_{\scr \tau L} \\ \tau^{\scr -}_{\scr L}\end{pmatrix}$
& 
$({\bf 1,2})_{\scr -\frac{1}{2}}$ & 
$\begin{matrix} \quad\! 0 \\ -1 \end{matrix}$ \\
\hline
$u^{\scr C}_{\scr L}$, $c^{\scr C}_{\scr L}$, $t^{\scr C}_{\scr L}$ & 
$({\bf \bar 3,1})_{\scr -\frac{2}{3}}$ & $-\frac{2}{3}$ &
 & & \\
\hline
$d^{\scr C}_{\scr L}$, $s^{\scr C}_{\scr L}$, $b^{\scr C}_{\scr L}$ & 
$({\bf \bar 3,1})_{\scr \frac{1}{3}}$ & $+\frac{1}{3}$ &
$e^{\scr +}_{\scr L}$, $\mu^{\scr +}_{\scr L}$, $\tau^{\scr +}_{\scr L}$ & 
$({\bf 1,1})_{\scr 1}$ & $+1$ \\
\hline
\end{tabular}
\end{center}
\caption{\label{SMFerm} Fermioninhalt des Standardmodells}
\end{table}

\noindent Es ist zu beachten, da"s im Fermionspektrum des SM keine
rechtsh"andigen Neutrinos vorkommen, da diese sich aufgrund ihrer
Farb- und Ladungsneutralit"at nach der $G_{\scr \textrm{SM}}$-Darstellung 
$({\bf 1,1})_{\scr 0}$ transformieren w"urden, also an keiner 
SM-Wechselwirkung teiln"ahmen. Dies steht im Einklang mit der
Tatsache, da"s rechtsh"andige Neutrinos und linksh"andige
Antineutrinos in Experimenten nicht beobachtet werden.

Die Eichbosonen, die als Austauschteilchen die
Wechselwirkung vermitteln, transformieren sich stets wie die
adjungierte Darstellung der Symmetriegruppe, im Falle des SM also
gem"a"s $({\bf 8,1})_{\scr 0} \oplus ({\bf 1,3})_{\scr 0} 
\oplus ({\bf 1,1})_{\scr 0}$.
Die Kopplungsst"arken der SM-Eichbosonen an die fermionischen Str"ome 
$j^{\mu}_{} = \bar \Psi \gamma^{\mu}_{} \Psi$ werden mit $g^{}_{\scr 3}$, 
$g^{}_{\scr 2}$ und $g'$ bezeichnet. Ferner besitzen 
Eichbosonen in Modellen mit nichtabelscher Symmetriegruppe auch eine 
Selbstwechselwirkung.
In Erweiterungen des SM durch Grand Unif\/ied-Theorien (GUTs) wird statt
$g'$ im allgemeinen mit $g^{}_{\scr 1} = \sqrt{5/3} \, g'$ gearbeitet,
da $g^{}_{\scr 1}$ die korrekte Normierung besitzt. Die Eigenschaften
der SM-Eichbosonen sind in Tabelle \ref{SMBos} zusammengefa"st.
\begin{table}[h]
\begin{center}
\begin{tabular}{|l|l|c|c|}
\hline
Bezeichnung & $G_{\scr \textrm{SM}}$ & Spin & Kopplung \\
\hline \hline
Gluonen $g$ \; [$SU(3)_{\scr C}$] & $({\bf 8,1})_{\scr 0}$ & 1 &
$g^{}_{\scr 3}$ \\
\hline
$W$-Bosonen \; [$SU(2)_{\scr L}$] & $({\bf 1,3})_{\scr 0}$ & 1 & 
$g^{}_{\scr 2}$ \\
\hline
$B$-Boson \; [$U(1)_{\scr Y}$] & $({\bf 1,1})_{\scr 0}$ & 1 & $g' =
\sqrt{3/5} \, g^{}_{\scr 1}$ \\
\hline
\end{tabular}
\end{center}
\caption{\label{SMBos} Eichbosonen des Standardmodells}
\end{table}
\section{Renormierung und laufende Kopplungen}
Ein wichtiges Kriterium f"ur die formale Konsistenz einer Eichtheorie ist ihre
Renormierbarkeit. Dies bedeutet, da"s bei der Berechnung von
physikalischen Prozessen in h"oheren Ordnungen der St"orungstheorie
nur endlich viele qualitativ verschiedene Divergenzen auftreten. Diese k"onnen
dann durch eine Redef\/inition der Modellparameter wie Kopplungen und
Massen absorbiert werden \cite{thv}.

Die Unendlichkeiten treten in Form von divergenten
Impulsintegralen auf. Durch Anwendung eines Regularisierungsverfahrens
werden die Integrale durch Ausdr"ucke ersetzt, die von einem neuen Parameter,
dem Regularisierungsparameter, abh"angen, und f"ur einen bestimmten
Wert desselben wieder die urspr"ungliche Gestalt annehmen. Bei der
dimensionalen Regularisierung zum Beispiel werden die Integrale statt
in vier in $D=4-2\varepsilon$ Impulsraumdimensionen gel"ost; die
Divergenzen gehen dann in Terme $\sim 1/\varepsilon$ "uber. Im
anschlie"senden Renormierungsproze"s werden die divergenten Parameter
$\mathcal{G}_0$ in der Lagrangedichte durch Einf"uhrung von
Renormierungskonstanten $Z$, welche die $1/\varepsilon$-Terme
aufnehmen, in die endlichen renormierten Gr"o"sen $\mathcal{G}_R$
umgewandelt: $\; \mathcal{G}_0 \equiv Z \, \mathcal{G}_R$.

Im Rahmen der Regularisierung wird aus formalen Gr"unden zwangsl"auf\/ig
eine freie Massenskala $\mu$ in die Theorie eingef"uhrt. Sowohl die 
Renormierungskonstanten als auch die renormierten Gr"o"sen h"angen von
dieser Skala ab; man spricht von laufenden Gr"o"sen. Die funktionalen 
Zusammenh"ange, welche die
Skalenabh"angigkeit der renormierten Gr"o"sen beschreiben, werden als
Renormierungsgruppengleichungen bezeichnet \cite{gml}. Sie sind immer
dann von Bedeutung, wenn physikalische Gr"o"sen bei verschiedenen
Massen- bzw. Energieskalen miteinander verglichen werden. Die in
dieser Arbeit verwendeten Renormierungsgruppengleichungen sind in
Anhang \ref{rge} angegeben.
Die Energieabh"angigkeit observabler Gr"o"sen ist experimentell
best"atigt; so hat die Kopplungsst"arke 
$\alpha^{}_{\scr \textrm{em}}=e_{}^2/4\pi$ der QED bei $\mu \approx 0$
den Betrag 1/137, bei $\mu \approx M_{\scr Z}$ ist sie etwa 1/129
gro"s \cite{pdg}.

Die Renormierbarkeit einer Eichtheorie kann durch das Auftreten
von Anomalien zerst"ort werden. Anomalien sind Symmetrien der
klassischen Lagrangedichte, die durch den Proze"s der Quantisierung
gebrochen werden \cite{abj}. In Modellen mit chiralen Fermionen 
"au"sert sich dies 
durch das Auftreten von linear divergenten Strahlungskorrekturen 
in Form von fermionischen Dreiecksdiagrammen mit einer ungeraden
Anzahl axialer Vertizes $\gamma_\mu\gamma_5$. Die Divergenzen
verletzen die Slavnov-Taylor-Identit"aten, deren G"ultigkeit
f"ur die vollst"an\-dige Renormierung der Theorie in allen Ordnungen der 
St"orungsrechnung ben"otigt wird.
Diese Beitr"age sind im wesentlichen proportional zu 
\begin{equation}
A^{}_{abc} = A^L_{abc} - A^R_{abc} =
\textrm{Tr} \, \big( \, \{ \, T^{}_a,T^{}_b \} \, T^{}_c\, \big)_L
- \textrm{Tr} \, \big( \, \{ \, T^{}_a,T^{}_b \} \, T^{}_c\, \big)_R
\; ,
\end{equation}
wobei die $T^{}_i$ die Generatoren der zur Symmetriegruppe geh"orenden
Lie-Algebra sind, und zwar in der Darstellung, nach der sich die
links- beziehungsweise rechtsh"andigen Fermionen transformieren 
\cite{ggan,gj}. Die Renormierbarkeit ist demnach sichergestellt, wenn 
\begin{equation} \label{anom1}
A^{L,R}_{abc} = \textrm{Tr} \, \big( \, \{ \, 
T^{}_a,T^{}_b \} \, T^{}_c \, \big)_{L,R} \stackrel{!}{=} 0
\end{equation}
gilt. Im Rahmen des SM f"uhrt (\ref{anom1}) auf die vier Bedingungen
\begin{eqnarray} \label{anom2}
\textrm{Tr} \, \big( \, \{ \, T^{}_a,T^{}_b \} \, 
T^{}_c \, \big) & = & 0 \; , \quad 
\textrm{Tr} \, \big( \, T^{}_a \, Y_{}^2 \, \big) \; = \; 0 \; , \\ 
\nonumber \textrm{Tr} \, \big( \, \{ \, T^{}_a,T^{}_b \} \, Y \,
\big) & = & 0 \; , \hspace{0.4cm}
\textrm{Tr} \, \big( \, Y_{}^3 \, \big) \; = \; 0
\end{eqnarray}
Die beiden oberen Gleichungen sind aufgrund der Spurfreiheit der
$SU(2)_{\scr L}$-Generatoren automatisch erf"ullt, die unteren beiden
wegen der SM-Zuordnung der Hyperladungen in (\ref{smfrep}). Das SM ist
also deshalb anomaliefrei, weil sich die Beitr"age der Quarks und der
Leptonen gerade aufheben.
\section{Symmetriebrechung und Massenerzeugung}
In Eichtheorien k"onnen prinzipiell zwei Arten von Massentermen
vorkommen. Wenn $\Psi$ ein Dirac-Spinor ist und $\Psi_{\scr L,R}$ seine links-
und rechtsh"andigen Komponenten bezeichnet, so hat der
Dirac-Massenterm die Gestalt (h.c. steht f"ur hermitesch konjugiert)
\begin{equation}
  m \, \bar \Psi \, \Psi = m \, ( \, \bar \Psi_{\scr L} \, \Psi_{\scr R}
 + \bar \Psi_{\scr R} \, \Psi_{\scr L} \, )
= m \, ( \, \bar \Psi_{\scr L} \, \Psi_{\scr R} + \textrm{h.c.} \, )
\end{equation}
Das ist derselbe Massenterm, der auch in der Dirac-Gleichung vorkommt.
Unter Verwendung von (\ref{ct}) kann man 
$\bar \Psi_{\scr R} \, \Psi_{\scr L}$ auch als 
$( \, \Psi^{\scr C}_{\scr L} \, )^T \, C \, \Psi^{}_{\scr L}$ 
schreiben. 

\noindent Desweiteren existieren die lorentzinvarianten Gr"o"sen 
\begin{eqnarray} \label{majmass1}
\dfrac{M_{\scr R}}{2} \, ( \, \bar \Psi^{\scr C}_{\scr L} \, \Psi_{\scr R}^{}
+ \textrm{h.c.} \, ) & = & 
\dfrac{M_{\scr R}}{2} \, ( \, \bar \Psi^{\scr C}_{\scr L} 
\, C \, ( \, \bar \Psi^{\scr C}_{\scr L} \, )^T
+ \textrm{h.c.} \, ) \quad \textrm{und} \quad \\ \label{majmass2}
\dfrac{M_{\scr L}}{2} \, ( \, \bar \Psi_{\scr L}^{} \, \Psi^{\scr C}_{\scr R}
+ \textrm{h.c.} \, ) & = & 
\dfrac{M_{\scr L}}{2} \, ( \, \bar \Psi_{\scr L}^{} 
\, C \, ( \, \bar \Psi^{}_{\scr L} \, )^T
+ \textrm{h.c.} \, )
\; ,
\end{eqnarray}
die man als Majorana-Massenterme bezeichnet, da sie nur f"ur 
Majorana-Teilchen def\/iniert sind. Letztere werden durch Spinoren $\Psi$ 
beschrieben, f"ur welche $\Psi^{\scr C} \equiv \Psi$ gilt, das hei"st
Teilchen und Antiteilchen sind identisch. Deswegen sind Majorana-Teilchen
zwangsl"auf\/ig elektrisch neutral.
Majorana-Massenterme werden sp"ater im Zusammenhang mit Neutrinomassen von 
Bedeutung sein.

Das Transformationsverhalten von Massentermen f"ur die geladenen
Fermionen im SM sieht folgenderma"sen aus:
\begin{eqnarray} \label{umass}
\textrm{($u,c,t$)-Quarks:} \hspace{1.1cm}
  ({\bf 3,2})_{\scr \frac{1}{6}} \otimes ({\bf \bar 3,1})_{\scr -\frac{2}{3}}
 & = & ({\bf 1,2})_{\scr -\frac{1}{2}} \oplus ({\bf 8,2})_{\scr
 -\frac{1}{2}} \\ \label{dmass}
\textrm{($d,s,b$)-Quarks:} \hspace{1.0cm} \,
  ({\bf 3,2})_{\scr \frac{1}{6}} \otimes ({\bf \bar 3,1})_{\scr \frac{1}{3}}
\;\; & = & ({\bf 1,2})_{\scr \frac{1}{2}} \oplus ({\bf 8,2})_{\scr
 \frac{1}{2}} \\ \label{emass}
\textrm{geladene Leptonen:} \quad
  ({\bf 1,2})_{\scr -\frac{1}{2}} \otimes ({\bf 1,1})_{\scr 1}
\; & = & ({\bf 1,2})_{\scr \frac{1}{2}}
\end{eqnarray}
Die Lagrangedichte des SM kann demnach keine Fermionmassenterme
enthalten, da diese nicht invariant unter 
$G_{\scr \textrm{SM}}$-Transformationen sind. Das steht aber im
Widerspruch zu der experimentellen Beobachtung massiver Teilchen.

Abgesehen davon wird in der Natur keineswegs die Symmetrie
des SM, sondern eine 
$SU(3)_{\scr C} \otimes U(1)_{\scr \textrm{em}}$-Sym\-me\-trie beobachtet.

Beide Probleme k"onnen durch die spontane Brechung lokaler
Eichsymmetrien, den sogenannten Higgs-Mechanismus, gel"ost werden
\cite{higgs}. Dazu werden dem Teilchenspektrum der Eichtheorie Spin-0-Teilchen
hinzugef"ugt. Im SM wird ein Doublett 
\begin{equation}
\Phi = \begin{pmatrix} \phi_{}^+ \\ \phi_{}^0 \end{pmatrix}
\end{equation}
dieser Higgs-Bosonen eingef"uhrt, wobei $\phi_{}^+$ und $\phi_{}^0$
komplexe Skalarfelder sind; sie liegen in der  
$G_{\scr \textrm{SM}}$-Darstellung $({\bf 1,2})_{\scr \frac{1}{2}}$. 
Das Potential f"ur $\Phi$ hat die Form
\begin{equation}
V(\Phi) = m^2 \, \Phi^{\dagger} \Phi + \lambda \, (\Phi^{\dagger} \Phi)^2
\end{equation}
und erzeugt f"ur $m^2 < 0$ einen nichtverschwindenden
Vakuumerwartungswert
\begin{equation}
\langle 0 | \Phi | 0 \rangle = \begin{pmatrix} 0 \\ 
\upsilon/\sqrt{2} \end{pmatrix} \quad \textrm{mit} \quad 
\upsilon = \sqrt{-\dfrac{m^2}{\lambda}}
\end{equation}
von $\Phi$. Den Wert von $\upsilon$ kann man in niedrigster Ordnung
aus der me"sbaren
Fermi-Konstanten $G_F$ und der Beziehung $\upsilon = (\sqrt{2} \, G_F)^{-1/2}$
bestimmen; er betr"agt $\upsilon = 246.22$ GeV.
Nun ist der Vakuumzustand nicht mehr 
$SU(2)_{\scr L} \otimes U(1)_{\scr Y}$-symmetrisch; die Symmetrie der
elektroschwachen Wechselwirkung wird spontan in die $U(1)_{\scr \textrm{em}}$ 
der QED gebrochen:
\begin{equation}
SU(3)_{\scr C} \otimes SU(2)_{\scr L} \otimes U(1)_{\scr Y}
\stackrel{\langle \Phi \rangle}{\longrightarrow} 
SU(3)_{\scr C} \otimes U(1)_{\scr \textrm{em}}
\end{equation}
Durch ihre Wechselwirkung mit den Higgs-Teilchen erhalten die
Eichbosonen Massenterme. Aus den vier 
$SU(2)_{\scr L} \otimes U(1)_{\scr Y}$-Bosonen entstehen so die
physikalischen Eichbosonen, das hei"st die Masseneigenzust"ande
\begin{eqnarray}
W_{}^{\pm} & = & \dfrac{1}{\sqrt{2}} \, ( \, W_1 \pm i \, W_2 \, ) \\
Z \quad & = & \sin{\theta_{\scr W}} \, B + \cos{\theta_{\scr W}} \, W_3 \\
A \quad & = & \cos{\theta_{\scr W}} \, B - \sin{\theta_{\scr W}} \, W_3
\end{eqnarray}
$\theta_{\scr W}$ ist der Weinberg-Winkel; f"ur ihn gilt
\begin{equation} \label{wangle}
  \sin{\theta_{\scr W}} = \dfrac{g'}{\sqrt{{g'}^{2}+g_{\scr 2}^2}} \; , \quad
  \tan{\theta_{\scr W}} = \dfrac{g'}{g_{\scr 2}^{}}
\end{equation}
$A$ ist das Eichboson der $U(1)_{\scr \textrm{em}}$, das masselose
Photon. Seine Kopplungskonstante, die Elementarladung $e$,
ergibt sich aus $e=g_{\scr 2}^{}\sin{\theta_{\scr W}}$. 
Die $W$- und $Z$-Bosonen erhalten die Massen 
\begin{equation}
M_{\scr W} = \dfrac{g_{\scr 2}^{}\upsilon}{2} \; , \quad 
M_{\scr Z} = \dfrac{M_{\scr W}}{\cos{\theta_{\scr W}}} \; ,
\end{equation}
w"ahrend die Gluonen masselos bleiben, da die Higgs-Teilchen
farbneutral sind. Tabelle \ref{Bomass} enth"alt die Werte
f"ur die Massen und den Weinberg-Winkel:
\begin{table}[h]
\begin{center}
\begin{tabular}{|l||c|c|c|}
\hline
Gr"o"se & $M_{\scr Z}$ & $M_{\scr W}$ & 
$\sin^2{\theta_{\scr W}}(M_{\scr Z})$ \\
\hline
Wert & $91.187 \pm 0.007$ GeV & $80.41 \pm 0.10$ GeV & $0.23124 \pm
0.00024$ \\
\hline
\end{tabular}
\end{center}
\caption[Werte f"ur $M_{\scr Z}$, $M_{\scr W}$ und 
$\sin^2{\theta_{\scr W}}(M_{\scr Z})$]{\label{Bomass} Werte f"ur 
$M_{\scr Z}$, $M_{\scr W}$ und $\sin^2{\theta_{\scr W}}(M_{\scr Z})$ 
\cite{pdg}}
\end{table}

\noindent Die experimentell me"sbaren Eichkopplungen der 
$SU(3)_{\scr C} \otimes U(1)_{\scr \textrm{em}}$-Theorie haben bei der
Skala $\mu=M_{\scr Z}$ folgende Werte ($\alpha^{}_{\scr 3}=g_{\scr 3}^2/4\pi$ 
und $\alpha^{}_{\scr \textrm{em}}=e_{}^2/4\pi$):
\begin{table}[h]
\begin{center}
\begin{tabular}{|l||c|c|}
\hline
Gr"o"se & $\alpha^{}_{\scr 3}(M_{\scr Z})$ & 
$\alpha^{}_{\scr \textrm{em}}(M_{\scr Z})$ \\
\hline
Wert & $0.119 \pm 0.002$ & $(128.88 \pm 0.09)^{-1}$ \\
\hline
\end{tabular}
\end{center}
\caption[Werte f"ur $\alpha^{}_{\scr 3}(M_{\scr Z})$ und 
$\alpha^{}_{\scr \textrm{em}}(M_{\scr Z})$]{\label{SMcpl1} Werte f"ur 
$\alpha^{}_{\scr 3}(M_{\scr Z})$ und 
$\alpha^{}_{\scr \textrm{em}}(M_{\scr Z})$ \cite{pdg}}
\end{table}

\noindent Daraus kann man mit Hilfe der Beziehungen
\begin{equation}
\alpha^{}_{\scr 1}(M_{\scr Z})= 
\dfrac{5 \, \alpha^{}_{\scr \textrm{em}}(M_{\scr Z})}
{3 \, \cos^2{\theta_{\scr W}}(M_{\scr Z})} \; , \quad
\alpha^{}_{\scr 2}(M_{\scr Z}) = 
\dfrac{\alpha^{}_{\scr \textrm{em}}(M_{\scr Z})}
{\sin^2{\theta_{\scr W}}(M_{\scr Z})} \; , \quad
\end{equation}
die Werte der SM-Gr"o"sen $\alpha^{}_{\scr 1,2}(M_{\scr Z})$
berechnen:
\begin{table}[h]
\begin{center}
\begin{tabular}{|l||c|c|}
\hline
Gr"o"se & $\alpha^{}_{\scr 1}(M_{\scr Z})$ & 
$\alpha^{}_{\scr 2}(M_{\scr Z})$ \\
\hline
Wert & $(59.447 \pm 0.060)^{-1}$ & $(29.802 \pm 0.052)^{-1}$ \\
\hline
\end{tabular}
\end{center}
\caption[Werte f"ur $\alpha^{}_{\scr 1}(M_{\scr Z})$
  und $\alpha^{}_{\scr 2}(M_{\scr Z})$]{\label{SMcpl2} Werte f"ur 
$\alpha^{}_{\scr 1}(M_{\scr Z})$ und $\alpha^{}_{\scr 2}(M_{\scr Z})$ 
\cite{pdg}}
\end{table}

\noindent Von den vier reellen Freiheitsgraden in $\Phi$ ist
nur einer physikalisch, er geh"ort zum elektrisch neutralen
Higgs-Boson $H$. Es ist das einzige SM-Teilchen, das experimentell
noch nicht nachgewiesen wurde. F"ur seine Masse $m_{\scr H} =
\sqrt{\lambda} \, \upsilon$ gibt es lediglich eine Untergrenze: 
$m_{\scr H} > 89.7$ GeV \cite{delphi}.

\noindent Die Fermionmassen werden durch Einf"uhrung von Yukawa-Kopplungen
zwischen den Fermionen und den Higgs-Teilchen realisiert. Diese
Yukawa-Terme k"onnen wegen (\ref{umass})-(\ref{emass}) und 
$\Phi \sim ({\bf 1,2})_{\scr \frac{1}{2}}$ eichinvariant konstruiert
werden und haben die Form
\begin{equation} \label{massterm1}
  \mathcal{L}_{\scr M} \; = \; \sum_{a,b=1}^{3} \, \Big( \,
\bar Q^{\scr a}_{\scr L} \, \tilde \Phi \, 
({\bf Y}^{}_u)_{\scr ab} \, u^{\scr b}_{\scr R}
+ \bar Q^{\scr a}_{\scr L} \, \Phi \, 
({\bf Y}^{}_d)_{\scr ab} \, d^{\scr b}_{\scr R}
+ \bar L^{\scr a}_{\scr L} \, \Phi \, 
({\bf Y}^{}_e)_{\scr ab} \, e^{\scr b}_{\scr R} \, \Big) 
\; + \; \textrm{h.c.}
\end{equation}
mit 
\begin{equation}
Q^{\scr a}_{\scr L} = \begin{pmatrix}  u^{\scr a}_{\scr L} \\
d^{\scr a}_{\scr L} \end{pmatrix} \; , \quad 
L^{\scr a}_{\scr L} = \begin{pmatrix}  \nu^{\scr a}_{\scr L} \\
e^{\scr a}_{\scr L} \end{pmatrix} \; , \quad 
\tilde \Phi = i \tau_2 \Phi^* = \begin{pmatrix}
\phi_{}^{0*} \\ -\phi_{}^-
\end{pmatrix} \sim ({\bf 1,2})_{\scr -\frac{1}{2}}
\end{equation}
Die Indizes $a$ und $b$ bezeichnen die Fermionfamilien 
($d^{\scr a}_{\scr L} \equiv s^{}_{\scr L}$ f"ur $a=2$), die Elemente
der (3$\times$3)-Matrizen ${\bf Y}^{}_i$ sind die
Yukawa-Kopplungen. 
Die Fermionmassen entstehen im Rahmen der spontanen Symmetriebrechung,
wenn $\Phi$ seinen Vakuumerwartungswert $\upsilon$ ausbildet. Dann
sind die Massenmatrizen durch 
\begin{equation} \label{yukmass}
{\bf M}^{}_u \; = \; \dfrac{1}{\sqrt{2}} \, \upsilon \, 
{\bf Y}^{}_u \; , \quad 
{\bf M}^{}_d \; = \; \dfrac{1}{\sqrt{2}} \, \upsilon \, 
{\bf Y}^{}_d \; , \quad 
{\bf M}^{}_e \; = \; \dfrac{1}{\sqrt{2}} \, \upsilon \, 
{\bf Y}^{}_e  
\end{equation}
gegeben; (\ref{massterm1}) geht "uber in
\begin{equation} \label{massterm2}
  \mathcal{L}_{\scr M} \; = \; \sum_{a,b=1}^{3} \, \Big( \,
({\bf M}^{}_u)_{\scr ab} \, \bar u^{\scr a}_{\scr L} \, u^{\scr b}_{\scr R}
+ ({\bf M}^{}_d)_{\scr ab} \, \bar d^{\scr a}_{\scr L} \, d^{\scr b}_{\scr R}
+ ({\bf M}^{}_e)_{\scr ab} \, \bar e^{\scr a}_{\scr L} \,e^{\scr b}_{\scr R} 
\, \Big)
\; + \; \textrm{h.c.}
\end{equation}
Da die Massenmatrizen beliebige komplexe (3$\times$3)-Matrizen 
sein k"onnen, sind die Fermionen in Tabelle \ref{SMFerm}, die
Eigenzust"ande der $SU(2)_{\scr L} \otimes U(1)_{\scr Y}$-Wechselwirkung, 
im allgemeinen nicht mehr mit den physikalischen Teilchen def\/inierter
Masse identisch. Letztere erh"alt man, wenn man die Massenmatrizen
durch biunit"are Transformationen diagonalisiert:
\begin{equation} \label{biuntr1}
{\bf U}^{\dagger}_{\scr L} \, {\bf M}^{}_u \, {\bf U}^{}_{\scr R} \; = \; 
{\bf M}^{\scr (D)}_u \; , \quad
{\bf D}^{\dagger}_{\scr L} \, {\bf M}^{}_d \, {\bf D}^{}_{\scr R} \; = \; 
{\bf M}^{\scr (D)}_d \; , \quad 
{\bf E}^{\dagger}_{\scr L} \, {\bf M}^{}_e \, {\bf E}^{}_{\scr R} \; = \; 
{\bf M}^{\scr (D)}_e
\end{equation}
Die nichtverschwindenden Elemente der Diagonalmatrizen ${\bf M}^{\scr (D)}_i$
sind die Fermionmassen; ihre Werte sind in Tabelle \ref{FermMass}
aufgelistet. 
\begin{table}[h]
\begin{center}
\begin{tabular}{|l||c|c|c|}
\hline Gr"o"se: & $m_u(M_{\scr Z})$ & $m_d(M_{\scr Z})$ & $m_s(M_{\scr Z})$ \\
\hline Wert: & 
$2.33 \, {+0.42 \atop -0.45}$ MeV & $4.69 \, {+0.60 \atop -0.66}$ MeV & 
$93.4 \, {+11.8 \atop -13.0}$ MeV \\
\hline \hline
Gr"o"se: & $m_c(M_{\scr Z})$ & $m_b(M_{\scr Z})$ & $m_t(M_{\scr Z})$ \\
\hline Wert: & $677 \, {+56 \atop -61}$ MeV & 
$3.00 \pm 0.11$ GeV & $181 \pm 13$ GeV \\
\hline \hline
Gr"o"se: & $m_e(M_{\scr Z})$ & $m_\mu(M_{\scr Z})$ & $m_\tau(M_{\scr Z})$ \\
\hline Wert: & $486.84727 \pm 0.00014$ keV & 
$102.75138 \pm 0.00033$ MeV & $1746.7 \pm 0.3$ MeV \\
\hline
\end{tabular}
\end{center}
\caption[Fermionmassen bei $M_{\scr Z}$]{\label{FermMass}
  Fermionmassen bei $M_{\scr Z}$ \cite{fusko}}
\end{table}

\noindent Die Yukawa-Kopplungen sind ebenso wie die Eichkopplungen renormierte
Gr"o"sen und h"angen somit von der Massenskala $\mu$ ab.
"Uber (\ref{yukmass}) und (\ref{biuntr1}) sind demnach auch die
Fermionmassen und -mischungen skalenabh"angig.

\noindent Die Transformationen (\ref{biuntr1}) liefern
den Zusammenhang zwischen den Eigenzust"anden der SM-Wechselwirkungen,
hier mit dem Index $(0)$ bezeichnet, und den Masseneigenzust"anden:
\begin{equation} \label{biuntr2}
u^{\scr a (0)}_{\scr L,R} \, = \, 
{\bf U}^{}_{\scr L,R} \, u^{\scr a}_{\scr L,R} \; , \quad
d^{\scr a (0)}_{\scr L,R} \, = \, 
{\bf D}^{}_{\scr L,R} \, d^{\scr a}_{\scr L,R} \; , \quad
e^{\scr a (0)}_{\scr L,R} \, = \, 
{\bf E}^{}_{\scr L,R} \, e^{\scr a}_{\scr L,R}
\end{equation}
Nun ist (\ref{biuntr2}) auch in den "ubrigen Termen der
Langrangedichte auszuf"uhren, in denen Fermionfelder vorkommen. 
W"ahrend die neutralen Str"ome, die an das $Z$-Boson und das Photon koppeln,
invariant unter (\ref{biuntr2}) sind, "andern sich die geladenen
schwachen Str"ome:
\begin{eqnarray} \label{cwc}
\bar u^{\scr a (0)}_{\scr L} \gamma^\mu d^{\scr a (0)}_{\scr L} \, W^+_\mu
& \Longrightarrow &
\bar u^{\scr a}_{\scr L} \gamma^\mu 
\big( {\bf U}^\dagger_{\scr L} {\bf D}^{}_{\scr L} \big)^{}_{\scr ab}
d^{\scr b}_{\scr L} \, W^+_\mu \\
\bar d^{\scr a (0)}_{\scr L} \gamma^\mu u^{\scr a (0)}_{\scr L} \, W^-_\mu
& \Longrightarrow &
\big( {\bf D}^\dagger_{\scr L} {\bf U}^{}_{\scr L} \big)^{}_{\scr ab}
\bar d^{\scr b}_{\scr L} \gamma^\mu 
u^{\scr a}_{\scr L} \, W^-_\mu
\end{eqnarray}
Im Experiment tritt nur die (unit"are) Kombination 
${\bf V} \equiv {\bf U}^\dagger_{\scr L} {\bf D}^{}_{\scr L}$ der
Mischungsmatrizen in Erscheinung; man nennt sie 
Cabibbo-Kobayashi-Maskawa-Matrix (CKM-Matrix) \cite{ckm}. 
Alle anderen Mischungen, insbesondere die rechtsh"andigen, 
sind im SM nicht observabel.
Die experimentellen Grenzen f"ur die Betr"age der Elemente
von ${\bf V}$ liegen bei \cite{pdg}:
\begin{equation} \label{ckmval}
|{\bf V}| \; = \; 
\left( \begin{array}{lll}
0.9745-0.9760 & 0.217\hspace{0.18cm}-0.224 & 0.0018-0.0045 \\
0.217\hspace{0.18cm}-0.224 & 0.9737-0.9753 & 0.036\hspace{0.18cm}-0.042 \\
0.004\hspace{0.18cm}-0.013 & 0.035\hspace{0.18cm}-0.042 & 0.9991-0.9994
\end{array} \right)
\end{equation}
Die in dieser Arbeit verwendete Parametrisierung f"ur ${\bf V}$ lautet:
\begin{equation} \label{ckmpar}
{\bf V} \; = \;
\begin{pmatrix}
C_{12} C_{31} & S_{12} C_{31} & S_{31} \, e^{-i\delta} \\
- S_{12} C_{23} - C_{12} S_{23} S_{31} \, e^{i\delta} &
C_{12} C_{23} - S_{12} S_{23} S_{31} \, e^{i\delta} &
S_{23} C_{31} \\
S_{12} S_{23} - C_{12} C_{23} S_{31} \, e^{i\delta} &
- C_{12} S_{23} - S_{12} C_{23} S_{31} \, e^{i\delta} &
C_{23} C_{31}
\end{pmatrix}
\end{equation}
mit $C_{ij} = \cos \theta_{ij}$ und $S_{ij} = \sin \theta_{ij}$. Als
numerische Werte f"ur die Winkel werden $\theta_{12}=0.223$, 
$\theta_{23}=0.039$ und $\theta_{31}=0.003$ benutzt. Ferner wird 
$\delta=0$ gew"ahlt, da ein Modell mit reellen Massenmatrizen (und
somit orthogonalen  Mischungsmatrizen) Gegenstand der Untersuchung
ist; auf das Problem der $CP$-Verletzung soll hier nicht n"aher
eingegangen werden. Die Parametrisierung (\ref{ckmpar}) mit $\delta=0$
wird auch f"ur alle Mischungsmatrizen in (\ref{biuntr1}) benutzt.

Die leptonischen Anteile der geladenen schwachen Str"ome "andern sich
im Falle masseloser Neutrinos nicht, da die linksh"andigen Neutrinos
derselben Transformation ${\bf E}^{}_{\scr L}$ wie die geladenen
Leptonen unterzogen werden k"onnen. 
\section{Grenzen des Standardmodells}
Das SM ist nicht nur mathematisch konsistent, sondern auch
ph"anomenologisch "uberaus erfolgreich. Viele seiner Vorhersagen
sind auf eindrucksvolle Weise experimentell best"atigt worden, zum
Beispiel die Existenz der dritten Fermiongeneration, der massiven
$W_{}^{\scr \pm}$- und $Z$-Bosonen und des neutralen schwachen
Stroms. Lediglich die beobachtete Baryonasymmetrie des Universums
\cite{rio} und die durch die Neutrinoexperimente der letzten Jahre implizierte
Existenz massiver Neutrinos \cite{bgg} lassen sich im Rahmen des SM nicht
befriedigend erkl"aren. Davon abgesehen ist die "Ubereinstimmung der
theoretischen Vorhersagen des SM mit den experimentellen Ergebnissen sehr gut 
\cite{Lang1}.

Es gibt allerdings auch eine Reihe von ungekl"arten theoretischen 
Fragestellungen, die stark darauf hindeuten, da"s das SM keine
wirklich fundamentale Theorie ist, sondern lediglich die ef\/fektive N"aherung
einer solchen f"ur niedrige Energien.

Zun"achst ist die Zahl der freien Parameter im SM, deren
Werte von der Theorie nicht vorhergesagt werden, sondern an
experimentelle Resultate angepa"st werden m"ussen, sehr gro"s:
\begin{itemize}
\item neun Massen der geladenen Fermionen
\item drei Winkel und eine Phase in der CKM-Matrix
\item drei Eichkopplungen
\item die Higgs-Masse und -kopplungskonstante
\item zwei $\theta$-Parameter in den $CP$-verletzenden
  Lagrangedichte-Termen $\sim \theta \, \textrm{Tr} \, 
F_{\scr \mu \nu}^{} \tilde F_{}^{\scr \mu \nu}$
\end{itemize}
Unter Ber"ucksichtigung massiver Neutrinos kommen drei  
Neutrinomassen sowie drei Winkel und drei Phasen in der leptonischen 
Mischungsmatrix hinzu.
Eine grundlegende Theorie sollte dagegen mit m"oglichst wenigen freien
Parametern auskommen.

\noindent Weitere Aspekte, die auf eine Theorie jenseits des SM
schlie"sen lassen, sind:
\begin{itemize}
\item Die Symmetriegruppe $G_{\scr \textrm{SM}}$ ist ein direktes
  Produkt, woraus die Existenz dreier voneinander unabh"angiger und 
  betragsm"a"sig stark unterschiedlicher Kopplungskonstanten folgt; eine
  Vereinheitlichung der Wechselwirkungen erfolgt nicht. Ebenfalls damit
  verbunden ist die komplizierte Darstellung (\ref{smfrep}), in der
  die Fermionen einer Familie liegen.
\item Die Zuordnung der Hyperladungen zu den Fermionen ist weitgehend
  willk"urlich und lediglich durch (\ref{hyp}) und (\ref{anom2}) 
  eingeschr"ankt. Dies liefert aber keine Erkl"arung f"ur die Quantisierung der
  elektrischen Ladung in Einheiten von $e/3$.
\item Die bis heute beobachteten Fermionen liegen in drei Familien,
  die hinsichtlich ihrer Quantenzahlen und ihrer Wechselwirkungen
  v"ollig identisch sind, sich aber bez"uglich ihrer Massen
  betr"achtlich unterscheiden.
\item Der Hauptgrund f"ur die Einordnung der linksh"andigen Fermionen in 
$SU(2)_{\scr  L}$-Doub\-letts und der rechtsh"andigen in Singuletts liegt im
ph"anomenologischen Erfolg dieses Ansatzes. 
\item Neutrinomassen k"onnen im Rahmen des SM zwar durch Einf"uhrung von
  nichtwechselwirkenden rechtsh"andigen Neutrinos konstruiert werden,
  aber es stellt sich dann die Frage, warum ihre Massen sehr viel
  kleiner als die der geladenen Fermionen sein sollten.
\item Der Higgs-Sektor des SM ist, sowohl was die Anzahl der
  Higgs-Teilchen als auch die Einordnung in Darstellungen von 
  $G_{\scr \textrm{SM}}$ angeht, weitgehend willk"urlich. Auch die
  Selbstkopplung sowie die Yukawa-Kopplungen und Massen der
  Higgs-Teilchen sind nicht festgelegt. 
\item Der QCD-$\theta$-Parameter ist extrem klein, $\theta \lesssim 
10_{}^{-9}$. Dies folgt direkt aus den experimentellen Grenzen
f"ur das Dipolmoment des Neutrons.
\item Die durch Einsteins Allgemeine Relativit"atstheorie beschriebene
  Gravitation l"a"st sich im Gegensatz zu den restlichen Wechselwirkungen
  nicht allein mit dem Prinzip der lokalen Eichinvarianz
  erkl"aren. Alle bisher entwickelten Eichtheorien der Gravitation
  haben sich als nichtrenormierbar erwiesen.
\end{itemize}
Zahlreiche Modif\/ikationen des minimalen SM sind konstruiert worden, um
einzelne dieser Schw"achen zu beheben. So kann das Problem der
$CP$-Verletzung in der QCD durch ein Modell mit zwei Higgs-Doubletts
und einer zus"atzlichen chiralen $U(1)$-Symmetrie gel"ost
werden \cite{pecc}, w"ahrend Modelle mit sogenannten horizontalen 
Symmetrien versuchen, die Massenhierarchie zwischen den 
Fermionfamilien zu erkl"aren \cite{frog}.

Im Gegensatz dazu gehen die Grand Unif\/ied-Theorien (GUTs), die im
n"achsten Kapitel diskutiert werden, "uber eine blo"se Erweiterung des
SM hinaus. Es wird vielmehr der Versuch gemacht, das SM in eine
umfassendere Theorie einzubetten. GUTs k"onnen einige, wenn auch nicht
alle, der oben erw"ahnten Schw"achen beseitigen und zum besseren 
Verst"andnis des SM beitragen.

%%% Local Variables: 
%%% mode: latex
%%% TeX-master: t
%%% End: 

\begin{fmffile}{gutfeyn}

\chapter{Grand Unif\/ied-Theorien}
\section{Grundidee und allgemeine Eigenschaften}
Das Hauptziel bei der Konstruktion von Grand\ Unif\/ied-Theorien (GUTs)
\cite{lang2,masie} besteht darin, die drei qualitativ und quantitativ 
verschiedenen Wechselwirkungen des SM in einer einzigen zu vereinheitlichen. 
Formal geschieht dies durch die Einbettung der SM-Symmetrie\-grup\-pe 
$G_{\scr \textrm{SM}}$ in eine einfache Lie-Gruppe 
$G \supset G_{\scr \textrm{SM}}$. Diese neue Theorie soll das SM als 
ef\/fektive Niederenergien"aherung enthalten, was durch spontane 
Symmetrie\-brechung in einem oder auch mehreren Schritten erreicht
werden kann. 

\begin{figure}[h]
%\begin{center}
\hspace{2cm}
\epsfig{file=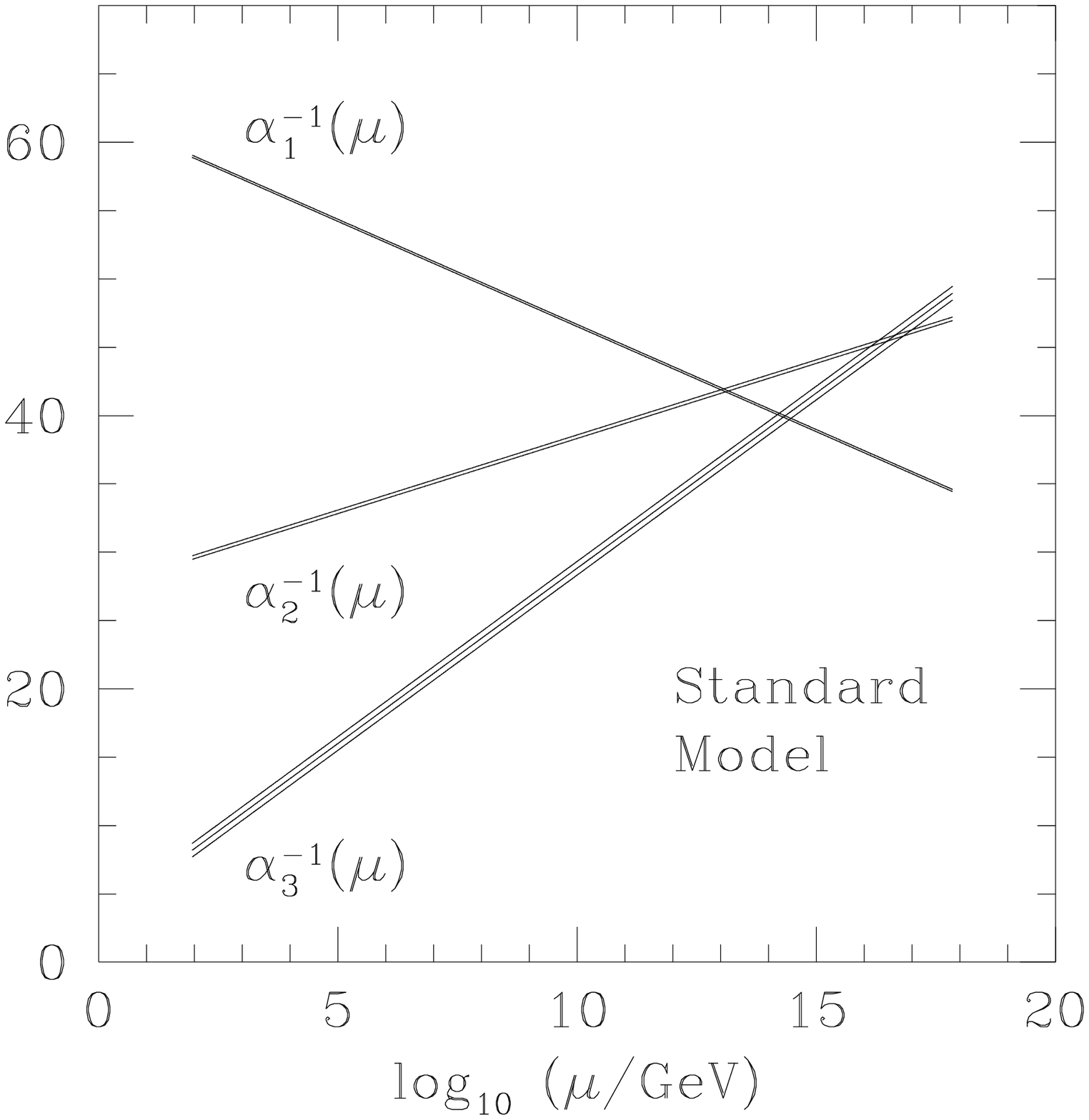,width=10cm}
\caption[Skalenabh"angigkeit der inversen Eichkopplungen 
des Standardmodells]{Skalenabh"angigkeit der inversen Eichkopplungen 
$1/\alpha^{}_{\scr i}(\mu)$ des SM (inklusive Fehlergrenzen) 
\cite{pic1} \label{smcoup}}
%\end{center}
\end{figure}

Motiviert wird dieser Ansatz in erster Linie durch die Skalenabh"angigkeit der
Eichkopplungen im SM \cite{gqw}. Integriert man die 
Renormierungsgruppengleichungen (\ref{smcrge1}-\ref{smcrge3}) f"ur die 
drei Gr"o"sen $\alpha^{}_{\scr i}=g^2_{\scr i}/4\pi$ \cite{rger} von
der Skala $M_{\scr Z}$ zu h"oheren Energien, erh"alt man das in Abbildung 
\ref{smcoup} gezeigte Verhalten.
%
%\psdraft
%\psfull
%
Die laufenden Kopplungen tref\/fen sich n"aherungsweise im Bereich 
$\mu\approx 10^{14}$-$10^{15} \, \textrm{GeV}$ und 
$\alpha^{}_{\scr i} \approx 1/40$. Da"s sie sich jedoch nachweislich
nicht in einem Punkt kreuzen, ist erst seit der Inbetriebnahme des
Teilchenbeschleunigers LEP am CERN 1989 und der damit verbundenen Steigerung
der Me"swertgenauigkeit f"ur $\alpha^{}_{\scr i}(M_{\scr Z})$ bekannt
\cite{deboer}. 

Der erste Schritt bei der Konstruktion eines GUT-Modells besteht in
der Wahl einer kompakten und einfachen Lie-Gruppe $G$. Man kann
prinzipiell auch halbeinfache Lie-Gruppen, also direkte Produkte
von einfachen Gruppen, verwenden, erh"alt dann aber f"ur jeden Faktor eine
separate Kopplungskonstante. Damit diese Kopplungen gleich sind und
eine Vereinheitlichung zustande kommt, m"ussen zus"atzliche Symmetrien 
eingef"uhrt werden. Deshalb sind einfache Gruppen vorzuziehen, da man
auf diese Weise eine Wechselwirkung mit genau einer Kopplung erh"alt;
die SM-Wechselwirkungen sind dann verschiedene Aspekte dieser Kraft.

Die einfachen Lie-Gruppen beziehungsweise -algebren sind
vollst"andig klassif\/iziert und ihre Eigenschaften und die ihrer
Darstellungen bekannt \cite{group,slans}; Tabelle \ref{Liealg} 
gibt eine "Ubersicht.
\begin{table}[h]
\begin{center}
\begin{tabular}{|l|c|c|c|}
\hline
Algebra & Rang & Ordnung & Dim(f) \\
\hline \hline
$SU(n+1)$, \hspace{0.2cm}$n\ge1$ & $n$ & $n(n+2)$ & $n+1$ \\
$SO(2n)$, \hspace{0.65cm}$n\ge4$ & $n$ & $n(2n-1)$ & $2n$ \\
$SO(2n+1)$, $n\ge2$ & $n$ & $n(2n+1)$ & $2n+1$ \\
$Sp(2n)$, \hspace{0.8cm}$n\ge3$ & $n$ & $n(2n+1)$ & $2n$ \\
$G_2$ & 2 & 14 & 7 \\
$F_4$ & 4 & 52 & 26 \\
$E_6$ & 6 & 78 & 27 \\
$E_7$ & 7 & 133 & 56 \\
$E_8$ & 8 & 248 & 248 \\
\hline
\end{tabular}
\end{center}
\caption{\label{Liealg} Klassif\/ikation und Eigenschaften einfacher 
Lie-Algebren}
\end{table}
Es gibt neben den vier unendlichen Reihen der klassischen
Lie-Gruppen die f"unf exzeptionellen Gruppen. Der Rang ist gleich der
maximalen Anzahl miteinander kommutierender Generatoren, die Ordnung
ist die Gesamtzahl der Generatoren und somit die Dimension der adjungierten
Darstellung, und Dim(f) gibt die Dimension der Fundamentaldarstellung
an.

%\noindent
Um eine geeignete Symmetriegruppe $G$ zu f\/inden, sollte diese
einer Reihe von Anforderungen gen"ugen:
\begin{itemize}
\item Zun"achst mu"s $G$ die SM-Symmetriegruppe $G_{\scr \textrm{SM}}$ als 
Untergruppe enthalten. Da der Rang von 
$SU(3)_{\scr C} \otimes SU(2)_{\scr L} \otimes U(1)_{\scr Y}$ gleich 4
ist, mu"s auch $G$ mindestens vom Rang 4 sein. Andererseits ist die
Forderung nach der Minimalit"at des Ranges der Eichgruppe sicher
sinnvoll, da mit dem Rang auch die Dimension der Darstellungen und
somit der Umfang des Teilchenspektrums zunimmt.
\item Die Gruppe $G$ sollte komplexe Darstellungen besitzen \cite{gmrs}
(eine Darstellung $\mathcal{R}$ hei"st komplex, wenn sie zu ihrer komplex
konjugierten Darstellung $\overline{\mathcal{R}}$ nicht "aquivalent
ist; ansonsten wird sie als reell bezeichnet). Diese Forderung ist 
aus folgendem Grund plausibel:

Sei $\mathcal{R}_{L,R}$ die Darstellung, nach der sich die links- 
beziehungsweise rechtsh"andigen Teilchen und Antiteilchen einer Fermionfamilie
transformieren. Dann gilt wegen 
$\Psi_{\scr R} = C \, ( \, \bar \Psi^{\scr C}_{\scr L} \, )^T$ die Beziehung
$\mathcal{R}_R=\overline{\mathcal{R}}_L$. Im SM ist die $\mathcal{R}_L$
entsprechende Darstellung (\ref{smfrep}) komplex, was die
Parit"atsverletzung und die Nichtexistenz 
$G_{\scr \textrm{SM}}$-invarianter Massenterme zur Folge hat. Um im 
Rahmen eines GUT-Modells eine reelle Darstellung $\mathcal{R}_L$
verwenden zu k"onnen, m"ussen in ihr neben den SM-Fermionen
zus"atzlich linksh"andige Fermionen liegen, die sich wie das komplex 
konjugierte von (\ref{smfrep}) transformieren. Diese neuen Teilchen
werden auch als Spiegelfermionen bezeichnet. Da sie in der Natur
jedoch nicht beobachtet werden, m"ussen ihre Massen w"ahrend des
Symmetriebrechungsschrittes entstehen, der auf das SM f"uhrt, was
aufgrund von Mischungen wiederum superschwere Massen auch f"ur die 
SM-Fermionen zur Folge hat. 
Dieses Problem l"a"st sich auf einfache Weise durch Verwendung
komplexer Darstellungen f"ur die Fermionen vermeiden. Von den
einfachen Lie-Gruppen enthalten jedoch nur 
\begin{equation} \label{cmplrep}
SU(n) \quad (n \ge 3) \; , \quad SO(4\,n+2) \quad (n \ge 2) 
\quad \textrm{und} \quad E_6
\end{equation}
komplexe Darstellungen \cite{mehta}, was die Wahlm"oglichkeiten f"ur $G$ stark
einschr"ankt.
\item Damit die Renormierbarkeit der Theorie sichergestellt ist, mu"s
sie anomaliefrei sein, also (\ref{anom1}) erf"ullen. W"ahrend die
orthogonalen Gruppen (au"ser $SO(6) \cong SU(4)$) \cite{ggan} 
und $E_6$ \cite{e6} automatisch anomaliefrei sind, gilt dies bei den
$SU(n)$-Gruppen nur f"ur bestimmte Kombinationen irreduzibler Darstellungen 
\cite{oku}. Im einfachsten Fall, der hier von Interesse ist,
n"amlich $SU(5)$, ist die (komplexe) Summe  
\begin{equation} \label{su5an}
\overline{{\bf 5}} \; \oplus \; {\bf 10} \quad \big[ \; \oplus {\bf 1} 
\; \equiv \; {\bf 16}_{SO(10)} \; \big]
\end{equation}
anomaliefrei.
\end{itemize}

\noindent Ber"ucksichtigt man bei der Wahl von $G$ diese Kriterien, so
verbleiben folgende M"oglichkeiten mit 4 $\le$ Rang($G$) $\le$ 6:
\begin{equation}
\textrm{Rang 4:} \;\; SU(5) \; ; \;\,
\textrm{Rang 5:} \;\; SU(6) \; , \; SO(10) \; ; \;\,
\textrm{Rang 6:} \;\; SU(7) \; , \; E_6
\end{equation}
Auf $SU(5)$, $SO(10)$ und $E_6$ basierende GUTs werden in den
n"achsten Abschnitten behandelt; $SU(6)$- und $SU(7)$-Theorien bieten
gegen"uber diesen Modellen keinerlei Vorteile, sind aber in einigen
Punkten wesentlich unhandlicher und deswegen weitgehend unbeachtet
geblieben.

Hat man auf diese Weise eine Wahl f"ur die Eichgruppe $G$ 
getrof\/fen, so sind noch folgende Schritte durchzuf"uhren:
\begin{itemize}
\item Die Einbettung von $G_{\scr \textrm{SM}}$ in $G$ und das Schema
\begin{equation} \label{ssbs}
G \; \stackrel{M^{}_{\scr U}}{\longrightarrow} \; 
G_1 \; \stackrel{M^{\scr (1)}_{\scr I}}{\longrightarrow} \; 
\cdots \; \stackrel{M^{\scr (k)}_{\scr I}}{\longrightarrow} \; 
G_{\scr \textrm{SM}} \; \stackrel{M^{}_{\scr Z}}{\longrightarrow} \; 
SU(3)_{\scr C} \otimes U(1)_{\scr \textrm{em}}
\end{equation}
f"ur die Symmetriebrechung in das SM sind anzugeben. 
Der einfachste Fall, die direkte Symmetriebrechung
$G \; \stackrel{M^{}_{\scr U}}{\longrightarrow} \; G_{\scr \textrm{SM}}$,
kommt allerdings wegen der Tatsache, da"s die laufenden SM-Kopplungen
sich nicht in einem Punkt tref\/fen (siehe Abbildung \ref{smcoup}), 
nicht in Frage.
\item F"ur jeden Symmetriebrechungsschritt in (\ref{ssbs}) m"ussen
geeignete Higgs-Darstellungen festgelegt und zugeh"orige Potentiale 
konstruiert werden, welche ihn realisieren k"on\-nen. Dabei ist darauf 
zu achten, da"s die Teilchen des SM nach wie vor erst im letzten
Schritt bei $M_{\scr Z}$ ihre Massen erhalten, w"ahrend alle
zus"atzlichen Teilchen sehr viel schwerer sind und deshalb mindestens 
Massen der Gr"o"senordnung $M^{\scr (k)}_{\scr I}$ haben. Dieser 
Sachverhalt wird auch als "`Survival Hypothesis"' \cite{surv} bezeichnet.
\item Die komplexe und anomaliefreie Darstellung, in der die Fermionen 
liegen sollen, ist auszuw"ahlen. Hierbei mu"s die Zerlegung der 
Fermiondarstellung unter $G_{\scr \textrm{SM}}$ die Zuordnung der korrekten
SM-Quantenzahlen gestatten, das hei"st die SM-Darstellung
(\ref{smfrep}) enthalten.
\end{itemize}
Ist das geschehen, so ist die formale Konstruktion der GUT
abgeschlossen. Die expliziten Eigenschaften und Vorhersagen des
Modells, wie zum Beispiel Vereinheitlichungsmasse $M_{\scr U}$ und -kopplung
$\alpha^{}_{\scr U}$, Fermionmassenrelationen oder die Rate des
Protonzerfalls und daraus eventuell resultierende Widerspr"uche zu
experimentellen Resultaten sind dann im Detail zu untersuchen.
\section{$SU(5)$}
\subsection{Konstruktion und Teilcheninhalt}
Die $SU(5)$-GUT \cite{su5} stellt gewisserma"sen den Prototyp dieser
Theorien dar. Wie oben gezeigt wurde, ist sie die einzige f"ur die 
Konstruktion einer GUT in Frage kommende Gruppe mit Rang($G$)=4.
Da die SM-Eichgruppe $G_{\scr \textrm{SM}}$ eine maximale Untergruppe
der $SU(5)$ ist, kann nur die direkte Symmetriebrechung 
\begin{equation} \label{sbsu5}
SU(5) \; \stackrel{M^{}_{\scr U}}{\longrightarrow} \; 
G_{\scr \textrm{SM}} \; \stackrel{M^{}_{\scr Z}}{\longrightarrow} \; 
SU(3)_{\scr C} \otimes U(1)_{\scr \textrm{em}}
\end{equation}
erfolgen, was wegen des durch Abbildung \ref{smcoup} verdeutlichten
SM-Kopplungsverhaltens f"ur eine realistische Theorie nicht in Frage kommt.
Dennoch sollen hier kurz die grundlegenden Eigenschaften der
$SU(5)$-GUT zusammengefa"st werden, da sie sich zum gro"sen Teil
zumindest qualitativ auch auf andere GUT-Modelle "ubertragen lassen.

\noindent Betrachtet man die Verzweigungen der beiden $SU(5)$-Darstellungen 
${\bf \bar 5}$ und ${\bf 10}$ unter $G_{\scr \textrm{SM}}$
\begin{equation} \label{fermsu5}
{\bf \bar 5} \longrightarrow ({\bf \bar 3,1})_{\scr \frac{1}{3}}
\oplus ({\bf 1,2})_{\scr -\frac{1}{2}} \; , \quad
{\bf 10} \longrightarrow ({\bf 3,2})_{\scr \frac{1}{6}} \oplus
({\bf \bar 3,1})_{\scr -\frac{2}{3}} \oplus
({\bf 1,1})_{\scr 1} \; ,
\end{equation}
stellt man fest, da"s die reduzible Darstellung 
$({\bf \bar 5} \oplus {\bf 10})$ gerade alle SM-Fermionen einer
Familie aufnehmen kann (siehe Tabelle \ref{SMFerm}). Ferner ist sie
komplex und anomaliefrei, erf"ullt also alle notwendigen Bedingungen.
Die gemeinsame Einbettung von Quarks und Leptonen in dieselbe
irreduzible Darstellung ist eine Eigenschaft aller GUTs und f"uhrt,
da die Eichtransformationen diese Teilchen miteinander mischen
k"onnen, zur Verletzung der Baryon- und Leptonzahlerhaltung.

An dieser Stelle wird ein wesentlicher Erfolg von GUTs deutlich,
n"amlich die Erkl"arung f"ur die Quantisierung der elektrischen
Ladung. Sie geh"ort zu den allgemeinen Eigenschaften von Eichtheorien
mit einfacher Symmetriegruppe, da die Eigenwerte der diagonalen
Generatoren solcher Gruppen im Gegensatz zu den Eigenwerten der
abelschen $U(1)$ stets diskret sind. Der Ladungsoperator $Q$ mu"s in
der Cartan-Unteralgebra von $G$
liegen, also eine Linearkombination der diagonalen Generatoren
sein. Da die Generatoren spurfrei sind, gilt dies auch f"ur
$Q$. Daraus folgt wiederum, da"s die Summe der Ladungen der in den
Darstellungen ${\bf \bar 5}$ und ${\bf 10}$ liegenden Fermionen
jeweils 0 sein mu"s: $Q(e_{}^-)=3\,Q(d)$ und $2\,Q(u)+Q(e_{}^-)+Q(d)=0$. 
Quarks haben deshalb drittelzahlige Ladungen, weil sie in drei Farben
auftreten, w"ahrend Leptonen farbneutral sind. 

Betrachtet man $G_{\scr \textrm{SM}}$ als Untergruppe von $SU(5)$,
sind die Hyperladungswerte in Tabelle \ref{SMFerm} mit einem Faktor
$\sqrt{5/3}$ umzunormieren. Wegen (\ref{wangle}), 
$g' = \sqrt{3/5} \, g^{}_{\scr 1}$ und 
$g^{}_{\scr 1}(M_{\scr U})=g^{}_{\scr 2}(M_{\scr U})$ gilt in GUTs mit
einfacher Symmetriegruppe stets $\sin^2{\theta_{\scr W}}(M_{\scr U})=3/8$; 
der Weinbergwinkel ist kein freier Parameter mehr. Der experimentelle
Wert $\sin^2{\theta_{\scr W}}(M_{\scr Z}) \approx 0.23$ mu"s durch
Integration der entsprechenden Renormierungsgruppengleichungen
reproduziert werden.

Die adjungierte Darstellung, nach der sich in Eichtheorien die
Eichbosonen transformieren, verzweigt sich bez"uglich $G_{\scr \textrm{SM}}$
gem"a"s
\begin{equation} \label{adjsu5}
{\bf 24} \longrightarrow 
({\bf 8,1})_{\scr 0} \oplus 
({\bf 1,3})_{\scr 0} \oplus 
({\bf 1,1})_{\scr 0} \oplus 
({\bf 3,2})_{\scr \frac{5}{6}} \oplus 
({\bf \bar 3,2})_{\scr -\frac{5}{6}}
\end{equation}
Die ersten drei Summanden entsprechen den SM-Eichbosonen, w"ahrend 
die Darstellungen $({\bf 3,2})_{\scr \frac{5}{6}}$ und 
$({\bf \bar 3,2})_{\scr -\frac{5}{6}}$ neue Bosonen enthalten, 
die mit $X$ und $Y$ beziehungsweise $\bar X$ und $\bar Y$ bezeichnet
werden. Da diese experimentell nicht beobachtet werden, m"ussen sie
Massen der Gr"o"senordnung $M_{\scr U}$ haben. 
\subsection{Protonzerfall}
Die $X$- und $Y$-Bosonen, welche die elektrischen Ladungen 
+4/3 und +1/3 besitzen, 
tragen sowohl eine Farbladung als auch schwachen Isospin und k"onnen 
deshalb Quarks und Leptonen ineinander "uberf"uhren. W"ahrend Baryon- 
und Leptonzahl $B,L$ globale $U(1)$-Symmetrien des (perturbativen) SM sind, 
werden sie in der $SU(5)$-Theorie wie
auch in allen anderen GUTs verletzt; allerdings bleibt $B-L$ erhalten.
Die direkte Konsequenz der Baryonzahlverletzung ist die Instabilit"at
von Proton und (gebundenem) Neutron. Der Nukleonenzerfall ist eine
wichtige und experimentell "uberpr"ufbare Vorhersage von GUTs, die
"uber das SM klar hinausgeht \cite{machk}. 

Der baryonzahlverletzende Teil der Lagrangedichte f"ur die
erste Fermionfamilie (unter Vernachl"assigung von Mischungen)
ergibt sich zu \cite{lang2}:
\begin{eqnarray} \nonumber
  \mathcal{L}_{\scr \Delta B \ne 0}
& = & \dfrac{g^{}_{\scr U}}{\sqrt{2}} \; \bar 
  X^{\scr \alpha}_{\scr \mu} \;
  \big( \varepsilon_{\scr \alpha \beta \gamma} \bar u_{\scr L}^{\scr C
    \gamma}  \gamma^{\scr \mu} u_{\scr L}^{\scr \beta} + \bar d_{\scr L
  \alpha}^{} \gamma^{\scr \mu} e^{\scr +}_{\scr L} 
   + \bar d_{\scr R \alpha}^{} \gamma^{\scr \mu} e^{\scr +}_{\scr R} \big)
\\
& + & \dfrac{g^{}_{\scr U}}{\sqrt{2}} \; 
  \bar Y^{\scr \alpha}_{\scr \mu} \; \big( 
  \varepsilon_{\scr \alpha \beta \gamma} \bar u_{\scr L}^{\scr C \gamma} 
  \gamma^{\scr \mu} d_{\scr L}^{\scr \beta}
  - \bar d_{\scr R \alpha}^{}  \gamma^{\scr \mu} \nu_{\scr e R}^{\scr C}
  - \bar u_{\scr L \alpha}^{} \gamma^{\scr \mu} e^{\scr +}_{\scr L}
  \big) \; + \; \textrm{h.c.}
\end{eqnarray}

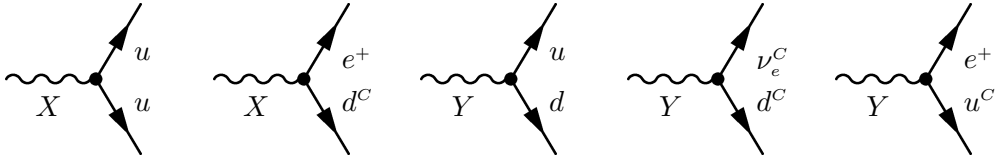
\begin{figure}[h]
\vspace{0.5cm}
\begin{minipage}{14cm}
\hspace{0.5cm}
\begin{fmfgraph*}(2,2)
\fmfleft{i} 
\fmfright{o1,o2}
\fmf{boson,width=1.0pt,label=$X$}{i,v} 
\fmf{fermion,width=1.0pt,label=$u$}{v,o1} 
\fmf{fermion,width=1.0pt,label=$u$}{v,o2}
\fmfdot{v}
\end{fmfgraph*}
\hspace{0.5cm}
\begin{fmfgraph*}(2,2)
\fmfleft{i} 
\fmfright{o1,o2}
\fmf{boson,width=1.0pt,label=$X$}{i,v} 
\fmf{fermion,width=1.0pt,label=$d^{\scr C}$}{v,o1} 
\fmf{fermion,width=1.0pt,label=$e^{\scr +}$}{v,o2}
\fmfdot{v}
\end{fmfgraph*}
\hspace{0.5cm}
\begin{fmfgraph*}(2,2)
\fmfleft{i} 
\fmfright{o1,o2}
\fmf{boson,width=1.0pt,label=$Y$}{i,v} 
\fmf{fermion,width=1.0pt,label=$d$}{v,o1} 
\fmf{fermion,width=1.0pt,label=$u$}{v,o2}
\fmfdot{v}
\end{fmfgraph*}
\hspace{0.5cm}
\begin{fmfgraph*}(2,2)
\fmfleft{i} 
\fmfright{o1,o2}
\fmf{boson,width=1.0pt,label=$Y$}{i,v} 
\fmf{fermion,width=1.0pt,label=$d^{\scr C}$}{v,o1} 
\fmf{fermion,width=1.0pt,label=$\nu_{\scr e}^{\scr C}$}{v,o2}
\fmfdot{v}
\end{fmfgraph*}
\hspace{0.5cm}
\begin{fmfgraph*}(2,2)
\fmfleft{i} 
\fmfright{o1,o2}
\fmf{boson,width=1.0pt,label=$Y$}{i,v} 
\fmf{fermion,width=1.0pt,label=$u^{\scr C}$}{v,o1} 
\fmf{fermion,width=1.0pt,label=$e^{\scr +}$}{v,o2}
\fmfdot{v}
\end{fmfgraph*}
\end{minipage}
\vspace{0.5cm}
\caption{Baryonzahlverletzende Vertizes der $X$- und $Y$-Bosonen \label{pdecv}}
\end{figure}

\noindent $\alpha$, $\beta$ und $\gamma$ sind $SU(3)_{\scr C}$-Farbindizes. 
Das entspricht den in Abbildung \ref{pdecv} aufgef"uhrten 
Wechselwirkungsvertizes.

\noindent Zwei typische Beispiele f"ur Feynmandiagramme von 
Protonzerfallsprozessen zeigt Abbildung \ref{pdecf}:

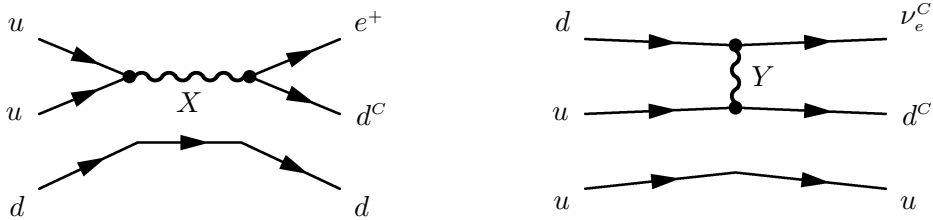
\begin{figure}[h]
\vspace{1cm}
\begin{minipage}{14cm}
\hspace{1cm}
\begin{fmfgraph*}(4,2)
\fmfstraight
\fmfleft{i1,i2,i3} \fmflabel{$u$}{i3} 
\fmflabel{$u$}{i2} \fmflabel{$d$}{i1}
\fmfright{o1,o2,o3} \fmflabel{$e^{\scr +}$}{o3} 
\fmflabel{$d^{\scr C}$}{o2} \fmflabel{$d$}{o1}
\fmf{fermion,width=1.0pt,l.s=left}{i3,v3}
\fmf{fermion,width=1.0pt}{i2,v3}
\fmf{boson,width=1.5pt,tension=1.5,label=$X$}{v3,v4}
\fmf{fermion,width=1.0pt}{v4,o3} 
\fmf{fermion,width=1.0pt}{v4,o2}
\fmffreeze
\fmf{fermion,width=1.0pt}{i1,v1}
\fmf{fermion,width=1.0pt}{v1,v2}
\fmf{fermion,width=1.0pt}{v2,o1}
\fmf{phantom,tension=0.7}{v1,v3}
\fmf{phantom,tension=0.7}{v2,v4}
\fmfdot{v3,v4}
\end{fmfgraph*}
\hspace{3cm}
\begin{fmfgraph*}(4,2)
\fmfstraight
\fmfleft{i1,i2,i3} \fmflabel{$d$}{i3} 
\fmflabel{$u$}{i2} \fmflabel{$u$}{i1}
\fmfright{o1,o2,o3} \fmflabel{$\nu_{\scr e}^{\scr C}$}{o3} 
\fmflabel{$d^{\scr C}$}{o2} \fmflabel{$u$}{o1}
\fmf{fermion,width=1.0pt}{i3,v3}
\fmf{fermion,width=1.0pt}{i2,v2}
\fmf{boson,width=1.5pt,tension=0.2,label=$Y$}{v2,v3}
\fmf{fermion,width=1.0pt}{v3,o3}
\fmf{fermion,width=1.0pt}{v2,o2}
\fmffreeze
\fmf{fermion,width=1.0pt}{i1,v1}
\fmf{fermion,width=1.0pt}{v1,o1}
\fmf{phantom,tension=0.5}{v1,v2}
\fmfdot{v2,v3}
\end{fmfgraph*}
\end{minipage}
\vspace{1cm}
\caption{Beispiele f"ur zum Protonzerfall beitragende Feynmandiagramme 
\label{pdecf}}
\end{figure}

\noindent Der zugrundeliegende Proze"s lautet stets Quark+Quark $\rightarrow$
Antiquark+Antilepton, was zu Zerfallskan"alen  Nukleon $\rightarrow$
Meson+Antilepton f"uhrt. Alle anderen Elementarprozesse sind aufgrund
von Phasenraumef\/fekten unterdr"uckt.
Unter der Annahme kleiner Mischungen ist 
$p \rightarrow \pi^{\scr 0} e^{\scr +}$ der dominante Kanal;
eine qualitative Absch"atzung der Lebensdauer gem"a"s
\begin{equation} \label{taupap}
\Gamma^{}_{\scr p} = \tau_{\scr p}^{-1} \approx
\alpha_{\scr U}^2 \, \dfrac{m^5_{\scr p}}{M^4_{\scr X,Y}}
\end{equation}
liefert mit $\alpha_{\scr U} \approx 1/40$ und 
$M_{\scr X,Y} \approx M_{\scr U} \approx 5 \cdot 10^{14}$ GeV f"ur
$\tau_{\scr p}$ den Wert $3 \cdot 10^{30}$ Jahre. Die neuesten
experimentellen Grenzen liegen bei 
$\tau_{\scr p} \ge 2.9 \cdot 10^{33}$ Jahre \cite{skpd}. 
Die zu kurze Lebensdauer des Protons war die erste Vorhersage der
$SU(5)$-GUT, die eindeutig im Widerspruch zu experimentellen Resultaten stand.
\subsection{Symmetriebrechung und Fermionmassen}
Um den ersten Symmetriebrechungsschritt in (\ref{sbsu5}) zu
verwirklichen, ben"otigt man eine Higgs-Darstellung $\Phi$, die ein
SM-Singulett enth"alt; dieses bildet dann den Vakuumerwartungswert (VEW)
aus. Der zweite Brechungsschritt bei $M_{\scr Z}$ erfordert eine
Darstellung $H$, in der ein SM-Doublett $({\bf 1,2})_{\scr \frac{1}{2}}$
liegt. Die einfachste Wahl hierf"ur lautet $\Phi={\bf 24}$ und $H={\bf 5}$
(siehe (\ref{fermsu5}) und (\ref{adjsu5})). Der VEW $\langle \Phi \rangle$ 
gibt allen Teilchen, die nicht zum SM-Spektrum geh"oren, Massen der
Gr"o"senordnung $M_{\scr U}$, w"ahrend $\langle H \rangle$ die
SM-Massen erzeugt.

Betrachtet man das Transformationsverhalten von
Fermionmassentermen in der $SU(5)$-Theorie, so ergeben sich die drei 
Tensorprodukte
\begin{eqnarray} \label{fms1}
\overline{{\bf 5}} \, \otimes \, \overline{{\bf 5}} \hspace{0.42cm} & = & 
\overline{{\bf 10}} \, \oplus \, \overline{{\bf 15}} \\ \label{fms2}
\overline{{\bf 5}} \, \otimes \, {\bf 10} \hspace{0.2cm} & = & 
{\bf 5} \, \oplus \, {\bf 45} \\ \label{fms3}
{\bf 10} \, \otimes \, {\bf 10} & = & \overline{{\bf 5}} \, \oplus \, 
\overline{{\bf 45}} \, \oplus \, \overline{{\bf 50}}
\end{eqnarray}
Daran erkennt man zun"achst, da"s keine Fermionmassen 
$\sim \langle \Phi \rangle$ entstehen k"onnen. Die Kopplung von $H$ an
den Term ${\bf \bar 5} \otimes {\bf 10}$ erzeugt Dirac-Massen f"ur die
$d$-Quarks und die geladenen Leptonen, jene an den Term 
${\bf 10} \otimes {\bf 10}$ Massen f"ur die $u$-Quarks. 
Majorana-Massen f"ur die Neutrinos entstehen nicht, da $H$ nicht an 
${\bf \bar 5} \otimes {\bf \bar 5}$ koppelt.

Bei $M_{\scr U}$ gilt die Beziehung ${\bf M}_d = {\bf M}_e$, w"ahrend 
${\bf M}_u$ von den anderen Massenmatrizen unabh"angig ist.
Vorhersagen von Relationen zwischen den fermionischen Massenmatrizen
geh"oren zu den typischen Eigenschaften von GUTs. In diesem Fall
ist die Vorhersage ${\bf M}_d(M_{\scr U}) = {\bf M}_e(M_{\scr U})$
nicht realistisch, da sich nach Ber"ucksichtigung der
Renormierungsef\/fekte die Beziehung 
${\bf M}_d(M_{\scr Z}) \approx 3 \, {\bf M}_e(M_{\scr Z})$ ergibt, was
im Widerspruch zu den Fermionmassen aus Tabelle \ref{FermMass} steht. 
Durch Einf"uhrung einer zus"atzlichen {\bf 45}-Higgs-Darstellung, die wegen 
(\ref{fms1}--\ref{fms3}) ebenfalls Fermionmassen erzeugen kann, ist es
m"oglich, die Relation zwischen ${\bf M}_d$ und ${\bf M}_e$ so zu
modif\/izieren, da"s sie zumindest qualitativ korrekte Ergebnisse 
liefert \cite{gjlk}.
\subsection{Das Hierarchieproblem}
Die $SU(5)$-GUT enth"alt zwei verschiedene Symmetriebrechungsskalen 
$M_{\scr U}$ und $M_{\scr Z}$. W"ah\-rend die SM-Teilchen nach wie vor 
Massen der Gr"o"senordnung $M_{\scr Z}$ haben sollen, m"ussen die zus"atzlich
eingef"uhrten Teilchen Massen $\sim M_{\scr U}$ besitzen. Das
Verh"altnis dieser Skalen ist extrem klein, 
$M_{\scr Z}/M_{\scr U} \approx 10^{-12}$. Diese Hierarchie "ubertr"agt
sich zwangsl"auf\/ig auf die Parameter im Higgs-Potential $V$, welches
die VEW und damit die Symmetriebrechung verursacht \cite{li}. 
Das vollst"andige Higgs-Potential beinhaltet neben
\begin{eqnarray}
V(\Phi) \, & = & -m_{\scr 1}^2 \, \textrm{Tr} \, \Phi^2
\, + \, \lambda_{\scr 1} \, \big( \, \textrm{Tr} \, \Phi^2 \, \big)^2
\, + \, \lambda_{\scr 2} \, \textrm{Tr} \, \Phi^4 \quad \textrm{und} \\
V(H) & = & -m_{\scr 2}^2 \, H^\dagger H \, + \, 
\lambda_{\scr 3} \, (H^\dagger H)^2  
\end{eqnarray}
auch Mischterme der Form
\begin{equation} \label{mixpot}
V(\Phi,H) \; = \; \lambda_{\scr 4} \, \textrm{Tr} \, \Phi^2 \, (H^\dagger H) 
\, + \, \lambda_{\scr 5} \, \big( \, H^\dagger \Phi^2 H \, \big)
\end{equation}
Der erste Brechungsschritt erfolgt, wenn $\Phi$ den VEW 
\begin{equation}
\langle \Phi \rangle = \upsilon_{\scr 1} \cdot \textrm{diag}(2,2,2,-3,-3)
\; \textrm{mit} \; \upsilon_{\scr 1}^2 = 
m_{\scr 1}^2/(60\lambda_{\scr 1}+14\, \lambda_{\scr 2}) \approx M_{\scr U}^2
\end{equation}
ausbildet. Die $X,Y$-Bosonen und die physikalischen Higgs-Teilchen in
$\Phi$ erhalten Massen $\sim M_{\scr U}$. Der zweite Schritt verl"auft
analog zum SM. Berechnet man jedoch die Massen der Teilchen in $H$, so
ergibt sich aufgrund der Potentialterme in (\ref{mixpot}):
\begin{equation}
({\bf 3,1})_H : m_{\scr T}^2 = -m_{\scr 2}^2
+(30\lambda_{\scr 4}+4\lambda_{\scr 5})\upsilon_{\scr 1}^2 \; , \;
({\bf 1,2})_H : m_{\scr D}^2 = -m_{\scr 2}^2
+(30\lambda_{\scr 4}+9\lambda_{\scr 5})\upsilon_{\scr 1}^2
\end{equation}
Sowohl $m_{\scr T}$ als auch $m_{\scr D}$ enthalten Terme 
$\sim \upsilon_{\scr 1}$,
was im Falle des Tripletts $({\bf 3,1})_H$ auch notwendig ist, da es ebenso
wie die $X,Y$-Bosonen Protonzerf"alle vermitteln kann; das SM-Doublett
 $({\bf 1,2})_H$ dagegen mu"s Massen $\sim M_{\scr Z}$ haben. Dies
 l"a"st sich wegen $|m_{\scr 2}/\upsilon_{\scr 1}| \approx 10^{-12}$ nur 
dann erreichen, wenn $(30\lambda_{\scr 4}+9\lambda_{\scr 5}) \lesssim
10^{-24}$ ist.

\begin{floatingfigure}{7cm}
\vspace{0.7cm}
\mbox{
\begin{fmfgraph*}(5,3)
\fmfstraight
\fmfleft{i1,i2} 
\fmflabel{$\Phi$}{i2} \fmflabel{$\Phi$}{i1}
\fmfright{o1,o2}
\fmflabel{$H$}{o2} \fmflabel{$H$}{o1}
\fmf{scalar,width=1.0pt}{i2,v1}
\fmf{scalar,width=1.0pt}{v1,i1}
\fmf{boson,left,width=1.0pt,tension=0.5,label=$X,,Y$}{v1,v2,v1}
\fmf{scalar,width=1.0pt}{o2,v2}
\fmf{scalar,width=1.0pt}{v2,o1}
\fmfdot{v1,v2}
\end{fmfgraph*}}
\vspace{0.5cm}
\caption{Zum Hierarchie\-pro\-blem bei\-tra\-gen\-de Strahlungskorrektur
\label{hierpr}}
\vspace{0.3cm}
\end{floatingfigure}

\noindent Die nicht sehr nat"urlich anmutende Feineinstellung der Parameter 
$\lambda_{\scr 4,5}$ mit einer Genauigkeit $\lesssim 10^{-24}$ mu"s
schlie"slich f"ur jede Ordnung der St"orungstheorie wiederholt
werden, da sonst Strahlungskorrekturen wie zum Beispiel die in
Abbildung \ref{hierpr} gezeigte Beitr"age h"oherer Ordnung zu $m_{\scr D}$
liefern, die wieder $\sim \upsilon_{\scr 1}$ sind \cite{gil}.

\noindent Dieses Hierarchieproblem taucht
grunds"atzlich in allen Theo\-rien auf, die zwei oder 
mehr Massenskalen stark unterschiedlicher Gr"o"senordnung
aufweisen und l"a"st sich zumindest im Rahmen nichtsupersymmetrischer 
GUTs nicht befriedigend l"osen. Deswegen geht man im allgemeinen von 
der G"ultigkeit der
sogenannten "`Extended Survival Hypothesis"' aus \cite{aguil}. Sie
besagt, da"s Higgs-Teilchen die gr"o"stm"ogliche Masse erhalten, die mit
dem Symmetriebrechungsschema des Modells vertr"aglich ist. Da in der
$SU(5)$-GUT das Triplett $({\bf 3,1})_H$ f"ur den Brechungsschritt bei 
$M_{\scr Z}$ ohne Bedeutung ist, hat es eine Masse $\sim M_{\scr U}$.
\subsection{Zusammenfassung}
Obwohl das $SU(5)$-Modell aufgrund der experimentellen Fakten
mittlerweile ausgeschlossen ist, besitzt es im Vergleich zum SM eine
Reihe von attraktiven Eigenschaften:
\begin{itemize}
\item Die SM-Wechselwirkungen werden vereinheitlicht; es gibt nur noch
  eine Kopplungskonstante $g_{\scr U}$ und eine Vorhersage f"ur
$\sin^2{\theta_{\scr W}}(M_{\scr U})$.
\item Die anomaliefreie Darstellung $({\bf \bar 5} \oplus {\bf 10})$ kann die
  Fermionen einer Familie aufnehmen. Obwohl nach wie vor reduzibel,
  hat sie eine einfachere Struktur als (\ref{smfrep}).
\item Die Einbettung der $U(1)_{\scr Y}$ in eine einfache Lie-Gruppe
  f"uhrt automatisch zur korrekten Quantisierung der elektrischen Ladung.
\item Die Massenmatrizen der Fermionen sind nicht mehr unabh"angig voneinander.
\end{itemize}
Dem stehen folgende Schwierigkeiten gegen"uber:
\begin{itemize}
\item Die fermionische Darstellung ist reduzibel.
\item Der Higgs-Sektor der Theorie ist wesentlich komplexer als im SM.
\item Die Existenz zweier stark unterschiedlicher Massenskalen f"uhrt
  zum Hierarchieproblem.
\item Die Probleme im Neutrinosektor sind weiterhin ungel"ost.
\item Auch in der $SU(5)$-GUT ist das Wechselwirkungsverhalten rechts-
  und linksh"andiger Fermionen verschieden.
\item Es gibt keine Erkl"arung f"ur die Existenz dreier
  Fermionfamilien und die Massenhierarchie zwischen ihnen.
\end{itemize}
Einige dieser Schw"achen k"onnen im Rahmen von GUTs mit gr"o"serer
Eichgruppe beseitigt werden.
\section{$SO(10)$}
\subsection{Spinordarstellungen der $SO(n)$-Gruppen}
Die $SO(10)$ ist die kleinste orthogonale Gruppe, die $G_{\scr \textrm{SM}}$ 
als Untergruppe enth"alt und komplexe Darstellungen besitzt. Da 
$SU(5) \otimes U(1)$ eine maximale Untergruppe der $SO(10)$ ist, sind
alle vorteilhaften Eigenschaften der $SU(5)$-GUT auch in
$SO(10)$-Theorien vorhanden.

Die orthogonalen Lie-Gruppen unterscheiden sich in einigen wichtigen 
Punkten von den unit"aren Gruppen \cite{group}. Alle irreduziblen 
Darstellungen der $SU(n)$ k"onnen durch Reduktion von Tensorprodukten 
der komplexen Fundamentaldarstellung
${\bf n}$ und der zu ihr komplex konjugierten Darstellung ${\bf \bar n}$ 
erhalten werden. Die orthogonalen Gruppen $SO(n)$ dagegen besitzen neben den 
Tensordarstellungen, welche man durch Produktbildung aus der reellen
Fundamentaldarstellung ${\bf n}$ konstruieren kann, die sogenannten
Spinordarstellungen, f"ur die das nicht gilt. Zu den
Spinordarstellungen der $SO(n)$ gelangt man "uber die komplexe 
Clif\/ford-Algebra 
\begin{equation} \label{cliff}
C_n: \quad \{ \, \Gamma_j \, , \, \Gamma_k \, \} \; = \; 2 \,
\delta_{jk} \, {\bf 1} \quad (j,k=1,\dots,n)
\end{equation}
Aus den Generatoren von $C_n$ kann man eine Darstellung f"ur die
Generatoren der $SO(n)$-Lie-Algebra konstruieren:
\begin{equation}
\Sigma_{jk} \; = \; \dfrac{i}{4} \, [ \, \Gamma_j \, , \, \Gamma_k \, ]
\end{equation}
Die so def\/inierten $\Sigma_{jk}$ erf"ullen die Vertauschungsrelation
\begin{equation}
[ \, \mathcal{M}_{ij} \, , \, \mathcal{M}_{kl} \, ] \; = \;
i \, \big( \, \delta_{ik} \mathcal{M}_{jl} - \delta_{il} \mathcal{M}_{jk} 
- \delta_{jk} \mathcal{M}_{il} + \delta_{jl} \mathcal{M}_{ik} \, \big) \; ,
\end{equation}
welche die $SO(n)$-Algebra def\/iniert ($\mathcal{M}_{ij} = -
\mathcal{M}_{ji} \; \textrm{mit} \; i,j=1,\dots,n$). Die Generatoren
$\mathcal{M}_{ij}$ in der Fundamentaldarstellung sind durch 
\begin{equation}
\big( \, \mathcal{M}_{ij} \, \big)_{ab} \; = \; -i \,
( \, \delta_{ia} \delta_{jb} - \delta_{ja} \delta_{ib} \, ) 
\end{equation}
gegeben. "Uber die Matrixdarstellungen der komplexen Clif\/ford-Algebra 
(\ref{cliff}) und somit der $\Sigma_{jk}$ kommt man zu den Spinoren, da diese 
die Elemente des entsprechenden Darstellungsraumes sind. Die Eigenschaften
der irreduziblen Spinordarstellungen der $SO(n)$ h"angen vom Wert von $n$ ab:
\begin{itemize}
\item Ist $n$ ungerade, so existiert eine reelle Spinordarstellung der
  Dimension $2^{(n-1)/2}$.
\item Ist $n$ gerade, so existieren zwei in"aquivalente Spinordarstellungen der
  Dimension $2^{(n/2-1)}$. Wenn $n/2$ gerade ist, so sind die beiden
  Darstellungen reell, w"ahrend sie f"ur ungerades $n/2$ komplex
  sind, wobei dann die eine das komplex Konjugierte der anderen ist.
\end{itemize}
Aus diesem Grunde verf"ugen nur die $SO(4n+2)$-Gruppen "uber komplexe
Darstellungen; im Falle der $SO(10)$ sind dies die ${\bf 16}$ und die 
${\bf \overline{16}}$.
\subsection{Teilcheninhalt und Symmetriebrechung} \label{tiusb}
Die Hauptmotivation f"ur die Konstruktion von $SO(10)$-GUTs
\cite{so10} wird deutlich, wenn man die Verzweigung der {\bf 16}
bez"uglich der Untergruppe $SU(5)$ betrachtet:
\begin{equation}
{\bf 16} \longrightarrow {\bf 10} \oplus {\bf \bar 5} \oplus {\bf 1}
\end{equation}
In einer {\bf 16} k"onnen demnach alle Fermionen einer Familie und ein
zus"atzliches  $SU(5)$-Singulett untergebracht werden. Dieses
Singulett besitzt exakt die Eigenschaften und Quantenzahlen des 
rechts\-h"an\-di\-gen Neutrinos (beziehungsweise linksh"andigen 
Antineutrinos), da es zur $G_{\scr \textrm{SM}}$-Dar\-stel\-lung 
$({\bf 1,1})_{\scr 0}$ identisch ist. 

Hier wird ein gro"ser Vorteil von $SO(10)$-Modellen gegen"uber der $SU(5)$-GUT
deutlich, denn alle Fermionen einer Generation inklusive eines linksh"andigen
Antineutrinos liegen in einer irreduziblen und komplexen Darstellung der
Eichgruppe. Ferner erkl"art die Anomaliefreiheit der orthogonalen Gruppen auf
nat"urliche Weise das Verschwinden der Anomalie der $SU(5)$-Darstellung 
${\bf 10} \oplus {\bf \bar 5}$. 

\noindent$SO(10)$-Modelle sind "uberdies zumindest im Eichboson- und 
Fermionsektor manifest $C$- und $P$-invariant. Da in der {\bf 16} die 
linksh"andigen Teilchen und Antiteilchen liegen, wird sie durch
Anwendung von $C$ auf sich selbst abgebildet. Unter $P$ wird die {\bf 16} auf 
die ${\bf \overline{16}}$ abgebildet, in welcher die rechtsh"andigen Teilchen 
und Antiteilchen untergebracht sind. $C$- und $P$-Verletzung kann entweder 
explizit im Higgs-Sektor oder durch die spontane Symmetriebrechung 
realisiert werden. 

Da der Rang der $SO(10)$ um eins gr"o"ser als der von $G_{\scr \textrm{SM}}$ 
ist, existiert eine Vielzahl von m"oglichen Symmetriebrechungsschemata. 
Die direkte Brechung in das SM ist dabei ebenso wie die Brechung nach
$SU(5) \, [\otimes U(1)]$ zwar m"oglich, aber wegen der in Abbildung 
\ref{smcoup} gezeigten SM-Kopplungsentwicklung ph"anomenologisch 
ausgeschlossen. Man ben"otigt demnach mindestens einen
Zwischenschritt, das hei"st
\begin{equation}
SO(10) \; \stackrel{M_{\scr U}}{\longrightarrow} \; 
G_{\scr I} \; \stackrel{M_{\scr I}}{\longrightarrow} \; 
G_{\scr \textrm{SM}} \; \stackrel{M_{\scr Z}}{\longrightarrow} \; 
SU(3)_{\scr C} \otimes U(1)_{\scr \textrm{em}}
\end{equation}
Es sind Modelle mit bis zu vier aufeinander folgenden intermedi"aren 
Symmetriegruppen konstruiert und ausf"uhrlich untersucht worden 
\cite{raj,deshp}, jedoch besitzen sie gegen"uber Modellen mit nur
einer solchen keine besonderen Vorz"uge. Tabelle \ref{symbr} gibt die 
verschiedenen M"oglichkeiten f"ur den Brechungsschritt 
$SO(10) \rightarrow G_{\scr I}$ bei $M_{\scr U}$ und die daf"ur zu 
verwendende Higss-Darstellung mit einem $G_{\scr I}$-Singulett an.
\begin{table}[h]
\begin{center}
\begin{tabular}{|l|c|}
\hline
Symmetriegruppe $G_{\scr I}$ & Higgs-Darstellung $\Phi$ \\
\hline \hline
$SU(4)_{\scr C} \otimes SU(2)_{\scr L} \otimes SU(2)_{\scr R}$ & {\bf 210} \\
$SU(4)_{\scr C} \otimes SU(2)_{\scr L} \otimes SU(2)_{\scr R} \otimes D$
& {\bf 54} \\
$SU(4)_{\scr C} \otimes SU(2)_{\scr L} \otimes U(1)_{\scr R}$ & {\bf 45} \\
$SU(3)_{\scr C} \otimes SU(2)_{\scr L} \otimes SU(2)_{\scr R}
\otimes U(1)_{\scr B-L}$ & {\bf 45}, {\bf 210} \\
$SU(3)_{\scr C} \otimes SU(2)_{\scr L} \otimes SU(2)_{\scr R}
\otimes U(1)_{\scr B-L} \otimes D$ & {\bf 210} \\
$SU(3)_{\scr C} \otimes SU(2)_{\scr L} \otimes U(1)_{\scr R} 
\otimes U(1)_{\scr B-L}$ & {\bf 210} \\
\hline
\end{tabular}
\end{center}
\caption{\label{symbr} Intermedi"are Symmetriegruppen in $SO(10)$-GUTs}
\end{table}

\noindent Die $D$-Parit"at \cite{dpar} ist eine diskrete Symmetrie, welche 
$SU(2)_{\scr L}$ und $SU(2)_{\scr R}$ vertauscht; sie erfordert ein
rechts-links-symmetrisches Teilchenspektrum. Wenn die Theorie also
Teilchen in einer $SU(2)_{\scr L} \otimes SU(2)_{\scr R}$-Darstellung
$(m,n)$ enth"alt, mu"s auch eine Darstellung $(n,m)$ vorhanden sein. 
Dies hat zur Folge, da"s die beiden Eichkopplungen $g_{\scr 2L}(\mu)$ 
und $g_{\scr 2R}(\mu)$ "uberall zwischen $M_{\scr U}$ und $M_{\scr I}$ 
gleich gro"s sind und die Parit"at erhalten ist. 

\noindent Welche Symmetriegruppe $G_{\scr I}$ durch die Brechung mittels
einer bestimmten Darstellung $\Phi$ realisiert wird, h"angt von den 
jeweiligen Werten der Parameter im Higgs-Potential $V(\Phi)$ ab 
\cite{acamp}. Die Symmetriebrechung bei $M_{\scr I}$ in das SM erfolgt 
in allen F"allen "uber einen VEW des $G_{\scr \textrm{SM}}$-Singuletts
einer $SO(10)$-Darstellung {\bf 126}.

In dieser Arbeit soll ein $SO(10)$-Modell mit intermedi"arer $SU(4)_{\scr C} 
\otimes SU(2)_{\scr L} \otimes SU(2)_{\scr R} \equiv 
G_{\scr \textrm{PS}}$-Symmetrie 
untersucht werden. Diese Gruppe geh"ort zu den maximalen Untergruppen
der $SO(10)$ und ist von Pati und Salam f"ur eine Erweiterung des SM
durch eine rechts-links-symmetrische Theorie vorgeschlagen worden \cite{ps}.
Allerdings hat sich gezeigt, da"s Modelle mit 
$G_{\scr \textrm{PS}} \otimes D$-Symmetrie f"ur $M_{\scr U}$ den Wert 
$\approx 10^{15}$ GeV liefern, welcher auf eine
zu kurze Protonlebensdauer f"uhrt \cite{deshp}. $G_{\scr \textrm{PS}}$-Modelle
ohne $D$-Parit"at besitzen diesen Nachteil nicht. Der $SU(2)_{\scr L}$-Faktor
ist identisch mit dem in $G_{\scr \textrm{SM}}$, die $SU(2)_{\scr R}$
ist das rechtsh"andige Gegenst"uck dazu, und die $SU(4)_{\scr C}$
schlie"slich ist eine erweiterte Farbgruppe mit der $B-L$-Quantenzahl
als "`vierter Farbe"'. In $SO(10)$-basierten Modellen ist $B-L$ bei Skalen 
$\mu \gtrsim M_{\scr I}$ eine lokale $U(1)$-Symmetrie, also Teil der
Eichgruppe, und keine globale Symmetrie wie in der $SU(5)$-GUT. Das hat
zur Folge, da"s auch Bosonen nichtverschwindendes $B-L$ besitzen k"onnen.

\begin{figure}[h]
\vspace{0.5cm}
\begin{minipage}{14cm}
\hspace{0.5cm}
\begin{fmfgraph*}(2,2)
\fmfleft{i} 
\fmfright{o1,o2}
\fmf{boson,width=1.0pt,label=$X'$}{i,v} 
\fmf{fermion,width=1.0pt,label=$d^{\scr C}$}{v,o1} 
\fmf{fermion,width=1.0pt,label=$d^{\scr C}$}{v,o2}
\fmfdot{v}
\end{fmfgraph*}
\hspace{0.5cm}
\begin{fmfgraph*}(2,2)
\fmfleft{i} 
\fmfright{o1,o2}
\fmf{boson,width=1.0pt,label=$X'$}{i,v} 
\fmf{fermion,width=1.0pt,label=$u$}{v,o1} 
\fmf{fermion,width=1.0pt,label=$\nu_{\scr e}^{\scr C}$}{v,o2}
\fmfdot{v}
\end{fmfgraph*}
\hspace{0.5cm}
\begin{fmfgraph*}(2,2)
\fmfleft{i} 
\fmfright{o1,o2}
\fmf{boson,width=1.0pt,label=$Y'$}{i,v} 
\fmf{fermion,width=1.0pt,label=$d^{\scr C}$}{v,o1} 
\fmf{fermion,width=1.0pt,label=$u^{\scr C}$}{v,o2}
\fmfdot{v}
\end{fmfgraph*}
\hspace{0.5cm}
\begin{fmfgraph*}(2,2)
\fmfleft{i} 
\fmfright{o1,o2}
\fmf{boson,width=1.0pt,label=$Y'$}{i,v} 
\fmf{fermion,width=1.0pt,label=$d$}{v,o1} 
\fmf{fermion,width=1.0pt,label=$\nu_{\scr e}^{\scr C}$}{v,o2}
\fmfdot{v}
\end{fmfgraph*}
\hspace{0.5cm}
\begin{fmfgraph*}(2,2)
\fmfleft{i} 
\fmfright{o1,o2}
\fmf{boson,width=1.0pt,label=$Y'$}{i,v} 
\fmf{fermion,width=1.0pt,label=$u$}{v,o1} 
\fmf{fermion,width=1.0pt,label=$e^{\scr -}$}{v,o2}
\fmfdot{v}
\end{fmfgraph*}
\end{minipage}
\vspace{0.5cm}
\caption{Baryonzahlverletzende Vertizes der $X'$- und $Y'$-Bosonen 
\label{pdecv2}}
\end{figure}
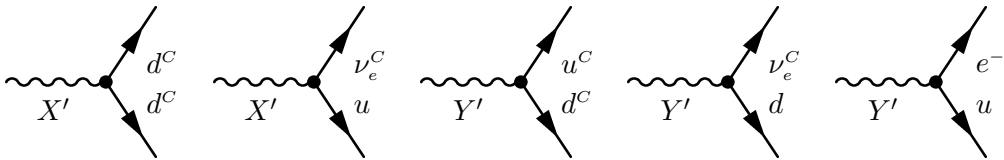

\noindent Die Fermiondarstellung {\bf 16} verzweigt sich bez"uglich 
$G_{\scr \textrm{PS}}$ und $SU(3)_{\scr C} \otimes SU(2)_{\scr L}$ gem"a"s
\begin{eqnarray} \label{lrsfs}
{\bf 16} & \longrightarrow & ({\bf 4,2,1}) \oplus ({\bf \bar 4,1,2})
\\ \nonumber
& \equiv & \begin{pmatrix} u^{}_1 & u^{}_2 & u^{}_3 & \nu^{}_{\scr e} \\ 
d^{}_1 & d^{}_2 & d^{}_3 & e_{}^- \end{pmatrix}_L
\oplus \begin{pmatrix} d^{\scr C}_1 & d^{\scr C}_2 & d^{\scr C}_3 & e_{}^+ \\
-u^{\scr C}_1 & -u^{\scr C}_2 & -u^{\scr C}_3 & -\nu^{\scr C}_{\scr e}
\end{pmatrix}_L \\
& \longrightarrow & ({\bf 3,2}) \oplus ({\bf 1,2})
\oplus 2 ({\bf \bar 3,1}) \oplus 2 ({\bf 1,1})
\end{eqnarray}
Der Operator der elektrischen Ladung ist durch 
$Q=T_{\scr 3L}+T_{\scr 3R}+(B-L)/2$ gegeben, die SM-Hyperladung durch 
$Y=T_{\scr 3R}+(B-L)/2$. Die adjungierte Darstellung {\bf 45}, in der
die Eichbosonen liegen, verzweigt sich bez"uglich $G_{\scr \textrm{PS}}$ wie
\begin{equation}
{\bf 45} \longrightarrow ({\bf 15,1,1}) \oplus ({\bf 1,3,1})
\oplus ({\bf 1,1,3}) \oplus ({\bf 2,2,6})
\end{equation}
Die ersten drei Faktoren entsprechen den $G_{\scr \textrm{PS}}$-Eichbosonen.
Die ({\bf 2,2,6}) enth"alt neben den aus der $SU(5)$-GUT bekannten
$X$- und $Y$-Bosonen die ebenfalls baryon- und leptonzahlverletzende 
Prozesse vermittelnden $X'$- und $Y'$-Bosonen (und deren Antiteilchen);
sie haben die Ladungen $+2/3$ beziehungsweise $-1/3$ und Massen der
Gr"o"senordnung $M_{\scr U}$. Abbildung \ref{pdecv2} zeigt die
zugeh"origen Wechselwirkungsvertizes. In der ({\bf 15,1,1}) liegen die acht 
Gluonen der QCD, ein an $B-L$ koppelndes neutrales Boson und die 
$SU(3)_{\scr C}$-Tripletts $X_{\scr 3}$ und $\bar X_{\scr 3}$ mit 
$Q(X_{\scr 3})=+2/3$. 

Der vollst"andige baryonzahlverletzende Teil der 
$SO(10)$-Lagrangedichte f"ur die erste Fermionfamilie und ohne
Mischungen lautet \cite{lang2}:
\begin{eqnarray} \nonumber
  \mathcal{L}_{\scr \Delta B \ne 0}
& = & \dfrac{g^{}_{\scr U}}{\sqrt{2}} \; \bar 
  X^{\scr \alpha}_{\scr \mu} \;
  \big( \varepsilon_{\scr \alpha \beta \gamma} \bar u_{\scr L}^{\scr C
    \gamma}  \gamma^{\scr \mu} u_{\scr L}^{\scr \beta} + \bar d_{\scr L
  \alpha}^{} \gamma^{\scr \mu} e^{\scr +}_{\scr L} 
   + \bar d_{\scr R \alpha}^{} \gamma^{\scr \mu} e^{\scr +}_{\scr R}
  \big) \\ \nonumber
& + & \dfrac{g^{}_{\scr U}}{\sqrt{2}} \; 
  \bar Y^{\scr \alpha}_{\scr \mu} \; \big( 
  \varepsilon_{\scr \alpha \beta \gamma} \bar u_{\scr L}^{\scr C \gamma} 
  \gamma^{\scr \mu} d_{\scr L}^{\scr \beta}
  - \bar d_{\scr R \alpha}^{}  \gamma^{\scr \mu} \nu_{\scr e R}^{\scr C}
  - \bar u_{\scr L \alpha}^{} \gamma^{\scr \mu} e^{\scr +}_{\scr L}
  \big) \\ \nonumber
& + & \dfrac{g^{}_{\scr U}}{\sqrt{2}} \; X^{' \scr \alpha}_{\scr \mu} \; \big( 
- \varepsilon_{\scr \alpha \beta \gamma} \bar d_{\scr L}^{\scr C \gamma}
  \gamma^{\scr \mu} d_{\scr L}^{\scr \beta}
  - \bar u_{\scr L \alpha}^{} \gamma^{\scr \mu} \nu^{\scr C}_{\scr e L} 
  - \bar u_{\scr R \alpha}^{} \gamma^{\scr \mu} \nu^{\scr C}_{\scr e R}
  \big) \\ \nonumber
& + & \dfrac{g^{}_{\scr U}}{\sqrt{2}} \; Y^{' \scr \alpha}_{\scr \mu} \; \big( 
  \varepsilon_{\scr \alpha \beta \gamma} \bar d_{\scr L}^{\scr C \gamma}
  \gamma^{\scr \mu} u_{\scr L}^{\scr \beta}
  - \bar d^{}_{\scr L \alpha} \gamma^{\scr \mu} \nu_{\scr e L}^{\scr C}
  - \bar u^{}_{\scr R \alpha} \gamma^{\scr \mu} e_{\scr R}^{\scr +}
  \big) \\ \nonumber
& + & \dfrac{g^{}_{\scr U}}{\sqrt{2}} \; X^{\scr \alpha}_{\scr 3\mu} \; \big( 
   \bar d^{}_{\scr L \alpha} \gamma^{\scr \mu} e^{\scr -}_{\scr L}
  + \bar d^{}_{\scr R \alpha} \gamma^{\scr \mu} e^{\scr -}_{\scr R}
  + \bar u^{}_{\scr L \alpha} \gamma^{\scr \mu} \nu^{}_{\scr e L} 
  + \bar u^{}_{\scr R \alpha} \gamma^{\scr \mu} \nu^{}_{\scr e R} 
  \big) \\ \label{bnvld} 
& + & \textrm{h.c.}
\end{eqnarray}
Durch direkten $X_{\scr 3}$-Austausch k"onnen keine 
Nukleonenzerf"alle vermittelt werden, was auch notwendig ist, da deren
Massen der Gr"o"senordnung $M_{\scr I}$ zu sehr gro"sen Zerfallsraten f"uhren
w"urden. Durch $X_{\scr 3}$-$X'$-Mischung entstehen in h"oheren Ordnungen
$B-L$-verletzende Nukleonzerf"alle, die aber gegen"uber den Prozessen
f"uhrender Ordnung durch zus"atzliche Faktoren $1/M_{\scr U}$ 
unterdr"uckt sind \cite{machk}.
\subsection{Fermionmassen} \label{smdlc}
Dirac-Massenterme f"ur die geladenen Fermionen und die Neutrinos haben
die Struktur $m ( \Psi^{\scr C}_{\scr L} )^T C \Psi^{}_{\scr L}$ 
und besitzen demnach das $SO(10)$-Transformationsverhalten
\begin{equation}
{\bf 16} \otimes {\bf 16} \; = \; {\bf 10} \oplus {\bf 120} \oplus {\bf 126}
\end{equation}
Die Fermionmassen und die Symmetriebrechung
$G_{\scr \textrm{SM}} \stackrel{M_{\scr Z}}{\longrightarrow}
SU(3)_{\scr C} \otimes U(1)_{\scr \textrm{em}}$
kommen also zustande, wenn die farblosen und elektrisch neutralen
Komponenten von Higgs-Teilchen in den Darstellungen ${\bf \overline{10}}$, 
${\bf \overline{120}}$ oder ${\bf \overline{126}}$ einen VEW der 
Gr"o"senordnung $M_{\scr Z}$ ausbilden.
Diese Darstellungen verzweigen sich folgenderma"sen unter 
$G_{\scr \textrm{PS}}$ (siehe Anhang \ref{brsor}):
\begin{eqnarray}
{\bf 10}  & \longrightarrow & ({\bf 1,2,2}) \oplus ({\bf 6,1,1}) \\
{\bf 120} & \longrightarrow & ({\bf 1,2,2}) \oplus ({\bf 15,2,2})
\oplus ({\bf 6,3,1}) \oplus ({\bf 6,1,3}) \oplus ({\bf 10,1,1}) 
\oplus ({\bf \overline{10},1,1}) \\
{\bf 126} & \longrightarrow & ({\bf 15,2,2}) \oplus ({\bf 10,3,1}) 
\oplus ({\bf \overline{10},1,3}) \oplus ({\bf 6,1,1})
\end{eqnarray}
Da die linksh"andigen Teilchen in ({\bf 4,2,1}) und die linksh"andigen
Antiteilchen in $({\bf \bar 4,1,2})$ liegen, wird die Erzeugung von
Dirac-Massen wegen 
\begin{equation}
({\bf 4,2,1}) \otimes ({\bf \bar 4,1,2}) = ({\bf 15,2,2}) \oplus ({\bf 1,2,2}) 
\end{equation}
durch die $SU(2)_{\scr L}$-Doubletts in den 
$G_{\scr \textrm{PS}}$-Darstellungen ({\bf 1,2,2}) der {\bf 10}
und {\bf 120} sowie denen in den ({\bf 15,2,2}) 
der {\bf 120} und {\bf 126} vermittelt. Ferner k"onnen aufgrund von
\begin{eqnarray} \label{majmas1}
({\bf 4,2,1}) \otimes ({\bf 4,2,1}) \otimes ({\bf \overline{10},3,1})
& = & ({\bf 1,1,1}) \oplus \; \dots \dots \\ \label{majmas2}
({\bf \bar 4,1,2}) \otimes ({\bf \bar 4,1,2}) \otimes ({\bf 10,1,3})
& = & ({\bf 1,1,1}) \oplus \; \dots \dots 
\end{eqnarray}
"uber den Term ${\bf 16} \otimes {\bf 16} \otimes {\bf \overline{126}}$ auch 
Majorana-Massen f"ur die links- und rechtsh"andigen Neutrinos
auftreten. Da der VEW des SM-Singuletts in ({\bf 10,1,3}) den
Symmetriebrechungsschritt $G_{\scr \textrm{PS}} 
\stackrel{M_{\scr I}}{\longrightarrow} G_{\scr \textrm{SM}}$
realisiert, sind die Majorana-Massen der rechtsh"andigen Neutrinos
zwangsl"auf\/ig von der Gr"o"senordnung $M_{\scr I}$. 

Die verschiedenen Yukawa-Kopplungen und VEW der {\bf 10}, {\bf 120}
und {\bf 126} sind in Tabelle \ref{hgs} zusammengefa"st (die Indizes
numerieren mehrere Darstellungen einer Art, wenn vorhanden).
\begin{table}[h]
\begin{center}
\begin{tabular}{|l||c|c|c|c|}
\hline
Higgs-Darstellung & $({\bf 1,2,2})_{\bf \scr 10}^{\scr (i)}$ & 
$({\bf 1,2,2})_{\bf \scr 120}^{\scr (j)}$ & 
$({\bf 15,2,2})_{\bf \scr 120}^{\scr (k)}$ & 
$({\bf 15,2,2})_{\bf \scr 126}^{\scr (l)}$ \\ \hline
Yukawa-Kopplungsmatrix & ${\bf Y}_{\bf \scr 10}^{\scr (i)}$ & 
${\bf Y}_{\bf \scr 120}^{\scr (j)}$ & ${\bf Y}_{\bf \scr 120}^{\scr (k)}$ & 
${\bf Y}_{\bf \scr 126}^{\scr (l)}$ \\ \hline
Vakuumerwartungswerte & $\upsilon^{\scr (i)}_u$, $\upsilon^{\scr (i)}_d$ & 
$\tilde \upsilon^{\scr (j)}_u$, $\tilde \upsilon^{\scr (j)}_d$ & 
$\tilde \omega^{\scr (k)}_u$, $\tilde \omega^{\scr (k)}_d$ & 
$\omega^{\scr (l)}_u$, $\omega^{\scr (l)}_d$ \\ \hline
\end{tabular}
\end{center}
\caption{Higgs-Kopplungen und Vakuumerwartungswerte in $SO(10)$-GUTs 
\label{hgs}}
\end{table}
Damit die beiden VEW, die zu jeder $G_{\scr \textrm{PS}}$-Darstellung
geh"oren, verschieden sein k"onnen, mu"s diese komplex sein. 
Das l"a"st sich durch Kombination zweier reeller Darstellungen 
$\mathcal{R}_1,\mathcal{R}_2$ zu $\mathcal{R}=\mathcal{R}_1+i\mathcal{R}_2$
erreichen.

Mit den in Tabelle \ref{hgs} festgelegten Bezeichnungen f"ur die 
Kopplungen und VEW (unter Vernachl"assigung der oberen Indizes)
ergeben sich f"ur die Fermionmassenmatrizen in
einer $SO(10)$-GUT folgende Identit"aten \cite{mass}:
\begin{eqnarray} \label{msrel1}
{\bf M}^{}_d \hspace{0.55cm} & = & 
\upsilon_d {\bf Y_{\scr 10}} + \omega_d {\bf Y_{\scr 126}}
+ (\tilde \upsilon_d + \tilde \omega_d) {\bf Y_{\scr 120}} \\
{\bf M}^{}_e \hspace{0.55cm} & = & 
\upsilon_d {\bf Y_{\scr 10}} - 3 \, \omega_d {\bf Y_{\scr 126}}
+ (\tilde \upsilon_d - 3 \, \tilde \omega_d) {\bf Y_{\scr 120}} \\
{\bf M}^{}_u \hspace{0.5cm} & = & 
\upsilon_u {\bf Y_{\scr 10}} + \omega_u {\bf Y_{\scr 126}}
+ (\tilde \upsilon_u + \tilde \omega_u) {\bf Y_{\scr 120}} \\ \label{msrel2a}
{\bf M}^{\scr (\textrm{Dir})}_\nu \hspace{0.1cm} & = & 
\upsilon_u {\bf Y_{\scr 10}} - 3 \, \omega_u {\bf Y_{\scr 126}}
+ (\tilde \upsilon_u - 3 \, \tilde \omega_u) {\bf Y_{\scr 120}} \\ 
\label{msrel2b}
{\bf M}^{\scr (\textrm{Maj})}_{\nu \scr{R}} & \sim & 
M_{\scr I} {\bf Y_{\scr 126}} \\ \label{msrel2}
{\bf M}^{\scr (\textrm{Maj})}_{\nu \scr{L}} & \sim & 
\dfrac{\omega_u^2}{M_{\scr I}} {\bf Y_{\scr 126}}
\end{eqnarray}
Die Beziehungen (\ref{msrel1}-\ref{msrel2}) sind in $SO(10)$-Theorien 
mit intermedi"arer $G_{\scr \textrm{PS}} \; [\otimes D]$-Sym\-me\-trie im 
gesamten Bereich oberhalb von $M_{\scr I}$ g"ultig, w"ahrend sie f"ur
die anderen $G_{\scr I}$ in Tabelle \ref{symbr} nur oberhalb von $M_{\scr U}$
gelten.

\noindent In (\ref{msrel1}-\ref{msrel2}) ist zu ber"ucksichtigen, da"s 
${\bf Y_{\scr 10}}$ und  ${\bf Y_{\scr 126}}$ symmetrisch sind, 
w"ahrend ${\bf Y_{\scr 120}}$ 
antisymmetrisch ist. Um Modelle mit asymmetrischen Massenmatrizen 
konstruieren zu k"onnen, mu"s also mindestens eine {\bf 120} an der 
Massenerzeugung beteiligt sein. Die Faktoren $(-3)$ sind
Clebsch-Gordan-Koef\/f\/izienten, welche auf der nichttrivialen
$SU(4)_{\scr C}$-Struktur der ({\bf 15,2,2}) beruhen.

\noindent Wird im einfachsten Fall lediglich eine (komplexe) {\bf 10} 
verwendet, gilt zwischen den Massenmatrize die Beziehung
\begin{equation} \label{massrel}
{\bf M}^{}_e \; = \; {\bf M}^{}_d \; \sim \; {\bf M}^{}_u \; = \; 
{\bf M}^{\scr (\textrm{Dir})}_\nu \; ,
\end{equation}
in der die $SU(5)$-Vorhersage ${\bf M}^{}_e = {\bf M}^{}_d$ enthalten
ist. Genauso wie dort f"uhrt (\ref{massrel}) aber zu falschen Werten
f"ur die Fermionmassen bei $M_{\scr Z}$. Verwendet man Modelle mit
komplizierterer Higgs-Struktur, kann man Massen und Mischungen der
Fermionen korrekt beschreiben.

Die beiden Majorana-Massenmatrizen der Neutrinos sind nur bis
auf konstante Vorfaktoren bekannt, die wiederum von der expliziten Form
des Higgs-Potentials und den Werten der Parameter in diesem abh"angen.
Da es f"ur Modelle mit nichtminimalem Higgs-Inhalt praktisch unm"oglich
ist, das zugeh"orige Potential im Detail zu analysieren, kann man
lediglich plausible Annahmen "uber die Vorfaktoren machen. Wir werden
im folgenden davon ausgehen, da"s sie betragsm"a"sig zwischen $1/100$
und $100$ liegen. Majorana-Massenmatrizen sind aufgrund der
Spinor-Struktur (\ref{majmass1}-\ref{majmass2}) der entsprechenden 
Massenterme stets symmetrisch. Im Rahmen von $SO(10)$-Modellen k"onnen
sie nur durch Kopplungen an {\bf 126}-Darstellungen erzeugt werden.

Die Majorana-Massen der linksh"andigen Neutrinos, die durch
einen VEW der $({\bf \overline{10},3,1})$ in (\ref{majmas1}) entstehen
k"onnen, m"ussen sehr klein sein, da sie sonst zu beobachtbaren
Ef\/fekten in durch neutrale schwache Str"ome vermittelten Prozessen
f"uhren w"urden. Die Konstruktion eines Potentials, in dem dieser VEW
verschwindet, st"o"st jedoch auf formale Probleme, denn in h"oheren
Ordnungen treten Divergenzen auf, die nur dann absorbiert werden
k"onnen, wenn der VEW der $({\bf \overline{10},3,1})$ ungleich Null
ist \cite{ggn}. Es kann allerdings gezeigt werden, da"s er von der
Gr"o"senordnung $\omega_u^2/M_{\scr I}$ sein mu"s \cite{muw} und somit
gegen"uber den Dirac-Massen um einen Faktor $\omega_u/M_{\scr I}$
unterdr"uckt ist und vernachl"assigt werden kann.

Bei Skalen unterhalb von $M_{\scr I}$ ist das Teilchenspektrum das des
SM. Das Higgs-Doublett im SM ist eine bestimmte Linearkombination aus den 
$SU(2)_{\scr L}$-Doubletts in den $({\bf 1,2,2})$- und 
$({\bf 15,2,2})$-Darstellungen, welche an der Massenerzeugung
mitwirken. Die "ubrigen Linearkombinationen haben Massen $\sim M_{\scr I}$
und treten im SM deshalb nicht in Erscheinung \cite{lhgs}.
\subsection{See-Saw-Mechanismus} \label{seesaw}
Die Massen der linksh"andigen Neutrinos m"ussen, wenn sie von Null
verschieden sind, sehr viel kleiner als die der geladenen Fermionen
sein. Die Obergrenzen aus den Experimenten zur direkten Messung von
Neutrinomassen liegen bei $m^{}_{\nu_e}<15$ eV (Untersuchung des 
$\beta$-Spektrums von Tritium), $m^{}_{\nu_\mu}<0.17$ MeV (aus
$\pi$-Zerf"allen) und $m^{}_{\nu_\tau}<18.2$ MeV (aus
$\tau$-Zerf"allen) \cite{pdg}. Kosmologische Argumente schr"anken die
Grenzen der Massenwerte weiter ein \cite{cosntr}; damit die kritische
Dichte des Universums nicht "uberschritten wird, mu"s f"ur die Summe
der Massen leichter stabiler Neutrinos $\sum_i m^{}_{\nu_i} < 30$ eV
gelten. Rechtsh"andige Neutrinos dagegen m"ussen sehr massiv sein, da sie 
experimentell nicht beobachtet werden.

Der See-Saw-Mechanismus \cite{sees} liefert in Modellen mit
rechtsh"andigen Neutrinos und einer Skala $M_{\scr I} \gg M_{\scr Z}$,
bei der die Leptonzahlerhaltung verletzt wird, eine nat"urliche
Erkl"arung f"ur die oben angegebenen Eigenschaften des 
Neutrinomassenspektrums. Seine Wirkungsweise soll zun"achst am
einfachen Fall einer Fermiongeneration verdeutlicht werden.

\noindent Der allgemeinste Massenterm f"ur das Neutrino $\nu$ lautet:
\begin{eqnarray} \nonumber
\mathcal{L}_{\scr M} & = & 
- \dfrac{1}{2} \, m^{\scr (\textrm{Maj})}_{\scr R} \, 
\bar \nu^{\scr C}_{\scr L} \nu^{}_{\scr R}
- \dfrac{1}{2} \, m^{\scr (\textrm{Maj})}_{\scr L} \, 
\bar \nu^{}_{\scr L} \nu^{\scr C}_{\scr R}
- m^{\scr (\textrm{Dir})}_{} \, \bar \nu^{}_{\scr L} \nu^{}_{\scr R}
+ \textrm{h.c.} \\ \label{nmm1}
& = & - \dfrac{1}{2} \, 
\begin{pmatrix} \bar \nu^{}_{\scr L} & \bar \nu^{\scr C}_{\scr L} 
\end{pmatrix} \, {\bf M} \,
\begin{pmatrix} \nu^{\scr C}_{\scr R} \\ \nu^{}_{\scr R} \end{pmatrix}
+ \textrm{h.c.}
\end{eqnarray}
mit
\begin{equation}
{\bf M} \; = \; \begin{pmatrix}
m^{\scr (\textrm{Maj})}_{\scr L} & m^{\scr (\textrm{Dir})}_{} 
\\ m^{\scr (\textrm{Dir})}_{} & m^{\scr (\textrm{Maj})}_{\scr R} \end{pmatrix}
\end{equation}
In $SO(10)$-GUTs mit intermedi"arer Symmetrie gelten die Relationen
\begin{equation}
\dfrac{m^{\scr (\textrm{Dir})}_{}}{m^{\scr (\textrm{Maj})}_{\scr R}} 
\, \ll \, 1 
\quad \textrm{und} \quad
\dfrac{m^{\scr (\textrm{Maj})}_{\scr L}}{m^{\scr (\textrm{Dir})}_{}} 
\, \ll \, 1 \; ,
\end{equation}
da, wie im vorigen Abschnitt erl"autert wurde, $m^{\scr (\textrm{Dir})}_{}$ 
dieselbe Gr"o"senordnung wie die Masse des positiv geladenen Quarks hat und 
$m^{\scr (\textrm{Maj})}_{\scr R}$ von der Ordnung der
$B-L$-brechenden Skala $M_{\scr I}$ ist; $m^{\scr (\textrm{Maj})}_{\scr L}$ 
wiederum ist gegen"uber $m^{\scr (\textrm{Dir})}_{}$ stark unterdr"uckt.

Die Diagonalisierung von {\bf M} f"uhrt (abgesehen von
Korrekturen $\sim m^{\scr (\textrm{Dir})}_{}/m^{\scr (\textrm{Maj})}_{\scr R}$)
auf die Masseneigenzust"ande 
\begin{equation}
\nu_{\scr 1} \; = \; \dfrac{1}{\sqrt{2}} \,
( \, \nu^{}_{\scr L} + \nu^{\scr C}_{\scr R} \, )
\quad \textrm{und} \quad
\nu_{\scr 2} \; = \; \dfrac{1}{\sqrt{2}} \, 
( \, \nu^{}_{\scr R} + \nu^{\scr C}_{\scr L} \, )
\end{equation}
und ihre Massen
\begin{eqnarray} \label{nmass1}
m_{\nu_1} & \approx & m^{\scr (\textrm{Maj})}_{\scr L}
- \left( \dfrac{m^{\scr (\textrm{Dir})}_{}}{m^{\scr (\textrm{Maj})}_{\scr R}}
\right) m^{\scr (\textrm{Dir})}_{} \; \approx \; 
- \left( \dfrac{m^{\scr (\textrm{Dir})}_{}}{m^{\scr (\textrm{Maj})}_{\scr R}}
\right) m^{\scr (\textrm{Dir})}_{} \\ \label{nmass2}
m_{\nu_2} & \approx & m^{\scr (\textrm{Maj})}_{\scr R}
+ \left( \dfrac{m^{\scr (\textrm{Dir})}_{}}{m^{\scr (\textrm{Maj})}_{\scr R}}
\right) m^{\scr (\textrm{Dir})}_{}
 \; \approx \; m^{\scr (\textrm{Maj})}_{\scr R}
\end{eqnarray}
Anstelle eines Dirac-Neutrinos $\nu=\nu_{\scr L}+\nu_{\scr R}$ erh"alt
man auf diese Weise zwei Majorana-Neutrinos $\nu^{}_{\scr 1}$ und 
$\nu^{}_{\scr 2}$ (es gilt $\nu_{\scr 1,2}^{\scr C}=\nu^{}_{\scr 1,2}$) 
mit Massen $m_{\nu_{1,2}}$, wobei $m_{\nu_1} \ll M_{\scr Z} \ll m_{\nu_2}$
ist. Das leichte Majorana-Neutrino $\nu^{}_{\scr 1}$ entspricht dem 
Neutrino des SM; das rechtsh"andige Antineutrino des SM ist identisch mit
$\nu^{}_{\scr 1R} \equiv \nu^{\scr C}_{\scr 1R}$. Das schwere Neutrino 
$\nu^{}_{\scr 2}$ ist aufgrund seiner Masse $m_{\nu_2} \sim M_{\scr I}$ 
nicht beobachtbar.

\noindent Aus (\ref{nmass1}) und (\ref{nmass2}) folgt die "`See-Saw"'-Relation
\begin{equation}
  m_{\nu_1} \cdot m_{\nu_2} \; \approx \; - m^{\scr (\textrm{Dir})}_{}
\cdot m^{\scr (\textrm{Dir})}_{} \; \lesssim \; - M^2_{\scr Z} \; ;
\end{equation}
je schwerer $\nu^{}_{\scr 2}$ ist, desto leichter wird $\nu^{}_{\scr 1}$.

\noindent Die Verallgemeinerung von (\ref{nmm1}) auf den Fall von $n$ 
Fermionfamilien liefert 
\begin{equation}
\mathcal{L}_{\scr M} \; = \;
- \dfrac{1}{2} \, 
\begin{pmatrix} \bar \nu^{}_{\scr L} & \bar \nu^{\scr C}_{\scr L} 
\end{pmatrix} \, {\bf M} \,
\begin{pmatrix} \nu^{\scr C}_{\scr R} \\ \nu^{}_{\scr R} \end{pmatrix}
+ \textrm{h.c.}
\end{equation}
mit $n$-dimensionalen Vektoren $\nu^{}_{\scr L,R}$ und der symmetrischen 
($2n \times 2n$)-Matrix
\begin{equation}
{\bf M} \; = \; \begin{pmatrix}
{\bf M}^{\scr (\textrm{Maj})}_{\nu \scr{L}} & 
{\bf M}^{\scr (\textrm{Dir})}_\nu \\
\big( {\bf M}^{\scr (\textrm{Dir})}_\nu \big)^{\scr T}
& {\bf M}^{\scr (\textrm{Maj})}_{\nu \scr{R}} 
\end{pmatrix}
\end{equation}
Hier sind die nichtverschwindenden Eintr"age von 
${\bf M}^{\scr (\textrm{Maj})}_{\nu \scr{L}}$ betragsm"a"sig sehr viel 
kleiner als die von ${\bf M}^{\scr (\textrm{Dir})}_\nu$, und diese sind
wiederum sehr viel kleiner als die von 
${\bf M}^{\scr (\textrm{Maj})}_{\nu \scr{R}}$ (siehe auch
(\ref{msrel2a}-\ref{msrel2})). Indem man {\bf M} auf blockdiagonale Form 
bringt, erh"alt man $n$ leichte Majorana-Neutrinos $\nu_{\scr i}$ mit
der symmetrischen Massenmatrix
\begin{equation} \label{seesmass}
{\bf M}_\nu \; \approx \; {\bf M}^{\scr (\textrm{Maj})}_{\nu \scr{L}}
- {\bf M}^{\scr (\textrm{Dir})}_\nu 
\big( {\bf M}^{\scr (\textrm{Maj})}_{\nu \scr{R}} \big)^{\scr -1}
\big( {\bf M}^{\scr (\textrm{Dir})}_\nu \big)^{\scr T}
\; \approx \; - {\bf M}^{\scr (\textrm{Dir})}_\nu 
\big( {\bf M}^{\scr (\textrm{Maj})}_{\nu \scr{R}} \big)^{\scr -1}
\big( {\bf M}^{\scr (\textrm{Dir})}_\nu \big)^{\scr T}\ \, ,
\end{equation}
der sogenannten See-Saw-Matrix, und $n$ schwere Majorana-Neutrinos $N_{\scr i}$
mit der Massenmatrix
\begin{equation}
{\bf M}_{\scr N} \; \approx \; {\bf M}^{\scr (\textrm{Maj})}_{\nu \scr{R}}
+ {\bf M}^{\scr (\textrm{Dir})}_\nu 
\big( {\bf M}^{\scr (\textrm{Maj})}_{\nu \scr{R}} \big)^{\scr -1}
\big( {\bf M}^{\scr (\textrm{Dir})}_\nu \big)^{\scr T}
\; \approx \; {\bf M}^{\scr (\textrm{Maj})}_{\nu \scr{R}}
\end{equation}
Die Diagonalisierung von ${\bf M}_\nu$ und ${\bf M}_{\scr N}$ 
\begin{equation} \label{ntrdiag}
{\bf N}^{\dagger}_{\scr L} \, {\bf M}_\nu \, {\bf N}^{}_{\scr L} \; = \; 
{\bf M}^{\scr (D)}_\nu \; , \quad
{\bf N}^{\dagger}_{\scr R} \, {\bf M}_{\scr N} \, {\bf N}^{}_{\scr R} \; = \; 
{\bf M}^{\scr (D)}_{\scr N}
\end{equation}
liefert f"ur $n=3$ die physikalischen Neutrinos $\nu_{\scr e}$, 
$\nu_{\scr \mu}$, $\nu_{\scr \tau}$ beziehungsweise $N_{\scr e}$, 
$N_{\scr \mu}$, $N_{\scr \tau}$ sowie deren Massen und Mischungen. 
Letztere haben nun Auswirkungen auf die leptonischen Anteile der 
geladenen schwachen Str"ome im SM, welche sich beim "Ubergang von
Wechselwirkungs- zu Masseneigenzust"anden analog zu (\ref{cwc}) gem"a"s
\begin{eqnarray}
\bar \nu^{\scr a (0)}_{\scr L} \gamma^\mu e^{\scr a (0)}_{\scr L} \, W^+_\mu
& \Longrightarrow &
\big( {\bf N}^\dagger_{\scr L} {\bf E}^{}_{\scr L} \big)^{}_{\scr ab}
\bar \nu^{\scr b}_{\scr L} \gamma^\mu 
e^{\scr a}_{\scr L} \, W^+_\mu \\
\bar e^{\scr a (0)}_{\scr L} \gamma^\mu \nu^{\scr a (0)}_{\scr L} \, W^-_\mu
& \Longrightarrow &
\bar e^{\scr a}_{\scr L} \gamma^\mu 
\big( {\bf E}^\dagger_{\scr L} {\bf N}^{}_{\scr L} \big)^{}_{\scr ab}
\nu^{\scr b}_{\scr L} \, W^-_\mu
\end{eqnarray}
"andern ($a$ und $b$ sind Familienindizes). Die unit"are Matrix 
${\bf U} \equiv {\bf E}^\dagger_{\scr L} {\bf N}^{}_{\scr L}$
entspricht der CKM-Matrix im Quarksektor und kann in
Neutrino-Oszillationsexperimenten gemessen werden. Die beobachtbaren
Ef\/fekte der Neutrinomassen und -mischungen werden im n"achsten Kapitel
erl"autert. 
\subsection{Zusammenfassung}
Im Vergleich zur $SU(5)$-GUT sind $SO(10)$-Modelle mit
intermedi"arer Symmetrie ph"anomenologisch sehr erfolgreich, und es
sind bis heute keine experimentellen Resultate bekannt, die sie
ausschlie"sen. Zus"atzlich zu den positiven Eigenschaften, welche
schon die $SU(5)$ besa"s, kommen hier folgende Vorz"uge: 
\begin{itemize}
\item Alle Fermionen einer Generation, einschlie"slich des
  rechtsh"andigen Neutrinos, liegen in einer irreduziblen Darstellung.
\item Die Eigenschaften der Neutrinos, insbesondere deren Massen und
  Mischungen, k"on\-nen durch den See-Saw-Mechanismus befriedigend
  erkl"art werden.
\item Die $SO(10)$ ist automatisch anomaliefrei, enth"alt $B-L$ als
  lokale Symmetrie und besitzt rechts-links-symmetrische Untergruppen.
\end{itemize}
Die Probleme, welche weder in $SO(10)$-Modellen noch in anderen
GUTs gel"ost werden k"onnen, sind im letzten Abschnitt zusammengefa"st.
\section{$E_6$}
Die $E_6$ \cite{e6,e6a} ist die einzige exzeptionelle Lie-Gruppe mit komplexen
Darstellungen. Zu den maximalen Untergruppen der $E_6$ geh"oren die f"ur die 
Symmetriebrechung nach $G_{\scr \textrm{SM}}$ in Frage kommenden
$SO(10) \otimes U(1)$ und $[ SU(3) ]^3$. Letztere kann man als 
$SU(3)_{\scr L} \otimes SU(3)_{\scr R} \otimes SU(3)_{\scr C}$-Symmetrie  
interpretieren.

Die Fermionen einer Generation liegen in der komplexen
Fundamentaldarstellung {\bf 27}, welche sich bez"uglich der $SO(10)$
und $[ SU(3) ]^3$ gem"a"s
\begin{eqnarray}
{\bf 27} & \longrightarrow & {\bf 16} \oplus {\bf 10} \oplus {\bf 1} \\
{\bf 27} & \longrightarrow & ({\bf \bar 3,3,1}) \oplus ({\bf 3,1,3}) 
\oplus ({\bf 1, \bar 3, \bar 3})
\end{eqnarray}
verzweigt. Im Gegensatz zu den $SU(5)$- und $SO(10)$-Modellen enth"alt die
$E_6$-GUT zus"atzliche Fermionen, die im SM nicht beobachtet werden
und somit superschwer sein m"ussen. In der {\bf 10} liegen ein
Quark $D$ mit Ladung $-1/3$, ein leptonisches $SU(2)_{\scr L}$-Doublett 
$(N,E)$ sowie die zugeh"origen Antiteilchen; das $SO(10)$-Singulett
wird mit $L$ bezeichnet. 

\noindent Die adjungierte Darstellung {\bf 78} der $E_6$ enth"alt die
Eichbosonen der Theorie. Die im Vergleich zur $SO(10)$ neu
hinzukommenden Eichbosonen koppeln sowohl an die superschweren als
auch an die SM-Fermionen und liefern zu Nukleonzerf"allen keine Beitr"age. 

\noindent Die Fermionmassen besitzen das $E_6$-Transformationsverhalten
\begin{equation}
{\bf 27} \otimes {\bf 27} \; = \; {\bf 27} \oplus {\bf 351} \oplus 
{\bf 351}' \, ,
\end{equation}
wobei die {\bf 351} und die ${\bf 351}'$ in"aquivalent sind. Unter
$SO(10)$ verzweigen sich diese beiden Darstellungen wie
\begin{eqnarray}
{\bf 351} \; & \longrightarrow & {\bf 1} \oplus {\bf 10} \oplus {\bf 126} 
\oplus \dots \dots \\
{\bf 351}' & \longrightarrow & {\bf 10} \oplus {\bf 120} \oplus \dots \dots 
\end{eqnarray}
Die {\bf 351} kann $E_6$ in die $SO(10)$ brechen, ohne da"s die
SM-Fermionen Massen der Gr"o"senordnung $M^{}_{\scr U}$ bekommen; in
diesem Brechungsschritt werden nur Massen f"ur die neuen Fermionen
erzeugt. Die Brechung $G_{\scr \textrm{SM}} \rightarrow
SU(3)_{\scr C} \otimes U(1)_{\scr \textrm{em}}$ und die Erzeugung der
SM-Massen werden durch die Higgs-Darstellung ${\bf \overline{27}}$ 
realisiert. Um asymmetrische Massenmatrizen erhalten zu k"onnen, mu"s
eine ${\bf 351}'$ vorhanden sein.

Die kleinste Darstellung, welche die Symmetriebrechung 
$E_6 \rightarrow [ SU(3) ]^3$ realisieren kann, ist die {\bf 650}. In
beiden Brechungsschemata ergeben sich Lebensdauern f"ur das Proton,
die weit oberhalb des experimentell zug"anglichen Bereichs liegen. 

Of\/fensichtlich ist die allgemeine Struktur von $E_6$-GUTs wesentlich
komplizierter als die von $SO(10)$-Modellen; insbesondere die Existenz
exotischer superschwerer Fermionen kommt neu hinzu. Dem steht als Vorteil
die M"oglichkeit gegen"uber, Fermionmassen und -mischun\-gen durch 
Strahlungskorrekturen zu erzeugen \cite{lukas}.
\section{Probleme und Grenzen von GUTs}
Wie in diesem Kapitel dargestellt wurde, k"onnen GUTs viele Schw"achen 
des SM erfolgreich beheben. Ferner liefern sie eine Reihe von zumindest
grunds"atzlich "uberpr"ufbaren Vorhersagen wie zum Beispiel die
Instabilit"at der Nukleonen oder die Neutrinomassen und -mischungen. 
Es gibt jedoch auch
Probleme, die im Rahmen von nichtsupersymmetrischen GUT-Modellen nicht
gel"ost werden k"onnen. Man kann sie in rein technische und fundamentale 
Probleme unterteilen.

Zu den ersteren geh"oren das schon erw"ahnte Hierarchieproblem und 
die daraus resultierenden Feineinstellungen verschiedener
Parameter. Diese Feineinstellungen sind bez"uglich
Strahlungskorrekturen instabil, so da"s sie f"ur jede Ordnung der
St"orungstheorie durchgef"uhrt werden m"ussen. Das 
$\theta_{\scr \textrm{QCD}}$-Problem 
($\theta_{\scr \textrm{QCD}} \lesssim 10_{}^{\scr -9}$) geh"ort
ebenfalls in diese Kategorie.

Ungel"oste fundamentale Fragestellungen sind die
Abwesenheit der Gravitationswechselwirkung in GUTs, eine fehlende 
Erkl"arung f"ur die Existenz dreier Fermionfamilien und ein 
hohes Ma"s an Unbestimmtheit
im Higgs-Sektor der Modelle. Auch der Ursprung der Massenhierarchie
der Fermionen kann nicht gekl"art werden, obwohl es in GUTs im
Gegensatz zum SM Beziehungen zwischen den Massenmatrizen der
verschiedenen Fermionarten gibt. 

Eine wirklich befriedigende L"osung dieser Probleme im Rahmen einer 
fundamentaleren Theorie als den hier geschilderten GUT-Modellen steht jedoch
bis heute aus.

\newpage

\mbox{ }

\end{fmffile}

%%% Local Variables: 
%%% mode: latex
%%% TeX-master: t
%%% End: 

\chapter{Neutrino-Oszillationen}
\section{Theoretische Grundlagen}
Neutrino-Oszillationen \cite{bgg,brrruno} k"onnen auftreten, wenn die 
Masseneigenzust"ande der Neutrinos nicht mit den 
Wechselwirkungseigenzust"anden "ubereinstimmen. 
Dies ist in $SO(10)$-Modellen mit intermedi"arer Symmetrie im
allgemeinen der Fall, da der See-Saw-Me\-cha\-nis\-mus aus Abschnitt 
\ref{seesaw} f"ur die leichten Majorana-Neutrinos $\nu_{\scr \alpha}$
($\alpha=e, \mu, \tau$) die "ublicherweise nichtdiagonale Massenmatrix 
(\ref{seesmass}) liefert. Wenn die Mischungsmatrix 
${\bf N}^{}_{\scr L} \ne {\bf 1}$
ist, sind die $\nu_{\scr \alpha}$ Linearkombinationen der Neutrinos 
$\nu_{\scr k}$ ($k$=1,2,3) mit den Massen $m_{\scr k}$, wobei 
$m_{\scr 1} \le m_{\scr 2} \le m_{\scr 3}$ gelten soll:
\begin{equation}
  \nu_{\scr \alpha L,R} \; = \; 
\sum^{\scr 3}_{\scr k=1} \, {\bf U}_{\scr \alpha k} \, \nu_{\scr k L,R}
\; , \quad {\bf U} \equiv {\bf E}^\dagger_{\scr L} {\bf N}^{}_{\scr L}
\end{equation}
Die schweren Neutrinos $N_{\scr \alpha}$ entkoppeln bei Skalen 
$\mu \sim M_{\scr I}$ und spielen f"ur die in diesem Kapitel
geschilderten Ph"anomene keine Rolle. 
\subsection{Vakuumoszillationen}
Zun"achst sollen Oszillationen im Vakuum betrachtet werden. Der
quantenmechanische Zustand eines zur Zeit $t=0$ in einem
elektroschwachen  Proze"s erzeugten relativistischen Neutrinos 
$\nu_{\scr \alpha}$ mit Impuls $p \gg m_{\scr k}$ ist durch
\begin{equation}
| \nu_{\scr \alpha}(x,t=0) \rangle \; = \;
\sum_{\scr k} \, {\bf U}_{\scr \alpha k} \, e^{ipx}
\, | \nu_{\scr k}(x,t=0) \rangle
\end{equation}
gegeben. Die Zeitentwicklung der Zust"ande $| \nu_{\scr k}(x,t=0) \rangle$ 
gem"a"s der Schr"odinger-Gleichung liefert unter Verwendung von
\begin{equation}
E = E(\nu_{\scr k}) = \sqrt{p^2+m_{\scr k}^2} \approx 
p+\dfrac{m_{\scr k}^2}{2\,p} \quad \textrm{und} \quad x \approx c t \quad 
(c \equiv 1)
\end{equation}
f"ur $| \nu_{\scr \alpha}(x) \rangle$ die Beziehung
\begin{equation}
| \nu_{\scr \alpha}(x) \rangle \; = \;
\sum_{\scr k} \, {\bf U}_{\scr \alpha k} \, e^{-i (m_{\scr k}^2 / 2E) x}
\, | \nu_{\scr k}(0) \rangle
\end{equation}
Dr"uckt man $| \nu_{\scr k}(x=0) \rangle$ als Linearkombination der  
$| \nu_{\scr \alpha}(x=0) \rangle$ aus, so ergibt sich 
\begin{equation}
| \nu_{\scr \alpha}(x) \rangle \; = \;
\sum_{\scr k,\beta} \, {\bf U}_{\scr \alpha k} \, {\bf U}^*_{\scr \beta k} 
\, e^{-i (m_{\scr k}^2 / 2E) x}
\, | \nu_{\scr \beta}(0) \rangle
\end{equation}
Mit $P(\nu_{\scr \alpha} \rightarrow \nu_{\scr \beta},L)$ sei die
Wahrscheinlichkeit bezeichnet, da"s ein Neutrino der Energie $E$,
welches bei $x=0$ als $\nu_{\scr \alpha}$ erzeugt wurde, in einer 
Entfernung $L$ von der Quelle als $\nu_{\scr \beta}$ detektiert wird. 
Dann gilt unter Ber"ucksichtigung der Unitarit"at von {\bf U} (die in
Modellen mit See-Saw-Mechanismus bis auf Korrekturen der Ordnung 
$M_{\scr Z}/M_{\scr I}$ gegeben ist):
\begin{equation} \label{pab1}
P(\nu_{\scr \alpha} \rightarrow \nu_{\scr \beta},L) \; = \; 
| \, \langle \nu_{\scr \beta}(L) | \nu_{\scr \alpha}(0) \rangle \, |^2 \; = \;
| \, \delta_{\scr \alpha \beta } + \sum_{\scr k=2}^{\scr 3} \, 
{\bf U}_{\scr \alpha k} \, {\bf U}^*_{\scr \beta k} \, 
[ \, e^{-i (\Delta m_{\scr k1}^2 / 2E) L}-1 \, ] \, |^2 
\end{equation}
Hierbei ist $\Delta m_{\scr k1}^2 \equiv m_{\scr k}^2 - m_{\scr 1}^2$. 
Aus der $CPT$-Invarianz folgen ferner die Beziehungen
\begin{equation}
P(\nu_{\scr \alpha} \rightarrow \nu_{\scr \beta},L) = 
P(\bar \nu_{\scr \beta} \rightarrow \bar \nu_{\scr \alpha},L) 
\quad \textrm{und} \quad
P(\nu_{\scr \alpha} \rightarrow \nu_{\scr \alpha},L) = 
P(\bar \nu_{\scr \alpha} \rightarrow \bar \nu_{\scr \alpha},L)
\end{equation}
Die "Ubergangswahrscheinlichkeit f"ur 
$\nu_{\scr \alpha} \rightarrow \nu_{\scr \beta}$ h"angt demnach von
den Elementen von {\bf U}, von zwei unabh"angigen Dif\/ferenzen der 
Massenquadrate und vom Parameter $L/E$ ab. Sie kann dann von Null
verschieden sein, wenn ${\bf U} \ne {\bf 1}$ ist und f"ur mindestens
ein $k$ $\Delta m_{\scr k1}^2 \gtrsim E/L$ gilt. Ist dies der Fall, so
wird ein Neutrinostrahl, welcher bei $x=0$ nur aus $\nu_{\scr \alpha}$ besteht,
an einem Ort $x=L$ auch Neutrinos der Art $\nu_{\scr \beta \ne \alpha}$
beinhalten. Ein nur f"ur den Nachweis von $\nu_{\scr \alpha}$
geeignetes Experiment wird dann ein Neutrinodef\/izit feststellen.

Die Bedeutung von (\ref{pab1}) wird besonders klar, wenn man den Fall 
zweier Neutrino-Arten betrachtet. Dann besitzt {\bf U} die einfache
Parametrisierung
\begin{equation}
{\bf U} \; = \; \begin{pmatrix}
\quad \cos{\theta} & \sin{\theta} \\ -\sin{\theta} & \cos{\theta} \end{pmatrix}
\end{equation}
({\bf U} kann zwar $CP$-verletzende Phasen enthalten, die aber 
auf Neutrino-Oszillationen keinen Einf\/lu"s haben) und es gilt:
\begin{eqnarray} \label{prob1}
P(\nu_{\scr \alpha} \rightarrow \nu_{\scr \beta},L) & = & 
\dfrac{1}{2} \, \sin^2{2\theta} \, 
\left( \, 1 - \cos{(\Delta m^2 L / 2 E)}\, \right) \quad (\alpha \ne \beta) 
\\ \label{prob2}
P(\nu_{\scr \alpha} \rightarrow \nu_{\scr \alpha},L) & = & 
P(\nu_{\scr \beta} \rightarrow \nu_{\scr \beta},L) \; = \;
1 - P(\nu_{\scr \alpha} \rightarrow \nu_{\scr \beta},L)
\end{eqnarray}
Die "Ubergangswahrscheinlichkeit (\ref{prob1}) ist eine periodische
Funktion von $L/E$; dieses Ph"a\-no\-men wird als Neutrino-Oszillation
bezeichnet. Die Amplitude der Oszillation ist gleich
$\sin^2{2\theta}$, h"angt also allgemein von den Eintr"agen von {\bf U} ab,
und die Oszillationsl"ange ist durch 
$L^{\scr \textrm{osc}}= 4 \pi E / \Delta m^2$
gegeben. Oszillationsef\/fekte k"onnen demnach beobachtet werden, sofern
$L \gtrsim L^{\scr \textrm{osc}}$ ist. 

Bei der Datenanalyse von Oszillationsexperimenten wird im allgemeinen
von nur zwei beteiligten Neutrino-Arten ausgegangen und 
(\ref{prob1}-\ref{prob2}) verwendet. Die graphische Auswertung erfolgt
dann in $\sin^2{2\theta}-\Delta m^2$-Diagrammen, wobei die Nichtbeobachtung 
von Oszillationen abh"angig von der Nachweisempf\/indlichkeit des Experiments 
bestimmte Bereiche des Parameterraums ausschlie"st, w"ahrend positive 
Resultate zu mehr oder weniger eng begrenzten erlaubten Parameterbereichen 
f"uhren. 

\noindent
Oszillationsexperimente k"onnen lediglich Aussagen "uber die
Gr"o"sen $|{\bf U}|$ und $\Delta m_{\scr jk}^2$ machen. Auf diese
Weise lassen sich die Verh"altnisse je zweier Neutrinomassen
bestimmen, nicht aber die Massenwerte selbst.
\subsection{Oszillationen in Materie}
Wenn Neutrinos Materie durchqueren, kann der sogenannte
Mikheyev-Smirnov-Wolfen\-stein-Ef\/fekt (MSW-Ef\/fekt)
\cite{msw1,msw2} auftreten, welcher das Oszillationsverhalten im
Vergleich zum Vakuumfall modif\/iziert. 
Ursache des MSW-Ef\/fekts ist die Tatsache, da"s in Materie f"ur die
Teilchendichten $N_i$ der geladenen Leptonen $N_e \gg N_{\mu,\tau} \approx 0$ 
gilt. W"ahrend alle Neutrinos die gleiche Wechselwirkung "uber
neutrale schwache Str"ome erfahren, wechselwirken nur die
Elektron-Neutrinos "uber geladene schwache Str"ome mit der Materie. 
Das f"uhrt zu einer Flavour-Asymmetrie in der Vorw"arts-Streuung von 
Neutrinos, die eine Phasenverschiebung zur Folge hat und sich somit
auf die Zeitentwicklung des Gesamtsystems auswirkt \cite{msw1}.

Eine formale Analyse dieses Sachverhalts liefert f"ur zwei
Neutrino-Arten wieder die Beziehungen (\ref{prob1}-\ref{prob2}), wobei
allerdings die Vakuumgr"o"sen $\sin^2{2\theta}$ und $L^{\scr \textrm{osc}}$
durch die neuen Gr"o"sen
\begin{eqnarray} \label{mswangle}
\sin^2{2\theta_{\scr M}} & = & \dfrac{\sin^2{2\theta}}
{( \, \cos{2\theta}-L^{\scr \textrm{osc}}_{}/L_{\scr 0} \, )^2 
+ \sin^2{2\theta}} \quad \textrm{und} \\
L^{\scr \textrm{osc}}_{\scr M} & = & 
\dfrac{L^{\scr \textrm{osc}}}{\sqrt{( \, \cos{2\theta}
 - L^{\scr \textrm{osc}}_{}/L_{\scr 0} \, )^2 + \sin^2{2\theta}}}
\end{eqnarray}
mit $L_{\scr 0} = 2 \pi / (\sqrt{2} \, G_{\scr F} N_e)$ ersetzt werden
m"ussen. Daraus ergeben sich folgende Spezialf"alle:
\begin{itemize}
\item $L^{\scr \textrm{osc}} \ll L_{\scr 0}$ (geringe Elektronendichte): 
\begin{equation}
\sin^2{2\theta_{\scr M}} \; \approx \; \sin^2{2\theta} \; , \quad
L^{\scr \textrm{osc}}_{\scr M} \; \approx \; L^{\scr \textrm{osc}}
\end{equation}
Die Materie-Ef\/fekte sind vernachl"assigbar klein.
\item $L^{\scr \textrm{osc}} \gg L_{\scr 0}$ (hohe Elektronendichte): 
\begin{equation}
\sin^2{2\theta_{\scr M}} \; \approx \; (L_{\scr 0}/L^{\scr \textrm{osc}})^2
\sin^2{2\theta} \; \approx \; 0 \; , \quad 
L^{\scr \textrm{osc}}_{\scr M} \; \approx \;  L_{\scr 0}
\end{equation}
Die Vakuumparameter sind stark unterdr"uckt; die Amplitude
ist verschwindend klein und die Oszillationsl"ange ergibt sich unabh"angig von 
$L^{\scr \textrm{osc}}$.
\item $L^{\scr \textrm{osc}}_{}/L_{\scr 0}=\cos{2\theta}$:
\begin{equation}
\sin^2{2\theta_{\scr M}} \; = \; 1 \; , \quad 
L^{\scr \textrm{osc}}_{\scr M} \; = \; 
L^{\scr \textrm{osc}} / \sin{2\theta}
\end{equation}
Die Oszillationsamplitude ist maximal, auch wenn der
entsprechende Vakuumwert $\sin^2{2\theta}$ sehr klein ist. 
Besitzt die Materieverteilung eine r"aumliche Ausdehnung der Gr"o"senordnung
$L^{\scr \textrm{osc}}_{\scr M}$, so kann die "Ubergangswahrscheinlichkeit 
durchaus $\lesssim 1$ sein. Dieses Resonanzverhalten von (\ref{mswangle}) wird
als MSW-Ef\/fekt bezeichnet 
\cite{msw2}. 
\end{itemize}
Die Materie-Ef\/fekte auf das Oszillationsverhalten sind besonders f"ur
den Fall der im Inneren der Sonne erzeugten Elektron-Neutrinos von Bedeutung.
\section{Experimentelle Situation}
Die Neutrino-Oszillationsexperimente lassen sich grob in zwei
Kategorien unterteilen. Die "`Appearence"'-Experimente untersuchen das
Erscheinen von $\nu_{\beta \ne \alpha}$ in einem Neutrinostrahl,
welcher bei seiner Entstehung ausschlie"slich aus
$\nu_{\alpha}$-Neutrinos bestand, w"ahrend
"`Disappearence"'-Experimente nach einer Verringerung des
$\nu_{\alpha}$-Flusses in einem solchen Strahl suchen. Weiterhin 
kann man die Experimente nach den Neutrinoquellen unterscheiden; es 
f\/inden Neutrinos aus Beschleunigern, Kernreaktoren und nat"urlichen
Quellen Verwendung. Ausf"uhrliche Referenzen zu den hier erw"ahnten 
Oszillationsexperimenten f\/inden sich in \cite{bgg}. 
Bis heute sind drei Indizien bekannt, die auf Neutrino-Oszillationen hindeuten.
\subsection{Sonnen-Neutrinos} \label{solntr}
Struktur und Dynamik der Sonne werden durch das Standard-Sonnenmodell
(SSM) \cite{ssm} beschrieben und gelten als gut verstanden. Die
Energie wird im Kern der Sonne "uber die thermonuklearen $pp$- und
CNO-Reaktionsketten erzeugt; der zugrundeliegende Fusionsproze"s ist
in beiden F"allen 
\begin{equation}
 4 \, p \longrightarrow \alpha + 2 \, e^+ + 2 \, \nu_e + 26.7 \, \textrm{MeV}
\end{equation}
Das Energiespektrum der Elektron-Neutrinos ist inhomogen und reicht je
nach erzeugender Reaktion von 100 keV bis zu 15 MeV. Da die
Unsicherheiten in den SSM-Vorhersagen "uber den Neutrinof\/lu"s relativ
klein sind, eignet sich dieser gut f"ur die Suche nach Oszillationen
$\nu_e \to \nu_{\mu,\tau}$. Alle bisher durchgef"uhrten Messungen
haben betr"achtliche Dif\/ferenzen zwischen tats"achlich
gemessenen Ereignissen $N_{\textrm{Exp}}$ und den aufgrund von
Monte-Carlo-Simulationen erwarteten Ereignissen $N_{\textrm{SSM}}$ 
festgestellt; die Resultate sind in Tabelle \ref{solex} dargestellt 
\cite{bgg}. 
\begin{table}[h]
\begin{center}
\begin{tabular}{|l|l|l|l|}
\hline
Experiment & Nachweisreaktion & Energiebereich (MeV)  &
$R=N_{\textrm{Exp}}/N_{\textrm{SSM}}$ \\
\hline \hline
Homestake & $\textrm{Cl}^{37} \, (\nu_e , e^-) \, \textrm{Ar}^{37}$ &
$E_\nu > 0.814$ & $0.33 \pm 0.06$ \\
GALLEX & $\textrm{Ga}^{71} \, (\nu_e , e^-) \, \textrm{Ge}^{71}$ &
$E_\nu > 0.233$ & $0.60 \pm 0.07$ \\
SAGE & $\textrm{Ga}^{71} \, (\nu_e , e^-) \, \textrm{Ge}^{71}$ &
$E_\nu > 0.233$ & $0.52 \pm 0.07$ \\
Kamiokande & $\nu_e e^- \rightarrow \, \nu_e e^-$ & $E_\nu > 7.5$ &
$0.54 \pm 0.07$ \\
Super-Kamiokande & $\nu_e e^- \rightarrow \, \nu_e e^-$ & $E_\nu > 6.5$ &
$0.47 \pm 0.08$ \\
\hline
\end{tabular}
\end{center}
\caption[Sonnen-Neutrino-Experimente und deren
  Resultate]{\label{solex} Sonnen-Neutrino-Experimente und deren
  Resultate f"ur $R=N_{\textrm{Exp}}/N_{\textrm{SSM}}$}
\end{table}

\noindent Homestake verwendet zum Nachweis 600\;t $\textrm{C}_2 \textrm{Cl}_4$,
GALLEX und SAGE bestehen aus 30\;t Galliumchlorid beziehungsweise 60\;t
metallischem Gallium, Kamiokande und das Nachfolge-Experiment Super-Kamiokande
sind Wasser-\v Cerenkovdetektoren.

Die Analyse der Daten f"uhrt zu dem Ergebnis, da"s sowohl
Vakuumoszillationen auf dem Weg zwischen der Sonne und der Erde als auch der
MSW-Ef\/fekt im Sonneninneren das Neutrinodef\/izit erkl"aren k"onnen;
%
%\psdraft
\begin{figure}[h]
\begin{center}
\subfigure[MSW-Ef\/fekt \label{mswpic}]
{\epsfig{file=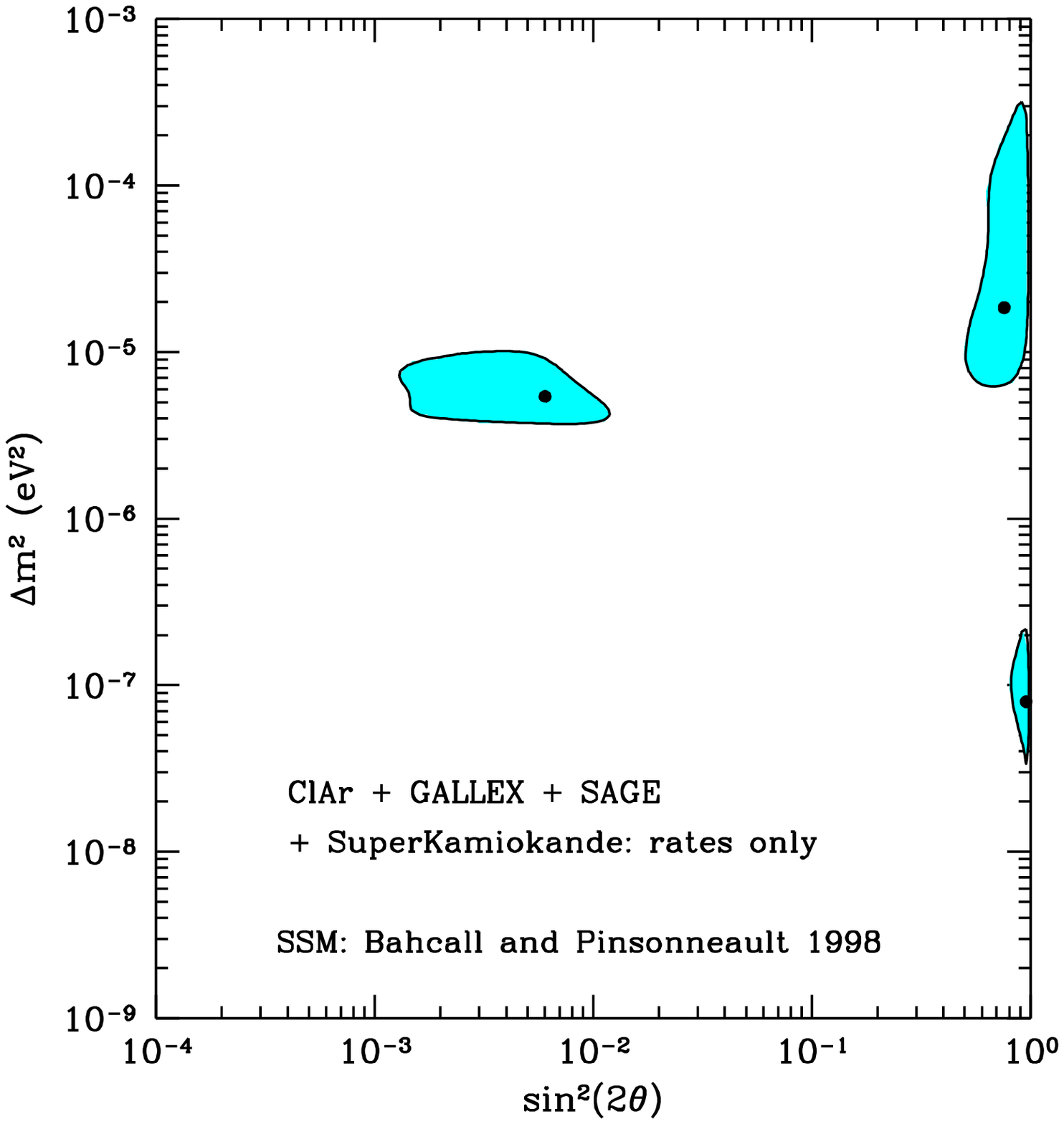,width=7.4cm}} 
\subfigure[Vakuumoszillationen \label{vacpic}]
{\epsfig{file=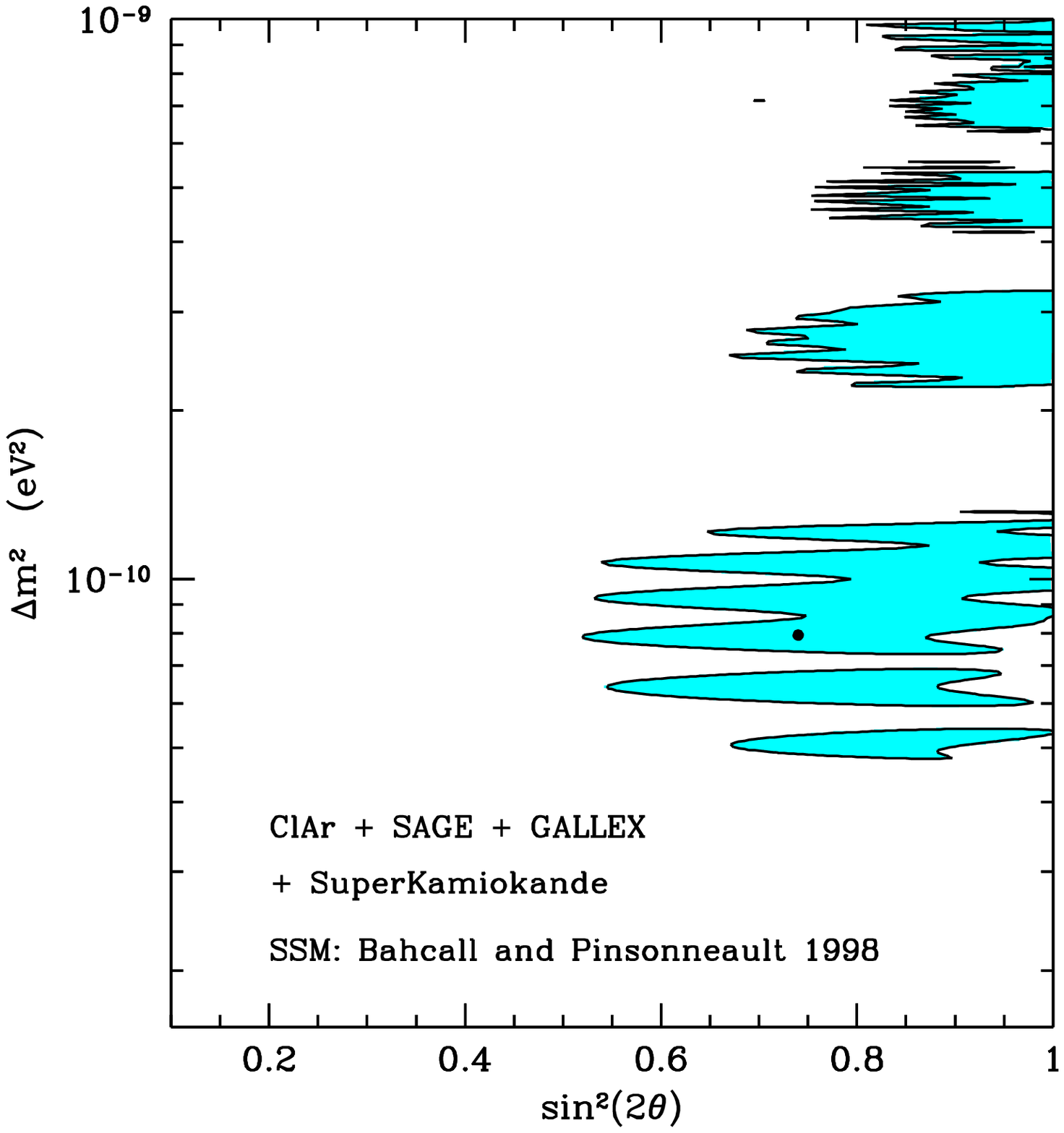,width=7.4cm}}
\caption[Anomalie der Sonnen-Neutrinos und m"ogliche L"osungen]
{Erlaubte Bereiche der Oszillationsparameter f"ur die L"osungen des 
Sonnen-Neutrinoproblems (99\,\% CL) \cite{snp} \label{snpic}}
\end{center}
\end{figure}
%\psfull
%
es kommen $\nu_e \to \nu_{\mu}$- und $\nu_e \to
\nu_{\tau}$-"Uberg"ange in Frage \cite{snp}. Die Abbildungen
\ref{mswpic} und 
\ref{vacpic} geben die erlaubten Bereiche (mit einem Conf\/idence
Level (CL) von 99\%) im $\sin^2{2\theta}-\Delta m^2$-Parameterraum an, 
wobei die Punkte den statistisch wahrscheinlichsten L"osungen entsprechen:
\begin{itemize}
\item eine MSW-L"osung mit gro"sem Mischungswinkel bei 
$\sin^2{2\theta} \approx 0.76$ und $\Delta m^2 \approx 2 \cdot 10^{-5} \,
\textrm{eV}^2$.
\item eine MSW-L"osung mit kleinem Mischungswinkel bei 
$\sin^2{2\theta} \approx 6 \cdot 10^{-3}$ und $\Delta m^2 \approx 5 \cdot
10^{-6} \, \textrm{eV}^2 $.
\item eine MSW-L"osung mit kleinem $\Delta m^2$ bei 
$\sin^2{2\theta} \approx 0.96$
und $\Delta m^2 \approx 8 \cdot 10^{-8} \, \textrm{eV}^2$. 
\item eine Vakuumoszillations-L"osung bei $\sin^2{2\theta} \approx 0.75$ 
und $\Delta m^2 \approx 8 \cdot 10^{-11} \, \textrm{eV}^2$. 
\end{itemize}
Die MSW-L"osung mit kleinem $\Delta m^2$ ist in ihrer Statistik wesentlich
unwahrscheinlicher als die anderen, da sie f"ur 95\,\% CL nicht mehr 
akzeptabel ist. Damit die Vakuumoszillations-L"osung das
Neutrinodef\/izit erkl"aren kann, ist eine nicht sehr nat"urlich
wirkende Feineinstellung zwischen der Oszillationsl"ange und dem
Sonne-Erde-Abstand erforderlich. Deswegen werden in dieser Arbeit nur
die ersten beiden L"osungen in Betracht gezogen.

\noindent Weitere Experimente zur Untersuchung des Sonnen-Neutrinodef\/izits 
sind in Vorbereitung. Der Nachfolger von GALLEX ist das unterirdische 
Gallium Neutrino Observatory (GNO) im Gran Sasso-Labor (Italien) mit
angestrebten 100\,t Ga im Jahre 2002. Borexino ist ein organischer
Fl"ussig-Szintillator mit einer Empf\/indlichkeit $E_\nu > 0.25$ MeV und
wird ebenfalls in Gran Sasso untergebracht. Das Sudbury Neutrino
Observatory (SNO) in Kanada wird ein Schwerwasser-\v Cerenkovdetektor
aus 1000\,t D$_2$O sein, welcher f"ur $E_\nu > 5$ MeV sensibel ist.
\subsection{Atmosph"arische Neutrinos} \label{atmntr}
Tref\/fen hochenergetische Protonen der kosmischen Strahlung auf die
obere Erdatmosph"are, so kollidieren sie mit den Kernen der Luftmolek"ule.
In diesen hadronischen Sto"sprozessen entstehen zahlreiche Pionen und Kaonen, 
die gem"a"s 
\begin{eqnarray} \nonumber
\pi^+ , \, K^+ \; \rightarrow \; \mu^+ + \nu_\mu \hspace{1.35cm} & & 
\pi^- , \, K^- \; \rightarrow \; \mu^- + \bar \nu_\mu \\ \label{decay}
\decay e^+ + \nu_e + \bar \nu_\mu & & \hspace{2.15cm}
\decay e^- + \bar \nu_e + \nu_\mu
\end{eqnarray}
zerfallen; die dabei erzeugten Neutrinos bezeichnet man als atmosph"arische 
Neutrinos. Wegen (\ref{decay}) erwartet man f"ur das zahlenm"a"sige 
Verh"altnis von Myon- zu Elektron-Neutrino-Ereignissen in einem
Detektor in guter N"aherung
\begin{equation} \label{ratm}
  (\mu/e) \; \equiv \; \dfrac{N(\nu_\mu+\bar \nu_\mu)}{N(\nu_e+\bar \nu_e)}
\; = \; 2
\end{equation}
Obwohl der exakte Wert von $(\mu/e)$ von der Neutrino-Energie abh"angt
und von 2 abweichen kann, erlauben detaillierte Analysen mit Hilfe von 
Monte-Carlo-Simulationen relativ pr"azise Vorhersagen. Diese bieten einen
weiteren Ansatzpunkt f"ur die Suche nach Neutrino-Oszillationen und
wurden vom Kamiokande-Experiment und dem gr"o"seren Nachfolger 
\begin{table}[b]
\begin{center}
\begin{tabular}{|l|l|l|}
\hline
Experiment & Energiebereich & 
$R = (\mu/e)_{\textrm{data}}/(\mu/e)_{\textrm{MC}}$ \\
\hline \hline
Kamiokande & $E_\nu < 1.33$ GeV & $0.60 \pm {0.06 \atop 0.05} \pm 0.05$ \\
Kamiokande & $E_\nu > 1.33$ GeV & $0.57 \pm {0.08 \atop 0.07} \pm 0.07$ \\
Super-Kamiokande & $E_\nu < 1.33$ GeV & $0.63 \pm 0.03 \pm 0.05$ \\
Super-Kamiokande & $E_\nu > 1.33$ GeV & $0.65 \pm 0.05 \pm 0.08$ \\
\hline
\end{tabular}
\end{center}
\caption[Resultate von Kamiokande und Super-Kamiokande]
{Resultate von Kamiokande \cite{kamio} und Super-Kamiokande
\cite{skamio} f"ur $R=\dfrac{(\mu/e)_{\textrm{data}}}{(\mu/e)_{\textrm{MC}}}$
\label{kamtab}}
\end{table}
Super-Kamiokande in Japan "uberpr"uft. Beides sind unterirdische 
Wasser-\v Cerenkovde\-tek\-to\-ren; Kamiokande bestand aus 3\,kt und 
Super-Kamiokande besteht aus 50\,kt reinen Wassers in 1000\,m
Tiefe. Als Nachweisreaktionen dienen
\begin{equation}
\nu_e + n \rightarrow e^- + p \, , \quad
\bar \nu_e + p \rightarrow e^+ + n \, , \quad
\nu_\mu + n \rightarrow \mu^- + p \, , \quad
\bar \nu_\mu + p \rightarrow \mu^+ + n
\end{equation}

\begin{figure}[th]
\begin{center}
\subfigure[Abh"angigkeit des von Super-Kamiokande gemessenen 
$(\mu/e)$-Def\/izits vom Zenit-Winkel $\theta$ \label{anpic1}]
{\epsfig{file=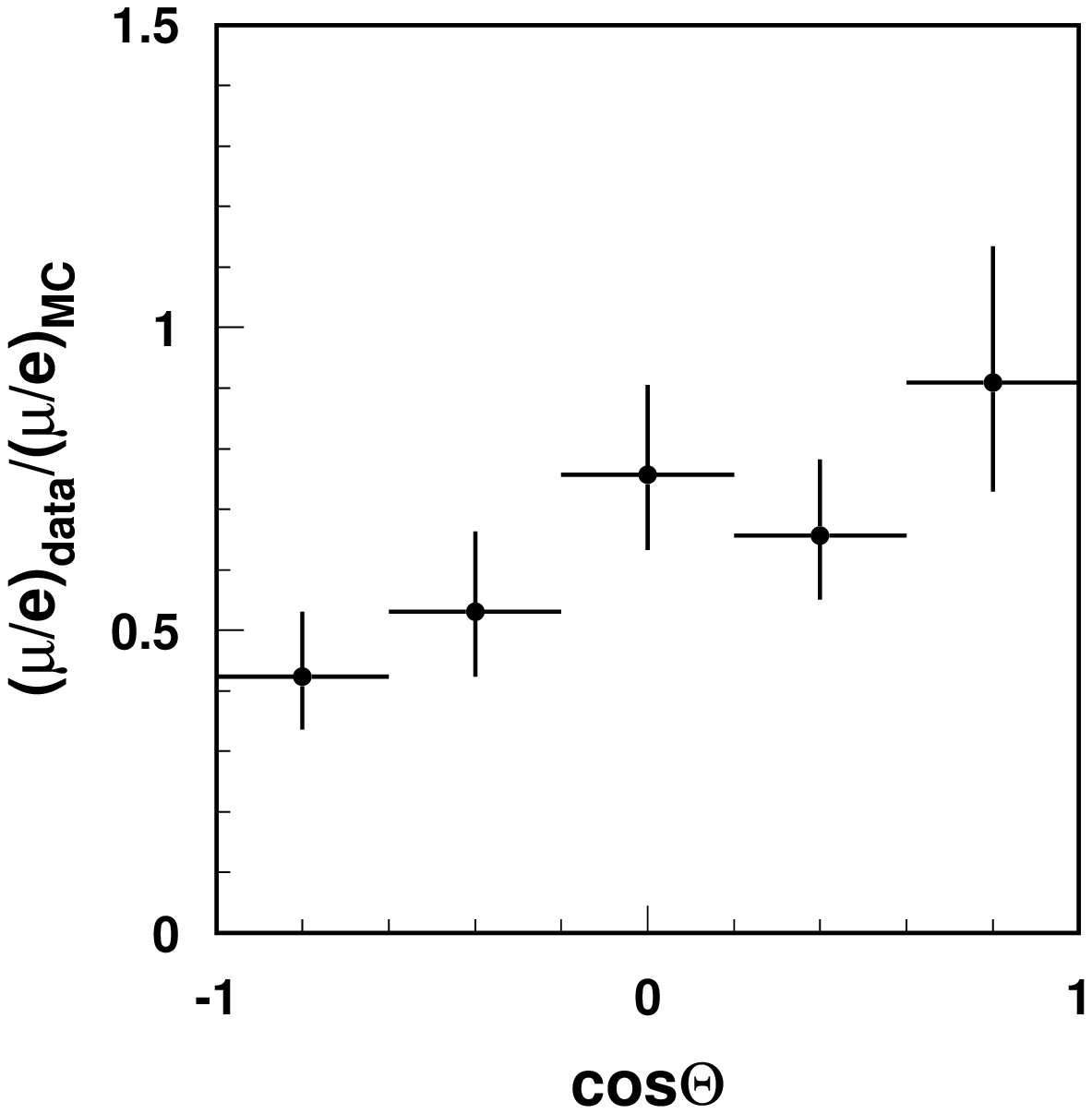,width=7.5cm}}
\subfigure[Erlaubte Bereiche der Oszillationsparameter f"ur Kamiokande 
und Super-Kamiokande \label{anpic2}]
{\epsfig{file=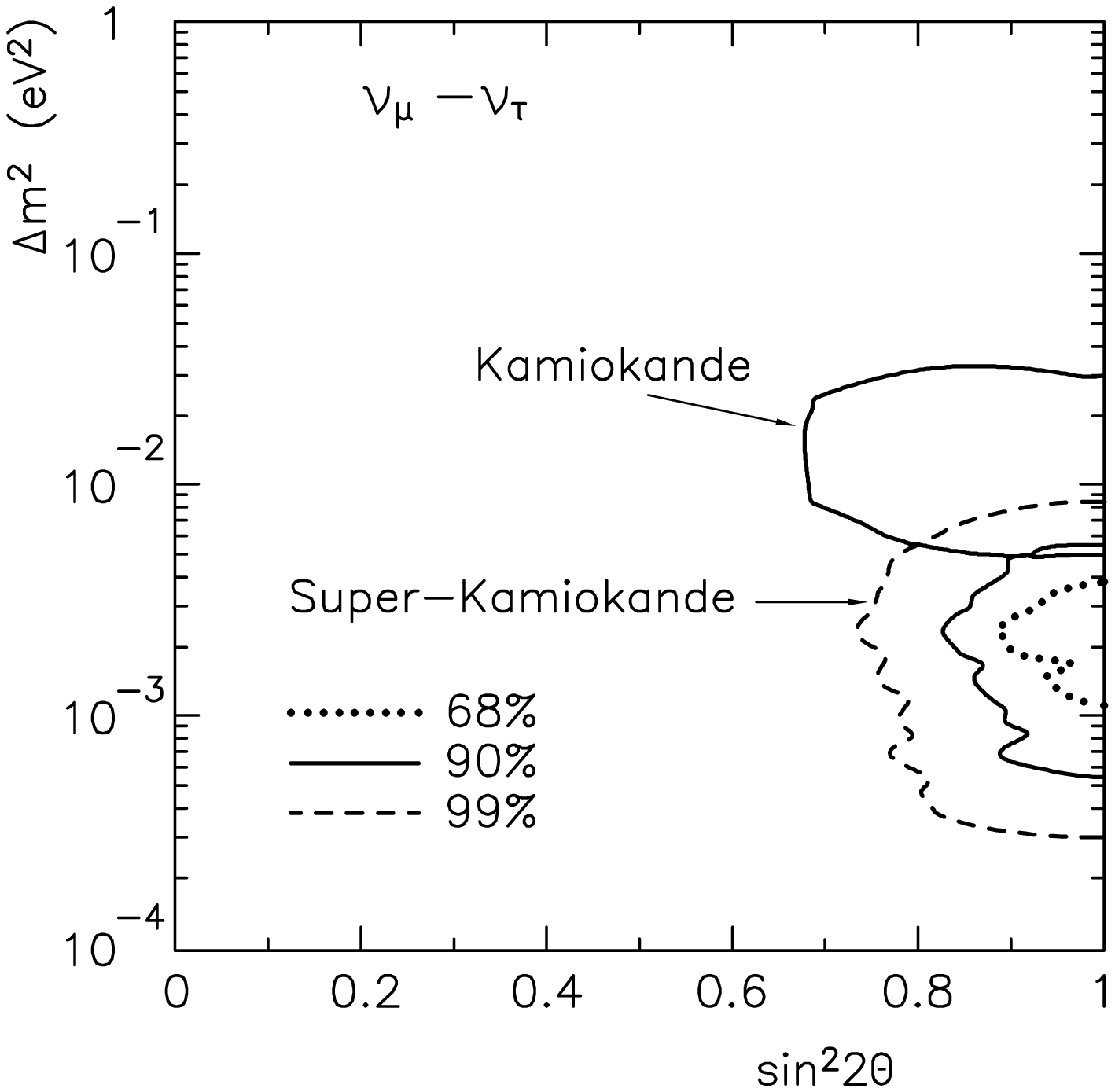,width=7cm}} 
\caption[Anomalie der atmosph"arischen Neutrinos und m"ogliche
L"osungen]{Anomalie der atmosph"arischen Neutrinos und m"ogliche
L"osungen durch $\nu_\mu-\nu_\tau$-Oszillationen \cite{kamio,skamio} 
\label{anpic}}
\end{center}
\end{figure}

\noindent 
Beide Experimente haben f"ur das Verh"altnis $R$ von gemessenem 
$(\mu/e)_{\textrm{data}}$ zu theoretisch erwartetem $(\mu/e)_{\textrm{MC}}$
erhebliche Abweichungen vom Wert 1 festgestellt, die auf einem Def\/izit von
$(\nu_\mu,\bar \nu_\mu)$-Ereignissen beruhen. Tabelle \ref{kamtab}
fa"st die Resultate zusammen, wobei der erste Fehler von $R$
statistischen und der zweite systematischen Ursprungs ist.

Eine m"ogliche Erkl"arung f"ur diese Anomalie k"onnen
$\nu_\mu-\nu_\tau$-Oszillationen liefern. Die Oszillation in 
Elektron-Neutrinos ist aufgrund der Resulate des
Reaktor-Experiments CHOOZ \cite{chooz} ausgeschlossen, welches nach
$\bar \nu_e \rightarrow \bar \nu_{\mu,\tau}$-"Uberg"angen sucht, im
Rahmen seines Me"sbereichs ($E_\nu \sim 3$ MeV, $L=1$ km) aber
keine solchen gefunden hat. 

Die Oszillationsl"osung wird durch die beobachtete Abh"angigkeit des 
$(\nu_\mu,\bar \nu_\mu)$-Def\/izits vom Zenit-Winkel $\theta$ (nicht zu
verwechseln mit dem gleichnamigen Mischungswinkel) unterst"utzt
(Abbildung \ref{anpic1}). W"ahrend die Neutrinos, welche den
Detektor senkrecht von oben erreichen ($\theta = 0$), eine Strecke von
$L \lesssim 20$ km zur"uckgelegt haben, m"ussen die den Detektor von
unten erreichenden die gesamte Erde durchqueren, das hei"st 
$L \approx 13000$ km f"ur $\theta = \pi$. Je gr"o"ser $\theta$ und
somit auch $L$ sind, desto mehr Myon-(Anti-)Neutrinos wandeln sich in 
Tau-(Anti-)Neutrinos um. 

Abbildung \ref{anpic2} zeigt den erlaubten Parameterbereich f"ur die 
Oszillationsl"osung; die wahrscheinlichste L"osung zur Erkl"arung der
Super-Kamiokande-Resultate liegt bei 
$\sin^2{2\theta} = 1.0$ und $\Delta m^2 = 2.2 \cdot 10^{-3}$ eV.
Die Annahme eines hierarchischen Massenschemas $m_1 \ll m_2 \ll m_3$
f"ur die Neutrinos f"uhrt zu 
$m_3 \approx \sqrt{\Delta m^2_{\textrm{atm}}} \sim 0.05$ eV, was
bedeuten w"urde, da"s Neutrinos zur dunklen Materie keinen
nennenswerten Beitrag liefern.

\noindent Unter Ber"ucksichtigung der 
$\nu_\mu \rightarrow \nu_{\tau}$-Oszillationsl"osung f"ur die Anomalie der 
atmosph"arischen Neutrinos ist das Sonnen-Neutrinoproblem demnach auf 
$\nu_e \rightarrow \nu_{\mu}$-Oszillationen zu\-r"uckzuf"uhren, da die
erlaubten Parameterbereiche in den Abbildungen \ref{snpic} und
\ref{anpic2} sich in keinem Fall "uberlappen.

Zur Erg"anzung von Super-Kamiokande, das weiterhin Daten aufnimmt, sind
mehrere sogenannte "`Long Baseline"'-Experimente geplant. Hierbei
sollen Neutrinostrahlen mit wohldef\/inierten Eigenschaften, die mittels
Sto"sprozessen an Teilchenbeschleunigern erzeugt werden, in einigen hundert 
km Entfernung detektiert werden. Erste Resulte sind jedoch nicht vor
2001 zu erwarten. 
\subsection{LSND und KARMEN}
\begin{figure}[ht]
\hspace{0.7cm}
%\begin{center}
\epsfig{file=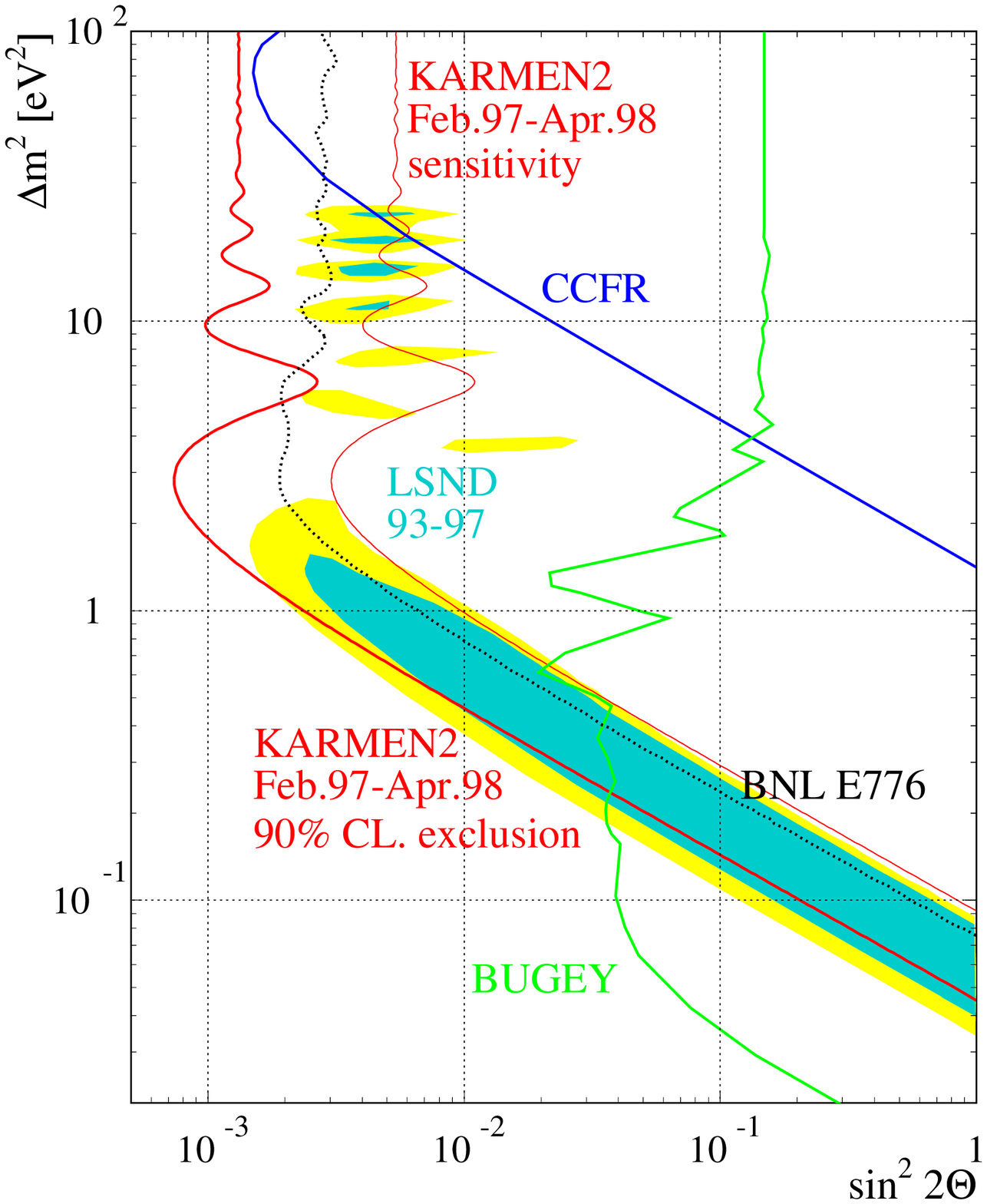,width=12cm}
\caption[Resultate der Beschleunigerexperimente LSND und KARMEN]
{Erlaubte Bereiche der Oszillationsparameter f"ur die Erkl"arung der 
LSND-Resultate und die durch KARMEN und weitere Experimente
ausgeschlossenen Parameterbereiche \cite{karmen} \label{karmpic}}
%\end{center}
\end{figure}

Das LSND-Beschleunigerexperiment in Los Alamos verwendet Neutrinos, die durch
das Auftref\/fen eines relativistischen Protonenstrahls auf ein
Wasser-Target entstehen. In den Sto"sprozessen werden $\pi^+$-Mesonen erzeugt,
welche anschlie"send zerfallen; gem"a"s (\ref{decay}) ent\-h"alt der
resultierende Neutrinostrahl keine Elektron-Antineutrinos. Im 30\,m
entfernten Fl"ussig-Szintillationsdetektor werden allerdings Photonen
nachgewiesen, welche aus der Reaktion 
$\bar \nu_e + p \rightarrow e^+ + n$ und nachfolgendem 
$e^+ + e^- \rightarrow \gamma$ beziehungsweise $n + p \rightarrow d + \gamma$ 
stammen. Eine m"ogliche Erkl"arung hierf"ur k"onnen 
$\bar \nu_\mu \rightarrow \bar \nu_e$-Oszillationen liefern. Der
erlaubte Parameterbereich ist in Abbildung \ref{karmpic} dargestellt, wobei
der dunkelgraue Bereich 90\,\% CL und der hellgraue 99\,\% CL kennzeichnet.

Eine Reihe von weiteren Reaktor- und Beschleunigerexperimenten haben 
$\bar \nu_\mu \rightarrow \bar \nu_e$-Oszil\-la\-tio\-nen untersucht, aber
im Rahmen ihrer Me"sbereiche keine entsprechenden Ereignisse
gefunden. Dies f"uhrt im Raum der Oszillationsparameter zu verbotenen
Bereichen, welche sich in Abbildung \ref{karmpic} rechts von den verschiedenen
Linien bef\/inden (90\,\% CL). Insbesondere das KARMEN-Experiment und
sein Nachfolger KARMEN\,2 am Rutherford-Appleton-Laboratorium, welche
im wesentlichen den gleichen Aufbau wie LSND haben (mit $L=17.6$ m), 
schlie"sen einen gro"sen Teil des LSND-Bereichs aus. Der Grund f"ur
diese widerspr"uchlichen Resultate ist bis heute nicht
bekannt. Allerdings wird das f"ur 2001 am Fermilab in Planung bef\/indliche
Szintillator-Experiment MiniBooNE \cite{boone} aufgrund einer deutlich 
verbesserten Statistik den gesamten von LSND erlaubten Parameterbereich 
untersuchen und somit die Frage nach 
$\bar \nu_\mu \rightarrow \bar \nu_e$-Oszil\-la\-tio\-nen
abschlie"send beantworten k"onnen.
\subsection{Analyse f"ur drei Neutrino-Arten} \label{ntranl}
In Tabelle \ref{sumntr} sind die Wertebereiche von $\Delta m^2$
zusammengefa"st, welche eine Oszillationsl"osung der jeweiligen
Neutrino-Anomalie erm"oglichen.
Da sich die zul"assigen Wertebereiche je zweier $\Delta m^2$ in keinem 
Fall "uberlappen und $n$ verschiedene Neutrinomassen 
$n-1$ unabh"angige $\Delta m^2$ liefern, k"onnen nur
dann alle drei Anomalien durch Neutrino-Oszillationen erkl"art werden,
wenn es vier leichte Neutrinos gibt. 
%\psdraft
%\psfull
Es sind verschiedene Modelle mit einem vierten, im Rahmen des SM nicht 
wechselwirkenden, Neutrino $\nu_s$ untersucht worden (Referenzen in 
\cite{bgg}); da aber die Resultate von LSND und KARMEN bisher
widerspr"uchlich sind, wird in der vorliegenden Arbeit die
LSND-Anomalie nicht weiter ber"ucksichtigt und von drei leichten
Neutrinos ausgegangen.
\begin{table}[h]
\begin{center}
\begin{tabular}{|l|l|l|}
\hline
Neutrino-Anomalie & $\Delta m^2$ (eV$^2$) & Oszillation \\
\hline \hline
Atmosph"arische Neutrinos & $(0.4-8.0) \cdot 10^{-3}$ & 
$\nu_\mu \longleftrightarrow \nu_\tau$ \\
Sonnen-Neutrinos (MSW-Ef\/fekt) & $(0.4-3.0) \cdot 10^{-5}$ & 
$\nu_e \longleftrightarrow \nu_{\mu,s}$ \\
Sonnen-Neutrinos (Vakuum-Osz.) & $(0.6-1.1) \cdot 10^{-10}$ & 
$\nu_e \longleftrightarrow \nu_{\mu,s}$ \\
LSND & $0.2-10$ & $\nu_\mu \longleftrightarrow \nu_e$ \\
\hline
\end{tabular}
\end{center}
\caption[Neutrino-Anomalien und erlaubte $\Delta m^2$-Bereiche]
{Neutrino-Anomalien und erlaubte $\Delta m^2$-Bereiche aus den 
Oszil\-la\-tions-Ana\-lysen \label{sumntr}}
\end{table}
Analysiert man alle Daten aus den Oszillationsl"osungen der
"ubrigen beiden Neutrinodef\/izite und aus den Experimenten, welche
keine Hinweise auf Oszillationen liefern, unter der Annahme dreier 
miteinander mischender leichter Neutrinos, so erh"alt man
je nach Erkl"arung der Sonnen-Neutrino-Anomalie folgende erlaubten
Bereiche f"ur die Massen und Mischungen \cite{bgg}:
\begin{eqnarray} \nonumber
\textrm{MSW-Ef\/fekt (kleiner Winkel):} \hspace{7.9cm} & & \\ \label{ntrcon1}
0.4 \cdot 10^{-5} \; \textrm{eV}^2
\; \lesssim \; \Delta m^2_{\scr \textrm{sun}} \;\, \lesssim \; 
1.2 \cdot 10^{-5} \; \textrm{eV}^2 \hspace{3cm} & & \\
0.4 \cdot 10^{-3} \; \textrm{eV}^2
\; \lesssim \; \Delta m^2_{\scr \textrm{atm}} \; \lesssim \; 
8.0 \cdot 10^{-3} \; \textrm{eV}^2 \hspace{3cm} & & \\
33 \; \lesssim \; 
\Delta m^2_{\scr \textrm{atm}}/\Delta m^2_{\scr \textrm{sun}}
\; \lesssim \; 2000 \hspace{5.15cm} & & \\ \label{ntrcon2}
|{\bf U}| \; = \; \left( \begin{array}{ccc}
\approx 1 & 0.03-0.05 & \ll 1 \\
0.02-0.05 & 0.71-0.87 & 0.49-0.71 \\
0.01-0.04 & 0.48-0.71 & 0.71-0.87
\end{array} \right) \hspace{1.6cm} & &
\end{eqnarray}
\begin{eqnarray} \nonumber
\textrm{MSW-Ef\/fekt (gro"ser Winkel):} \hspace{8cm} & & \\ \label{ntrcon3}
0.8 \cdot 10^{-5} \; \textrm{eV}^2
\; \lesssim \; \Delta m^2_{\scr \textrm{sun}} \;\, \lesssim \; 
3.0 \cdot 10^{-5} \; \textrm{eV}^2 \hspace{3cm} & & \\
0.4 \cdot 10^{-3} \; \textrm{eV}^2
\; \lesssim \; \Delta m^2_{\scr \textrm{atm}} \; \lesssim \; 
8.0 \cdot 10^{-3} \; \textrm{eV}^2 \hspace{3cm} & & \\
13 \; \lesssim \; 
\Delta m^2_{\scr \textrm{atm}}/\Delta m^2_{\scr \textrm{sun}}
\; \lesssim \; 1000 \hspace{5.15cm} & & \\  \label{ntrcon4}
|{\bf U}| \; = \; \left( \begin{array}{ccc}
0.87-0.94 & 0.35-0.49 & \ll 1 \\
0.25-0.43 & 0.61-0.82 & 0.49-0.71 \\
0.17-0.35 & 0.42-0.66 & 0.71-0.87
\end{array} \right) \hspace{1.6cm} & & 
\end{eqnarray}
\begin{eqnarray} \nonumber
\textrm{Vakuumoszillationen:} \hspace{9.3cm} & & \\
0.6 \cdot 10^{-10} \; \textrm{eV}^2
\; \lesssim \; \Delta m^2_{\scr \textrm{sun}} \;\, \lesssim \; 
1.1 \cdot 10^{-10} \; \textrm{eV}^2 \hspace{2.7cm} & & \\
0.4 \cdot 10^{-3} \; \textrm{eV}^2
\; \lesssim \; \Delta m^2_{\scr \textrm{atm}} \; \lesssim \; 
6.0 \cdot 10^{-3} \; \textrm{eV}^2 \hspace{3cm} & & \\
3 \cdot 10^{6} \; \lesssim \; 
\Delta m^2_{\scr \textrm{atm}}/\Delta m^2_{\scr \textrm{sun}}
\; \lesssim \; 10^{8} \hspace{4.75cm} & & \\
|{\bf U}| \; = \; \left( \begin{array}{ccc}
0.71-0.88 & 0.48-0.71 & \ll 1 \\
0.34-0.61 & 0.50-0.76 & 0.51-0.71 \\
0.24-0. 50 & 0.36-0.62 & 0.71-0.86
\end{array} \right) \hspace{1.6cm} & & 
\end{eqnarray}
Ein realistisches Massenmodell f"ur die Fermionen mu"s demnach
auf Neutrino-Eigen\-schaf\-ten f"uhren, welche mit einem der drei
obigen F"alle innerhalb der Grenzen konsistent sind.

%%% Local Variables: 
%%% mode: latex
%%% TeX-master: t
%%% End: 

\chapter{Das $SO(10)$-Massenmodell}
\section{Bestimmung der Symmetriebrechungsskalen} \label{essb}
Gegenstand dieser Arbeit ist die detaillierte Analyse eines Modellansatzes
f"ur die fermionischen Massenmatrizen im Rahmen einer nichtsupersymmetrischen 
$SO(10)$-GUT mit dem Brechungsschema 
\begin{equation}
SO(10) \; \stackrel{M_{\scr U}}{\longrightarrow} \; 
G_{\scr I} \; \stackrel{M_{\scr I}}{\longrightarrow} \; 
G_{\scr \textrm{SM}} \; \stackrel{M_{\scr Z}}{\longrightarrow} \; 
SU(3)_{\scr C} \otimes U(1)_{\scr \textrm{em}}
\end{equation}
Als intermedi"are Symmetriegruppe $G_{\scr I}$ sollen zun"achst sowohl 
$G_{\scr \textrm{PS}}$ als auch $G_{\scr \textrm{PS}} \otimes D$ in Frage
kommen. Deren Vorteil gegen"uber Symmetriegruppen, welche
$SU(3)_{\scr C} \otimes U(1)_{\scr B-L}$ enthalten, besteht in der
Vereinheitlichung von Quarks und Leptonen bereits bei der Skala 
$M_{\scr I}$. Das hat unter anderem zur Folge, da"s die 
$SO(10)$-Relationen (\ref{msrel1}-\ref{msrel2}) zwischen den 
verschiedenen Massenmatrizen 
nicht nur oberhalb von $M_{\scr U}$, sondern auch im Energiebereich
zwischen $M_{\scr I}$ und $M_{\scr U}$ gelten. Ferner resultiert aus
dem links-rechts-symmetrischen Fermionspektrum (\ref{lrsfs}) auf
nat"urliche Weise die Existenz rechtsh"andiger Neutrinos.

Gem"a"s Tabelle \ref{symbr} wird die Symmetriebrechung bei 
$M_{\scr U}$ durch eine {\bf 210} beziehungsweise {\bf 54} realisiert; 
die Brechung bei $M_{\scr I}$ erfolgt in beiden F"allen durch das
SM-Singulett einer $({\bf 10,1,3})_{126}$. Bez"uglich der Massen 
der Higgs-Teilchen wird
von der G"ultigkeit der "`Extended Survival Hypothesis"' \cite{aguil} 
ausgegangen, das hei"st nur diejenigen Komponenten von 
$SO(10)$-Higgs-Darstellungen besitzen Massen $m \ll M_{\scr U}$,
welche f"ur die Symmetriebrechungen bei $M_{\scr I}$ und $M_{\scr Z}$
sowie f"ur die Erzeugung der Fermionmassen erforderlich sind. 
Higgs-Teilchen mit Massen der Gr"o"senordnung $M_{\scr I}$ k"onnen 
demnach in folgenden Darstellungen liegen:
\begin{itemize}
\item in der $({\bf 10,1,3})_{126}$, da sie die Symmetriebrechung bei 
$M_{\scr I}$ realisiert.
\item in $({\bf 1,2,2})_{10/120}$ und $({\bf 15,2,2})_{120/126}$, da
  diese f"ur die Massenerzeugung der Fermionen in Frage kommen.
\item in Modellen mit $G_{\scr \textrm{PS}} \otimes D$-Symmetrie
  in der $({\bf \overline{10},3,1})_{126}$, welche aufgrund der 
  $D$-Parit"at das Gegenst"uck zur $({\bf 10,1,3})_{126}$ bildet 
  (siehe Abschnitt \ref{tiusb}).
\end{itemize}
Wie bereits in Abschnitt \ref{smdlc} erl"autert wurde, ist das 
SM-Higgs-Doublett eine Linearkombination der $SU(2)_{\scr L}$-Doubletts in 
den $({\bf 1,2,2})$- und $({\bf 15,2,2})$-Darstellungen. 

Um ein realistisches $SO(10)$-Massenmodell konstruieren zu
k"onnen, ist wegen
(\ref{massrel}) ein Higgs-Spektrum erforderlich, welches "uber den
einfachsten Fall einer komplexen {\bf 10} hinausgeht. W"ahrend im
Bereich unterhalb von $M_{\scr I}$ der Teilcheninhalt stets der des SM
mit einem Higgs-Doublett sein soll, ist die Anzahl der f"ur
die Erzeugung der Fermionmassen relevanten 
$G_{\scr \textrm{PS}}$-Higgs-Darstellungen 
$({\bf 1,2,2})_{10/120}$ und $({\bf 15,2,2})_{120/126}$ mit Massen 
$\sim M_{\scr I}$ zun"achst nicht weiter festgelegt und kann an die
konkreten Anforderungen des Modells angepa"st werden. Der Umfang des
Teilchenspektrums bei $M_{\scr I}$ hat allerdings direkte Auswirkungen auf die
Skalenabh"angigkeit der $G_{\scr \textrm{PS}} [\otimes
D]$-Eichkopplungen, wie man an den Renormierungsgruppengleichungen 
(\ref{rgeps1}-\ref{rgeps3}) erkennt. Bedeutsam wird dieser Einf\/lu"s,
wenn man die ph"anomenologischen Eigenschaften des Massenmodells wie zum
Beispiel die Zerfallsraten der Nukleonen berechnen will.
Dazu m"ussen n"amlich die Werte der Symmetriebrechungsskalen $M_{\scr U,I}$
und die der verschiedenen Eichkopplungen bei diesen Skalen bekannt sein,
und um diese quantitativ bestimmen zu k"onnen, ist eine numerische 
Integration der Renormierungsgruppengleichungen f"ur die Kopplungen 
sowohl des SM (\ref{smcrge1}-\ref{smcrge3}) als auch des 
$G_{\scr \textrm{PS}} [\otimes D]$-Modells (\ref{rgeps1}-\ref{rgeps3})
erforderlich.

Im folgenden werden zun"achst $M_{\scr U,I}$ und die Kopplungswerte
f"ur beide intermedi"aren Symmetriegruppen in Abh"angigkeit vom
Higgs-Spektrum numerisch bestimmt, wobei das verwendete Verfahren dem 
in \cite{deshp} entspricht. Dort ist der minimale Fall mit einer {\bf 10}
untersucht worden; in \cite{aguil} wurde der Einf\/lu"s des 
Higgs-Spektrums auf die $\beta$-Funktionen der $G_{\scr I}$-Kopplungen 
analysiert, ohne jedoch die Renormierungsgruppengleichungen numerisch 
zu l"osen.
\subsection{$SU(4)_{\scr C} \otimes SU(2)_{\scr L} 
\otimes SU(2)_{\scr R}$-Modell} \label{sbsps}
Die Parameter $\Delta_{R,L}$ in (\ref{rgeps1}-\ref{rgeps3}) haben hier
die festen Werte $\Delta_R=1$ und $\Delta_L=0$, die Werte von $N_{1}$
und $N_{15}$ k"onnen frei vorgegeben werden. Unter Ber"ucksichtigung
des Prinzips, da"s die Anzahl der Higgs-Teilchen bei gegebenem
Massenmodell grunds"atzlich so klein wie m"oglich gew"ahlt werden
soll, erscheint eine Untersuchung der F"alle $N_{1}=1,\dots,5$ und 
$N_{15}=1,2,3$ sinnvoll. Die Vorgehensweise sieht dann wie folgt aus:
\begin{itemize}
\item F"ur $M_{\scr I}$ wird ein Wert vorgegeben, der sinnvollerweise zwischen
  $M_{\scr Z}$ und $M_{\scr \textrm{Planck}} \approx 10^{19}$ GeV liegen kann. 
\item Ausgehend von den bekannten Werten der SM-Eichkopplungen bei $M_{\scr Z}$
(siehe Tabelle \ref{vwsmk}) werden die Renormierungsgruppengleichungen
(\ref{smcrge1}-\ref{smcrge3}) unter Vernachl"assigung der Yukawabeitr"age
von $M_{\scr Z}$ bis $M_{\scr I}$ integriert.
\item Aus den nun bekannten SM-Kopplungen bei $M_{\scr I}$ werden die
  Kopplungen des $G_{\scr \textrm{PS}}$-Modells bei dieser Skala
  berechnet. Die Anschlu"sbedingungen lauten
  \begin{eqnarray} \label{psmc1}
    \alpha^{-1}_{\scr 4C}(M_{\scr I}) & = & \alpha^{-1}_{\scr 3}(M_{\scr I})
    + \frac{1}{12\pi} \\ \label{psmc2}
    \alpha^{-1}_{\scr 2L}(M_{\scr I}) & = & \alpha^{-1}_{\scr 2}(M_{\scr I}) 
    \\ \label{psmc3}
    \alpha^{-1}_{\scr 2R}(M_{\scr I}) & = & 
    \frac{5}{3} \, \alpha^{-1}_{\scr 1}(M_{\scr I}) - 
    \frac{2}{3} \, \alpha^{-1}_{\scr 3}(M_{\scr I}) + \frac{1}{3\pi}
  \end{eqnarray}
  Die Faktoren 5/3 und 2/3 in (\ref{psmc3}) stammen aus der korrekten 
  Normierung der $G_{\scr \textrm{PS}}$-Generatoren im Rahmen der
  "ubergeordneten $SO(10)$ \cite{masie}, w"ahrend die Korrekturen
  $1/(12\pi)$ und $1/(3\pi)$ auf Schwellenef\/fekte zur"uckzuf"uhren
  sind \cite{hall}. Diese Ef\/fekte modif\/izieren die "`naiven"'
  Anschlu"sbedingungen in der zweiten Ordnung der St"orungsrechnung.
  Vernachl"assigt man logarithmische Korrekturen $\sim \ln(m_i/M_{\scr I,U})$, 
  wobei $m_i$ f"ur die in der Regel nicht exakt bekannten Massen
  aller Teilchen steht, welche ihre Massen durch die Symmetriebrechung 
  bei $M_{\scr I,U}$ erhalten, k"onnen die Anschlu"sbedingungen allgemein als
  \begin{equation}
    \alpha^{-1}_{\scr j}(M^{}_{\scr I/U}) - \dfrac{1}{12\pi} \,
    S_2({\bf \mathcal{G}}_{j}) \; = \; \alpha^{-1}_{\scr k}(M^{}_{\scr I/U}) 
    - \dfrac{1}{12\pi} \, S_2({\bf \mathcal{G}}_{k})
  \end{equation}
  geschrieben werden \cite{deshp}. $S_2({\bf \mathcal{G}}_{j})$ 
  bezeichnet hier den Dynkin-Index der adjungierten Darstellung der 
  Eichgruppe $G_j$ (siehe Anhang \ref{gtapp}). In Abschnitt
  \ref{thresh} werden die Ursachen von Schwellenkorrekturen und ihr 
  Einf\/lu"s auf die Bestimmung der Symmetriebrechungsskalen
  ausf"uhrlich diskutiert.
\item  Ausgehend von den $G_{\scr \textrm{PS}}$-Kopplungen bei
  $M_{\scr I}$ integriert man die Renormierungsgruppengleichungen 
  (\ref{rgeps1}-\ref{rgeps3}) von $M_{\scr I}$ zu h"oheren Energien
  und "uberpr"uft, ob eine Skala $\mu \equiv M_{\scr U} > M_{\scr I}$
  existiert, bei welcher sich die drei Kopplungen unter Ber"ucksichtigung
  der ebenfalls durch Schwellenef\/fekte modif\/izierten
  Anschlu"sbedingungen 
  \begin{eqnarray}
    \alpha^{-1}_{\scr U}(M_{\scr U}) & = & \alpha^{-1}_{\scr 4C}(M_{\scr U})
    + \frac{1}{3\pi} \\
    & = & \alpha^{-1}_{\scr 2L}(M_{\scr U}) + \frac{1}{2\pi} \\
    & = & \alpha^{-1}_{\scr 2R}(M_{\scr U}) + \frac{1}{2\pi}
  \end{eqnarray}
  in einem Punkt tref\/fen. Ist das der Fall, so sind die Brechungsskalen
  $M_{\scr I}$ und $M_{\scr U}$ sowie die Werte der verschiedenen 
  Kopplungskonstanten bei $M_{\scr I,U}$ bekannt.
\end{itemize}
Auf diese Weise ist f"ur jedes der in Frage kommenden Higgs-Spektren
der gesamte erlaubte Parameterbereich f"ur $M_{\scr I}$ zu
untersuchen; dabei besteht durchaus die M"oglichkeit, da"s keine 
Vereinheitlichung erreicht werden kann. 

Die Ergebnisse sind in Anhang \ref{sukfps}
zusammengefa"st. Man erkennt zun"achst, da"s f"ur $N_{15}=3$ und 
$N_{1}=1,\dots,5$ sowie $N_{15}=2$ und $N_{1}=1,2,3$ keine Vereinheitlichung
stattf\/indet. In den "ubrigen untersuchten F"allen sind folgende Eigenschaften
erkennbar:
\begin{itemize}
\item Bei festem $N_{15}$ sind $M_{\scr U}$ und $\alpha_{\scr U}$ umso 
kleiner und $M_{\scr I}$ umso gr"o"ser, je gr"o"ser $N_{1}$ ist.
\item Bei festem $N_{1}$ sind $M_{\scr U}$ und $\alpha_{\scr U}$ umso 
gr"o"ser und $M_{\scr I}$ umso kleiner, je gr"o"ser $N_{15}$ ist.
\end{itemize}
Eine qualitative Absch"atzung der Protonlebensdauer gem"a"s (\ref{taupap})
f"uhrt unter Ber"ucksichtigung der experimentellen Grenzen
mit $\alpha_{\scr U} \approx 1/35 \dots 1/20$ zu der Einschr"ankung
$M_{\scr U} \gtrsim 3 \cdot 10^{15}$ GeV f"ur die Vereinheitlichungsskala.
Deswegen sind die Modelle mit $(N_1,N_{15})=(4,1)$ und $(5,1)$ wegen
ihrer relativ kleinen Werte von $M_{\scr U}$ im Vergleich zu den
"ubrigen vom ph"anomenologischen Standpunkt her eher ungeeignet.
\subsection{$SU(4)_{\scr C} \otimes SU(2)_{\scr L} \otimes SU(2)_{\scr R}
\otimes D$-Modell} \label{sbspsd}
Die Parameter $\Delta_{R,L}$ haben in diesem Fall wegen des
zwangsl"auf\/ig links-rechts-symmetri\-schen Teilcheninhalts beide den Wert 1.
Im Vergleich zum $G_{\scr \textrm{PS}}$-Modell ergibt sich hier ein
wesentlicher Unterschied in der Vorgehensweise, der auf den
Anschlu"sbedingungen bei $M_{\scr I}$ beruht. Zus"atzlich zu 
(\ref{psmc1}-\ref{psmc3}) gilt in Modellen mit $D$-Parit"at n"amlich 
$\alpha^{}_{\scr 2L}(\mu) = \alpha^{}_{\scr 2R}(\mu)$ f"ur alle $\mu$ mit
$M^{}_{\scr I} \le \mu \le M^{}_{\scr U}$. Das f"uhrt auf
\begin{eqnarray} \label{psmcd1}
\alpha^{-1}_{\scr 4C}(M^{}_{\scr I})
& = & \alpha^{-1}_{\scr 3}(M^{}_{\scr I}) + \dfrac{1}{12\pi} \\ \label{psmcd2}
\alpha^{-1}_{\scr 2L}(M^{}_{\scr I})
& = & \alpha^{-1}_{\scr 2}(M^{}_{\scr I}) \quad = \quad
\alpha^{-1}_{\scr 2R}(M^{}_{\scr I}) \\ \label{psmcd3}
0 & = & \alpha^{-1}_{\scr 2}(M^{}_{\scr I}) - 
\dfrac{5}{3} \alpha^{-1}_{\scr 1}(M^{}_{\scr I}) 
+ \dfrac{2}{3} \alpha^{-1}_{\scr 3}(M^{}_{\scr I}) - \dfrac{1}{3\pi}
\end{eqnarray}
Die letzte Gleichung liefert eine Einschr"ankung f"ur die SM-Kopplungen,
welche bei genau einer Skala $M^{}_{\scr I}$ erf"ullt ist. In \cite{noh}
ist gezeigt worden, da"s dieser Wert f"ur $M^{}_{\scr I}$ von den
physikalischen Eigenschaften des Modells bei Skalen $\mu > M^{}_{\scr I}$ 
v"ollig unabh"angig ist; insbesondere hat das Higgs-Spektrum keinen
Einf\/lu"s auf $M^{}_{\scr I}$ und die Kopplungskonstanten bei 
$\mu = M^{}_{\scr I}$. In Tabelle \ref{emdp} sind die Werte dieser
Gr"o"sen angegeben. Es f"allt auf, da"s der Wert von $M^{}_{\scr I}$
um etwa zwei bis drei Gr"o"senordnungen "uber den entsprechenden
Werten im $G_{\scr \textrm{PS}}$-Modell liegt.

Das Problem reduziert sich also auf die Bestimmung von $M^{}_{\scr U}$
und $\alpha_{\scr U}$ in Abh"angigkeit vom Teilcheninhalt. Von den
zwei bekannten Kopplungen bei $M^{}_{\scr I}$ ausgehend werden 
(\ref{rgeps1}-\ref{rgeps2}) bis zu der Skala $\mu \equiv M^{}_{\scr U}$ 
integriert, bei welcher $\alpha_{\scr 4C}$ und $\alpha_{\scr 2L}$ die 
Bedingungen
\begin{eqnarray}
\alpha^{-1}_{\scr U}(M_{\scr U}) & = & \alpha^{-1}_{\scr 4C}(M_{\scr U})
+ \frac{1}{3\pi} \\
& = & \alpha^{-1}_{\scr 2L}(M_{\scr U}) + \frac{1}{2\pi}
\end{eqnarray}
erf"ullen. Die Resultate sind in Tabelle \ref{skpsd} zusammengefa"st,
es sind folgende Tendenzen in den L"osungen erkennbar:
\begin{itemize}
\item Bei festem $N_{15}$ sind $M_{\scr U}$ und $\alpha_{\scr U}$ umso 
kleiner, je gr"o"ser $N_{1}$ ist.
\item Bei festem $N_{1}$ sind $M_{\scr U}$ und $\alpha_{\scr U}$ umso 
gr"o"ser, je gr"o"ser $N_{15}$ ist.
\end{itemize}
F"ur alle untersuchten F"alle f\/indet eine Vereinheitlichung der
Kopplungen statt, allerdings bei vergleichsweise kleinen 
$M^{}_{\scr U}$-Werten von $ \approx (1 - 2) \cdot 10^{15}$ GeV. Das
f"uhrt zu Vorhersagen f"ur die Lebensdauern der Nukleonen,
die den experimentellen Grenzen widersprechen, weshalb 
$G_{\scr \textrm{PS}} \otimes D$ als intermedi"are Symmetriegruppe 
ausscheidet. Im folgenden ist $G_{\scr I}=G_{\scr \textrm{PS}}$; der
Higgs-Inhalt des Modells wird sp"ater aus den Anforderungen des
Ansatzes f"ur die Massenmatrizen unter Ber"ucksichtigung der Resultate
in Abschnitt \ref{sbsps} bestimmt.
\section{Schwellenkorrekturen} \label{thresh}
In diesem Abschnitt wird auf den Ursprung der Schwellenkorrekturen und
deren Ef\/fekte auf die Bestimmung von Symmetriebrechungsskalen eingegangen.

Wird eine auf der Gruppe $G_{\scr J}$ beruhende
Eichsymmetrie bei einer Skala $M$ spontan in die zur Untergruppe 
$G_{\scr j}$ geh"orige Symmetrie gebrochen, ist damit formal der "Ubergang
von der vollen Eichtheorie zu einer bei Energien $\mu < M$ g"ultigen
ef\/fektiven Theorie verbunden \cite{wnbrg}. Diese ef\/fektive
Niederenergie-N"aherung erh"alt man durch Ausintegration der schweren
Freiheitsgrade aus dem Wirkungsfunktional. Wenn $\Phi$ die Felder
bezeichnet, die zu Teilchen mit Massen der Gr"o"senordnung $M$
geh"oren, und $\phi$ f"ur die auch unterhalb von $M$ masselosen
Felder steht, erh"alt man die Wirkung $S_{\scr j}[\phi]$ der
ef\/fektiven Theorie aus der urspr"unglichen Wirkung $S_{\scr J}[\phi,\Phi]$
durch Funktionalintegration "uber die schweren Felder:
\begin{equation} \label{funcint}
\exp(i\,S_{\scr j}[\phi]) \; = \; \int [d\Phi] \exp(i\,S_{\scr J}[\phi,\Phi])
\end{equation}
Hierbei ist die Invarianz von $S_{\scr j}[\phi]$ unter 
$G_{\scr j}$-Transformationen zu fordern. Um die Integration in 
(\ref{funcint}) ausf"uhren zu k"onnen, mu"s in der Wirkung 
$S_{\scr J}[\phi,\Phi]$ ein Eichf\/ixierungsterm 
$f_{\scr J}(\phi,\Phi)$ eingef"uhrt werden. Nach der 
Ausintegration der schweren Felder verbleibt von $f_{\scr J}(\phi,\Phi)$ 
jedoch ein Rest, der die $G_{\scr j}$-Eichinvarianz der neuen Wirkung 
$S_{\scr j}[\phi]$  verletzt. Dieses Problem kann durch eine Redef\/inition der
Kopplungskonstanten und der Eichfelder in $S_{\scr j}$ behoben
werden. Als Konsequenz "andert sich die naive Anschlu"sbedingung 
$\alpha^{}_{\scr j}(M) \; = \; \alpha^{}_{\scr J}(M)$ in
\begin{equation} \label{mcwte}
\alpha^{-1}_{\scr j}(M) \; = \; \alpha^{-1}_{\scr J}(M) 
- \lambda_{\scr j}(M) \; ,
\end{equation}
wobei $\lambda_{\scr j}(M)$ in erster Ordnung der St"orungsrechung durch
\begin{eqnarray} \nonumber
\lambda_{\scr j}(M) & = & \dfrac{1}{12\pi} \, \Big( \, 
S_2({\bf \mathcal{R}}^{\scr (G)}_{j})
- 21 \, S_2({\bf \mathcal{R}}^{\scr (G)}_{j}) \cdot \ln(m_{\scr G}/M) \\
& & + \, \Lambda \sum_{\scr S} S_2({\bf \mathcal{R}}^{\scr (S)}_{j}) 
\cdot \ln(m_{\scr S}/M)
+ 8 \, \sum_{\scr F} S_2({\bf \mathcal{R}}^{\scr (F)}_{j}) 
\cdot \ln(m_{\scr F}/M) \, \Big)
\end{eqnarray}
gegeben ist \cite{hall}. Hierbei ist ${\bf \mathcal{R}}^{\scr (G)}_{j}$ 
die Darstellung von $G_{\scr j}$, nach welcher die im Rahmen der
Symmetriebrechung massiv gewordenen Eichbosonen von $G_{\scr J}$ 
transformieren. Weiterhin steht
${\bf \mathcal{R}}^{\scr (F,S)}_{j}$ f"ur die 
${\bf \mathcal{G}}_{j}$-Darstellun\-gen, 
in denen die Fermionen beziehungsweise Higgs-Teilchen mit Massen 
$m_{\scr F,S} \sim M$ liegen; $S_2$ ist der Dynkin-Index dieser 
Darstellungen (siehe Tabelle \ref{grp}). Der Operator $\Lambda$
projeziert die Goldstone-Bosonen aus dem Higgs-Spektrum heraus. 
$\lambda_{\scr j}(M)$ hat seinen physikalischen Ursprung in den 
Strahlungskorrekturen zum Propagator der
Eichbosonen der ef\/fektiven Theorie durch die schweren Teilchen.

\noindent In der Analyse des letzten Abschnitts wurde bei den
Anschlu"sbedingungen lediglich der erste, nicht von den Massen abh"angige,
Term von $\lambda_{\scr j}$ in abgewandelter Form benutzt. 
Das entspricht der "ublicherweise
gemachten Annahme, da"s die durch Symmetriebrechung bei einer Skala $M$
erzeugten Teilchenmassen sich nur minimal von $M$ unterscheiden und
die logarithmischen Terme vernachl"assigt werden k"onnen.

Hier soll nun der Einf\/lu"s der zu den Skalaren geh"origen
logarithmischen Terme auf die Bestimmung
der Werte der Symmetriebrechungsskalen $M_{\scr I}$ und $M_{\scr U}$ 
untersucht werden, wenn man f"ur die Argumente der Logarithmen einen
Wertebereich von 0.1 bis 10 zul"a"st. Das ist angesichts der
Komplexit"at des Higgs-Potentials in Modellen mit nichtminimalem
Higgs-Inhalt und der damit verbundenen gro"sen Anzahl unbekannter
Koef\/f\/izienten der Gr"o"senordnung 1 durchaus plausibel. 
Das Verfahren lehnt sich an \cite{lmpr} an, wo dies
f"ur $SO(10)$-Modelle mit minimalem Higgs-Inhalt durchgef"uhrt
wurde. Beitr"age durch Fermionen gibt es bei $M_{\scr U,I}$ nicht, da
die rechtsh"andigen Neutrinos als einzige schwere Teilchen
SM-Singuletts sind. Die Eichbosonbeitr"age werden
vernachl"assigt, da der Higgs-Inhalt der Theorie deutlich
umfangreicher als der Eichboson-Inhalt ist.

Die f"ur die Berechnung der $\lambda_{\scr j}$-Koef\/f\/izienten
erforderlichen Dynkin-Indizes der einzelnen Higgs-Darstellungen sind
in Tabelle \ref{grp} angegeben. Mit ihrer Hilfe lassen sich die 
$\lambda_{\scr j}^{\scr I,U}$ f"ur das hier diskutierte Modell
bestimmen; Anhang \ref{threshnr} enth"alt alle relevanten
Resultate. Der Einf\/lu"s der Schwellenkorrekturen auf die Vorhersagen
von $M_{\scr U,I}$ wird durch folgende Beziehungen beschrieben \cite{noh}:
\begin{eqnarray}
\Delta \ln(M_{\scr U}/M_{\scr Z}) & = & 
\dfrac{ K_\lambda A_{\scr I} - J_\lambda B_{\scr I} }
{ A_{\scr U} B_{\scr I} - A_{\scr I} B_{\scr U} } \\
\Delta \ln(M_{\scr I}/M_{\scr Z}) & = & \dfrac
{ J_\lambda B_{\scr U} - K_\lambda A_{\scr U} }
{ A_{\scr U} B_{\scr I} - A_{\scr I} B_{\scr U} }
\end{eqnarray}
Die Gr"o"sen $A_{\scr I,U}$ und $B_{\scr I,U}$ sind Linearkombinationen der 
f"uhrenden Koef\/f\/izienten der Eichkopplungs-$\beta$-Funktionen
\begin{eqnarray}
A_{\scr U} & = & 2\,\beta_{\scr 4C}-\beta_{\scr 2L}-\beta_{\scr 2R} \\
B_{\scr U} & = & -\dfrac{2}{3}\,\beta_{\scr 4C}+\dfrac{5}{3}\,\beta_{\scr 2L}
-\beta_{\scr 2R} \\
A_{\scr I} & = & \dfrac{8}{3}\,\beta_{\scr 3}-\beta_{\scr 2}
-\dfrac{5}{3}\,\beta_{\scr 1}-A_{\scr U} \\
B_{\scr I} & = & \dfrac{5}{3}\,(\beta_{\scr 2}-\beta_{\scr 1})-B_{\scr U}
\end{eqnarray}
und k"onnen Anhang \ref{rge} entnommen werden, $J_\lambda$ und
$K_\lambda$ sind als
\begin{eqnarray}
J_\lambda & = & 2 \pi \, \big( \,
-2\,\lambda_{\scr 4C}^{\scr U}
+\lambda_{\scr 2L}^{\scr U}
+\lambda_{\scr 2R}^{\scr U} 
-\dfrac{8}{3}\,\lambda_{\scr 3C}^{\scr I}
+\lambda_{\scr 2L}^{\scr I}
+\dfrac{5}{3}\,\lambda_{\scr 1Y}^{\scr I}
\, \big) \\
K_\lambda & = & 2 \pi \, \big( \,
\dfrac{2}{3}\,\lambda_{\scr 4C}^{\scr U}
-\dfrac{5}{3}\,\lambda_{\scr 2L}^{\scr U}
+\lambda_{\scr 2R}^{\scr U}
-\dfrac{5}{3}\,\lambda_{\scr 2L}^{\scr I}
+\dfrac{5}{3}\,\lambda_{\scr 1Y}^{\scr I}
\, \big)
\end{eqnarray}
def\/iniert. Dabei ist $\lambda_{\scr j}^{\scr I,U}$ als Summe der
Beitr"age der einzelnen Higgs-Darstellungen zu verstehen.

\noindent Um daraus quantitative Resultate zu erhalten, sei an 
dieser Stelle das
Higgs-Spektrum des noch zu konstruierenden Massenmodells
vorweggenommen. Es wird sich abgesehen von $\Delta_R=1$ und
$\Delta_L=0$ zu $(N_1,N_{15})=(4,2)$ ergeben, woraus sich f"ur die durch
Schwellenef\/fekte verursachten Unsicherheiten in den Vorhersagen f"ur 
$M_{\scr U,I}$
\begin{eqnarray} \nonumber
\Delta \ln(M_{\scr U}/M_{\scr Z}) & = & 
- \dfrac{42}{215} \, \eta^{}_{\scr ({\bf 1,2,2})}
+ \dfrac{24}{215} \, \eta^{}_{\scr ({\bf 15,2,2})}
+ \dfrac{141}{430} \, \eta^{}_{\scr ({\bf 10,1,3})} \\
& & + \dfrac{12}{215} \, \eta^{}_{\scr {\bf 210}}
- \dfrac{17}{43} \, \eta^{}_{\scr {\bf 126}}
+ \dfrac{12}{215} \, \eta^{}_{\scr {\bf 120}}
+ \dfrac{24}{215} \, \eta^{}_{\scr {\bf 10}} \\ \nonumber
\Delta \ln(M_{\scr I}/M_{\scr Z}) & = &
\dfrac{7}{43} \, \eta^{}_{\scr ({\bf 1,2,2})}
- \dfrac{4}{43} \, \eta^{}_{\scr ({\bf 15,2,2})}
- \dfrac{44}{43} \, \eta^{}_{\scr ({\bf 10,1,3})} \\
& & - \dfrac{2}{43} \, \eta^{}_{\scr {\bf 210}}
+ \dfrac {50}{43} \, \eta^{}_{\scr {\bf 126}}
- \dfrac {2}{43} \, \eta^{}_{\scr {\bf 120}}
- \dfrac{4}{43} \, \eta^{}_{\scr {\bf 10}}
\end{eqnarray}
ergibt. $\eta^{}_{\scr {\bf \mathcal{R}}}$ steht f"ur 
$\ln(m_{\scr {\bf \mathcal{R}}}/M_{\scr U})$ 
(${\bf \mathcal{R}}={\bf 10},{\bf 120},{\bf 126},{\bf 210}$)
beziehungsweise $\ln(m_{\scr {\bf \mathcal{R}}}/M_{\scr I})$ 
(${\bf \mathcal{R}}=({\bf 1,2,2}),({\bf 15,2,2}),({\bf 10,1,3})$), 
wobei vereinfachend davon ausgegangen wird, da"s alle Teilchen in
einer Darstellung ${\bf \mathcal{R}}$, welche ihre Massen durch den
Symmetriebrechungsschritt bei $M_{\scr U,I}$ erhalten, dieselbe Masse 
haben. L"a"st man nun, wie oben angek"undigt, f"ur die
$\eta^{}_{\scr {\bf \mathcal{R}}}$ Werte zwischen $\ln(0.1)$ und
$\ln(10)$ zu, erh"alt man folgende Maximalbetr"age f"ur 
$\Delta \ln(M_{\scr U,I}/M_{\scr Z})$:
\begin{eqnarray}
| \, \Delta \ln(M_{\scr U}/M_{\scr Z}) \, |_{\textrm{max}} & = & 
\dfrac{539}{430} \cdot \ln(10) \; = \; 2.89 \\
| \, \Delta \ln(M_{\scr I}/M_{\scr Z}) \, |_{\textrm{max}} & = & 
\dfrac{113}{43} \cdot \ln(10) \; = \; 6.05
\end{eqnarray}
Demnach k"onnen die Schwellenkorrekturen im vorliegenden Fall
aufgrund ihres Einf\/lusses auf die Anschlu"sbedingungen gem"a"s (\ref{mcwte}) 
den Wert von $M_{\scr U}$ um einen Faktor $\lesssim 10^{\pm 1.25}$ 
und den von $M_{\scr I}$ um einen Faktor $\lesssim 10^{\pm 2.63}$
modif\/izieren. 

Obwohl hier lediglich eine Absch"atzung des Maximalef\/fekts
durchgef"uhrt wurde, sollte bei den Angaben "uber die Werte von 
Symmetriebrechungsskalen der m"ogliche Einf\/lu"s der unbekannten 
Higgs-Massen nicht vergessen werden.
\section{Der Ansatz f"ur die Massenmatrizen}
Im Rahmen des SM ist die Form der fermionischen Massenmatrizen in
keiner Weise eingeschr"ankt. Es besteht aber allgemein die
"Uberzeugung, da"s eine fundamentalere Theorie als das SM existiert,
welche auch den Fermionsektor einschlie"slich der Struktur der
Massenmatrizen erkl"aren kann. In Ermangelung einer solchen Theorie
ist bis heute eine Vielzahl von ph"anomenologisch motivierten
Ans"atzen f"ur die Massenmatrizen vorgeschlagen worden. Sie zeichnen
sich durch bestimmte Symmetrieeigenschaften wie zum Beispiel 
Hermitezit"at und sogenannte Texturen, das hei"st Nullen als
Matrixeintr"age an bestimmten Stellen, aus. Der Grundgedanke besteht
darin, die Anzahl der freien Parameter in den Matrizen kleiner als die 
Zahl der zu reproduzierenden observablen Massen und Mischungen zu
machen, um Vorhersagen "uber Beziehungen zwischen diesen Gr"o"sen zu
erhalten und somit m"oglicherweise etwas "uber die Eigenschaften der
zugrundeliegenden Theorie zu erfahren. 

\noindent Der wohl popul"arste dieser Ans"atze geht auf Fritzsch zur"uck
\cite{frtz} und basiert auf zwei Annahmen: Die Quark-Massenmatrizen sind
hermitesch und haben die "`Nearest Neighbour Interaction"' (NNI)-Form.
Letzteres bedeutet, da"s nur das schwerste Fermion einer Art seine
Masse direkt "uber die Yukawakopplung erh"alt; die Massen der leichten 
Teilchen kommen durch Wechselwirkungen respektive Mischungen 
mit ihren n"achsten Nachbarn in
der Massenmatrix zustande. F"ur zwei und drei Generationen haben die
Matrizen demnach folgende Gestalt:
\begin{equation} \label{frtzsch}
{\bf M} \; = \; \left( \begin{array}{lll}
0 & A \\ A^* & B
\end{array} \right) \quad \textrm{beziehungsweise} \quad
{\bf M} \; = \; \left( \begin{array}{lll}
0 & A & 0 \\ A^* & 0 & B \\ 0 & B^* & C
\end{array} \right)
\end{equation}
Wendet man den Fritzsch-Ansatz auf die Quark-Massenmatrizen im Falle zweier
Generationen an, erh"alt man f"ur den Cabibbo-Winkel die mit $\phi
\sim \pi/2$ sehr gute Vorhersage
\begin{equation}
\theta_{\scr C} \; \approx \; | \, \sqrt{m_d/m_s}- e^{i\,\phi}
 \sqrt{m_u/m_c} \, |
\end{equation}
Ferner l"a"st sich mit $|A| \ll |B|$ auch die Massenhierarchie erkl"aren.

Problematisch ist der Ansatz jedoch f"ur drei Fermiongenerationen, da
es keine L"osungen gibt, welche gleichzeitig die gro"se $t$-Quarkmasse
und den kleinen Wert f"ur das CKM-Matrixelement $V_{cb}$ realisieren
k"onnen \cite{kang}. Daran "andert sich auch nichts, wenn man den 
Fritzsch-Ansatz in eine $SO(10)$-GUT ohne \cite{georgna} oder mit
Supersymmetrie \cite{basha1} einbettet.

Eine M"oglichkeit, dieses Problem zu umgehen, k"onnte darin bestehen,
hermitesche Massenmatrizen mit einer anderen Struktur als
(\ref{frtzsch}) zu verwenden. In \cite{rrr} ist eine systematische
Analyse aller Kombinationen von hermiteschen Quark-Massenmatrizen mit 
insgesamt $\ge 5$ unabh"angigen Null-Eintr"agen durchgef"uhrt worden,
aber auch dort sind keine wirklich "uberzeugenden L"osungen gefunden
worden. 

Es deutet also einiges darauf hin, da"s ein ph"anomenologisch
erfolgreiches Massenmodell nichthermitesche Matrizen enthalten mu"s. 
Hier hat sich das Interesse auf supersymmetrische GUTs
konzentriert, wobei zur Massenerzeugung h"auf\/ig von
nichtrenormierbaren Operatoren Gebrauch gemacht wird, welche aus einer 
noch fundamentaleren Theorie stammen sollen; \cite{berezh} bietet
einen "Uberblick "uber solche Ans"atze in SUSY-GUTs. 

Modelle mit nichthermiteschen Massenmatrizen, welche die
NNI-Form besitzen, haben im Rahmen supersymmetrischer $SU(5)$-
\cite{rossi} und $SO(10)$-Theorien \cite{basha2} vergleichsweise gute
Resultate auch im Neutrinosektor geliefert. Da"s ein solcher Ansatz
auch in einem $SO(10)$-Modell ohne Supersymmetrie erfolgreich sein
kann, wird in dieser Arbeit gezeigt werden. 
Ein fr"uherer Versuch in dieser Richtung war 
aufgrund der zu einfachen Struktur des verwendeten Higgs-Spektrums 
(eine ${\bf 10}$ und eine ${\bf 120}$) gescheitert \cite{georgna2}.

Ein weiteres Argument zugunsten nichthermitescher NNI-Matrizen liefert
\cite{branco}. Dort ist gezeigt worden, da"s ein NNI-Ansatz f"ur die
Quark-Massenmatrizen im SM keine physikalischen Konsequenzen beinhaltet,
sofern die Zahl der Fermiongenerationen nicht gr"o"ser als vier
ist. Im SM kann man beide Matrizen unabh"angig von deren Ausgangsgestalt
durch Transformationen auf NNI-Form bringen, ohne da"s
sich observable Gr"o"sen "andern. Der Grund daf"ur liegt in der
Tatsache, da"s von den vier Mischungsmatrizen im Quarksektor 
${\bf U}^{}_{\scr L,R}$ und ${\bf D}^{}_{\scr L,R}$ aus 
(\ref{biuntr1}) nur die Kombination 
${\bf V} \equiv {\bf U}^\dagger_{\scr L} {\bf D}^{}_{\scr L}$
physikalisch relevant ist; insbesondere die rechtsh"andigen Mischungen
sind experimentell nicht beobachtbar. Es handelt sich beim NNI-Ansatz im SM 
also streng genommen nicht um einen Ansatz, sondern um eine spezielle
Wahl der Basis im Raum der schwachen Eigenzust"ande. Wird der
NNI-Ansatz jedoch in Theorien jenseits des SM eingebettet, ergeben
sich daraus direkte Konsequenzen, die zumindest prinzipiell im
Experiment "uberpr"uft werden k"onnen. So beeinf\/lussen alle
Mischungsmatrizen in (\ref{biuntr1}) die Verzweigungsraten der 
Nukleonenzerf"alle, und Beziehungen wie (\ref{msrel1}-\ref{msrel2})
liefern Eigenschaften des Neutrinosektors aus den Massenmatrizen der
geladenen Fermionen.

Ausgangspunkt der nun folgenden "Uberlegungen wird die Annahme sein,
da"s die NNI-Form der Dirac-Massenmatrizen aus den
Symmetrieeigenschaften einer wirklich
grundlegenden Theorie der Elementarteilchen folgt. Diese nicht n"aher
bekannte Theorie soll als Niederenergie-N"aherung eine 
nichtsupersymmetrische $SO(10)$-GUT besitzen, welche wiederum "uber
eine intermedi"are $G_{\scr \textrm{PS}}$-Symmetrie in das SM
gebrochen wird. F"ur die fermionischen Massenmatrizen wird 
bei Skalen $\mu \gtrsim M_{\scr I}$ der Ansatz 
\begin{equation} \label{mmans}
{\bf M} \; = \; \left( \begin{array}{lll}
0 & A & 0 \\ B & 0 & C \\ 0 & D & E
\end{array} \right)
\end{equation}
gemacht, wobei der Einfachheit halber nur reelle Eintr"age betrachtet
werden. Das hat auf die Nukleonenzerf"alle und Neutrinoeigenschaften,
welche die zentralen Gegenst"ande der Untersuchung darstellen,
keine nennenswerten Auswirkungen; lediglich auf das Problem der
$CP$-Verletzung kann nicht eingegangen werden. Es sollte jedoch keine
Schwierigkeiten bereiten, mit Hilfe komplexer Yukawa-Kopplungen die
beobachtete $CP$-Verletzung zu reproduzieren.

Eine g"angige Methode, bestimmte Texturen in den Massenmatrizen zu
realisieren, besteht darin, eine globale $U(1)$-Familiensymmetrie
zus"atzlich zur Eichsymmetrie zu verwenden \cite{pqs}. Die drei
$SO(10)$-Fermiondarstellungen transformieren sich dann gem"a"s ${\bf 16}_j
\rightarrow \exp(i \alpha_j \theta){\bf 16}_j$, wobei $j$ der
Generationsindex ist und $\alpha_j$ die paarweise verschiedenen
Ladungen bez"uglich der $U(1)$ bezeichnet. Besitzt nun eine 
Higgs-Darstellung ${\bf \Phi}_k ={\bf 10},{\bf120}$ oder ${\bf 126}$ das 
$U(1)$-Trans\-for\-ma\-tionsverhalten 
${\bf \Phi}_k \rightarrow \exp(i \beta_k \theta) {\bf \Phi}_k$, so
sind nur solche Yukawa-Terme ${\bf 16}_i {\bf \bar \Phi}_k {\bf 16}_j$ 
$U(1)$-invariant, welche die Bedingung $\alpha_i + \alpha_j = \beta_k$ 
erf"ullen. Da die Eintr"age der Massenmatrizen gem"a"s
\begin{equation}
{\bf M} \; \sim \; \left( \begin{array}{lll}
\alpha_1+\alpha_1 & \alpha_1+\alpha_2 & \alpha_1+\alpha_3 \\
\alpha_1+\alpha_2 & \alpha_2+\alpha_2 & \alpha_2+\alpha_3 \\ 
\alpha_1+\alpha_3 & \alpha_2+\alpha_3 & \alpha_3+\alpha_3
\end{array} \right)
\end{equation}
bestimmte $U(1)$-Ladungen besitzen, kann man die NNI-Form realisieren,
indem nur Higgs-Darstellungen ${\bf \Phi}_k$ mit den Ladungen
$\beta = \alpha_1+\alpha_2,\alpha_2+\alpha_3$ und $\alpha_3+\alpha_3$
verwendet werden. Dabei ist es m"oglich, $\alpha_1+\alpha_2 = 2 \alpha_3$ 
zu w"ahlen, was die Anzahl der ben"otigten Darstellungen weiter
reduziert. Die $U(1)$-Familiensymmetrie wird bei $M_{\scr I}$ spontan
gebrochen, da dort die massenerzeugenden Higgs-Darstellungen $({\bf 1,2,2})$
und $({\bf 15,2,2})$ selbst massiv werden. Unterhalb von $M_{\scr I}$ werden
aufgrund von Renormierungsef\/fekten im allgemeinen auch die Eintr"age
ungleich Null sein, welche gem"a"s (\ref{mmans}) verschwinden.

F"ur die Yukawa-Kopplungsmatrizen kommen dann bei Skalen 
$\mu \gtrsim M_{\scr I}$ unter Verwendung der Bezeichnungsweisen in 
Tabelle \ref{hgs} folgende M"oglichkeiten in Frage:
\begin{eqnarray} \nonumber
{\bf Y}_{\bf \scr 10}^{\scr (1)} = \begin{pmatrix}
0 & x_1 & 0 \\ x_1 & 0 & 0 \\ 0 & 0 & \tilde x_1 \end{pmatrix} \; ; \quad
{\bf Y}_{\bf \scr 126}^{\scr (1)} = \begin{pmatrix}
0 & y_1 & 0 \\ y_1 & 0 & 0 \\ 0 & 0 & \tilde y_1 \end{pmatrix} \; ; \quad
{\bf Y}_{\bf \scr 120}^{\scr (1)} = \begin{pmatrix}
0 & z_1 & 0 \\ -z_1 & 0 & 0 \\ 0 & 0 & 0 \end{pmatrix} \; ; & & \\ \label{ycm}
{\bf Y}_{\bf \scr 10}^{\scr (2)} = \begin{pmatrix}
0 & 0 & 0 \\ 0 & 0 & x_2 \\ 0 & x_2 & 0 \end{pmatrix} \hspace{0.28cm} ; \quad
{\bf Y}_{\bf \scr 126}^{\scr (2)} = \begin{pmatrix}
0 & 0 & 0 \\ 0 & 0 & y_2 \\ 0 & y_2 & 0 \end{pmatrix} \hspace{0.22cm} ; \quad
{\bf Y}_{\bf \scr 120}^{\scr (2)} = \begin{pmatrix}
0 & 0 & 0 \\ 0 & 0 & z_2 \\ 0 & -z_2 & 0 \end{pmatrix} \hspace{0.25cm} & &
\end{eqnarray}
Das Higgs-Spektrum des Modells soll genau eine {\bf 126} enthalten,
welche die Symmetriebrechung bei $M_{\scr I}$ realisiert. Mehr als
eine {\bf 126} zu verwenden ist nicht verboten, aber unn"otig und w"urde
die Vorhersagekraft insbesondere im Neutrinosektor betr"achtlich reduzieren.
Damit der See-Saw-Mechanismus gem"a"s (\ref{seesmass}) anwendbar ist,
mu"s die Majorana-Massen\-ma\-trix der rechts\-h"an\-di\-gen Neutrinos 
(\ref{msrel2b}) invertierbar sein. Aus diesem Grunde ist die Kopplung 
der {\bf 126} durch ${\bf Y}_{\bf \scr 126}^{\scr (1)}$ aus
(\ref{ycm}) bestimmt.

Um die Asymmetrie gem"a"s (\ref{mmans}) erzeugen zu k"onnen, sind beide
{\bf 120}-Darstellungen mit den Kopplungsmatrizen 
${\bf Y}_{\bf \scr 120}^{\scr (1)}$ und 
${\bf Y}_{\bf \scr 120}^{\scr (2)}$ erforderlich. "Ubernimmt man auch
beide m"oglichen {\bf 10}-Darstellungen aus (\ref{ycm}) in den Higgs-Inhalt des
Modells und beteiligt s"amtliche $({\bf 1,2,2})_{10/120}$- und 
$({\bf 15,2,2})_{120/126}$-Komponenten an der Massenerzeugung der
Fermionen bei $M_{\scr Z}$, das hei"st gibt ihnen Massen $\sim M_{\scr I}$, 
so erh"alt man $(N_1,N_{15})=(4,3)$. F"ur diesen Fall f\/indet
gem"a"s der Analyse in Abschnitt \ref{sbsps} keine
Vereinheitlichung statt, so da"s eine der $({\bf 15,2,2})$ Massen der
Gr"o"senordnung $M_{\scr U}$ besitzen mu"s. Die Werte der Skalen und
Kopplungen f"ur den Fall $N_1=4$ und $N_{15}=2$ kann man Tabelle
\ref{skps} entnehmen; dabei erscheint $M_{\scr U} \approx 1.3 \cdot 10^{16}$ 
GeV bez"uglich der Konsequenz f"ur die Lebensdauer des Protons
sinnvoll zu sein. Um entscheiden zu k"onnen, welche der $({\bf 15,2,2})$ 
f"ur die Massenerzeugung die geringste Bedeutung hat,
werden zun"achst die Massenmatrizen unter Verwendung der vier 
$({\bf 1,2,2})_{10/120}$ und drei $({\bf 15,2,2})_{120/126}$ konstruiert.
Deren allgemeine Gestalt sieht analog zu (\ref{msrel1}-\ref{msrel2}) 
bei Energien $\mu \gtrsim M_{\scr I}$ dann folgenderma"sen aus:
\begin{eqnarray} \label{mmsrel1} \nonumber
{\bf M}^{}_d \hspace{0.55cm} & = & 
\upsilon_d^{\scr (1)} {\bf Y}_{\bf \scr 10}^{\scr (1)}
+ \upsilon_d^{\scr (2)} {\bf Y}_{\bf \scr 10}^{\scr (2)}
+ \omega_d^{\scr (1)} {\bf Y}_{\bf \scr 126}^{\scr (1)} \\ & & 
+ \; (\tilde \upsilon_d^{\scr (1)} + \tilde \omega_d^{\scr (1)}) 
{\bf Y}_{\bf \scr 120}^{\scr (1)}
+ (\tilde \upsilon_d^{\scr (2)} + \tilde \omega_d^{\scr (2)}) 
{\bf Y}_{\bf \scr 120}^{\scr (2)} \\ \nonumber
{\bf M}^{}_e \hspace{0.55cm} & = & 
\upsilon_d^{\scr (1)} {\bf Y}_{\bf \scr 10}^{\scr (1)}
+ \upsilon_d^{\scr (2)} {\bf Y}_{\bf \scr 10}^{\scr (2)}
- 3 \, \omega_d^{\scr (1)} {\bf Y}_{\bf \scr 126}^{\scr (1)} \\ & &
+ \; (\tilde \upsilon_d^{\scr (1)} - 3 \, \tilde \omega_d^{\scr (1)}) 
{\bf Y}_{\bf \scr 120}^{\scr (1)}
+ (\tilde \upsilon_d^{\scr (2)} - 3 \, \tilde \omega_d^{\scr (2)}) 
{\bf Y}_{\bf \scr 120}^{\scr (2)} \\ \nonumber
{\bf M}^{}_u \hspace{0.5cm} & = & 
\upsilon_u^{\scr (1)} {\bf Y}_{\bf \scr 10}^{\scr (1)}
+ \upsilon_u^{\scr (2)} {\bf Y}_{\bf \scr 10}^{\scr (2)}
+ \omega_u^{\scr (1)} {\bf Y}_{\bf \scr 126}^{\scr (1)} \\ & &
+ \ (\tilde \upsilon_u^{\scr (1)} + \tilde \omega_u^{\scr (1)}) 
{\bf Y}_{\bf \scr 120}^{\scr (1)}
+ (\tilde \upsilon_u^{\scr (2)} + \tilde \omega_u^{\scr (2)}) 
{\bf Y}_{\bf \scr 120}^{\scr (2)} %\\ 
\end{eqnarray}
\begin{eqnarray}\label{mmsrel2a} \nonumber
{\bf M}^{\scr (\textrm{Dir})}_\nu \hspace{0.1cm} & = & 
\upsilon_u^{\scr (1)} {\bf Y}_{\bf \scr 10}^{\scr (1)}
+ \upsilon_u^{\scr (2)} {\bf Y}_{\bf \scr 10}^{\scr (2)}
- 3 \, \omega_u^{\scr (1)} {\bf Y}_{\bf \scr 126}^{\scr (1)} \\ & &
+ \ (\tilde \upsilon_u^{\scr (1)} - 3 \, \tilde \omega_u^{\scr (1)}) 
{\bf Y}_{\bf \scr 120}^{\scr (1)}
+ (\tilde \upsilon_u^{\scr (2)} - 3 \, \tilde \omega_u^{\scr (2)}) 
{\bf Y}_{\bf \scr 120}^{\scr (2)} \\ 
\label{mmsrel2b}
{\bf M}^{\scr (\textrm{Maj})}_{\nu \scr{R}} & \sim & 
M_{\scr I} {\bf Y}_{\bf \scr 126}^{\scr (1)} 
\; \equiv \; M_{\scr R} {\bf Y}_{\bf \scr 126}^{\scr (1)} \\ \label{mmsrel2}
{\bf M}^{\scr (\textrm{Maj})}_{\nu \scr{L}} & \sim & 
\dfrac{\omega_u^{\scr (1)2}}{M_{\scr I}} {\bf Y}_{\bf \scr 126}^{\scr (1)}
\; \equiv \; M_{\scr L} {\bf Y}_{\bf \scr 126}^{\scr (1)}
\; ; \quad M_{\scr L} \ll \omega_u^{\scr (1)}
\end{eqnarray}
F"ur die einzelnen nichtverschwindenden Matrixelemente gilt unter
Verwendung von (\ref{ycm}) dann:
\begin{eqnarray} \label{dmm12}
({\bf M}_d)_{\scr 12} & = & 
x_1 \upsilon^{\scr (1)}_d + y_1 \omega^{\scr (1)}_d 
+ z_1 (\tilde \upsilon^{\scr (1)}_d + \tilde \omega^{\scr (1)}_d) 
\\ \label{dmm21}
({\bf M}_d)_{\scr 21} & = & 
x_1 \upsilon^{\scr (1)}_d + y_1 \omega^{\scr (1)}_d 
- z_1 (\tilde \upsilon^{\scr (1)}_d + \tilde \omega^{\scr (1)}_d) 
\\ \label{dmm23}
({\bf M}_d)_{\scr 23} & = & x_2 \upsilon^{\scr (2)}_d 
+ z_2 (\tilde \upsilon^{\scr (2)}_d + \tilde \omega^{\scr (2)}_d) 
\\ \label{dmm32}
({\bf M}_d)_{\scr 32} & = & x_2 \upsilon^{\scr (2)}_d 
- z_2 (\tilde \upsilon^{\scr (2)}_d + \tilde \omega^{\scr (2)}_d) \\ \nonumber
({\bf M}_d)_{\scr 33} & = & \tilde x_1 \upsilon^{\scr (1)}_d 
+ \tilde y_1 \omega^{\scr (1)}_d \\ \label{dmm33}
       & = & \Big( \dfrac{\tilde x_1}{x_1} \Big) x_1 \upsilon^{\scr (1)}_d
           + \Big( \dfrac{\tilde y_1}{y_1} \Big) y_1 \omega^{\scr (1)}_d
\\ \nonumber & & \\ \nonumber
({\bf M}_e)_{\scr 12} & = & 
x_1 \upsilon^{\scr (1)}_d - 3 \, y_1 \omega^{\scr (1)}_d 
+ z_1 (\tilde \upsilon^{\scr (1)}_d -3 \, \tilde \omega^{\scr (1)}_d) 
\\ \label{emm12}
& = & ({\bf M}_d)_{\scr 12} -4 \, y_1 \omega^{\scr (1)}_d 
-4 \, z_1 \tilde \omega^{\scr (1)}_d \\ \nonumber
({\bf M}_e)_{\scr 21} & = & 
x_1 \upsilon^{\scr (1)}_d - 3 \, y_1 \omega^{\scr (1)}_d 
- z_1 (\tilde \upsilon^{\scr (1)}_d -3 \, \tilde \omega^{\scr (1)}_d) 
\\ \label{emm21}
& = & ({\bf M}_d)_{\scr 21} -4 \, y_1 \omega^{\scr (1)}_d 
+4 \, z_1 \tilde \omega^{\scr (1)}_d \\ \nonumber
({\bf M}_e)_{\scr 23} & = & 
x_2 \upsilon^{\scr (2)}_d + z_2 (\tilde \upsilon^{\scr (2)}_d 
-3 \, \tilde \omega^{\scr (2)}_d) \\ \label{emm23}
& = & ({\bf M}_d)_{\scr 23} -4 \, z_2\tilde \omega^{\scr (2)}_d 
\\ \nonumber
({\bf M}_e)_{\scr 32} & = & 
x_2 \upsilon^{\scr (2)}_d - z_2 (\tilde \upsilon^{\scr (2)}_d 
-3 \, \tilde \omega^{\scr (2)}_d) \\ \label{emm32}
& = & ({\bf M}_d)_{\scr 32} +4 \, z_2\tilde \omega^{\scr (2)}_d \\ \nonumber
({\bf M}_e)_{\scr 33} & = & \tilde x_1 \upsilon^{\scr (1)}_d 
- 3 \, \tilde y_1 \omega^{\scr (1)}_d \\ \label{emm33}
& = & ({\bf M}_d)_{\scr 33} 
-4 \, \Big( \dfrac{\tilde y_1}{y_1} \Big) y_1  \omega^{\scr (1)}_d
\\ \nonumber & & \\ \label{umm12}
({\bf M}_u)_{12} & = & x_1 \upsilon^{\scr (1)}_u + y_1 \omega^{\scr (1)}_u 
+ z_1 (\tilde \upsilon^{\scr (1)}_u + \tilde \omega^{\scr (1)}_u) 
\\ \label{umm21}
({\bf M}_u)_{21} & = & x_1 \upsilon^{\scr (1)}_u + y_1 \omega^{\scr (1)}_u 
- z_1 (\tilde \upsilon^{\scr (1)}_u + \tilde \omega^{\scr (1)}_u) 
\\ \label{umm23}
({\bf M}_u)_{23} & = & x_2 \upsilon^{\scr (2)}_u 
+ z_2 (\tilde \upsilon^{\scr (2)}_u + \tilde \omega^{\scr (2)}_u) 
\\ \label{umm32}
({\bf M}_u)_{32} & = & x_2 \upsilon^{\scr (2)}_u 
- z_2 (\tilde \upsilon^{\scr (2)}_u + \tilde \omega^{\scr (2)}_u) \\ \nonumber
({\bf M}_u)_{33} & = & \tilde x_1 \upsilon^{\scr (1)}_u 
+ \tilde y_1 \omega^{\scr (1)}_u \\ \label{umm33}
       & = & \Big( \dfrac{\tilde x_1}{x_1} \Big) x_1 \upsilon^{\scr (1)}_u
           + \Big( \dfrac{\tilde y_1}{y_1} \Big) y_1 \omega^{\scr (1)}_u
\end{eqnarray}
\begin{eqnarray} \nonumber
({\bf M}^{\scr (\textrm{Dir})}_\nu)_{12} & = & 
x_1 \upsilon^{\scr (1)}_u - 3 \, y_1 \omega^{\scr (1)}_u 
+ z_1 (\tilde \upsilon^{\scr (1)}_u -3 \, \tilde \omega^{\scr (1)}_u) 
\\ \label{nmm12}
& = & ({\bf M}_u)_{12} -4 \, y_1 \omega^{\scr (1)}_u 
-4 \, z_1 \tilde \omega^{\scr (1)}_u \\ \nonumber
({\bf M}^{\scr (\textrm{Dir})}_\nu)_{21} & = & 
x_1 \upsilon^{\scr (1)}_u - 3 \, y_1 \omega^{\scr (1)}_u 
- z_1 (\tilde \upsilon^{\scr (1)}_u -3 \, \tilde \omega^{\scr (1)}_u) 
\\ \label{nmm21}
& = & ({\bf M}_u)_{21} -4 \, y_1 \omega^{\scr (1)}_u 
+4 \, z_1 \tilde \omega^{\scr (1)}_u \\ \nonumber 
({\bf M}^{\scr (\textrm{Dir})}_\nu)_{23} & = & 
x_2 \upsilon^{\scr (2)}_u + z_2 (\tilde \upsilon^{\scr (2)}_u 
-3 \, \tilde \omega^{\scr (2)}_u) \\ \label{nmm23}
& = & ({\bf M}_u)_{23} -4 \, z_2\tilde \omega^{\scr (2)}_u \\ \nonumber
({\bf M}^{\scr (\textrm{Dir})}_\nu)_{32} & = & 
x_2 \upsilon^{\scr (2)}_u - z_2 (\tilde \upsilon^{\scr (2)}_u 
-3 \, \tilde \omega^{\scr (2)}_u) \\ \label{nmm32}
& = & ({\bf M}_u)_{32} +4 \, z_2\tilde \omega^{\scr (2)}_u \\ \nonumber
({\bf M}^{\scr (\textrm{Dir})}_\nu)_{33} & = & 
\tilde x_1 \upsilon^{\scr (1)}_u - 3 \, \tilde y_1 \omega^{\scr (1)}_u 
\\ \label{nmm33}
& = & ({\bf M}_u)_{33} -4 \, \Big( \dfrac{\tilde y_1}{y_1} \Big) 
y_1 \omega^{\scr (1)}_u
\end{eqnarray}
Eine der $({\bf 15,2,2})_{120/126}$ nicht an der Massenerzeugung
teilnehmen zu lassen ist gleichbedeutend mit der Tatsache, da"s entweder
$\omega^{\scr (1)}_{u,d}$, $\tilde \omega^{\scr (1)}_{u,d}$
oder $\tilde \omega^{\scr (2)}_{u,d}$ Null sein m"ussen. Wenn man bei
fester Parameterzahl zun"achst
die gr"o"stm"ogliche Freiheit bez"uglich der Gestalt der einzelnen 
Matrizen haben
m"ochte, n"amlich f"ur alle $j,k$ $({\bf M}_d)_{jk} \ne ({\bf M}_e)_{jk}$ und 
$({\bf M}_u)_{jk} \ne ({\bf M}^{\scr (\textrm{Dir})}_\nu)_{jk}$
w"ahlen zu k"onnen, f"uhrt das wegen der Gleichungen 
(\ref{emm12}-\ref{emm33}) und (\ref{nmm12}-\ref{nmm33}) 
zwangsl"auf\/ig zu der Wahl $\tilde \omega^{\scr (1)}_{u,d} \equiv 0$.

Damit ist der Ansatz f"ur die Massenmatrizen der Fermionen
vollst"andig festgelegt.
Alle Matrix\-eintr"age setzen sich aus 14 voneinander unabh"angigen freien 
$SO(10)$-Higgs-Parametern in Form von Produkten aus
Vakuumerwartungswerten und Yukawa-Kopplun\-gen zusammen. Dazu kommt noch
die Massenskala $M_{\scr R}$ der schweren Neutrinos,
welche "uber den See-Saw-Mechanismus (\ref{seesmass}) auch als
Vorfaktor $(1/M_{\scr R})$ in der See-Saw-Matrix erscheint und somit
die Betr"age der leichten Neutrinomassen festlegt. 

Dem stehen 18 zu reproduzierende observable Gr"o"sen gegen"uber,
n"amlich die 12 Quark- und Leptonmassen sowie jeweils drei Mischungswinkel in 
{\bf V} und {\bf U}. Me"sbar sind, zumindest bisher, nicht
die eigentlichen Massen der leichten Neutrinos, sondern "uber die
Oszillationsexperimente nur die beiden Parameter $\Delta m_{\scr 32}^2$ 
und $\Delta m_{\scr 21}^2$ (siehe Kapitel\,3). Dem entspricht in
gewisser Weise die Tatsache, da"s der Wert von $M_{\scr R}$ im
Rahmen von GUT-Massenmodellen mit See-Saw-Mechanismus 
aufgrund des zu komplizierten vollst"andigen Higgs-Potentials nicht
exakt berechenbar ist. Wie schon in Abschnitt \ref{smdlc} erl"autert 
wurde, l"a"st sich lediglich die qualitative Relation 
$M_{\scr R} \sim M_{\scr I}$ angeben, der genaue Wert von $M_{\scr R}$
mu"s an die experimentellen Resultate angepa"st werden.

Es k"onnen demnach drei Vorhersagen gemacht werden, die im
vorliegenden Fall alle im Neutrinosektor liegen. Allein aufgrund der
Parameterzahl ist eine m"ogliche Erkl"arung der Anomalien von Sonnen- und 
atmosph"arischen Neutrinos gem"a"s Abschnitt \ref{ntranl} als absolut
nichttrivial zu betrachten. Die Zahl der freien Higgs-Parameter mag im 
Vergleich zu "alteren Massenmodellen gro"s erscheinen; daf"ur werden 
in dem hier vorgeschlagenen Modell aber auch alle Massen der geladenen 
Fermionen sowie die CKM-Mischungswinkel bis auf Rundungsfehler korrekt 
wiedergegeben, w"ahrend bisher Resultate $\pm 10 \%$ vom
tats"achlichen Wert schon als erfolgreich galten. Ferner wird zur 
Massenerzeugung auf keine Mechanismen zur"uckgegrif\/fen,
die, wie zum Beispiel nichtrenormierbare Operatoren, au"serhalb der 
zugrundegelegten $SO(10)$-Theorie liegen.
\section{Numerische L"osung des Massenmodells} \label{numsolmm}
Ausgangspunkt f"ur den ersten Schritt in der numerischen Bestimmung der 
Massenmatrizen und Mischungen ist das Gleichungssystem
\begin{eqnarray} \nonumber
{\bf U}^{\dagger}_{\scr L} \, {\bf M}^{}_u \, {\bf U}^{}_{\scr R} & = &
{\bf M}^{\scr (D)}_u \; , \quad
{\bf D}^{\dagger}_{\scr L} \, {\bf M}^{}_d \, {\bf D}^{}_{\scr R} \ = \
{\bf M}^{\scr (D)}_d \; , \\ \label{soeq}
{\bf E}^{\dagger}_{\scr L} \, {\bf M}^{}_e \, {\bf E}^{}_{\scr R} \ & = &
{\bf M}^{\scr (D)}_e \; , \quad
{\bf U}^\dagger_{\scr L} {\bf D}^{}_{\scr L} \ = \ {\bf V} \; ,
\end{eqnarray}
welches, um die $SO(10)$-Beziehungen (\ref{mmsrel1}-\ref{mmsrel2})
ausnutzen zu k"onnen, bei $\mu=M_{\scr I}$ gel"ost werden mu"s. Auf den
rechten Seiten der Gleichungen stehen dann die physikalischen Fermionmassen
und die CKM-Matrix bei $M_{\scr I}$. Bekannt sind deren Werte
zun"achst nur bei $M_{\scr Z}$ (siehe (\ref{ckmval}) und Tabelle 
\ref{FermMass}). Um mit Hilfe der Renormierungsgruppengleichungen des
SM aus Anhang \ref{rgeA} trotz Unkenntnis der Massenmatrizen bei
$M_{\scr Z}$ die entsprechenden Werte bei $M_{\scr I}$ zu erhalten,
werden zwei vereinfachende Annahmen gemacht:
\begin{itemize}
\item Die Skalenabh"angigkeit der CKM-Matrix {\bf V} ist sehr
  klein. Das ist in \cite{olech} best"atigt worden.
\item Die Teilchenmassen bei $M_{\scr I}$ sind in guter N"aherung
  unabh"angig von der expliziten Form der Massenmatrizen bei $M_{\scr Z}$. Das
  ist zumindest dann korrekt, wenn $|({\bf M}^{}_u)_{33}|$ sehr viel
  gr"o"ser als alle anderen Eintr"age in ${\bf M}^{}_{u,d,e}$ ist.
\end{itemize}
Inwieweit obige Annahmen im vorliegenden Fall berechtigt sind, wird
sich sp"ater zeigen, wenn die Renormierungsgruppengleichungen der dann 
bekannten Yukawa-Matrizen von $M_{\scr I}$ nach $M_{\scr Z}$
integriert und diese bei $M_{\scr Z}$ diagonalisiert werden.

Nun kann man die Yukawa-Matrizen bei $M_{\scr Z}$ diagonal ansetzen
und die Gleichungen (\ref{smcrge1}-\ref{rgehgs}) nach 
$M_{\scr I} = 6.14 \cdot 10^{10}$ GeV
integrieren; die Resultate sind in Tabelle \ref{FermMass2} zusammengefa"st.
\begin{table}[h]
\begin{center}
\begin{tabular}{|l||c|c|c|}
\hline Gr"o"se: & $m_u(M_{\scr I})$ & $m_d(M_{\scr I})$ & $m_s(M_{\scr I})$ \\
\hline Wert: & $1.16$ MeV & $2.38$ MeV & $47.4$ MeV \\
\hline \hline
Gr"o"se: & $m_c(M_{\scr I})$ & $m_b(M_{\scr I})$ & $m_t(M_{\scr I})$ \\
\hline Wert: & $337.6$ MeV & $1360$ MeV & $101.2$ GeV \\
\hline \hline
Gr"o"se: & $m_e(M_{\scr I})$ & $m_\mu(M_{\scr I})$ & $m_\tau(M_{\scr I})$ \\
\hline Wert: & $513$ keV & $108.14$ MeV & $1838.3$ MeV \\
\hline
\end{tabular}
\end{center}
\caption{\label{FermMass2} Fermionmassen bei $M_{\scr I}$}
\end{table}

\noindent In die Berechnungen geht auch die experimentell nicht
bekannte Higgs-Selbstkopplung $\lambda$ ein, welche in niedrigster
Ordnung "uber
\begin{equation}
m^2_{\scr H} \; = \; \lambda \, \upsilon^2 \; ; \quad 
\upsilon \; = \; 246.2 \; \textrm{GeV}
\end{equation}
mit der Masse $m_{\scr H}$ des SM-Higgs-Bosons zusammenh"angt. Es mu"s
also eine Annahme f"ur $\lambda(M_{\scr Z})$ gemacht werden, wobei sich
zeigt, da"s $\lambda(M_{\scr Z})=0.5$ ein sinnvoller Wert zu sein
scheint. F"ur $\lambda(M_{\scr Z}) \lesssim 0.4$ f"uhrt die
Integration von (\ref{smcrge1}-\ref{rgehgs}) n"amlich auf 
$\lambda(M_{\scr I})<0$, w"ahrend f"ur $\lambda(M_{\scr Z}) \gtrsim 0.6$ 
der Wert von $\lambda(M_{\scr I})$ gr"o"ser als 1 wird, was die
Anwendbarkeit der St"orungsrechnung gef"ahrdet; dieses Ph"anomen ist
in \cite{fusko} erw"ahnt worden. $\lambda(M_{\scr Z})=0.5$
w"urde in f"uhrender Ordnung einer Higgs-Masse von 
$m_{\scr H} \approx 174$ GeV entsprechen, was mit den experimentellen
Grenzen in \cite{delphi} vereinbar ist.

Damit sind die Matrizen auf den rechten Seiten von (\ref{soeq})
bestimmt. Auf den linken Seiten stehen 31 unbekannte Gr"o"sen:
\begin{itemize}
\item In den drei Massenmatrizen ${\bf M}_f$ ($f=u,d,e$) die 14 
$SO(10)$-Higgs-Parameter aus (\ref{dmm12}-\ref{umm33}). Es sind
jedoch nur 13 voneinander unabh"angig, da zwei lediglich in der Kombination 
$z_2 (\tilde \upsilon^{\scr (2)}_u + \tilde \omega^{\scr (2)}_u)$ vorkommen;
um $z_2 \tilde \upsilon^{\scr (2)}_u$ und $z_2\tilde \omega^{\scr (2)}_u$ 
separat bestimmen zu k"onnen, mu"s die Dirac-Massenmatrix
der Neutrinos bekannt sein. Dies wird aber erst nach dem n"achsten Schritt der
Fall sein.
\item Gem"a"s der Parametrisierung (\ref{ckmpar}) jeweils drei Winkel 
$\theta^{\scr (f)}_{\scr L,Rij}$ ($ij=12,23,31$) in den
sechs Mischungsmatrizen, also insgesamt 18 Mischungswinkel.
\end{itemize}
Dem stehen 30 nichtlineare Bestimmungsgleichungen gegen"uber, jeweils 
neun aus den ersten drei Beziehungen von (\ref{soeq}) und drei 
unabh"angige Gleichungen aus der letzten.
Um das Problem numerisch behandeln zu k"onnen, mu"s also einer der
freien Parameter vorgegeben werden. Hierf"ur wird der Quotient 
$\tilde y_1/y_1$ ausgew"ahlt, welcher die Form der Majorana-Matrix 
(\ref{mmsrel2b}) bis auf einen Vorfaktor eindeutig festlegt:
\begin{equation}
{\bf M}^{\scr (\textrm{Maj})}_{\nu \scr{R}} \; = \; M_{\scr R} \,
\begin{pmatrix}
0 & y_1 & 0 \\ y_1 & 0 & 0 \\ 0 & 0 & \tilde y_1 \end{pmatrix}
\; = \; y_1 \, M_{\scr R} \, \begin{pmatrix}
0 & 1 & 0 \\ 1 & 0 & 0 \\ 0 & 0 & \tilde y_1/y_1 \end{pmatrix}
\end{equation}
Als m"oglicher Wertebereich f"ur $\tilde y_1/y_1$ erscheint aufgrund
der Massenhierarchie der Fermionen 
$1 \lesssim |\tilde y_1/y_1| \lesssim 1000$ plausibel.
Bevor jedoch die L"osungen und ihre Eigenschaften behandelt werden,
sei an dieser Stelle auf die Bedeutung der $SO(10)$-spezif\/ischen
Beziehungen (\ref{mmsrel1}-\ref{mmsrel2a}) hingewiesen. 

Wie man an (\ref{dmm12}-\ref{emm33}) erkennt, sind bei
bekanntem ${\bf M}_d$ in ${\bf M}_e$ noch drei Freiheitsgrade
enthalten, n"amlich $y_1 \omega^{\scr (1)}_d$, $\tilde y_1/y_1$ und 
$z_2\tilde \omega^{\scr (2)}_d$. Da die Leptonmassen sich auch bei $M_{\scr I}$
deutlich von den Massen der $d$-Quarks unterscheiden, m"ussen diese 
Parameter derart festgelegt werden, da"s ${\bf M}_e$ die korrekten
Teilchenmassen liefert. Dann sind die $SO(10)$-Higgs-Parameter im
$d$-$e$-Sektor eindeutig bestimmt:
\begin{eqnarray} \label{parmrel1}
x_1 \upsilon^{\scr (1)}_d & = & 
\dfrac{1}{8} \, \big( \, 3 \, ({\bf M}_d)_{12}+3 \, ({\bf M}_d)_{21}
+({\bf M}_e)_{12}+({\bf M}_e)_{21} \, \big) \\
y_1 \omega^{\scr (1)}_d & = & \dfrac{1}{8} \, \big( \, ({\bf M}_d)_{12}
+({\bf M}_d)_{21}-({\bf M}_e)_{12}-({\bf M}_e)_{21} \, \big) \\
z_1 \tilde \upsilon^{\scr (1)}_d & = & \dfrac{1}{2} \, \big( \, 
({\bf M}_d)_{12}-({\bf M}_d)_{21} \, \big)
\; = \; \dfrac{1}{2} \, \big( \, ({\bf M}_e)_{12}-({\bf M}_e)_{21} \, \big) \\
\tilde x_1 \upsilon^{\scr (1)}_d & = & \dfrac{1}{4} \, 
\big( \, 3 \, ({\bf M}_d)_{33}+({\bf M}_e)_{33} \, \big) \\
\tilde y_1 \omega^{\scr (1)}_d & = & \dfrac{1}{4} \, 
\big( \, ({\bf M}_d)_{33}-({\bf M}_e)_{33} \, \big) \\
x_2 \upsilon^{\scr (2)}_d & = & 
\dfrac{1}{2} \big( \, ({\bf M}_d)_{23} + ({\bf M}_d)_{32}\, \big) 
\; = \; \dfrac{1}{2} \big( \, ({\bf M}_e)_{23} + ({\bf M}_e)_{32}\, \big) \\
z_2 \tilde \upsilon^{\scr (2)}_d & = & \dfrac{1}{8} \, \big( \, 
3 \, ({\bf M}_d)_{23}-3 \, ({\bf M}_d)_{32}+({\bf M}_e)_{23}
-({\bf M}_e)_{32} \, \big) \\
z_2 \tilde \omega^{\scr (2)}_d & = & \dfrac{1}{8} \, \big( \, 
({\bf M}_d)_{23}-({\bf M}_d)_{32}-({\bf M}_e)_{23}+({\bf M}_e)_{32} \, \big) \\
\dfrac{\tilde x_1}{x_1} & = & 2 \, \dfrac{3 \, ({\bf M}_d)_{33}
+({\bf M}_e)_{33}}
{3 \, ({\bf M}_d)_{12}+3 \, ({\bf M}_d)_{21}+({\bf M}_e)_{12}
+({\bf M}_e)_{21}} \\
\dfrac{\tilde y_1}{y_1} & = & 2 \, \dfrac{({\bf M}_d)_{33}-({\bf M}_e)_{33}}
{({\bf M}_d)_{12}+({\bf M}_d)_{21}-({\bf M}_e)_{12}-({\bf M}_e)_{21}} \\
\dfrac{\tilde \upsilon^{\scr (1)}_d}{\tilde \omega^{\scr (1)}_d} & = & 
\dfrac{3 \, ({\bf M}_d)_{12}-3 \, ({\bf M}_d)_{21}+({\bf M}_e)_{12}
-({\bf M}_e)_{21}}
{({\bf M}_d)_{12}-({\bf M}_d)_{21}-({\bf M}_e)_{12}+({\bf M}_e)_{21}} \\
\dfrac{\tilde \upsilon^{\scr (2)}_d}{\tilde \omega^{\scr (2)}_d} & = & 
\dfrac{3 \, ({\bf M}_d)_{23}-3 \, ({\bf M}_d)_{32}+({\bf M}_e)_{23}
-({\bf M}_e)_{32}}
{({\bf M}_d)_{23}-({\bf M}_d)_{32}-({\bf M}_e)_{23}+({\bf M}_e)_{32}}
\end{eqnarray}
Ist nun auch ${\bf M}_u$ bekannt, enth"alt die Dirac-Massenmatrix
der Neutrinos ${\bf M}^{\scr (\textrm{Dir})}_\nu$ gem"a"s
(\ref{umm12}-\ref{nmm33}) relativ zu ${\bf M}_u$ nur noch den
Freiheitsgrad $z_2\tilde \omega^{\scr (2)}_u$, da $\tilde y_1/y_1$ aus
dem $d$-$e$-Sektor bestimmt ist. Auch $y_1 \omega^{\scr (1)}_u$ ist
durch die Kenntnis von ${\bf M}_u$ bereits eindeutig festgelegt, wie
die zweite der folgenden Gleichungen zeigt:
\begin{eqnarray}
x_1 \upsilon^{\scr (1)}_u & = & 
\dfrac{ -\dfrac{\tilde y_1}{2y_1} \big(({\bf M}_u)_{12}+({\bf M}_u)_{21}\big)
+ ({\bf M}_u)_{33} }
{ \dfrac{\tilde x_1}{x_1} - \dfrac{\tilde y_1}{y_1} } \\
y_1 \omega^{\scr (1)}_u & = & 
\dfrac{ \dfrac{\tilde x_1}{2x_1} \big(({\bf M}_u)_{12}+({\bf M}_u)_{21}\big) 
- ({\bf M}_u)_{33} }
{ \dfrac{\tilde x_1}{x_1} - \dfrac{\tilde y_1}{y_1} } \\
z_1 \tilde \upsilon^{\scr (1)}_u & = & \dfrac{1}{2} \, \big( \, 
({\bf M}_u)_{12}-({\bf M}_u)_{21} \, \big) 
\; = \; \dfrac{1}{2} \, \big( \, ({\bf M}^{\scr (\textrm{Dir})}_\nu)_{12}
-({\bf M}^{\scr (\textrm{Dir})}_\nu)_{21} \, \big) \\
\tilde x_1 \upsilon^{\scr (1)}_u & = & \Big( \dfrac{\tilde x_1}{x_1} \Big)
\, x_1 \upsilon^{\scr (1)}_u \\
\tilde y_1 \omega^{\scr (1)}_u & = & \Big( \dfrac{\tilde y_1}{y_1} \Big)
\, y_1 \omega^{\scr (1)}_u \\
x_2 \upsilon^{\scr (2)}_u & = & 
\dfrac{1}{2} \big( \, ({\bf M}_u)_{23} + ({\bf M}_u)_{32}\, \big) 
\; = \; \dfrac{1}{2} \big( \, ({\bf M}^{\scr (\textrm{Dir})}_\nu)_{23} 
+ ({\bf M}^{\scr (\textrm{Dir})}_\nu)_{32}\, \big)
\end{eqnarray}
\begin{eqnarray}
z_2 \tilde \upsilon^{\scr (2)}_u & = & \dfrac{1}{8} \, \big( \, 
3 \, ({\bf M}_u)_{23}-3 \, ({\bf M}_u)_{32}
+({\bf M}^{\scr (\textrm{Dir})}_\nu)_{23}
-({\bf M}^{\scr (\textrm{Dir})}_\nu)_{32} \, \big) \\ \label{parmrel2}
z_2 \tilde \omega^{\scr (2)}_u & = & \dfrac{1}{8} \, \big( \, 
({\bf M}_u)_{23}-({\bf M}_u)_{32}-({\bf M}^{\scr (\textrm{Dir})}_\nu)_{23}
+({\bf M}^{\scr (\textrm{Dir})}_\nu)_{32} \, \big)
\end{eqnarray}
Anders ausgedr"uckt, wenn man einen Wert f"ur $\tilde y_1/y_1$ vorgibt
und L"osungen f"ur die Massenmatrizen der geladenen Fermionen
f\/indet, welche (\ref{soeq}) erf"ullen, so ist der Neutrinosektor der
Theorie bis auf die beiden Parameter $y_1 M_{\scr R}$ und 
$z_2\tilde \omega^{\scr (2)}_u$ eindeutig festgelegt. Das wiederum deutet
schon den zweiten Schritt der Analyse an, n"amlich die Suche nach Werten f"ur
$\tilde y_1/y_1$ und $z_2\tilde \omega^{\scr (2)}_u$, welche eine
Oszillationsl"osung f"ur die Anomalien der Sonnen- und atmosph"arischen 
Neutrinos gem"a"s \ref{ntranl} erm"oglichen.

Die Suche nach L"osungen von (\ref{soeq}) f"ur vorgegebene Werte von 
$\tilde y_1/y_1$ mit $1 \lesssim |\tilde y_1/y_1| \lesssim 1000$
liefert folgende Resultate:
\begin{itemize}
\item F"ur $|\tilde y_1/y_1| \lesssim 5$ und $|\tilde y_1/y_1| \gtrsim 500$
  k"onnen keine L"osungen gefunden werden.
\item F"ur $5 \lesssim |\tilde y_1/y_1| \lesssim 500$ existieren L"osungen.
  Es gibt bei gegebenem $\tilde y_1/y_1$ bis zu sechs 
  (bei $20 \lesssim \tilde y_1/y_1 \lesssim 25$) verschiedene S"atze von 
  Quark-Massenmatrizen und zu jedem dieser Paare jeweils zwei M"oglichkeiten
  f"ur die Massenmatrix der geladenen Leptonen, welche (\ref{soeq})
  erf"ullen. Letzteres resultiert aus der Tatsache, da"s es zu
  jeder Matrix ${\bf M}_d$ zwei Werte des Parameters 
  $z_2\tilde \omega^{\scr (2)}_d$ gibt, welche die
  korrekten Leptonmassen liefern; der Parameter $y_1 \omega^{\scr (1)}_d$ 
  dagegen ist in beiden F"allen gleich gro"s.
\end{itemize}
Es werden nur ganzzahlige Werte f"ur $\tilde y_1/y_1$ untersucht, da
die L"osungen, sofern sie existieren, stetig von $\tilde y_1/y_1$
abh"angen. Man erh"alt mit dem Ansatz (\ref{mmsrel1}-\ref{mmsrel2}) f"ur die
Massenmatrizen demnach eine Vielzahl von L"osungen f"ur die Matrizen
der geladenen Fermionen, welche die beobachteten Massen und
CKM-Mischungen liefern. Nun mu"s der Neutrinosektor in die Analyse mit
einbezogen werden. 

Es verbleiben zwei Parameter, welche aus den nun bekannten Massenmatrizen
der geladenen Fermionen gem"a"s den Beziehungen 
(\ref{parmrel1}-\ref{parmrel2}) nicht berechnet werden k"onnen, n"amlich 
$y_1 M_{\scr R}$ und $z_2\tilde \omega^{\scr (2)}_u$. Ersterer kann im
Rahmen des Massenmodells nicht exakt vorhergesagt werden, sondern mu"s
am Ende an die Resultate der Oszillationsexperimente angepa"st werden,
wobei $M_{\scr R} \sim M_{\scr I}$ gelten sollte. Bei vorgegebenem 
$\tilde y_1/y_1$ ist demnach $z_2\tilde \omega^{\scr (2)}_u$ der
einzige noch freie Parameter in der Dirac-Massenmatrix der Neutrinos;
die Majorana-Matrix ist bis auf den globalen Vorfaktor $y_1 M_{\scr R}$ 
festgelegt. Die Aufgabe besteht also darin, zu untersuchen, ob f"ur
diejenigen Werte von $\tilde y_1/y_1$, welche L"osungen von (\ref{soeq})
liefern, der Parameter $z_2\tilde \omega^{\scr (2)}_u$ so gew"ahlt
werden kann, da"s die Massen und Mischungen der leichten Neutrinos
ph"anomenologisch sinnvoll sind, das hei"st den experimentellen
Grenzen in (\ref{ntrcon1}-\ref{ntrcon4}) gen"ugen.
Da das Fermionspektrum des Modells
drei Neutrino-Arten enth"alt, k"onnen durch Oszillationsl"osungen zwei der
drei in Kapitel\,3 diskutierten Neutrinoprobleme erkl"art werden; das sollen 
aufgrund der im Vergleich zu den LSND-Resultaten deutlich st"arkeren Evidenz
die Anomalien der Sonnen- und atmosph"arischen Neutrinos
sein. Wegen der Argumentation in Abschnitt \ref{solntr} wird f"ur
das Sonnen-Neutrinodef\/izit nur der MSW-Ef\/fekt ber"ucksichtigt.

\begin{figure}[h]
\begin{center}
% GNUPLOT: LaTeX picture with Postscript
\begingroup%
  \makeatletter%
  \newcommand{\GNUPLOTspecial}{%
    \@sanitize\catcode`\%=14\relax\special}%
  \setlength{\unitlength}{0.1bp}%
\begin{picture}(3600,2160)(0,0)%
\special{psfile=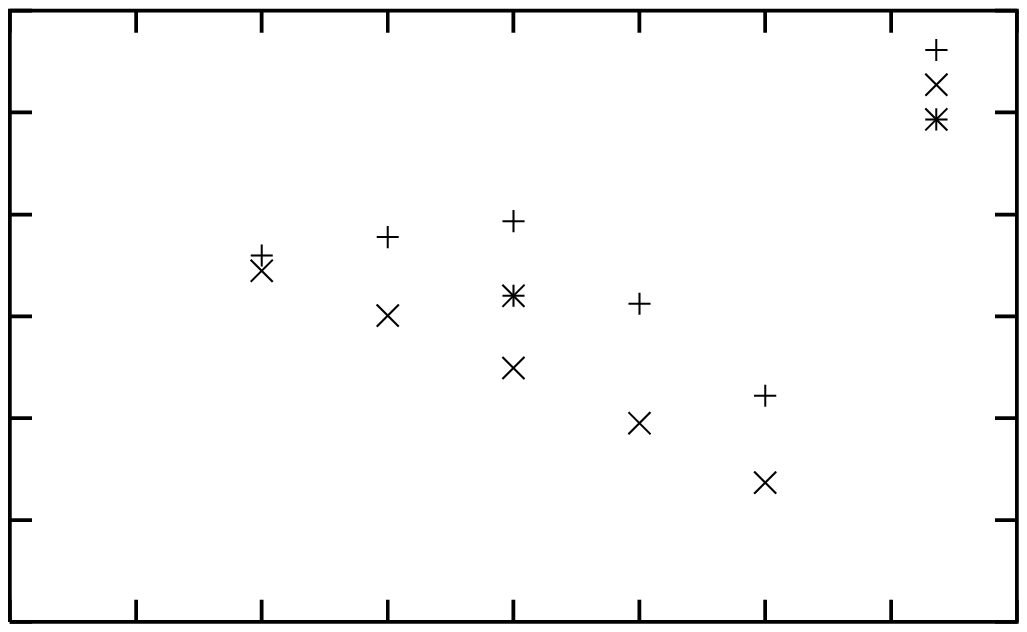 llx=0 lly=0 urx=720 ury=504 rwi=7200}
\put(3037,1747){\makebox(0,0)[r]{\small{Untersuchte L"osung}}}%
\put(3037,1847){\makebox(0,0)[r]{\small{Minimalwerte f"ur $4\,z_2\tilde\omega^{\scriptscriptstyle (2)}_u$}}}%
\put(3037,1947){\makebox(0,0)[r]{\small{Maximalwerte f"ur $4\,z_2\tilde\omega^{\scriptscriptstyle (2)}_u$}}}%
\put(2000,50){\makebox(0,0){$\tilde y_1/y_1$}}%
\put(100,1180){%
\special{ps: gsave currentpoint currentpoint translate
270 rotate neg exch neg exch translate}%
\makebox(0,0)[b]{\shortstack{$4\,z_2\tilde\omega^{\scriptscriptstyle (2)}_u$ (MeV)}}%
\special{ps: currentpoint grestore moveto}%
}%
\put(3450,200){\makebox(0,0){23}}%
\put(3088,200){\makebox(0,0){22}}%
\put(2725,200){\makebox(0,0){21}}%
\put(2363,200){\makebox(0,0){20}}%
\put(2000,200){\makebox(0,0){19}}%
\put(1638,200){\makebox(0,0){18}}%
\put(1275,200){\makebox(0,0){17}}%
\put(913,200){\makebox(0,0){16}}%
\put(550,200){\makebox(0,0){15}}%
\put(500,2060){\makebox(0,0)[r]{-11000}}%
\put(500,1767){\makebox(0,0)[r]{-12000}}%
\put(500,1473){\makebox(0,0)[r]{-13000}}%
\put(500,1180){\makebox(0,0)[r]{-14000}}%
\put(500,887){\makebox(0,0)[r]{-15000}}%
\put(500,593){\makebox(0,0)[r]{-16000}}%
\put(500,300){\makebox(0,0)[r]{-17000}}%
\end{picture}%
\endgroup
 
\caption{MSW-L"osung 1 (gro"se Mischung) im 
$\tilde y_1/y_1$-$z_2\tilde \omega^{\scr (2)}_u$-Parameterraum \label{mod1}} 
\end{center}
\end{figure}

\begin{figure}
\begin{center}
% GNUPLOT: LaTeX picture with Postscript
\begingroup%
  \makeatletter%
  \newcommand{\GNUPLOTspecial}{%
    \@sanitize\catcode`\%=14\relax\special}%
  \setlength{\unitlength}{0.1bp}%
\begin{picture}(3600,2160)(0,0)%
\special{psfile=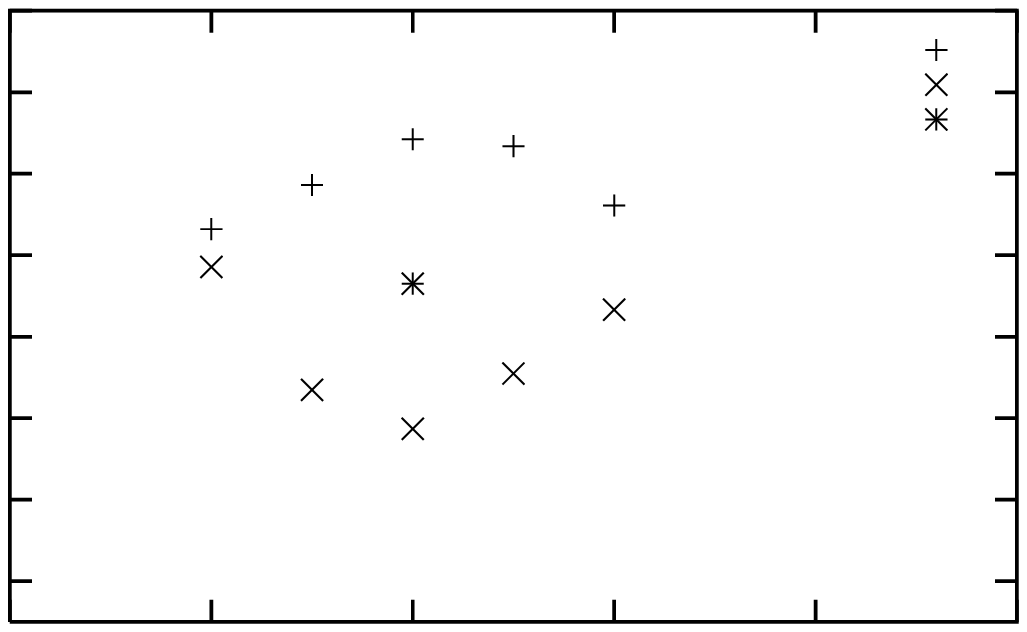 llx=0 lly=0 urx=720 ury=504 rwi=7200}
\put(3037,1747){\makebox(0,0)[r]{\small{Untersuchte L"osung}}}%
\put(3037,1847){\makebox(0,0)[r]{\small{Minimalwerte f"ur $4\,z_2\tilde\omega^{\scriptscriptstyle (2)}_u$}}}%
\put(3037,1947){\makebox(0,0)[r]{\small{Maximalwerte f"ur $4\,z_2\tilde\omega^{\scriptscriptstyle (2)}_u$}}}%
\put(2000,50){\makebox(0,0){$\tilde y_1/y_1$}}%
\put(100,1180){%
\special{ps: gsave currentpoint currentpoint translate
270 rotate neg exch neg exch translate}%
\makebox(0,0)[b]{\shortstack{$4\,z_2\tilde\omega^{\scriptscriptstyle (2)}_u$ (MeV)}}%
\special{ps: currentpoint grestore moveto}%
}%
\put(3450,200){\makebox(0,0){30}}%
\put(2870,200){\makebox(0,0){28}}%
\put(2290,200){\makebox(0,0){26}}%
\put(1710,200){\makebox(0,0){24}}%
\put(1130,200){\makebox(0,0){22}}%
\put(550,200){\makebox(0,0){20}}%
\put(500,2060){\makebox(0,0)[r]{-30000}}%
\put(500,1825){\makebox(0,0)[r]{-32000}}%
\put(500,1591){\makebox(0,0)[r]{-34000}}%
\put(500,1356){\makebox(0,0)[r]{-36000}}%
\put(500,1121){\makebox(0,0)[r]{-38000}}%
\put(500,887){\makebox(0,0)[r]{-40000}}%
\put(500,652){\makebox(0,0)[r]{-42000}}%
\put(500,417){\makebox(0,0)[r]{-44000}}%
\end{picture}%
\endgroup
 
\caption{MSW-L"osung 2a (kleine Mischung) im 
$\tilde y_1/y_1$-$z_2\tilde \omega^{\scr (2)}_u$-Parameterraum \label{mod2a}}
\end{center}
\end{figure}

\begin{figure}
\begin{center}
% GNUPLOT: LaTeX picture with Postscript
\begingroup%
  \makeatletter%
  \newcommand{\GNUPLOTspecial}{%
    \@sanitize\catcode`\%=14\relax\special}%
  \setlength{\unitlength}{0.1bp}%
\begin{picture}(3600,2160)(0,0)%
\special{psfile=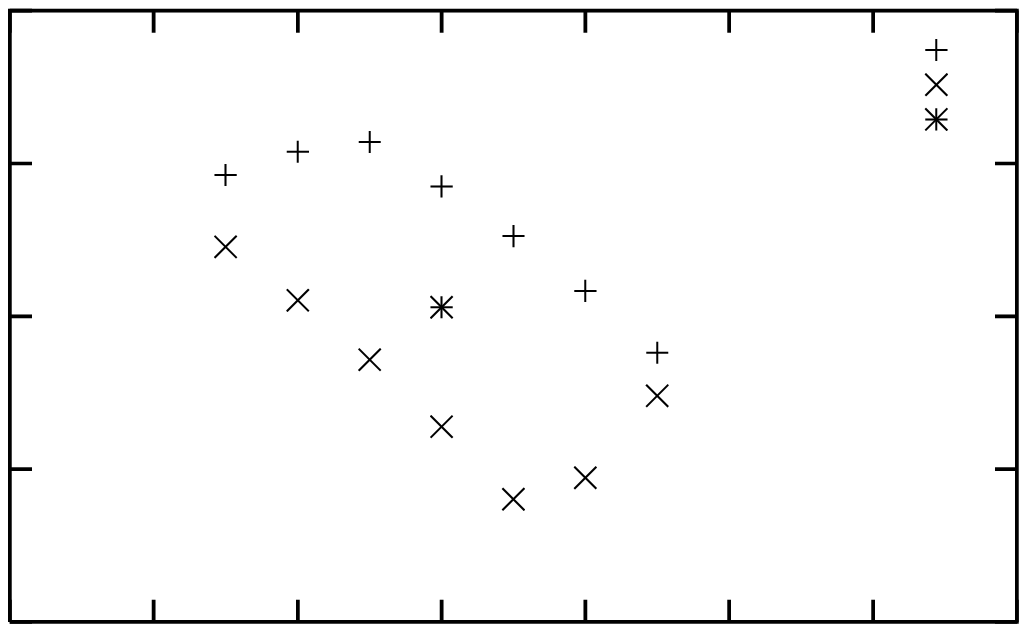 llx=0 lly=0 urx=720 ury=504 rwi=7200}
\put(3037,1747){\makebox(0,0)[r]{\small{Untersuchte L"osung}}}%
\put(3037,1847){\makebox(0,0)[r]{\small{Minimalwerte f"ur $4\,z_2\tilde\omega^{\scriptscriptstyle (2)}_u$}}}%
\put(3037,1947){\makebox(0,0)[r]{\small{Maximalwerte f"ur $4\,z_2\tilde\omega^{\scriptscriptstyle (2)}_u$}}}%
\put(2000,50){\makebox(0,0){$\tilde y_1/y_1$}}%
\put(100,1180){%
\special{ps: gsave currentpoint currentpoint translate
270 rotate neg exch neg exch translate}%
\makebox(0,0)[b]{\shortstack{$4\,z_2\tilde\omega^{\scriptscriptstyle (2)}_u$ (MeV)}}%
\special{ps: currentpoint grestore moveto}%
}%
\put(3450,200){\makebox(0,0){-10}}%
\put(3036,200){\makebox(0,0){-12}}%
\put(2621,200){\makebox(0,0){-14}}%
\put(2207,200){\makebox(0,0){-16}}%
\put(1793,200){\makebox(0,0){-18}}%
\put(1379,200){\makebox(0,0){-20}}%
\put(964,200){\makebox(0,0){-22}}%
\put(550,200){\makebox(0,0){-24}}%
\put(500,2060){\makebox(0,0)[r]{-30000}}%
\put(500,1620){\makebox(0,0)[r]{-35000}}%
\put(500,1180){\makebox(0,0)[r]{-40000}}%
\put(500,740){\makebox(0,0)[r]{-45000}}%
\put(500,300){\makebox(0,0)[r]{-50000}}%
\end{picture}%
\endgroup
 
\caption{MSW-L"osung 2b (kleine Mischung) im 
$\tilde y_1/y_1$-$z_2\tilde \omega^{\scr (2)}_u$-Parameterraum \label{mod2b}}
\end{center}
\end{figure}

\noindent Konkret sieht die Vorgehensweise wie folgt aus:
\begin{itemize}
\item F"ur festes $\tilde y_1/y_1$ wird der gesamte plausibel
  erscheinende Wertebereich f"ur $z_2\tilde \omega^{\scr (2)}_u$
  untersucht. Im folgenden wird $|z_2\tilde \omega^{\scr (2)}_u|
  \lesssim m_t(M_{\scr I})$ gew"ahlt.
\item Bei gegebenem $z_2\tilde \omega^{\scr (2)}_u$-Wert werden
  die Dirac-, Majorana- und See-Saw-Massenma\-tri\-zen der Neutrinos
  bestimmt; letztere wird gem"a"s (\ref{ntrdiag}) diagonalisiert.
\item Auf diese Weise erh"alt man die Massen $m_{\nu_i}$ der leichten 
  Neutrinos bis auf den Vorfaktor $1/(y_1 M_{\scr R})$, es wird
  $|m_{\nu_3}|\ge|m_{\nu_2}|\ge|m_{\nu_1}|$ sowie
  $\Delta m^2_{\scr \textrm{atm}} \equiv m^2_{\nu_3}-m^2_{\nu_2}$ 
  und $\Delta m^2_{\scr \textrm{sun}} \equiv m^2_{\nu_2}-m^2_{\nu_1}$
  angenommen. Die leptonische Mischungsmatrix 
  ${\bf U} \equiv {\bf E}^\dagger_{\scr L} {\bf N}^{}_{\scr L}$ ist
  damit ebenfalls bekannt.
\item Nun wird "uberpr"uft, ob die so erhaltenen Neutrinoeigenschaften
  den Einschr"ankungen (\ref{ntrcon1}-\ref{ntrcon4}) gen"ugen. In
  Unkenntnis der Massenskala $M_{\scr R}$ werden folgende Forderungen
  an die L"osungen gestellt:
  \begin{eqnarray} \label{numcons1}
  |({\bf U})_{13}| \le 0.05 \hspace{5.33cm} & & \\
  0.49 \le |({\bf U})_{23}| \le 0.71 \hspace{5.33cm} & & \\ \nonumber
  0.03 \le |({\bf U})_{12}| \le 0.05 \quad 
  \textrm{(MSW mit kleiner Mischung)} & & \\
  \textrm{oder} \quad 0.35 \le |({\bf U})_{12}| \le 0.49 \quad 
  \textrm{(MSW mit gro"ser Mischung)} \hspace{0.08cm} & & \\ \label{numcons2}
  50 \le \Delta m^2_{\scr \textrm{atm}}/\Delta m^2_{\scr \textrm{sun}} 
  \equiv (m^2_{\nu_3}-m^2_{\nu_2})/(m^2_{\nu_2}-m^2_{\nu_1})
  \le 1000 & &
\end{eqnarray}
\end{itemize}
Es zeigt sich, da"s drei Bereiche im 
$\tilde y_1/y_1$-$z_2\tilde \omega^{\scr (2)}_u$-Parameterraum existieren, in
denen L"osungen von (\ref{soeq}) liegen, welche zudem im Neutrinosektor die 
Forderungen (\ref{numcons1}-\ref{numcons2}) erf"ullen. Die Abbildungen 
\ref{mod1}, \ref{mod2a} und \ref{mod2b} zeigen die Lage der entsprechenden
Parameterbereiche, wobei zu jedem untersuchten 
ganzzahligen $\tilde y_1/y_1$-Wert
die Minimal- und Maximalwerte von $z_2\tilde \omega^{\scr (2)}_u$
angegeben sind, die zu Neutrinoeigenschaften f"uhren, welche den 
Einschr"ankungen (\ref{numcons1}-\ref{numcons2}) gen"ugen. 
Ein Bereich liefert MSW-L"osungen f"ur das Sonnen-Neutrinodef\/izit 
mit gro"sem Mischungswinkel, die anderen beiden Bereiche liefern 
MSW-L"osungen mit kleiner Mischung.


Aus jedem der drei Bereiche wird nun eine repr"asentative
L"osung ausgew"ahlt, die genauer untersucht werden soll. Deren Werte
f"ur $\tilde y_1/y_1$ und $z_2\tilde \omega^{\scr (2)}_u$ werden so
gew"ahlt, da"s die zugeh"origen Punkte ungef"ahr in der Mitte der
erlaubten Parameterbereiche liegen, da sowohl die bestehenden als auch
die neu hinzukommenden Neutrinoexperimente die Grenzen in 
(\ref{ntrcon1}-\ref{ntrcon4}) weiter einschr"anken werden. Tabelle
\ref{mmsols} fa"st die Parameterwerte f"ur die drei Beispielmodelle zusammen.
\begin{table}[h]
\begin{center}
\begin{tabular}{|l|l|r|c|}
\hline
L"osung & MSW-Ef\/fekt & $\tilde y_1/y_1$ & 
$z_2\tilde \omega^{\scr (2)}_u$ (MeV) \\
\hline \hline
Modell\,1 & gro"se Mischung & $19$ & $-3450$ \\
Modell\,2a & kleine Mischung & $24$ & $-9175$ \\
Modell\,2b & kleine Mischung & $-18$ & $-9925$ \\
\hline
\end{tabular}
\end{center}
\caption{\label{mmsols} Werte von $\tilde y_1/y_1$ und 
$z_2\tilde \omega^{\scr (2)}_u$ f"ur die drei Beispiell"osungen}
\end{table}

Die numerischen Resultate f"ur die untersuchten Beispielmodelle sind 
in Anhang \ref{appmm} zusammengefa"st. Zun"achst gibt Abschnitt 
\ref{appmm1} Auskunft "uber die nichtverschwindenden Eintr"age aller 
Massenmatrizen sowie die
Fermionmassen und -mischungen bei $M_{\scr I}$. Daraus lassen sich
mit Hilfe von (\ref{parmrel1}-\ref{parmrel2}) die Werte der 
$SO(10)$-Higgs-Parameter berechnen; die Resultate enth"alt Abschnitt 
\ref{appmm2}. Um die Werte dieser Gr"o"sen bei $M_{\scr Z}$ zu
erhalten, m"ussen die Renormierungsgruppengleichungen
(\ref{smcrge1}-\ref{rgessn}) des SM einschlie"slich derjenigen f"ur die
See-Saw-Massenmatrix von $M_{\scr I}$ nach $M_{\scr Z}$ integriert
werden. Dabei werden auch die Matrixeintr"age ungleich Null, die bei
$M_{\scr I}$ wegen des Ansatzes (\ref{mmans}) verschwinden. Die
Elemente der Massenmatrizen der geladenen Fermionen und leichten
Neutrinos sowie deren Massen und Mischungen bei $M_{\scr Z}$ sind in
Abschnitt \ref{appmm3} aufgelistet. Insbesondere die Mischungswinkel
werden bei der Berechnung der Nukleonzerfallsraten im n"achsten
Kapitel von gro"ser Bedeutung sein. Aus ihnen lassen sich auch die
CKM-Matrix {\bf V} und die leptonische Mischungsmatrix {\bf U}
bestimmen; die Ergebnisse f"ur die drei betrachteten Modelle enth"alt 
Abschnitt \ref{appmm4}.

Obwohl die Massenmatrizen in den einzelnen F"allen durchaus
verschieden sind, lassen sich folgende gemeinsame Eigenschaften der L"osungen
feststellen:
\begin{itemize}
\item Die Winkel der CKM-Matrix in der hier f"ur alle
  Mischungsmatrizen verwendeten Parametrisierung (\ref{ckmpar}) sind
  mit $\theta_{12}=0.223$, $\theta_{23}=0.039$ und $\theta_{31}=0.003$ 
  angesetzt worden, was innerhalb der experimentellen Grenzen
  liegt. Es zeigt sich, da"s die linksh"andigen Mischungswinkel der
  Quarks $\theta^{\scr (u)}_{\scr Ljk}$ und $\theta^{\scr (d)}_{\scr Ljk}$ 
  betragsm"a"sig zum Teil deutlich gr"o"ser als die CKM-Winkel
  sind. Besonders auf\/f"allig ist das im Fall von 
  $\theta_{12} \approx \theta^{\scr (d)}_{\scr L12} 
  - \theta^{\scr (u)}_{\scr L12}$. Demnach kann man von der Gr"o"se
  der observablen CKM-Winkel nicht zwangsl"auf\/ig auf die der linksh"andigen
  Quarkmischungen schlie"sen, wie es h"auf\/ig getan wird.
\item Alle Modelle enthalten betragsm"a"sig gro"se Mischungswinkel mit 
  Betr"agen zwischen $0.5$ und $1.0$; so ist zum Beispiel 
  $\theta^{\scr (d)}_{\scr R23}$ in allen F"allen sehr gro"s.
  Gro"se rechtsh"andige Mischungen sind insofern von besonderem
  Interesse, als sie im Rahmen des SM keine experimentell
  beobachtbaren Auswirkungen haben, in GUT-Modellen wie dem
  vorliegenden aber durchaus relevant sind. Unter anderem h"angen die 
  Verzweigungsraten der Nukleonenzerf"alle stark von den
  Fermionmischungen ab, wie man im n"achsten Kapitel sehen wird.
\item In den L"osungen ist $({\bf M}_u)_{\scr 33} \approx m_t$ und
  somit betragsm"a"sig deutlich gr"o"ser als alle "ubrigen Matrixeintr"age,
  $({\bf M}_d)_{\scr 33}$ und $({\bf M}_e)_{\scr 33}$ dagegen
  sind in ${\bf M}_d$ beziehungsweise ${\bf M}_e$ nicht allein dominant.
  Die $({\bf M}_f)_{\scr 12}$-$({\bf M}_f)_{\scr 21}$- und 
  $({\bf M}_f)_{\scr 23}$-$({\bf M}_f)_{\scr 32}$-Asymmetrien 
  in den Massenmatrizen sind in allen F"allen stark ausgepr"agt.
\item Die zu Beginn des Abschnitts gemachten Annahmen "uber die
  Skalenabh"angigkeit der Fermionmassen und CKM-Mischungen sind
  gerechtfertigt. Alle Werte f"ur die Massen bei $M_{\scr Z}$, welche
  die drei Modelle durch Integration der Renormierungsgruppengleichungen
  liefern, ergeben sich abgesehen von Rundungsfehlern in sehr guter 
  "Ubereinstimmung mit den urspr"unglich angesetzten experimentellen Gr"o"sen.
  Auch die aus den L"osungen numerisch bestimmten CKM-Matrizen
  gen"ugen den experimentellen Grenzen. Lediglich
  die Skalenabh"angigkeit von $\theta^{\scr (d)}_{\scr L23}$ 
  und $\theta^{\scr (d)}_{\scr L31}$ ist nicht v"ollig zu
  vernachl"assigen; es gilt 
  $\theta^{\scr (d)}_{\scr L23}(M_{\scr I}) > 
  \theta^{\scr (d)}_{\scr L23}(M_{\scr Z})$ und 
  $\theta^{\scr (d)}_{\scr L31}(M_{\scr I}) <
  \theta^{\scr (d)}_{\scr L31}(M_{\scr Z})$. 
  Das wiederum f"uhrt innerhalb der Grenzen zu
  vergleichsweise gro"sen Werten f"ur $({\bf V})_{\scr 33}$, w"ahrend 
  $({\bf V})_{\scr 23}$ und $({\bf V})_{\scr 32}$ betragsm"a"sig 
  am unteren Ende des erlaubten Bereichs liegen. 
  Man k"onnte diese Schw"ache jedoch ohne weiteres
  beheben, indem man die nun bekannte Skalenabh"angigkeit auf den
  rechten Seiten von (\ref{soeq}) ber"ucksichtigt. Da die Ef\/fekte aber
  sehr klein sind, wird hier darauf verzichtet.
\end{itemize}
Die drei untersuchten Beispiell"osungen beschreiben den Sektor der
geladenen Fermionen in sehr guter "Ubereinstimmung mit den experimentellen
Resultaten f"ur die Massen und Mischungen. Das ist jedoch lediglich
ein notwendiges Kriterium daf"ur, da"s der Ansatz (\ref{mmans}) als
sinnvoll zu betrachten ist. Auf die Neutrinoeigenschaften der
einzelnen L"osungen, welche die eigentlichen Vorhersagen des
Massenmodells darstellen, wird im n"achsten Abschnitt eingegangen.

Die in diesem Kapitel verwendeten numerischen Algorithmen sind im 
wesentlichen \cite{numrec} entnommen. Zur Integration der
Renormierungsgruppengleichungen wurde eine auf der Runge-Kutta-Methode
vierter Ordnung basierende Routine mit adaptiver
Schrittweitensteuerung benutzt. Dabei wird der durch die endliche Schrittweite
erzeugte Fehler abgesch"atzt und unterhalb einer vorzugebenden
Schranke gehalten. Diese Schranke wurde so festgelegt, da"s die Fehler
aufgrund der Integration deutlich kleiner als die experimentellen
Unsicherheiten sind. In den $\beta$-Funktionskoef\/f\/izienten zweiter
Ordnung der Yukawa-Matrizen und der Higgs-Kopplung wurde die N"aherung
benutzt, nur die betragsm"a"sig gr"o"sten Eintr"age der
Yukawa-Matrizen zu verwenden und alle "ubrigen auf Null zu setzen.

Die Diagonalisierung der Massenmatrizen gem"a"s
(\ref{biuntr1}) entspricht der numerischen L"o\-sung eines nichtlinearen
Gleichungssystems und wurde mit Hilfe der multidimensionalen 
Newton-Raphson-Methode durchgef"uhrt. 
\section{Eigenschaften der leichten Neutrinos}
Die Massen und Mischungen der leichten Neutrinos bei $M_{\scr Z}$ 
sind in Tabelle \ref{ntrchar} zusammengefa"st, welche einen Ausschnitt
aus Tabelle \ref{parmz} darstellt. 
\begin{table}[h]
\begin{center}
\begin{tabular}{|c|r|r|r|}
\hline
Parameter & Wert in Modell\, 1 & Wert in Modell\,2a & Wert in Modell\,2b \\
\hline \hline
$(y_1 M_{\scr R}/M_{\scr I}) \cdot m_{\nu_1}$ 
& $-0.0245$ eV & $-8.73 \cdot 10^{-4}$ eV & $1.16 \cdot 10^{-3}$ eV \\
$(y_1 M_{\scr R}/M_{\scr I}) \cdot m_{\nu_2}$ 
& $0.0876$ eV & $0.355 \,$ eV & $-0.467$ eV \\
$(y_1 M_{\scr R}/M_{\scr I}) \cdot m_{\nu_3}$ 
& $-2.402$ eV & $-3.031$ eV & $4.365$ eV \\
\hline
$\theta^{\scr (\nu)}_{\scr 12}$ & $-0.487$ & $-0.050$ & $0.051$ \\
$\theta^{\scr (\nu)}_{\scr 23}$ & $0.205$ & $0.506$ & $0.496$ \\
$\theta^{\scr (\nu)}_{\scr 31}$ & $0.004$ & $0.003$ & $-0.003$ \\
\hline
$m_{\nu_2}/m_{\nu_1}$ & $-3.57$ & $-406.3$ & $-401.2$ \\
$m_{\nu_3}/m_{\nu_2}$ & $-27.43$ & $-8.54$ & $-9.35$ \\
$\left( \dfrac{m_{\nu_3}^2-m_{\nu_2}^2}{m_{\nu_2}^2-m_{\nu_1}^2} \right)$ 
& $815.4$ & $71.9$ & $86.4$ \\
\hline
\end{tabular}
\end{center}
\caption{\label{ntrchar} Massen und Mischungswinkel der leichten
  Neutrinos bei $M_{\scr Z}$}
\end{table}

\noindent Da sich die Massen der geladenen Leptonen zwischen 
$\mu=M_{\scr Z}$ und $\mu/M_{\scr Z} \approx 0$ kaum "andern,
kann man das in noch st"arkerem Ma"se von den Neutrinomassen
erwarten, da die Neutrinos nur schwach wechselwirken. Vergleicht man 
die Tabellen \ref{parmz} einerseits und \ref{parmiglf} sowie \ref{parmintr}
andererseits, so erkennt man, da"s die Mischungswinkel aller Fermionen 
zwischen $M_{\scr I}$ und $M_{\scr Z}$ nur minimal skalenabh"angig sind.
Im folgenden wird deshalb von $\theta^{\scr (f)}_{\scr L,Rjk}(M_{\scr Z}) = 
\theta^{\scr (f)}_{\scr L,Rjk}(m_p)$ ausgegangen.

In Tabelle \ref{ntrchar} erkennt man zun"achst die starke "Ahnlichkeit der
Eigenschaften von Modell\,2a und 2b. Beide realisieren eine
Oszillationsl"osung des Sonnen-Neutrinodef\/izits "uber den MSW-Ef\/fekt
mit kleinen Mischungen, w"ahrend Modell\,1 auf den MSW-Ef\/fekt mit
gro"ser Mischung f"uhrt. Von den linksh"andigen Mischungen der
geladenen Leptonen ist lediglich $\theta^{\scr (e)}_{\scr L23}$ 
in Modell\,1 betragsm"a"sig deutlich gr"o"ser als Null. In den anderen
beiden F"allen kommt die gro"se (23)-Mischung in {\bf U}, welche die
Anomalie der atmosph"arischen Neutrinos erkl"art, allein durch
die reine Neutrinomischung zustande. 

\noindent F"ur $(m_{\nu_3}^2-m_{\nu_2}^2)/(m_{\nu_2}^2-m_{\nu_1}^2) \equiv 
\Delta m^2_{\scr \textrm{atm}}/\Delta m^2_{\scr \textrm{sun}}$ waren
aufgrund der experimentellen Grenzen Werte zwischen 50 und 1000
zugelassen worden. Hier f"allt auf, da"s der Wert in Modell\,1 nahe
an der oberen Schranke ist, w"ahrend die Resultate in den Modellen
2a,b deutlich im unteren Teil des erlaubten Bereichs liegen. 
Dem steht die durch Abbildung \ref{mswpic} begr"undete Erwartung
gegen"uber, bei festem $\Delta m^2_{\scr \textrm{atm}}$ f"ur den
MSW-Ef\/fekt mit gro"ser Mischung den im Vergleich zum MSW-Ef\/fekt mit 
kleiner Mischung kleineren Wert f"ur 
$\Delta m^2_{\scr \textrm{atm}}/\Delta m^2_{\scr \textrm{sun}}$
zu f\/inden.

Tabelle \ref{yemr} gibt die Werte von $(y_1 M_{\scr R}/M_{\scr I})$ und 
$y_1 M_{\scr R}$ an, welche ben"otigt werden, um die erhaltenen
Vorhersagen f"ur $(m_{\nu_3}^2-m_{\nu_2}^2)$ und 
$(m_{\nu_2}^2-m_{\nu_1}^2)$ an die experimentellen Grenzen 
f"ur $\Delta m^2_{\scr \textrm{atm}}$ und $\Delta m^2_{\scr \textrm{sun}}$
aus (\ref{ntrcon1}-\ref{ntrcon4}) anzupassen. 
\begin{table}[h]
\begin{center}
\begin{tabular}{|l|r|r|r|}
\hline
Gr"o"se & Wert in Modell\,1 & Wert in Modell\,2a & Wert in Modell\,2b \\
\hline \hline
$M_{\scr I}$ (GeV) & $6.14 \cdot 10^{10}$ & $6.14 \cdot 10^{10}$
& $6.14 \cdot 10^{10}$ \\
$(y_1 M_{\scr R}/M_{\scr I})$ & $25-30$ & $100-150$ & $135-215$ \\
$y_1 M_{\scr R}$ (GeV) & $(1.54-1.84) \cdot 10^{12}$
& $(6.14-9.21) \cdot 10^{12}$ & $(8.29-13.20) \cdot 10^{12}$ \\
\hline
$m_{\nu_1}$ (eV)& $-(8.2-9.8) \cdot 10^{-4}$
& $-(5.8-8.7) \cdot 10^{-6}$ & $(5.4-8.6) \cdot 10^{-6}$ \\
$m_{\nu_2}$ (eV) & $(2.9-3.5) \cdot 10^{-3}$
& $(2.4-3.6) \cdot 10^{-3}$ & $-(2.2-3.5) \cdot 10^{-3}$ \\
$m_{\nu_3}$ (eV) & $-(8.0-9.6) \cdot 10^{-2}$ 
& $-(2.0-3.0) \cdot 10^{-2}$ & $(2.0-3.2) \cdot 10^{-2}$ \\
$m_{\nu_2}^2-m_{\nu_1}^2$ (eV$^2$) & $(7.7-11.3) \cdot 10^{-6}$
& $(5.8-13.0) \cdot 10^{-6}$ & $(4.8-12.3) \cdot 10^{-6}$ \\
$m_{\nu_3}^2-m_{\nu_2}^2$ (eV$^2$) & $(6.4-9.2) \cdot 10^{-3}$ 
& $(3.9-8.9) \cdot 10^{-4}$ & $(4.0-10.1) \cdot 10^{-4}$ \\
\hline
\end{tabular}
\end{center}
\caption[Werte von $y_1 M_{\scr R}$ und daraus resultierende Neutrinomassen]
{\label{yemr} Werte von $(y_1 M_{\scr R}/M_{\scr I})$ und 
$y_1 M_{\scr R}$ und daraus resultierende Neutrinomassen}
\end{table}
Sie sind angesichts der erwarteten Relation 
$M_{\scr R} \sim M_{\scr I}$ vergleichsweise gro"s, da $y_1$ als 
Yukawa-Kopplung $\lesssim 1$ sein sollte. Andererseits ist 
$(y_1 M_{\scr R}/M_{\scr I}) \sim 100$ auch 
keineswegs ausgeschlossen, da der exakte Wert von $M_{\scr R}$
von zahlreichen Parametern im nicht explizit bekannten Higgs-Potential
abh"angt. Diese sollten zwar von der Gr"o"senordnung 1 sein, 
aber es ist durchaus m"oglich, da"s der Proportionalit"atskoef\/f\/izient
in $M_{\scr R} \sim M_{\scr I}$ aus einer nichttrivialen Kombination
der Parameter besteht und im Bereich $1/100$ bis $100$ liegt. 
"Uberdies ist in Abschnitt \ref{thresh} eine Unsicherheit
$\lesssim 10^{\pm 2.6} \approx 400$ in $M_{\scr I}$ aufgrund von
Schwellenef\/fekten berechnet worden. Die Vernachl"assigung der
massenabh"angigen Schwellenkorrekturen bei der Bestimmung von $M_{\scr I}$ 
in Abschnitt \ref{essb} k"onnte somit theoretisch zu einem um zwei
Gr"o"senordnungen zu kleinen Wert gef"uhrt haben. Eine Kombination
dieser beiden M"oglichkeiten ist demnach def\/initiv in der Lage, 
$(y_1 M_{\scr R}/M_{\scr I}) \sim 100$ zu realisieren.

Wie schon weiter oben erw"ahnt wurde, liegen im Modell\,1 die Werte f"ur
$m_{\nu_2}^2-m_{\nu_1}^2$ im unteren Teil des f"ur 
$\Delta m^2_{\scr \textrm{sun}}$ zul"assigen Bereichs, w"ahrend jene f"ur 
$m_{\nu_3}^2-m_{\nu_2}^2$ im Rahmen der experimentellen Grenzen von 
$\Delta m^2_{\scr \textrm{atm}}$ relativ gro"s sind. In den Modellen\,2a,b 
beobachtet man ein gegenteiliges Verhalten dieser Gr"o"sen.

Vergleicht man die Modellvorhersagen f"ur die leptonische
Mischungsmatrix {\bf U} (siehe Anhang \ref{appmm4}) mit den Grenzen aus 
(\ref{ntrcon1}-\ref{ntrcon4}), so erkennt man eine gute "Ubereinstimmung
zwischen beiden; alle Eintr"age von {\bf U} liegen innerhalb der
erlaubten Bereiche. Auf\/f"allig ist bei allen drei Modellen die
Tatsache, da"s die (23)-Mischung in {\bf U} vergleichsweise klein
ist. Insbesondere in Modell\,1 ist $|({\bf U})_{23}|=0.504$ nahe an der
unteren Schranke 0.49, w"ahrend die $|({\bf U})_{23}|$-Werte in den
Modellen\,2a,b ($0.536$ beziehungsweise $0.552$) etwas gr"o"ser sind.
Da die existierenden ebenso wie die neu hinzukommenden 
Neutrinoexperimente die Grenzen in (\ref{ntrcon1}-\ref{ntrcon4}) 
weiter einschr"anken werden, ist Modell\,1 in diesem Sinne weniger
"uberzeugend als die anderen beiden. Das hier vorgeschlagene
$SO(10)$-Massenmodell deutet also auf eine MSW-L"osung mit kleiner
Mischung hin. Sollte der MSW-Ef\/fekt mit kleiner Mischung durch Experimente
eindeutig als Ursache f"ur das Def\/izit der Sonnen-Neutrinos best"atigt
werden, wird es schwierig sein, eines der Modelle\,2a und 2b anhand
der Eigenschaften im Neutrinosektor als in der Natur realisiert
zu identif\/izieren. Dazu mu"s man auf die Vorhersagen bez"uglich der
Nukleonzerfallsraten zur"uckgreifen, welche im n"achsten Kapitel
berechnet werden.

Die Existenz von leichten Majorana-Neutrinos hat eine weitere
"uberpr"ufbare Konsequenz, n"amlich die sogenannten neutrinolosen
doppelten $\beta$-Zerf"alle. Sie kommen durch den Austausch eines 
Majorana-Neutrinos zwischen zwei zerfallenden Neutronen zustande, was
zu Kernreaktionen der Form 
$X^{\scr A}_{\scr Z} \rightarrow X^{\scr A}_{\scr Z+2} + 2 \, e^-$
f"uhrt. Die experimentelle Nichtbeobachtung solcher Prozesse liefert
die Einschr"ankung \cite{baud}:
\begin{equation}
\langle m_\nu \rangle \; \equiv \; \sum_i 
\big( \, ({\bf U})^2_{1i} m_{\nu_i} \, \big) 
\; \le \; 0.2 \, \textrm{eV}
\end{equation}
Alle hier untersuchten L"osungen erf"ullen diese Bedingung.
\section{Baryonasymmetrie durch Neutrinozerf"alle}
Obwohl nicht Gegenstand dieser Arbeit, soll hier kurz auf die
M"oglichkeit hingewiesen werden, die Baryonasymmetrie des Universums
im Rahmen einer $SO(10)$-GUT mit See-Saw-Mechanismus durch den Zerfall 
der schweren Majorana-Neutrinos zu erkl"aren. Einen aktuellen "Uberblick
zu diesem Thema liefert \cite{rio}.

Das beobachtete Universum scheint wesentlich mehr Materie als
Antimaterie zu enthalten. Wenn $n^{}_b$, $n_{\bar b}$ und 
$n_\gamma$ die mittleren Teilchenzahldichten der Baryonen, 
Antibaryonen und Photonen im Universum bezeichnen, ergibt sich f"ur
die Gr"o"se $\eta \equiv (n^{}_b-n_{\bar b})/n_\gamma$ aus 
kosmologischen "Uberlegungen der Wert 
$4 \cdot 10^{-10} \lesssim \eta \lesssim 7 \cdot 10^{-10}$
\cite{rio,kotu}. Wenn man davon ausgeht, da"s $\eta$ bei der Entstehung  
des Universums Null war, so kann die Kosmologie die heute beobachtete
Baryonasymmetrie nicht erkl"aren. Dazu mu"s auf die Physik der
Elementarteilchen zur"uckgegrif\/fen werden.

Wie in \cite{sakh} gezeigt wurde, sind f"ur die Entstehung der 
Baryonasymmetrie in einem urspr"unglich symmetrischen Universum 
drei notwendige Kriterien zu erf"ullen, n"amlich die 
Existenz baryonzahlverletzender 
Wechselwirkungen, die Verletzung von $C$ und $CP$ sowie das
Vorhandensein eines thermischen Ungleichgewichts. Die
Baryonzahlverletzung ist naheliegenderweise notwendig, um $\eta \ne 0$
zu erm"oglichen, und ist sowohl in $SO(10)$- als auch 
$G_{\scr \textrm{PS}}$-Modellen enthalten; lediglich die Kombination
$B-L$ ist Teil der Eichsymmetrie und somit eine Erhaltungsgr"o"se.
$C$- und $CP$-Verletzung sind erforderlich, da sich zeigen l"a"st,
da"s andernfalls die Rate der Prozesse, welche Baryonen erzeugen,
genauso gro"s wie die der antibaryonerzeugenden Prozesse ist. W"ahrend die
Eichboson- und Fermionsektoren von $SO(10)$-GUTs $C$- und
$CP$-invariant sind, lassen sich diese Symmetrien durch Einf"uhrung
komplexer Higgs-Kopplungen explizit brechen. Schlie"slich ist ein
thermisches Ungleichgewicht erforderlich, da der Baryonzahloperator
$B$ bei einer Temperatur $T$ im Gleichgewicht den Erwartungswert
\begin{eqnarray} \nonumber
\langle B \rangle_T & = & \textrm{Tr} \big( \, e^{-\beta H} B \, \big)
\; = \; \textrm{Tr} \big( \, [\mathcal{CPT}][\mathcal{CPT}]^{-1} 
e^{-\beta H} B \, \big) \\
& = & \textrm{Tr} \big( \, e^{-\beta H} 
[\mathcal{CPT}]^{-1} B [\mathcal{CPT}] \, \big) 
\; = \; -\textrm{Tr} \big( \, e^{-\beta H} B \, \big) 
\; \stackrel{!}{=} \; 0
\end{eqnarray}
besitzt, vorausgesetzt, die Theorie ist $CPT$-invariant. Ob sich ein System
im thermischen Gleichgewicht bef\/indet, l"a"st sich nur durch L"osung
seiner Boltzmann-Gleichungen feststellen. In erster N"aherung kann man 
aber sagen, da"s ein Ungleichgewicht vorliegt, wenn die Expansionsrate 
des Universums gr"o"ser als die Raten der baryon- und
leptonzahlverletzenden Wechselwirkungen ist. Das sollte bei
Temperaturen $T \gtrsim M_{\scr I}$ der Fall sein.

Somit ist im vorliegenden $SO(10)$-Modell die Entstehung einer
$B+L$-Asymmetrie bei Energien $\mu \gtrsim M_{\scr I}$
m"oglich. Allerdings existieren auch im SM baryon- und 
leptonzahlverletzende Prozesse, welche $B-L$ erhalten und auf 
nichtperturbativen Ef\/fekten beruhen. W"ahrend n"amlich die klassische
Lagrangedichte des SM invariant unter globalen $U(1)_{\scr B,L}$-Symmetrien 
ist, besitzen die zu diesen Symmetrien geh"orenden Str"ome in der
quantisierten Theorie Anomalien \cite{sph1}. Das f"uhrt zu $B+L$-verletzenden 
Prozessen, welche bei $T \approx 0$ durch einen Faktor
$\exp(-2\pi/\alpha_{\scr 2})$ unterdr"uckt sind und vernachl"assigt
werden k"onnen. Bei Temperaturen $10^2$ GeV $\lesssim T \lesssim$ $10^{12}$ 
GeV jedoch sind sie in Gegenwart von statischen topologischen
Feldkonf\/igurationen, den sogenannten Sphaleronen, stark genug, 
um eine vorhandene $B+L$-Asymmetrie vollst"andig auszuwaschen 
\cite{sph2}. Andererseits sind sie in der Lage, eine bereits
vorhandene $B-L$-Asymmetrie teilweise in eine $B$-Asymmetrie gem"a"s
\begin{equation} \label{lptg}
B(T \lesssim M_{\scr Z}) \; = \; \dfrac{32+4\,n_{\scr H}}{98+13\,n_{\scr H}} 
\, (B-L)
\end{equation}
umzuwandeln.

Nun erhalten im Rahmen der $B-L$-Symmetriebrechung bei $M_{\scr I}$
die schweren Neutrinos ihre Massen. Aufgrund ihrer Majorana-Natur
k"onnen sie "uber leptonzahlverletzende Prozesse
\begin{equation}
N_{\scr R} \longrightarrow l^{}_{\scr L}+\bar \phi \; , \quad
N_{\scr R} \longrightarrow l^{\scr C}_{\scr L}+\phi
\end{equation}
sowohl in Leptonen als auch in Antileptonen zerfallen;
$\phi$ bezeichnet ein Higgs-Teilchen. Die notwendige $CP$-Verletzung
wird durch komplexe Kopplungen zwischen $N_{\scr R}$, $l^{}_{\scr L}$
und $\bar \phi$ realisiert. Auf diese Weise erh"alt man eine reine
Lepton- und somit eine $B-L$-Asymmetrie, die gem"a"s (\ref{lptg}) in
die beobachtete Baryonasymmetrie des Universums konvertiert werden
kann \cite{lepg}. Damit die leptonzahlverletzenden Zerf"alle der
schweren Neutrinos im erforderlichen thermischen Ungleichgewicht
stattf\/inden, mu"s die zugeh"orige Zerfallsrate kleiner als die 
Expansionsrate des Universums bei dieser Temperatur sein. Das l"a"st
sich qualitativ als Bedingung 
\begin{equation} \label{lgcs}
m_\nu \; \lesssim \; \left( \dfrac{M}{10^{10} \, \textrm{GeV}} 
\right)^{-1/2} \cdot 4 \; \textrm{eV}
\end{equation}
f"ur die Massen $m_\nu$ der leichten Neutrinos schreiben,
wobei $M \equiv$ Min$\{M_{\scr I},10^{12}$ GeV$\}$ ist \cite{ritro}.
Im hier untersuchten Massenmodell ist (\ref{lgcs}) of\/fensichtlich
erf"ullt.

\newpage

\mbox{}

%%% Local Variables: 
%%% mode: latex
%%% TeX-master: t
%%% End: 

\chapter{Zerfallsraten der Nukleonen}
Wie in Kapitel\,2 dargestellt wurde, geh"ort die Instabilit"at der
Nukleonen zu den bemerkenswertesten Vorhersagen von GUTs. Die
Bestimmung der partiellen und totalen Zerfallsraten im Rahmen des hier
untersuchten $SO(10)$-Massenmodells wird Gegenstand dieses Kapitels
sein. Dazu sind zwei wesentliche Schritte notwendig:
\begin{itemize}
\item Zun"achst mu"s die vollst"andige ef\/fektive Lagrangedichte 
  $\mathcal{L}_{\scr \textrm{ef\/f}}$ der durch Eichboson\-austausch
  vermittelten Nukleonenzerf"alle in der Basis der Masseneigenzust"ande
  bestimmt werden. Das geschieht analog zur Herleitung der
  Lagrangedichte der Fermi-Theorie aus jener der elektroschwachen
  Wechselwirkung.  
\item Weiterhin mu"s man die hadronischen "Ubergangsmatrixelemente 
  berechnen. Hierzu sind in der Vergangenheit das nichtrelativistische
  Quark-Modell \cite{jarl,karl,gavel}, das Bag-Modell \cite{bag} und 
  die chirale St"orungsrechnung \cite{chir} verwendet
  worden. Urspr"ungliche Diskrepanzen zwischen den Resultaten der
  chiralen St"orungsrechnung und denen der Quark-Modelle sind in 
  \cite{isg} beseitigt worden; \cite{luch} gibt eine "Ubersicht "uber
  die mit den verschiedenen Verfahren erzielten Ergebnisse. In dieser
  Arbeit wird analog zu \cite{karl} und \cite{gavel} vorgegangen.
\end{itemize}
Prinzipiell existieren auch Nukleonzerfallsprozesse aufgrund von
Higgs-Austausch. Diese sind aber selbst dann vernachl"assigbar, wenn
das Higgs-Teilchen eine Masse der Gr"o"senordnung $M_{\scr I}$ hat, da
die "Ubergangswahrscheinlichkeit proportional zur vierten Potenz der
sehr kleinen Yukawa-Kopplungen an die drei leichten Quarks ist.
\section{Die ef\/fektive Lagrangedichte}
Ausgangspunkt ist der baryonzahlverletzende Anteil der
$SO(10)$-Lagrangedichte (\ref{bnvld}). Verallgemeinert man ihn auf drei
Fermiongenerationen und ersetzt die Wechselwirkungs- durch die
Masseneigenzust"ande analog zu (\ref{biuntr2}), kann man die 
vollst"andige ef\/fektive Lagrangedichte der
Vier-Fermion-Wechselwirkungen berechnen, welche die
eichbosonvermittelten Nukleonenzerf"alle beschreibt. 
F"ur den Fall zweier Generationen ist das in \cite{hara} durchgef"uhrt
worden. Unter Benutzung der Fierz-Identit"aten
\begin{eqnarray}
\big( \, \bar \Psi^{}_{\scr 1L} \gamma_\mu \Psi^{}_{\scr 2L} \, \big)
\big( \, \bar \Psi^{}_{\scr 3L} \gamma^\mu \Psi^{}_{\scr 4L} \, \big)
& = &
\big( \, \bar \Psi^{}_{\scr 1L} \gamma_\mu \Psi^{}_{\scr 4L} \, \big)
\big( \, \bar \Psi^{}_{\scr 3L} \gamma^\mu \Psi^{}_{\scr 2L} \, \big)
\; , \\
\bar \Psi^{}_{\scr 1L} \gamma_\mu \Psi^{}_{\scr 2L} & = &
- \bar \Psi^{\scr C}_{\scr 2R} \gamma_\mu \Psi^{\scr C}_{\scr 1R}
\end{eqnarray}
erh"alt man als Resultat:
\begin{eqnarray} \nonumber
\mathcal{L}_{\scr \textrm{ef\/f}}
& = & A^{}_1 \, 
\big( \varepsilon_{\scr \alpha \beta \gamma} \bar u_{\scr L}^{\scr C
    \gamma}  \gamma^{\scr \mu} u_{\scr L}^{\scr \beta} \big)
\big( \bar e^{\scr +}_{\scr L} \gamma_{\scr \mu} d_{\scr L}^{\scr \alpha} \big)
\; + \; A^{}_2 \, 
\big( \varepsilon_{\scr \alpha \beta \gamma} \bar u_{\scr L}^{\scr C
    \gamma}  \gamma^{\scr \mu} u_{\scr L}^{\scr \beta} \big)
\big( \bar e^{\scr +}_{\scr R} \gamma_{\scr \mu} d_{\scr R}^{\scr \alpha} \big)
\\ \nonumber
& + & A^{}_3 \, 
\big( \varepsilon_{\scr \alpha \beta \gamma} \bar u_{\scr L}^{\scr C
    \gamma}  \gamma^{\scr \mu} u_{\scr L}^{\scr \beta} \big)
\big( \bar \mu^{\scr +}_{\scr L} \gamma_{\scr \mu} d_{\scr L}^{\scr
  \alpha} \big) 
\; + \; A^{}_4 \, 
\big( \varepsilon_{\scr \alpha \beta \gamma} \bar u_{\scr L}^{\scr C
    \gamma}  \gamma^{\scr \mu} u_{\scr L}^{\scr \beta} \big)
\big( \bar \mu^{\scr +}_{\scr R} \gamma_{\scr \mu} d_{\scr R}^{\scr \alpha} 
\big) \\ \nonumber
& + & A^{}_5 \, 
\big( \varepsilon_{\scr \alpha \beta \gamma} \bar u_{\scr L}^{\scr C
    \gamma}  \gamma^{\scr \mu} u_{\scr L}^{\scr \beta} \big)
\big( \bar e^{\scr +}_{\scr L} \gamma_{\scr \mu} s_{\scr L}^{\scr \alpha} \big)
\; + \; A^{}_6 \, 
\big( \varepsilon_{\scr \alpha \beta \gamma} \bar u_{\scr L}^{\scr C
    \gamma}  \gamma^{\scr \mu} u_{\scr L}^{\scr \beta} \big)
\big( \bar e^{\scr +}_{\scr R} \gamma_{\scr \mu} s_{\scr R}^{\scr \alpha} 
\big) \\ \nonumber
& + & A^{}_7 \, 
\big( \varepsilon_{\scr \alpha \beta \gamma} \bar u_{\scr L}^{\scr C
    \gamma}  \gamma^{\scr \mu} u_{\scr L}^{\scr \beta} \big)
\big( \bar \mu^{\scr +}_{\scr L} \gamma_{\scr \mu} 
s_{\scr L}^{\scr \alpha} \big) 
\; + \; A^{}_8 \, 
\big( \varepsilon_{\scr \alpha \beta \gamma} \bar u_{\scr L}^{\scr C
    \gamma}  \gamma^{\scr \mu} u_{\scr L}^{\scr \beta} \big)
\big( \bar \mu^{\scr +}_{\scr R} \gamma_{\scr \mu} s_{\scr R}^{\scr \alpha} 
\big) \\ \nonumber
& + & A^{}_9 \, 
\big( \varepsilon_{\scr \alpha \beta \gamma} 
\bar u_{\scr L}^{\scr C \gamma} \gamma^{\scr \mu} d_{\scr L}^{\scr \beta} 
\big)
\big( \bar \nu_{\scr e R}^{\scr C} \gamma_{\scr \mu} d_{\scr R}^{\scr
  \alpha}
\big)
\; + \; A^{}_{10} \, 
\big( \varepsilon_{\scr \alpha \beta \gamma} 
\bar u_{\scr L}^{\scr C \gamma} \gamma^{\scr \mu} d_{\scr L}^{\scr \beta} 
\big)
\big( \bar \nu_{\scr \mu R}^{\scr C} \gamma_{\scr \mu} d_{\scr
  R}^{\scr \alpha} \big)
\\ \nonumber
& + & A^{}_{11} \, 
\big( \varepsilon_{\scr \alpha \beta \gamma} 
\bar u_{\scr L}^{\scr C \gamma} \gamma^{\scr \mu} d_{\scr L}^{\scr \beta} 
\big)
\big( \bar \nu_{\scr e R}^{\scr C} \gamma_{\scr \mu} s_{\scr
  R}^{\scr \alpha} \big)
\; + \; A^{}_{12} \, 
\big( \varepsilon_{\scr \alpha \beta \gamma} 
\bar u_{\scr L}^{\scr C \gamma} \gamma^{\scr \mu} d_{\scr L}^{\scr \beta} 
\big)
\big( \bar \nu_{\scr \mu R}^{\scr C} \gamma_{\scr \mu} s_{\scr R}^{\scr
  \alpha} \big)
\\ \nonumber
& + & A^{}_{13} \, 
\big( \varepsilon_{\scr \alpha \beta \gamma} 
\bar u_{\scr L}^{\scr C \gamma} \gamma^{\scr \mu} s_{\scr L}^{\scr \beta} 
\big)
\big( \bar \nu_{\scr e R}^{\scr C} \gamma_{\scr \mu} d_{\scr R}^{\scr
  \alpha} \big)
\; + \; A^{}_{14} \, 
\big( \varepsilon_{\scr \alpha \beta \gamma} 
\bar u_{\scr L}^{\scr C \gamma} \gamma^{\scr \mu} s_{\scr L}^{\scr \beta} 
\big)
\big( \bar \nu_{\scr \mu R}^{\scr C} \gamma_{\scr \mu} d_{\scr
  R}^{\scr \alpha} \big)
\\ \nonumber
& + & A^{}_{15} \, 
\big( \varepsilon_{\scr \alpha \beta \gamma} 
\bar u_{\scr L}^{\scr C \gamma} \gamma^{\scr \mu} d_{\scr L}^{\scr \beta} 
\big)
\big( \bar \nu_{\scr \tau R}^{\scr C} \gamma_{\scr \mu} d_{\scr R}^{\scr
  \alpha} \big)
\; + \; A^{}_{16} \, 
\big( \varepsilon_{\scr \alpha \beta \gamma} 
\bar u_{\scr L}^{\scr C \gamma} \gamma^{\scr \mu} d_{\scr L}^{\scr \beta} 
\big)
\big( \bar \nu_{\scr \tau R}^{\scr C} \gamma_{\scr \mu} s_{\scr
  R}^{\scr \alpha} \big)
\\ \nonumber
& + & A^{}_{17} \, 
\big( \varepsilon_{\scr \alpha \beta \gamma} 
\bar u_{\scr L}^{\scr C \gamma} \gamma^{\scr \mu} s_{\scr L}^{\scr \beta} 
\big)
\big( \bar \nu_{\scr \tau R}^{\scr C} \gamma_{\scr \mu} d_{\scr R}^{\scr
  \alpha} \big) \\ \nonumber
& + & \; \textrm{(\, Terme mit zwei $s$-Quarks \,)}
\\ \nonumber
& + & \; \textrm{(\, Terme mit $c$- ,$b$- und $t$-Quarks \,)}
\\ \nonumber
& + & \; \textrm{(\, Terme mit $\bar \tau^{\scr +}_{\scr L,R}$ und 
$\bar \nu^{\scr C}_{\scr e,\mu,\tau L}$ \,)}
\\ \label{eldfnd}
& + & \; \textrm{(\, h.c.\,)} 
\end{eqnarray}
Die $A_i$-Koef\/f\/izienten sind in Anhang \ref{aicel} aufgef"uhrt und h"angen
au"ser von den Eichbosonmassen $M_{\scr X,Y}$ und $M_{\scr X',Y'}$, 
f"ur welche $M_{\scr X,Y} = M_{\scr X',Y'} \approx M_{\scr U}$ angenommen wird,
von $g_{\scr U}$ und den Elementen der fermionischen Mischungsmatrizen 
${\bf U}_{\scr R,L}$, ${\bf D}_{\scr R,L}$, 
${\bf E}_{\scr R,L}$ und ${\bf N}_{\scr L}$ ab. Diese Gr"o"sen
sind bereits im vorigen Kapitel bestimmt worden, so da"s alle $A_i$
bekannt sind. ${\bf N}_{\scr R}$ hat keinen Einf\/lu"s auf die $A_i$, 
da sie die Massenmatrix der schweren Neutrinos
diagonalisiert, welche als Zerfallsprodukte der Nukleonen nicht
vorkommen. An dieser Stelle wird klar, warum die Fermionmischungen f"ur die
Verzweigungsraten der Nukleonenzerf"alle von gro"ser Bedeutung sind. 
\section{Hadronische "Ubergangsmatrixelemente} \label{hmtrxel}
Die hier verwendete Methode zur Berechnung der "Ubergangsmatrixelemente 
entspricht dem sogenannten "`recoil"'-Modell in \cite{karl}. Grundlage
ist ein Quark-Modell mit $SU(6)$-Spin-Flavour-Symmetrie; die Quarks
treten in drei Flavours $u$, $d$ und $s$ sowie zwei Spinzust"anden 
$\uparrow$ und $\downarrow$ auf. Alle sechs unabh"angigen
Quarkzust"ande haben dieselbe Masse $m$, was eine einheitliche Masse
f"ur die Mesonen zur Folge hat. Die $SU(6)$-Symmetrie wird sp"ater bei der
Berechnung der Zerfallsraten durch einen Phasenraumfaktor explizit gebrochen.

\noindent Es wird in einem Bezugssystem gearbeitet, in dem das
Nukleon, bevor es zerf"allt, im Ursprung des Systems in Ruhe ist. 
Die Spineinstellungen $\uparrow \; \equiv +1/2$ beziehungsweise 
$\downarrow \; \equiv -1/2$ werden
relativ zur $z$-Achse angegeben. Der Zerfall f"uhrt stets auf ein
Antilepton, welches sich in positive  $z$-Richtung bewegt, w"ahrend
das Meson einen R"ucksto"s in Richtung der negativen $z$-Achse
erh"alt. Das entstehende Antilepton wird als relativistisch angenommen
und es gilt $l^{\scr C}_{\scr R} \equiv l^{\scr C}_{} \! \uparrow $ und
$l^{\scr C}_{\scr L} \equiv l^{\scr C}_{} \! \downarrow $. 

Die Vorgehensweise zur Berechnung der Matrixelemente soll kurz am Beispiel von
$\langle \epi|\mathcal{L}_{\scr \textrm{ef\/f}}|p \rangle$ erl"au\-tert
werden, f"ur Details sei auf \cite{karl} verwiesen. Aufgrund der
Drehimpulserhaltung l"a"st sich die "Ubergangswahrscheinlichkeit
zun"achst gem"a"s 
\begin{equation} \label{matre}
|\langle \epi|\mathcal{L}_{\scr \textrm{ef\/f}}|p \rangle|^2 
\; \longrightarrow \; 
|\langle e_{\scr R}^{\scr +} \pi^0|\mathcal{L}_{\scr \textrm{ef\/f}}
|p \! \uparrow \rangle|^2 \; + \;
|\langle e_{\scr L}^{\scr +} \pi^0|\mathcal{L}_{\scr \textrm{ef\/f}}
|p \! \downarrow \rangle|^2
\end{equation}
zerlegen. Eine weitere Zerlegung ist m"oglich, indem der Zustand 
$|\pi^0\rangle$ durch den Spin-, Flavour- und Farbanteil seiner 
Wellenfunktion ausgedr"uckt wird:
\begin{equation} \label{piosf}
|\pi^0 \rangle \; = \; 1/\sqrt{12} \, \delta_{\scr \rho \sigma} \, |
u^{\scr C \rho} \!\! \uparrow u^{\scr \sigma} \! \downarrow 
- u^{\scr C \rho} \!\! \downarrow u^{\scr \sigma} \! \uparrow
- d^{\scr C \rho} \!\! \uparrow d^{\scr \sigma} \! \downarrow 
+ d^{\scr C \rho} \!\! \downarrow d^{\scr \sigma} \! \uparrow \rangle
\end{equation}
Dadurch ergeben sich die Matrixelemente der f"ur den Zerfall 
$p \rightarrow \epi$ relevanten Elementarprozesse:
\begin{eqnarray} \nonumber
\langle e_{\scr R}^{\scr +} u^{\scr C} \!\! \uparrow 
u \! \downarrow |\mathcal{L}_{\scr \textrm{ef\/f}}
|p \! \uparrow \rangle & , &
\langle e_{\scr R}^{\scr +} u^{\scr C} \!\! \downarrow 
u \! \uparrow |\mathcal{L}_{\scr \textrm{ef\/f}}
|p \! \uparrow \rangle \; , \\ \label{mtrelp}
\langle e_{\scr R}^{\scr +} d^{\scr C} \!\! \uparrow 
d \! \downarrow |\mathcal{L}_{\scr \textrm{ef\/f}}
|p \! \uparrow \rangle & , &
\langle e_{\scr R}^{\scr +} d^{\scr C} \!\! \downarrow 
d \! \uparrow |\mathcal{L}_{\scr \textrm{ef\/f}}
|p \! \uparrow \rangle
\end{eqnarray}
Die Matrixelemente mit $e_{\scr L}^{\scr +}$ im Endzustand erh"alt man analog.
Auch das Proton kann gem"a"s
\begin{equation}
|p \! \uparrow \rangle \; = \; 1/\sqrt{18} \,
\varepsilon_{\scr \alpha \beta \gamma}
\, | ( \, u^{\scr \alpha} \! \uparrow d^{\scr \beta} \! \downarrow 
- u^{\scr \alpha} \! \downarrow d^{\scr \beta} \! \uparrow \, )
u^{\scr \gamma} \! \uparrow \rangle
\end{equation}
in seine Spin-, Flavour- und Farbanteile
zerlegt werden.
Die Ortsanteile der Wellenfunktionen werden sp"ater ber"ucksichtigt,
da sie in guter N"aherung f"ur alle Zerfallsprozesse gleich gro"s sind.
Die Spin-Flavour-Anteile der Mesonwellenfunktionen k"onnen
in Anhang \ref{wfm} gefunden werden. 

Weiterhin identif\/iziert man die f"ur die Bestimmung von (\ref{matre})
relevanten Terme in $\mathcal{L}_{\scr \textrm{ef\/f}}$ als
\begin{itemize}
\item $A^{}_2 \, \big( \varepsilon_{\scr \alpha \beta \gamma} 
\bar u_{\scr L}^{\scr C \gamma} \gamma^{\scr \mu} 
u_{\scr L}^{\scr \beta} \big) \big( \bar e^{\scr +}_{\scr R} 
\gamma_{\scr \mu} d_{\scr R}^{\scr \alpha} \big)$ f"ur
$\langle e_{\scr R}^{\scr +} \pi^0|\mathcal{L}_{\scr \textrm{ef\/f}}
|p \! \uparrow \rangle$
\item
$A^{}_1 \, \big( \varepsilon_{\scr \alpha \beta \gamma} 
\bar u_{\scr L}^{\scr C \gamma} \gamma^{\scr \mu}
u_{\scr L}^{\scr \beta} \big) \big( \bar e^{\scr +}_{\scr L} 
\gamma_{\scr \mu} d_{\scr L}^{\scr \alpha} \big)$ f"ur 
$\langle e_{\scr L}^{\scr +} \pi^0|\mathcal{L}_{\scr \textrm{ef\/f}}
|p \! \downarrow \rangle$
\end{itemize}
Im n"achsten Schritt werden nun alle Quark-
und Antiquarkfelder sowohl in den Zustandsfunktionen 
$|p \rangle$ und $\langle \pi^0|$ als auch in den
Lagrangedichtetermen in Form von Erzeugungs- und
Vernichtungsoperatoren sowie Dirac-Spinoren {\bf u} und {\bf v}
ausgedr"uckt. F"ur ein Quark $q$ beziehungsweise Antiquark $q^{\scr C}$ 
erh"alt man auf diese Weise
\begin{equation}
q \; \sim \; \hat a(q) {\bf u} + \hat b^\dagger(q) {\bf v} \; , \quad
q^{\scr C} \; \sim \; \hat a(q) {\bf v} + \hat b^\dagger(q) {\bf u}
\; ,
\end{equation}
wobei der Operator $\hat a(q)$ ein $q$-Quark vernichtet und der Operator 
$\hat b^\dagger(q)$ ein $q^{\scr C}$-Antiquark erzeugt; die Spinoren 
{\bf u} und {\bf v} mit 
\begin{equation}
(\slash{p} - m) \, {\bf u} \; = \; 0 \; , \quad 
(\slash{p} + m) \, {\bf v} \; = \; 0
\end{equation}
sind in \cite{bjo} def\/iniert. Man erh"alt auf diese Weise
umfangreiche Ausdr"ucke f"ur die Gr"o"sen in (\ref{mtrelp}), welche sich
jedoch betr"achtlich vereinfachen, wenn man unter Ausnutzung der
Antikom\-mu\-ta\-tor-Eigenschaften 
\begin{eqnarray}
& & \{\hat a(q),\hat a^\dagger(\tilde q)\} \; = \; 
\{\hat b(q),\hat b^\dagger(\tilde q)\} \; \sim \; \delta_{q \tilde q} \\
& & \{\hat a(q),\hat a(\tilde q)\} \; = \; 
\{\hat b(q),\hat b(\tilde q)\} \; = \; 
\{\hat a^\dagger(q),\hat a^\dagger(\tilde q)\} \; = \; 
\{\hat b^\dagger(q),\hat b^\dagger(\tilde q)\} \; = \; 0 \\
& & \{\hat a(q),\hat b(\tilde q)\} \; = \; 
\{\hat a^\dagger(q),\hat b(\tilde q)\} \; = \; 
\{\hat a(q),\hat b^\dagger(\tilde q)\} \; = \; 
\{\hat a^\dagger(q),\hat b^\dagger(\tilde q)\} \; = \; 0
\end{eqnarray}
die Vernichtungsoperatoren auf den Vakuumzustand $|0\rangle$ und die
Erzeugungsoperatoren auf den Zustand $\langle0|$ wirken l"a"st und
dadurch f"ur die meisten Terme Null erh"alt. In den
nichtverschwindenden Beitr"agen verbleiben die
Spinor-Amplituden, welche noch berechnet werden m"ussen.

Im "`recoil"'-Modell wird vom "Ubergang zweier statischer Quarks
derselben Masse $m$ in ein relativistisches Antilepton und ein
Antiquark, welches einen R"ucksto"simpuls $p$ mit $(p/m=3/4)$ besitzt,
ausgegangen. Die Spinoren der Antileptonen sind demnach durch
\begin{eqnarray}
l^{\scr C}_{\scr R} \; = \; \dfrac{1}{\sqrt{2}} \, \begin{pmatrix} 
\chi \! \uparrow \\ \chi \! \uparrow \end{pmatrix} & , &
l^{\scr C}_{\scr L} \; = \; \dfrac{1}{\sqrt{2}} \, \begin{pmatrix} 
\chi \! \downarrow \\ -\chi \! \downarrow \end{pmatrix}
\end{eqnarray}
und die der statischen Quarks durch
\begin{equation}
{\bf u} \! \uparrow \; = \; \begin{pmatrix} 
\chi \! \uparrow \\ 0 \end{pmatrix} \; , \quad
{\bf u} \! \downarrow \; = \; \begin{pmatrix} 
\chi \! \downarrow \\ 0 \end{pmatrix} \; , \quad
{\bf v} \! \uparrow \; = \; \begin{pmatrix} 
0 \\ -\chi \! \downarrow \end{pmatrix} \; , \quad
{\bf v} \! \downarrow \; = \; \begin{pmatrix} 
0 \\ \chi \! \uparrow \end{pmatrix}
\end{equation}
gegeben. Die nichtstatischen Antiquarks im Endzustand werden durch die
Spinoren  
\begin{eqnarray} \nonumber
{\bf u} \! \uparrow \; = \; \dfrac{1}{\sqrt{10}} \, \begin{pmatrix} 
3 \, \chi \! \uparrow \\ -\chi \! \uparrow \end{pmatrix} & , &
{\bf u} \! \downarrow \; = \; \dfrac{1}{\sqrt{10}} \, \begin{pmatrix} 
3 \, \chi \! \downarrow \\ \chi \! \downarrow \end{pmatrix} \; , \\
{\bf v} \! \uparrow \; = \; \dfrac{-1}{\sqrt{10}} \, \begin{pmatrix} 
\chi \! \downarrow \\ 3 \, \chi \! \downarrow \end{pmatrix} & , &
{\bf v} \! \downarrow \; = \; \dfrac{1}{\sqrt{10}} \, \begin{pmatrix} 
-\chi \! \uparrow \\ 3 \, \chi \! \uparrow \end{pmatrix}
\end{eqnarray}
beschrieben. Nun lassen sich die Spinor-Amplituden bestimmen; Tabelle
\ref{spamp} enth"alt alle nichtverschwindenden Amplituden und deren
Werte. 
\begin{table}[h]
\begin{center}
\begin{tabular}{|c||c|c|c|}
\hline
Amplitude & 
$(\bar {\bf u} \! \uparrow \gamma_\mu {\bf u} \! \uparrow)
(\bar l^{\scr C}_{\scr R} \gamma^\mu {\bf u}\!\uparrow)$ & 
$(\bar {\bf u} \! \downarrow \gamma_\mu {\bf u} \! \uparrow)
(\bar l^{\scr C}_{\scr R} \gamma^\mu {\bf u}\!\downarrow)$ & 
$(\bar {\bf v} \! \downarrow \gamma_\mu {\bf v} \! \downarrow)
(\bar l^{\scr C}_{\scr R} \gamma^\mu {\bf u}\!\uparrow)$ \\
\hline
Wert & $2/\sqrt{5}$ & $1/\sqrt{5}$ & $2/\sqrt{5}$ \\
\hline \hline
Amplitude & 
$(\bar {\bf v} \! \uparrow \gamma_\mu {\bf v} \! \downarrow)
(\bar l^{\scr C}_{\scr R} \gamma^\mu {\bf u}\!\downarrow)$ & 
$(\bar {\bf v} \! \uparrow \gamma_\mu {\bf u} \! \uparrow)
(\bar l^{\scr C}_{\scr R} \gamma^\mu {\bf v}\!\uparrow)$ & 
$(\bar {\bf v} \! \uparrow \gamma_\mu {\bf u} \! \uparrow)
(\bar l^{\scr C}_{\scr R} \gamma^\mu {\bf v}\!\downarrow)$ \\
\hline
Wert & $-2/\sqrt{5}$ & $-2/\sqrt{5}$ & $-1/\sqrt{5}$ \\
\hline
\end{tabular}
\end{center}
\caption{Spinor-Amplituden der hadronischen Matrixelemente \label{spamp}}
\end{table}
Die Amplituden zu den Zerf"allen mit linksh"andigen
Antileptonen im Endzustand erh"alt man durch Ersetzen von 
$\bar l^{\scr C}_{\scr R}$ mit $\bar l^{\scr C}_{\scr L}$ und
Vertauschen von $\uparrow$ und $\downarrow$.

Damit kann man die Matrixelemente (\ref{mtrelp}) berechnen, und aus
diesen lassen sich unter Ber"ucksichtigung der Spin-Flavour-Struktur
(\ref{piosf}) des Mesons wiederum die "Ubergangswahrscheinlichkeiten
in (\ref{matre}) bestimmen. Analog verf"ahrt man f"ur alle 
relevanten Zerfallskan"ale; die Tabelle \ref{aepnd} gibt die Resultate
f"ur die Amplituden der elementaren Zerfallsprozesse an, w"ahrend Tabelle
\ref{trprob} die "Ubergangswahrscheinlichkeiten f"ur die Zerf"alle mit
physikalischen Endzust"anden beinhaltet. Zu beachten ist die Tatsache,
da"s die Amplituden der einzelnen Elementarprozesse von
unterschiedlichen $A_i$-Koef\/f\/izienten gem"a"s (\ref{eldfnd})
begleitet werden.
\section{Berechnung der Zerfallsraten}
F"ur die partielle Zerfallsrate eines bestimmten Prozesses
$\textrm{Nukleon} \rightarrow \textrm{Meson} + \textrm{Antilepton}$
gilt die Beziehung
\begin{equation}
\Gamma^{}_j \; = \; \dfrac{1}{16\pi} \, m^2_{\scr \textrm{Nukl}} \,
\rho_{\scr j} \, |S|^2 \, |\mathcal{A}_{}|^2 \,
\Big( |\mathcal{A}^{}_{\scr L}|^2 \sum_l |A^{}_{l} \mathcal{M}^{}_{l}|^2
+ |\mathcal{A}^{}_{\scr R}|^2 \sum_r |A^{}_{r} \mathcal{M}^{}_{r}|^2 
\Big) \; , \label{decform} 
\end{equation}
welche man aus der allgemeinen Bestimmungsgleichung f"ur $\Gamma$ in
Zwei-Teilchen-Zerf"allen unter Ber"ucksichtigung der spezif\/ischen
Annahmen des "`recoil"'-Modells erh"alt. 
(\ref{decform}) ist "aquivalent zu der in \cite{gavel} gegebenen
Gleichung f"ur $\Gamma^{}_j$ und unterscheidet sich von dieser
lediglich darin, da"s der Phasenraumfaktor und die
"Ubergangsmatrixelemente aus \cite{karl} verwendet wurden, welche
direkt proportional zu den in \cite{gavel} benutzten sind. Weiterhin ist in
(\ref{decform}) die Abh"angigkeit von $\alpha_{\scr U}$ und den
Eichbosonmassen in den Koef\/f\/izienten $A_i$ aus (\ref{eldfnd}) enthalten. 
Die verschiedenen Gr"o"sen in (\ref{decform}) sind wie folgt 
def\/iniert:
\begin{itemize}
\item
$\mathcal{M}^{}_{l}$ und $\mathcal{M}^{}_{r}$ sind die f"ur den jeweiligen
Zerfallsproze"s relevanten hadronischen "Ubergangsmatrixelemente,
wobei sich die Indizes $l$ und $r$ auf die Chiralit"at des Antileptons
im Endzustand beziehen. Die Summation ber"ucksichtigt die Tatsache,
da"s f"ur einige Zerf"alle zwei Lagrangedichteterme verantwortlich
sind. Die Bestimmung der Matrixelemente war Gegenstand des letzten Abschnitts.
\item
$A^{}_{l}$ und $A^{}_{r}$ sind die zu $\mathcal{M}^{}_{l}$
beziehungsweise $\mathcal{M}^{}_{r}$ geh"orenden Koef\/f\/izienten aus 
der ef\/fektiven Lagrangedichte. Gem"a"s (\ref{eldfnd}) treten alle
Matrixelemente der elementaren Zerfallsprozesse in Kombination mit
einem bestimmten $A_i$-Koef\/f\/izient auf. Tabelle \ref{trprob} 
gibt die Werte f"ur die unabh"angigen $|A^{}_{r} \mathcal{M}^{}_{r}|^2$ 
an; alle dort nicht explizit aufgef"uhrten Gr"o"sen lassen sich durch
Symmetrie"uberlegungen aus diesen herleiten.
\item
$\mathcal{A}, \; \mathcal{A}^{}_{\scr L}$ und $\mathcal{A}^{}_{\scr  R}$
sind Faktoren, welche aus der Renormierung der Vier-Fermion-Operatoren
in der ef\/fektiven Lagrangedichte resultieren. W"ahrend n"amlich die 
Berechnung der hadronischen Matrixelemente $\mathcal{M}^{}_{r,l}$
f"ur Energieskalen $\mu \approx m_p$
durchgef"uhrt wird, ist die eigentliche baryonzahlverletzende
Wechselwirkung bei Skalen $\mu \gtrsim M^{}_{\scr U}$ wirksam. Das
f"uhrt zu Renormierungsef\/fekten, welche auf Strahlungskorrekturen zu
den Vier-Fermion-Operatoren durch die $G_{\scr \textrm{PS}}$- und 
SM-Eichbosonen beruhen. Eine formale Analyse dieses Ph"anomens
im Rahmen einer $SU(5)$-GUT ist in \cite{dan} zu f\/inden, w"ahrend 
\cite{bucc2} die Renormierungsef\/fekte f"ur den Fall einer 
$SO(10)$-Theorie untersucht. F"ur das in dieser Arbeit untersuchte
Modell sind die Faktoren durch
\begin{eqnarray}
\mathcal{A}^{}_{\scr L} & = & \Big( 
\dfrac{\alpha^{}_{\scr 1}(M^{}_{\scr Z})}{\alpha^{}_{\scr 1}(M^{}_{\scr I})} 
\Big)_{}^{-\frac{23}{82}} \\
\mathcal{A}^{}_{\scr R} & = & \Big( 
\dfrac{\alpha^{}_{\scr 1}(M^{}_{\scr Z})}{\alpha^{}_{\scr 1}(M^{}_{\scr I})} 
\Big)_{}^{-\frac{11}{82}} \\ \nonumber
\mathcal{A} \hspace{0.2cm} & = & \Big( 
\dfrac{\alpha^{}_{\scr 4C}(M^{}_{\scr I})}{\alpha^{}_{\scr 4C}(M^{}_{\scr U})} 
\Big)_{}^{-\frac{5}{8}}
\Big( \dfrac{\alpha^{}_{\scr 2L}(M^{}_{\scr I})}{\alpha^{}_{\scr 2L}
(M^{}_{\scr U})} \Big)_{}^{-\frac{27}{100}}
\Big( \dfrac{\alpha^{}_{\scr 2R}(M^{}_{\scr I})}{\alpha^{}_{\scr 2R}
(M^{}_{\scr U})} \Big)_{}^{-\frac{3}{20}} 
\Big( \dfrac{\alpha^{}_{\scr 2}(M^{}_{\scr Z})}{\alpha^{}_{\scr 2}
(M^{}_{\scr I})} \Big)_{}^{\frac{27}{38}}
\\ & &  \hspace{-0.16cm} \cdot \,
\Big( \dfrac{\alpha^{}_{\scr 3}(M^{}_{\scr Z})}{\alpha^{}_{\scr 3}
(M^{}_{\scr I})} \Big)_{}^{\frac{2}{7}}
\Big( \dfrac{\alpha^{}_{\scr 3}(m^{}_b)}
{\alpha^{}_{\scr 3}(M^{}_{\scr Z})} \Big)_{}^{\frac{6}{23}}
\Big( \dfrac{\alpha^{}_{\scr 3}(m^{}_c)}
{\alpha^{}_{\scr 3}(m^{}_b)} \Big)_{}^{\frac{6}{25}}
\Big( \dfrac{\alpha^{}_{\scr 3}(1 \; {\textrm{GeV}})}
{\alpha^{}_{\scr 3}(m^{}_c)} \Big)_{}^{\frac{2}{9}}
\end{eqnarray}
gegeben. In die Exponenten gehen die anomalen Dimensionen der
Vier-Fermion-Operatoren und die f"uhrenden Koef\/f\/izienten der
jeweiligen Eichkopplungs-$\beta$-\-Funk\-tionen ein.
Die Eichkopplungen bei $M^{}_{\scr Z}$ sind in Tabelle
\ref{vwsmk} zu f\/inden, diejenigen bei $M^{}_{\scr U}$ und $M^{}_{\scr I}$
wurden im vorigen Kapitel bestimmt; Anhang \ref{sukfps} 
enth"alt die numerischen Resultate. Die Werte von
$\alpha^{}_{\scr 3}$ bei Energien $\mu < M^{}_{\scr Z}$ sind 
\cite{buras2} entnommen:
\begin{equation}
\alpha^{}_{\scr 3}(1 \; {\textrm{GeV}}) \; = \; 0.544 \; , \quad
\alpha^{}_{\scr 3}(m^{}_c) \; = \; 0.412 \; , \quad
\alpha^{}_{\scr 3}(m^{}_b) \; = \; 0.226
\end{equation}
Der Einf\/lu"s der Skalenabh"angigkeit von 
$\alpha^{}_{\scr \textrm{em}}$ wird vernachl"assigt, da er
viel kleiner als der von $\alpha^{}_{\scr 3}$ ist 
($\alpha^{-1}_{\scr \textrm{em}}(M^{}_{\scr Z}) \approx 129$, 
$\alpha^{-1}_{\scr \textrm{em}}(0) \approx 137$). 
Unter Verwendung aller Kopplungswerte erh"alt man
\begin{equation}
|\mathcal{A}^{}_{\scr L}|^2 \; = \; 1.155 \; , \quad
|\mathcal{A}^{}_{\scr R}|^2 \; = \; 1.071 \; , \quad
|\mathcal{A}|^2 \; = \; 23.59
\end{equation}
\item
$|S|_{}^2 = \langle 
\Psi^s_{\scr \textrm{Nucl}}
(\vec r^{}_{\scr 1},\vec r^{}_{\scr 2},\vec r^{}_{\scr 3}) \, | \, 
\delta(\vec r^{}_{\scr 1}-\vec r^{}_{\scr 2}) \, | \,
\Psi^s_{\scr \textrm{Nucl}}
(\vec r^{}_{\scr 1},\vec r^{}_{\scr 2},\vec r^{}_{\scr 3})
\rangle$ 
steht f"ur die Wahrscheinlichkeit, zwei Valenzquarks des Nukleons in
einem Raumpunkt zu f\/inden. Das ist erforderlich, damit der
Elementarproze"s $qq \rightarrow q_{}^c l_{}^c$ stattf\/inden kann, da
die zugrundeliegende Wechselwirkung extrem kurzreichweitig ist.
$\Psi^s_{\scr \textrm{Nucl}}$ ist der Raumanteil der Wellenfunktion
des Nukleons. Der r"aumliche Anteil der Mesonwellenfunktion wird
wie in \cite{gavel} vernachl"assigt; die Spin-, Flavour- und
Farbanteile der Wellenfunktionen sind bereits in die Berechnung von 
$\mathcal{M}^{}_{r,l}$ eingegangen.
F"ur $|S|_{}^2$ wird der Wert 0.012 GeV$^3$ verwendet \cite{kroll}.
\item
$\rho_{\scr j} \equiv (1-\chi_j^2)(1-\chi_j^4)$ mit
$\chi_j^{}=m^{}_{\scr \textrm{Meson}}/m^{}_{\scr \textrm{Nukl}}$
ist ein Phasenraumfaktor, welcher die $SU(6)$-Spin-Flavour-Symmetrie
des im letzten Abschnitt verwendeten Quarkmodells explizit bricht. Er
ber"ucksichtigt die in der Natur existierenden unterschiedlichen
Massen der Mesonen in den
Endzust"anden der Zerfallsprozesse und ihren Einf\/lu"s auf die
Raten. Tabelle \ref{phspf} enth"alt die relevanten Werte von
$\rho_{\scr p}$ und $\rho_{\scr n}$.
\end{itemize}
Damit sind alle Gr"o"sen in (\ref{decform}) bekannt und die partiellen
sowie totalen Zerfallsraten der Nukleonen k"onnen berechnet
werden. S"amtliche Resultate sowohl f"ur die untersuchten Modelle als
auch f"ur den Fall verschwindender Fermionmischungen sind in den
Tabellen \ref{pdrt}-\ref{tdrnt} im Anhang zusammengefa"st. 

Man erkennt die generelle Tendenz, da"s der Einf\/lu"s der Mischungen zu 
einer deutlichen Unterdr"uckung des Zerfallskanals 
$p,n \rightarrow e^{\scr +} X$ f"uhrt, w"ahrend die Kan"ale 
$p,n \rightarrow \mu^{\scr +} X$ und $p,n \rightarrow \nu^{\scr C} X$ 
im Vergleich zum Fall verschwindender Mischungen bevorzugt werden. 
Besonders auf\/f"allig ist das bei den Zerf"allen 
$p,n \rightarrow \mu^{\scr +} \pi$, $p,n \rightarrow \nu^{\scr C} K$
und $p \rightarrow \mu^{\scr +} \omega$. Insofern unterscheiden sich die
Verzweigungsraten zum Teil betr"achtlich von denen nichtsupersymmetrischer
GUT-Modelle mit kleinen Mischungen, welche eine deutliche Dominanz von 
$p,n \rightarrow e^{\scr +} X$-Zerf"allen vorhersagen. SUSY-GUTs
dagegen bevorzugen Zerf"alle mit $K$-Mesonen im Endzustand \cite{luch}.

Zwischen den Verzweigungsraten der drei analysierten Modelle
bestehen teilweise deutlich ausgepr"agte Unterschiede, was man auch 
an den in Tabelle \ref{rocdr} dargestellten Verh"altnissen partieller 
Raten erkennen kann.
\begin{table}[h]
\begin{center}
\begin{tabular}{|c|c|c|c|c|}
\hline
Gr"o"se & ohne Mischungen & \; Modell\,1 \; & \; Modell\,2a \; 
& \; Modell\,2b \; \\
\hline \hline
$\dfrac{\Gamma(p \rightarrow \ek)}{\Gamma(p \rightarrow \epi)}$ 
& 0 & 0.145 & 0.104 & 0.162 \\ \hline
$\dfrac{\Gamma(p \rightarrow \mpi)}{\Gamma(p \rightarrow \muk)}$ 
& 0 & 3.27 \; & 6.33 \; & 3.11 \; \\ \hline
$\dfrac{\Gamma(p \rightarrow \nu^{\scr C}_{} K^+)}
{\Gamma(p \rightarrow \nu^{\scr C}_{} \pi^+)}$ 
& 0.003 & 0.098 & 0.064 & 0.157 \\ \hline
$\dfrac{\Gamma(p \rightarrow \epi)}
{\Gamma(p \rightarrow \nu^{\scr C}_{} \pi^+)}$ 
& 1.040 & 0.618 & 0.693 & 0.993 \\
\hline \hline
$\dfrac{\Gamma(n \rightarrow \mpin)}{\Gamma(n \rightarrow \epin)}$ 
& 0 & 0.394 & 0.225 & 0.200 \\ \hline
$\dfrac{\Gamma(n \rightarrow \mron)}{\Gamma(n \rightarrow \eron)}$ 
& 0 & 0.387 & 0.225 & 0.195 \\ \hline
$\dfrac{\Gamma(n \rightarrow \nu^{\scr C}_{} K^0)}
{\Gamma(n \rightarrow \nu^{\scr C}_{} \pi^0)}$ 
& 0.113 & 0.469 & 0.347 & 0.722 \\ \hline
$\dfrac{\Gamma(n \rightarrow \epin)}
{\Gamma(n \rightarrow \nu^{\scr C}_{} \pi^0)}$ 
& 4.16 \; & 2.48 \; & 2.78 \; & 3.96 \; \\
\hline
\end{tabular}
\end{center}
\caption{Verh"altnisse ausgew"ahlter partieller Nukleonzerfallsraten 
\label{rocdr}}
\end{table}
W"ahrend die Modelle 2a und 2b anhand ihrer Eigenschaften im
Neutrinosektor praktisch nicht unterscheidbar waren, zeigen sich in
ihren Verzweigungsraten Dif\/ferenzen, welche durchaus experimentell
zug"anglich sein sollten, sofern in Zukunft Nukleonenzerf"alle 
beobachtet werden. Die totalen Zerfallsraten der einzelnen Modelle 
sind nahezu gleich und f"uhren zu Lebensdauern im Bereich von 
$\tau_{p,n} \approx (3-4) \cdot 10^{34}$ Jahren.

Die in den Tabellen \ref{pdrt}-\ref{tdrnt} dargelegten Resultate
dieser Untersuchung sind echte Vorhersagen des verwendeten
$SO(10)$-Massenmodells und verdeutlichen auf eindrucksvolle Weise den
Einf\/lu"s der Fermionmischungen auf die Verzweigungsraten der
Nukleonenzerf"alle.

Bei allen Rechnungen ist stets mit den Mittelwerten der verwendeten Gr"o"sen
gearbeitet worden, ohne deren Fehler explizit zu ber"ucksichtigen. Das
ist insofern begr"undet, als die Hauptquellen der auftretenden Unsicherheiten
in den Einf\/l"ussen der Schwellenkorrekturen und den modellspezif\/ischen
N"aherungen bei der Bestimmung der hadronischen Matrixelemente zu
sehen sind. Gerade diese sind aber im Gegensatz zu den bekannten
Me"sungenauigkeiten in den SM-Kopplungen und den Fermionmassen und
-mischungen nur schwer abzusch"atzen. F"ur die Schwellenef\/fekte und
deren Einlu"s auf die Werte der Symmetriebrechungsskalen ist
das in Abschnitt \ref{thresh} versucht worden, w"ahrend \cite{lang3}
die Auswirkung der Unsicherheit in den "Ubergangsmatrixelementen auf die
Lebensdauer des Protons mit einem Faktor $10^{\pm 0.7}$ ansetzt. 
Es kann allerdings davon ausgegangen werden, da"s die Ef\/fekte dieser 
Unsicherheiten in erster Linie die totalen Zerfallsraten beeinf\/lussen, 
die Verzweigungsraten dagegen deutlich weniger betref\/fen. Letztere
stellen aber die wirklich relevanten Vorhersagen des Massenmodells dar.
\section{Experimenteller Status}
Es bleibt zu pr"ufen, ob die im letzten Abschnitt gemachten
Vorhersagen bez"uglich der partiellen und totalen Zerfallsraten der
Nukleonen mit den experimentellen Grenzen f"ur diese Gr"o"sen
vertr"aglich sind. 

Die Experimente, welche nach Nukleonenzerf"allen suchen, lassen sich in
zwei Kategorien einteilen. Wasser-\v Cerenkovdetektoren wie IMB\,3
\cite{imb} und Kamiokande beziehungsweise Super-Kamiokande \cite{skam} 
sind besonders zum Nachweis von $p,n \rightarrow e^{\scr +} \pi$ 
geeignet, w"ahrend die
Eisenkalorimeter-Spurdetektoren wie Fr\'ejus \cite{frej} und Soudan\,2 
\cite{soud} f"ur Zerf"alle mit $K$-Mesonen und $\mu^+$ im Endzustand 
sensitiv sind. Keines der Experimente hat bis heute eindeutige Hinweise auf
die Instabilit"at der Nukleonen liefern k"onnen, so da"s lediglich
obere Grenzen f"ur die Zerfallsraten existieren. 
\begin{table}[h]
\begin{center}
\begin{tabular}{|l|c|c|}
\hline
Zerfallsproze"s & Super-Kamiokande \cite{skpd} &
Part.\,Data\,Group\,1998 \cite{pdg} \\
\hline \hline
$p \; \rightarrow \; \epi$ & 29 \hspace{0.33cm} & 5.5 \, \\
$p \; \rightarrow \; \et$  & 11 \hspace{0.33cm} & 1.4 \, \\
$p \; \rightarrow \; \mpi$ & 23 \hspace{0.33cm} & 2.7 \, \\
$p \; \rightarrow \; \muk$ & 4.0 & 1.2 \, \\
$p \; \rightarrow \; \mt$  & 7.8 & 0.69 \\
$p \; \rightarrow \; \nu^{\scr C}_{} K^+$ & 6.8 & 1.0 \, \\
$n \; \rightarrow \; \nu^{\scr C}_{} \eta$ & 5.6 & 0.54 \\
\hline
\end{tabular}
\end{center}
\caption[Aktuelle Resultate von Super-Kamiokande f"ur die
unteren Grenzen der Nukleonlebensdauern]
{Aktuelle Resultate von Super-Kamiokande f"ur die
unteren Grenzen der Nukleonlebensdauern in einigen Zerfallskan"alen 
(alle Werte in $10^{32}$ Jahren) \label{ndsumt}}
\end{table}
Die in \cite{pdg} angegebenen Werte f"ur diese Grenzen sind in 
Tabelle \ref{exptab} aufgef"uhrt, seitdem ver"of\/fentlichte Resultate 
von Super-Kamiokande zeigt Tabelle \ref{ndsumt}. Der Vergleich mit den
Modellvorhersagen in den Tabellen \ref{pdrt}-\ref{tdrnt} f"uhrt zu dem
Schlu"s, da"s diese mit allen experimentellen Grenzen vertr"aglich sind.

Die Grenzwerte f"ur die Lebensdauern bez"uglich bestimmter 
Zerfallsprozesse wie zum Beispiel $p \rightarrow \epi,\mpi$ erreichen
gem"a"s Tabelle \ref{ndsumt} bereits den Bereich einiger 
$10^{33}$ Jahre, was lediglich eine Gr"o"senordnung unter den in
dieser Arbeit gemachten Vorhersagen
liegt. Letztere werden also in absehbarer Zeit direkt "uberpr"ufbar
sein, da Super-Kamiokande weiterhin Daten aufnimmt und Experimente wie
das auf einem Detektor aus f\/l"ussigem Argon basierende ICARUS
\cite{icar} neu hinzukommen werden.

\newpage

\mbox{ }

%%% Local Variables: 
%%% mode: latex
%%% TeX-master: t
%%% End: 

\chapter*{Zusammenfassung und Ausblick}
\addcontentsline{toc}{chapter}{Zusammenfassung und Ausblick}

\pagestyle{myheadings}\markboth{ZUSAMMENFASSUNG UND AUSBLICK}{ZUSAMMENFASSUNG 
UND AUSBLICK}

\thispagestyle{plain}

Gegenstand dieser Untersuchung war ein Modell f"ur die fermionischen 
Massenmatrizen auf der Grundlage einer $SO(10)$-Theorie, welche 
"uber eine intermedi"are Pati-Salam-Symmetrie in das
Standardmodell gebrochen wird. Der gew"ahlte Ansatz bestand in einer
asymmetrischen "`Nearest Neighbour Interaction"'-Form der Matrizen,
die durch eine globale $U(1)$-Familiensymmetrie realisiert wird.
Dieser beinhaltet im Rahmen des Standardmodells keine physikalischen 
Konsequenzen, da die rechtsh"an\-digen Mischungen dort nicht beobachtbar sind.
In Theorien jenseits des Standardmodells jedoch haben alle Fermionmischungen 
betr"achtlichen Einf\/lu"s auf observable Gr"o"sen wie die Zerfallsraten der 
Nukleonen. 

Nach der systematischen Bestimmung der Symmetriebrechungsskalen und 
Eichkopplungen in Abh"angigkeit vom Teilcheninhalt des Modells sowie 
einer Absch"atzung der Schwellenef\/fekte wurden
die Massenmatrizen unter Ber"ucksichtigung der $SO(10)$-spezif\/i\-schen
Beziehungen zwischen diesen konstruiert. Anschlie"send sind
alle L"osungen des Modells numerisch bestimmt worden, welche neben den 
bekannten Parametern im Sektor der geladenen Fermionen auch 
ph"anomenologisch sinnvolle Neutrinoeigenschaften liefern. Es wurden 
insgesamt drei L"osungen des Modells gefunden, welche in der Lage sind, 
die Anomalien der Sonnen- und atmosph"arischen Neutrinos durch 
Oszillationen zu erkl"aren. Zwei davon deuten auf eine 
L"osung des Sonnen-Neutrinoproblems 
durch den MSW-Ef\/fekt mit kleiner Mischung hin, w"ahrend die dritte den
MSW-Ef\/fekt mit gro"ser Mischung beinhaltet. Bemerkenswert ist, da"s 
alle L"osungen auf leptonische (23)-Mischungen f"uhren, die im Rahmen 
des zur Erkl"arung der Anomalie atmosph"arischer Neutrinos erlaubten
Parameterbereichs vergleichsweise klein sind. Da die (23)-Mischung 
desjenigen Modells, welches den MSW-Ef\/fekt mit gro"ser Mischung
realisiert, nahe an der unteren Grenze des zul"assigen Bereichs liegt und
letzterer durch die Oszillationsexperimente in naher Zukunft weiter
eingeschr"ankt werden wird, sind die anderen beiden L"osungen als
realistischer zu betrachten. Im Rahmen des hier analysierten
Massenmodells erscheint eine L"osung des Sonnen-Neutrinoproblems durch 
den MSW-Ef\/fekt mit kleiner Mischung als wahrscheinlich.

Alle gefundenen L"osungen besitzen mehrere betragsm"a"sig gro"se
Mischungen im Sektor der geladenen Fermionen. Dabei handelt es sich
sowohl um im Standardmodell nicht observable rechtsh"andige als auch
um linksh"andige Mischungswinkel. Demnach kann man von der Gr"o"se
der CKM-Winkel nicht zwangsl"auf\/ig auf die der linksh"andigen
Fermionmischungen schlie"sen, wie es h"auf\/ig getan wird.
Die teilweise gro"sen Mischungen f"uhren zu charakteristischen 
Verzweigungsraten der Nukleonenzerf"alle, welche von denen
bei verschwindenden Mischungen deutlich abweichen. Generell l"a"st
sich eine Unterdr"uckung der Kan"ale mit $e^+$ im Endzustand
feststellen, w"ahrend Kan"ale mit $\mu^+$ und $\nu^{\scr C}$ bevorzugt 
werden. Die drei L"osungen sind aufgrund ihrer Vorhersagen f"ur die
Verzweigungsraten auch untereinander unterscheidbar. Die berechneten 
Lebensdauern der Nukleonen betragen
$\tau_{p,n} \approx (3-4) \cdot 10^{34}$ Jahre, was je nach
Zerfallskanal lediglich ein bis zwei Gr"o"senordnungen "uber den
experimentellen Grenzen liegt. Damit sind die Modellvorhersagen in
absehbarer Zeit direkt "uberpr"ufbar.

Der Einfachheit halber sind alle Massenmatrizen als reell angenommen
worden, da das Problem der $CP$-Verletzung nicht Gegenstand der
Arbeit war. Abgesehen von der Standardmethode, die $CP$-Verletzung
"uber komplexe Yukawa-Kopplungen zu realisieren, bieten Grand
Unif\/ied-Theorien aufgrund ihres erweiterten Eichboson- und
Higgs-Sektors daf"ur verschiedene weitere M"oglichkeiten \cite{ynir}.

Weiterhin wurde auf die Verwendung von Supersymmetrie verzichtet, da
supersymmetrische Theorien wegen des wesentlich umfangreicheren 
Teilchenspektrums zahlreiche zus"atzliche Parameter enthalten, 
deren Werte weitgehend
unbestimmt sind. Das hat naturgem"a"s negative Auswirkungen auf die 
Vorhersagekraft der Modelle. Auch eine f"ur die Anwendbarkeit des
See-Saw-Mechanismus vorteilhafte intermedi"are Skala im Bereich
$10^{10}-10^{12}$ GeV ist in supersymmetrischen Theorien nicht
auf nat"urliche Weise realisierbar, da sich die Eichkopplungen des 
supersymmetrischen Standardmodells im Gegensatz zum 
nichtsupersymmetrischen Fall genau in einem Punkt tref\/fen. 
Die Abwesenheit jeglicher experimenteller Hinweise
auf eine bei Energien im TeV-Bereich gebrochene Supersymmetrie in der 
Natur \cite{giud} ist ein weiterer Schwachpunkt, mit dem
supersymmetrische Theorien bisher behaftet sind.
Schlie"slich deutet auch die Existenz nichtsupersymmetrischer 
Niederenergie-Vakua in der M-Theorie \cite{iban} darauf hin, da"s
Supersymmetrie keine zwingende Eigenschaft von Theorien jenseits des
Standardmodells ist.

Trotz dieser Schw"achen im ph"anomenologischen Bereich ist die
Supersymmetrie ein formal "uberaus ansprechendes Konzept, und es 
w"are interessant, das hier diskutierte Massenmodell in eine
supersymmetrische Grand Unif\/ied-Theorie einzubetten und die daraus
resultierenden Konsequenzen zu untersuchen. Auch eine Analyse des
Modells hinsichtlich seiner M"oglichkeiten zur Erkl"arung der
$CP$-Verletzung und der Baryonasymmetrie des Universums w"are
sicherlich sinnvoll.

Auf die generellen Probleme von nichtsupersymmetrischen Grand 
Unif\/ied-Theorien wurde in Kapitel\,2 eingegangen. Das 
Hierarchieproblem und die Nichtber"ucksichtigung der Gravitation sind 
wohl die fundamentalsten Schwachpunkte dieser Modelle.
Das Hierarchieproblem kann ebenso wie das Problem der Divergenzen 
zumindest prinzipiell durch eine bei Energien von etwa 1 TeV gebrochene 
globale Supersymmetrie gel"ost werden. Bei solchen Modellen stellt
sich aber die Frage, auf welche Weise die Supersymmetrie gebrochen
wird. Sowohl die spontane Brechung als auch die explizite Brechung
durch Lagrangedichteterme deuten letztendlich auf eine noch
fundamentalere Theorie hin. Ein Kandidat f"ur eine solche ist die
Supergravitationstheorie, welche auf einer lokalen Supersymmetrie
basiert und die Gravitationswechselwirkung mit einschlie"st. Auf 
Supergravitation basierende Modelle sind allerdings nicht renormierbar
und k"onnen wiederum nur ef\/fektive N"aherungen einer grundlegenden
Theorie sein. Den zur Zeit zweifellos popul"arsten und vielleicht sogar
einzigen ernstzunehmenden Ansatz daf"ur stellen die
Superstringtheorien und das "ubergeordnete Modell der M-Theorie dar. 
Sie werden in zehn beziehungsweise elf Raumzeitdimensionen formuliert,
was eine Kompaktif\/izierung der "uberz"ahligen Dimensionen notwendig
macht, um vierdimensionale ef\/fektive Theorien und m"oglicherweise
"uberpr"ufbare Voraussagen erhalten zu k"onnen. In der "au"serst
eingeschr"ankten Vorhersagekraft f"ur den Bereich experimentell
zug"anglicher Energien besteht die Schw"ache dieser Modelle.

Es ist bis heute keine wirklich fundamentale Theorie bekannt, welche die
Eigenschaften der Fermionen erkl"art. Aus diesem Grunde kommt der
Untersuchung ef\/fektiver Massenmodelle nach wie vor gro"se Bedeutung
zu, da sie vielleicht einen Hinweis auf die zugrundeliegende Symmetrie
der Natur liefern.

\newpage

\mbox{}

\newpage

\pagestyle{headings}

\thispagestyle{plain}

\begin{appendix}
\chapter{Gruppentheoretischer Anhang} \label{gtapp}
Die in diesem Anhang zusammengefa"sten Eigenschaften ausgew"ahlter 
irreduzibler Darstellungen sind den Tabellen in \cite{slans} entnommen; 
dort wird ferner eine detaillierte Beschreibung des f"ur Anwendungen
in der Elementarteilchenphysik relevanten gruppentheoretischen Formalismus 
gegeben.
\section{Casimir-Operator und Dynkin-Index}
Die Generatoren einer Lie-Algebra erf"ullen unabh"angig von der
expliziten Darstellung die Vertauschungsrelationen
\begin{equation}
[ \, T_a,T_b \, ] \; = \; i \, f_{abc} \, T_c \quad ,
\end{equation}
wobei die $f_{abc}$ die Strukturkonstanten der Algebra sind. Der
quadratische Casimir-Opera\-tor $\hat{C}_2 \equiv \sum_a T_a T_a$ kommutiert
mit allen Generatoren und ist somit bez"uglich jeder irreduziblen Darstellung
${\bf \mathcal{R}}$ der Liegruppe $G$ ein Vielfaches der Einheitsmatrix; der
zugeh"orige Eigenwert wird mit $C_2({\bf \mathcal{R}})$ bezeichnet:
\begin{equation}
\hat{C}_2({\bf \mathcal{R}}) \, {\bf \mathcal{R}} \; = \; 
C_2({\bf \mathcal{R}}) \, {\bf \mathcal{R}}
\end{equation}
Der Dynkin-Index $S_2({\bf \mathcal{R}})$ einer irreduziblen Darstellung 
${\bf \mathcal{R}}$ ist "uber
\begin{equation}
\textrm{Tr}^{}_{{\bf \mathcal{R}}}(\,T_a T_b\,) \; = \; 
S_2({\bf \mathcal{R}}) \, \delta_{ab}
\end{equation}
def\/iniert. Zwischen $C_2({\bf \mathcal{R}})$ und $S_2({\bf \mathcal{R}})$ 
gilt die Beziehung 
\begin{equation}
d({\bf \mathcal{G}}) \, S_2({\bf \mathcal{R}}) \; = \; 
d({\bf \mathcal{R}}) \, C_2({\bf \mathcal{R}}) \quad ,
\end{equation}
wobei $d({\bf \mathcal{R}})$ die Dimension von ${\bf \mathcal{R}}$ bezeichnet;
${\bf \mathcal{G}}$ steht f"ur die adjungierte Darstellung von $G$. F"ur die
$SU(N)$-Gruppen gelten die Normierungen 
$C_2({\bf \mathcal{G}}) = S_2({\bf \mathcal{G}}) = N$ und $S_2(N) = 1/2$,
w"ahrend f"ur die $U(1)_{\scr Y}$ $C_2({\bf \mathcal{R}}) = 
S_2({\bf \mathcal{R}}) = Y^2$ und $C_2({\bf \mathcal{G}}) = 
S_2({\bf \mathcal{G}}) = 0$ 
ist.

In der folgenden Tabelle sind $S_2({\bf \mathcal{R}})$ und 
$C_2({\bf \mathcal{R}})$ einiger irreduzibler Darstellungen verschiedener
Lie-Gruppen aufgelistet.
\begin{table}[h]
\begin{center}
\begin{tabular}{|r|r|r||r|r|r||r|r|r|}
\hline
${\bf \mathcal{R}}$ & $S_2({\bf \mathcal{R}})$ & $C_2({\bf \mathcal{R}})$ &
${\bf \mathcal{R}}$ & $S_2({\bf \mathcal{R}})$ & $C_2({\bf \mathcal{R}})$ &
${\bf \mathcal{R}}$ & $S_2({\bf \mathcal{R}})$ & $C_2({\bf \mathcal{R}})$ \\
\hline \hline
{\bf 10} & 1 & 9/2 & {\bf 4} & 1/2 & 15/8 & 
{\bf 3} & 1/2 & 4/3 \\
{\bf 16} & 2 & 45/8 & {\bf 6} & 1 & 5/2 & 
{\bf 6} & 5/2 & 10/3 \\
{\bf 45} & 8 & 8 & {\bf 10} & 3 & 9/2 & 
{\bf 8} & 3 & 3 \\
{\bf 54} & 12 & 10 & {\bf 15} & 4 & 4 & 
{\bf 10} & 15/2 & 6 \\
{\bf 120} & 28 & 21/2 & {\bf 20} & 13/2 & 39/8 & 
{\bf 15} & 10 & 16/3 \\
{\bf 126} & 35 & 25/2 & {\bf 20'}\!\!& 8 & 6 & & & \\
{\bf 144} & 34 & 85/8 & & & & & & \\
{\bf 210} & 56 & 12 & & & & & & \\
{\bf 210'}\!\!& 77 & 33/2 & & & & & & \\
\hline
\end{tabular}
\end{center}
\caption[Dynkin-Indizes und Eigenwerte des quadratischen
Casimir-Operators ausgew"ahlter irreduzibler Darstellungen]
{Dynkin-Indizes $S_2({\bf \mathcal{R}})$ und Eigenwerte
$C_2({\bf \mathcal{R}})$ des quadratischen Casimir-Operators f"ur
ausgew"ahlte irreduzible Darstellungen ${\bf \mathcal{R}}$ der einfachen 
Lie-Gruppen $SO(10)$, $SU(4)$ und $SU(3)$ (von links nach rechts) \label{grp}}
\end{table}
\section{Verzweigungsregeln f"ur $SO(10)$-Darstellungen} \label{brsor}
\begin{eqnarray} \nonumber
SO(10) & \supset & SU(4)_{\scr C} \otimes SU(2)_{\scr L} \otimes 
SU(2)_{\scr R} \\
{\bf 10} & = & ({\bf 1,2,2}) \oplus ({\bf 6,1,1}) \\
{\bf 16} & = & ({\bf 4,2,1}) \oplus ({\bf \bar 4,1,2}) \\
{\bf 45} & = & ({\bf 15,1,1}) \oplus ({\bf 1,3,1}) 
\oplus ({\bf 1,1,3}) \oplus ({\bf 6,2,2}) \\
{\bf 54} & = & ({\bf 1,1,1}) \oplus ({\bf 1,3,3})
\oplus ({\bf 20',1,1}) \oplus ({\bf 6,2,2}) \\ \nonumber
{\bf 120} & = & ({\bf 1,2,2}) \oplus ({\bf 10,1,1}) 
\oplus ({\bf \overline{10},1,1}) \oplus ({\bf 6,3,1}) \\
& & \oplus \; ({\bf 6,1,3}) \oplus ({\bf 15,2,2}) \\
{\bf 126} & = & ({\bf 6,1,1}) \oplus ({\bf \overline{10},3,1}) 
\oplus ({\bf 10,1,3}) \oplus ({\bf 15,2,2}) \\ \nonumber
{\bf 210} & = & ({\bf 1,1,1}) \oplus ({\bf 15,1,1}) 
\oplus ({\bf 6,2,2}) \oplus ({\bf 15,3,1}) 
\oplus ({\bf 15,1,3}) \\
& & \oplus \; ({\bf 10,2,2}) \oplus ({\bf \overline{10},2,2})
\end{eqnarray}
%
%Die Darstellungen {\bf 16} und {\bf 126} sind komplex.
%
\section{Verzweigungsregeln f"ur $SU(4)$-Darstellungen}
\begin{eqnarray} \nonumber
SU(4)_{\scr C} & \supset & SU(3)_{\scr C} \otimes U(1)_{\scr B-L} \\
{\bf 4} & = & {\bf 1}_{\scr -1} \oplus {\bf 3}_{\scr 1/3} \\
{\bf 6} & = & {\bf 3}_{\scr -2/3} \oplus {\bf \bar 3}_{\scr 2/3} \\
{\bf 10} & = & {\bf 1}_{\scr -2} \oplus {\bf 3}_{\scr -2/3} 
\oplus {\bf 6}_{\scr 2/3} \\
{\bf 15} & = & {\bf 1}_{\scr 0} \oplus {\bf 3}_{\scr 4/3} 
\oplus {\bf \bar 3}_{\scr -4/3} \oplus {\bf 8}_{\scr 0} \\
{\bf 20} & = & {\bf 3}_{\scr 1/3} \oplus {\bf \bar 3}_{\scr 5/3} 
\oplus {\bf \bar 6}_{\scr 1/3} \oplus {\bf 8}_{\scr -1}
\end{eqnarray}
%
%Die Darstellungen {\bf 4}, {\bf 10} und {\bf 20} sind komplex.
%
\section{Schwellenkorrektur-Koef\/f\/izienten} \label{threshnr}
Hier ist zu ber"ucksichtigen, da"s die $SO(10)$-Darstellungen {\bf 210} 
und {\bf 54} als reell, die an der Massenerzeugung beteiligten Darstellungen
{\bf 10}, {\bf 120} und {\bf 126} dagegen als komplex angenommen
werden. Hat die Darstellung einer Produktgruppe $G_1 \otimes G_2 \otimes G_3$ 
die Gestalt 
$({\bf {\bf \mathcal{R}}_1,{\bf \mathcal{R}}_2,{\bf \mathcal{R}}_3})$, so
sind die zugeh"origen Schwellenkorrektur-Koef\/f\/izienten durch
\begin{equation}
(\lambda_1,\lambda_2,\lambda_3) = \big( \,
d({\bf \mathcal{R}_2}) \, d({\bf \mathcal{R}_3}) \, S_2({\bf \mathcal{R}_1}),
d({\bf \mathcal{R}_1}) \, d({\bf \mathcal{R}_3}) \, S_2({\bf \mathcal{R}_2}),
d({\bf \mathcal{R}_1}) \, d({\bf \mathcal{R}_2}) \, S_2({\bf \mathcal{R}_3})
\, \big)
\end{equation}
gegeben. Bei komplexen Darstellungen kommt noch ein Faktor 2 hinzu.

\noindent F"ur die Berechnung der $\lambda_{\scr 1Y}^{\scr I}$ ist die
korrekte GUT-Normierung der Hyperladung verwendet worden, was die
$\sqrt{3/5}$-Faktoren erkl"art.
\begin{table}[h]
\begin{center}
\begin{tabular}{|c|c|c|}
\hline
$SO(10)$-Darstellung & $G_{\scr \textrm{PS}}$-Darstellung & 
$(\lambda_{\scr 4C}^{\scr U},\lambda_{\scr 2L}^{\scr U},
\lambda_{\scr 2R}^{\scr U})$ \\ \hline \hline
{\bf 210} & ({\bf 1,1,1}) & (0,0,0) \\
& ({\bf 15,1,1}) & (4,0,0) \\
& ({\bf 6,2,2}) & (4,6,6) \\
& ({\bf 15,3,1}) & (12,30,0) \\
& ({\bf 15,1,3}) & (12,0,30) \\
& ({\bf 10,2,2}) & (12,10,10) \\
& ({$\bf \overline{10}$}{\bf,2,2}) & (12,10,10) \\
\hline
{\bf 54} & ({\bf 1,1,1}) & (0,0,0) \\
& ({\bf 1,3,3}) & (0,6,6) \\
& ({\bf 20',1,1}) & (8,0,0) \\
& ({\bf 6,2,2}) & (4,6,6) \\
\hline
{\bf 126} & ({\bf 6,1,1}) & (2,0,0) \\
& ({$\bf \overline{10}$}{\bf,3,1}) & (18,40,0) \\
& ({\bf 10,1,3}) & (18,0,40) \\
& ({\bf 15,2,2}) & (32,30,30) \\
\hline
{\bf 120} & ({\bf 1,2,2}) & (0,2,2) \\
& ({\bf 10,1,1}) & (6,0,0) \\
& ({$\bf \overline{10}$}{\bf,1,1}) & (6,0,0) \\
& ({\bf 6,3,1}) & (6,24,0) \\
& ({\bf 6,1,3}) & (6,0,24) \\
& ({\bf 15,2,2}) & (32,30,30) \\
\hline
{\bf 10} & ({\bf 1,2,2}) & (0,2,2) \\
& ({\bf 6,1,1}) & (2,0,0) \\
\hline
\end{tabular}
\end{center}
\caption{Koef\/f\/izienten f"ur Schwellenkorrekturen durch Higgs-Teilchen
  bei $M_{\scr U}$}
\end{table}

\begin{table}[h]
\begin{center}
\begin{tabular}{|c|c|c|}
\hline
$G_{\scr \textrm{PS}}$-Darstellung & $G_{\scr \textrm{SM}}$-Darstellung & 
$(\lambda_{\scr 3C}^{\scr I},\lambda_{\scr 2L}^{\scr I},
\lambda_{\scr 1Y}^{\scr I})$ \\ \hline \hline
({\bf 1,2,2}) & ({\bf 1,2,}$+1/2\sqrt{3/5}$) & (0,1,3/5) \\
& ({\bf 1,2,}$-1/2\sqrt{3/5}$) & (0,1,3/5) \\
({\bf 15,2,2}) & ({\bf 1,2,}$+1/2\sqrt{3/5}$) & (0,1,3/5) \\
& ({\bf 1,2,}$-1/2\sqrt{3/5}$) & (0,1,3/5) \\
& ({\bf 3,2,}$+7/6\sqrt{3/5}$) & (2,3,49/5) \\
& ({\bf 3,2,}$+1/6\sqrt{3/5}$) & (2,3,1/5) \\
& ({\bf \=3,2,}$-1/6\sqrt{3/5}$) & (2,3,1/5) \\
& ({\bf \=3,2,}$-7/6\sqrt{3/5}$) & (2,3,49/5) \\
& ({\bf 8,2,}$+1/2\sqrt{3/5}$) & (12,8,24/5) \\
& ({\bf 8,2,}$-1/2\sqrt{3/5}$) & (12,8,24/5) \\
({$\bf \overline{10}$}{\bf,3,1}) & ({\bf 1,3,}$+\sqrt{3/5}$) & (0,4,18/5) \\
& ({\bf \=3,3,}$+1/3\sqrt{3/5}$) & (3,12,6/5) \\
& ({\bf \=6,3,}$-1/3\sqrt{3/5}$) & (15,24,12/5) \\
({\bf 10,1,3}) & ({\bf 1,1,}0) & (0,0,0) \\
& ({\bf 1,1,}$-\sqrt{3/5}$) & (0,0,6/5) \\
& ({\bf 1,1,}$-2\sqrt{3/5}$) & (0,0,24/5) \\
& ({\bf 3,1,}$+2/3\sqrt{3/5}$) & (1,0,8/5) \\
& ({\bf 3,1,}$-1/3\sqrt{3/5}$) & (1,0,2/5) \\
& ({\bf 3,1,}$-4/3\sqrt{3/5}$) & (1,0,32/5) \\
& ({\bf 6,1,}$+4/3\sqrt{3/5}$) & (5,0,64/5) \\
& ({\bf 6,1,}$+1/3\sqrt{3/5}$) & (5,0,4/5) \\
& ({\bf 6,1,}$-2/3\sqrt{3/5}$) & (5,0,16/5) \\
\hline
\end{tabular}
\end{center}
\caption{Koef\/f\/izienten f"ur Schwellenkorrekturen durch Higgs-Teilchen
  bei $M_{\scr I}$}
\end{table}
\vspace*{5cm}
\newpage
\chapter{Renormierungsgruppengleichungen} \label{rge}
In \cite{Machvgn} sind allgemeine Ausdr"ucke f"ur die
Renormierungsgruppengleichungen von Quantenfeldtheorien in 
Zweischleifenordnung entwickelt worden, aus denen sich auch die
hier verwendeten Gleichungen herleiten lassen. Die 
Renormierungsgruppengleichungen f"ur das Standardmodell sind 
in \cite{rger} und die der See-Saw-Massenmatrix der Neutrinos 
in \cite{chanpl} ausf"uhrlich analysiert worden.
\section{Standardmodell} \label{rgeA}
Alle Gleichungen bis auf die der See-Saw-Massenmatrix sind in
Zweischleifenordnung angegeben.
Als Variable wird im folgenden $t = \ln(\mu/M^{}_{\scr Z})$ verwendet,
$n_{\scr H}$ gibt die Anzahl der komplexen Higgs-Doubletts an (in
dieser Arbeit wird stets $n_{\scr H}=1$ angenommen).
Die Def\/inition der Yukawa-Kopplungsmatrizen lautet 
\begin{equation}
{\bf Y}^{}_i(\mu) = \dfrac{\sqrt{2}}{\upsilon} \; {\bf M}^{}_i(\mu) \quad
(i=u,d,e) \; ,
\end{equation}
wobei ${\bf M}^{}_i(\mu)$ die skalenabh"angigen fermionischen
Massenmatrizen bezeichnet; $\upsilon/\sqrt{2} = 174.1 \; \textrm{GeV}$
ist der Vakuumerwartungswert des Higgs-Feldes. Desweiteren werden die
nachfolgend aufgef"uhrten Def\/initionen benutzt:
\begin{eqnarray}
{\bf H}^{}_{i} & = & {\bf Y}^{}_i {\bf Y}^{\dagger}_i \quad
(i=u,d,e)
\\
Y^{}_2 & = & \textrm{Tr} 
\big( 3\,{\bf H}^{}_u+3\,{\bf H}^{}_d+{\bf H}^{}_e \big) 
\\ \nonumber
Y^{}_4 & = & \Big( \dfrac{17}{20} g^{2}_{\scr 1}
+ \dfrac{9}{4} g^{2}_{\scr 2} + 8 g^{2}_{\scr 3}
\Big) \, \textrm{Tr} \big( {\bf H}^{}_u \big) \\ \nonumber
& + & \Big( \dfrac{1}{4} g^{2}_{\scr 1}
+ \dfrac{9}{4} g^{2}_{\scr 2} + 8 g^{2}_{\scr 3}
\Big) \, \textrm{Tr} \big( {\bf H}^{}_d \big) \\
& + & \Big( \dfrac{3}{4} g^{2}_{\scr 1}+ 
\dfrac{3}{4} g^{2}_{\scr 2} \Big) \, 
\textrm{Tr} \big( {\bf H}^{}_e \big)
\\
H^{}_2 & = & \textrm{Tr} 
\big( 3\,{\bf H}^2_u+3\,{\bf H}^2_d+{\bf H}^2_e \big) 
\\
\chi^{}_4 & = & \dfrac{9}{4} \textrm{Tr} \big(
3\,{\bf H}^2_u+3\,{\bf H}^2_d+{\bf H}^2_e 
- \dfrac{2}{3}{\bf H}^{}_u {\bf H}^{}_d \big) 
\end{eqnarray}
\subsection[Renormierungsgruppengleichungen f"ur die Eichkopplungen]
{Renormierungsgruppengleichungen f"ur die Eichkopplungen}
\begin{eqnarray} \nonumber
\dfrac{d}{dt} \, \alpha^{}_{\scr 1} & = & 
-\dfrac{1}{2\pi} \Big( -4-\dfrac{1}{10}n_{\scr H} \Big) \,
\alpha^{2}_{\scr 1} \\ \nonumber
& & -\dfrac{1}{8\pi^2}
\Big[
 \Big(-\dfrac{19}{5}-\dfrac{9}{50}n_{\scr H} \Big) \, \alpha^{}_{\scr 1}
+\Big(-\dfrac{9}{5}-\dfrac{9}{10}n_{\scr H} \Big) \, \alpha^{}_{\scr 2}
+\Big(-\dfrac{44}{5} \Big) \, \alpha^{}_{\scr 3}
\Big] \, \alpha^{2}_{\scr 1} \\ \label{smcrge1}
& & -\dfrac{1}{32\pi^3} \textrm{Tr} \Big(
 \dfrac{17}{10} {\bf H}^{}_u
+\dfrac{1}{2} {\bf H}^{}_d
+\dfrac{3}{2} {\bf H}^{}_e \Big)
\, \alpha^{2}_{\scr 1} 
\\ \nonumber
\dfrac{d}{dt} \, \alpha^{}_{\scr 2} & = & 
-\dfrac{1}{2\pi} \Big( \dfrac{10}{3}-\dfrac{1}{6}n_{\scr H} \Big) 
\, \alpha^{2}_{\scr 2} \\ \nonumber
& & -\dfrac{1}{8\pi^2}
\Big[
 \Big(-\dfrac{3}{5}-\dfrac{3}{10}n_{\scr H} \Big) \, \alpha^{}_{\scr 1}
+\Big(-\dfrac{11}{3}-\dfrac{13}{6}n_{\scr H} \Big) \, \alpha^{}_{\scr 2}
+\big(-12 \big) \, \alpha^{}_{\scr 3}
\Big] \, \alpha^{2}_{\scr 2} \\ \label{smcrge2}
& & -\dfrac{1}{32\pi^3} \textrm{Tr} \Big(
 \dfrac{3}{2} {\bf H}^{}_u
+\dfrac{3}{2} {\bf H}^{}_d
+\dfrac{1}{2} {\bf H}^{}_e \Big)
\, \alpha^{2}_{\scr 2} 
\\ \nonumber
\dfrac{d}{dt} \, \alpha^{}_{\scr 3} & = & 
-\dfrac{1}{2\pi} \, 7 \, \alpha^{2}_{\scr 3} \\ \nonumber
& & -\dfrac{1}{8\pi^2}
\Big[
 \Big(-\dfrac{11}{10} \Big) \, \alpha^{}_{\scr 1}
+\Big(-\dfrac{9}{2} \Big) \, \alpha^{}_{\scr 2}
+ 26 \, \alpha^{}_{\scr 3}
\Big] \, \alpha^{2}_{\scr 3} \\ \label{smcrge3}
& & -\dfrac{1}{32\pi^3} \textrm{Tr} \big(
  2 \, {\bf H}^{}_u
+ 2 \, {\bf H}^{}_d \big)
\, \alpha^{2}_{\scr 3}
\end{eqnarray}
\subsection[Renormierungsgruppengleichungen f"ur die Yukawa-Matrizen]
{Renormierungsgruppengleichungen f"ur die Yukawa-Matrizen}
\begin{eqnarray} \label{rgeyuk}
\dfrac{d}{dt} \, {\bf Y}^{}_i & = & \Big( \, \dfrac{1}{16 \pi^2}
\beta^{\scr (1)}_i + \dfrac{1}{(16 \pi^2)^2}
\beta^{\scr (2)}_i \, \Big) \, {\bf Y}^{}_i \quad (i=u,d,e)
\\ \nonumber & & \\
\beta^{\scr (1)}_u & = & 
\dfrac{3}{2} \big( {\bf H}^{}_u - {\bf H}^{}_d \big) + Y^{}_2
- \Big( \dfrac{17}{20}g_{\scr 1}^{2} + \dfrac{9}{4}g_{\scr 2}^{2}  
+ 8 g_{\scr 3}^{2} \Big)
\\
\beta^{\scr (1)}_d & = & 
\dfrac{3}{2} \big( {\bf H}^{}_d - {\bf H}^{}_u \big) + Y^{}_2
- \Big( \dfrac{1}{4}g_{\scr 1}^{2} + \dfrac{9}{4}g_{\scr 2}^{2}  
+ 8 g_{\scr 3}^{2} \Big)
\\
\beta^{\scr (1)}_e & = & 
\dfrac{3}{2} {\bf H}^{}_e + Y^{}_2
- \Big( \dfrac{9}{4} g_{\scr 1}^{2} + \dfrac{9}{4} g_{\scr 2}^{2} \Big)
\\ \nonumber & & \\ \nonumber
\beta^{\scr (2)}_u & = & 
\dfrac{3}{2} {\bf H}^2_u - {\bf H}^{}_u {\bf H}^{}_d
- \dfrac{1}{4} {\bf H}^{}_d {\bf H}^{}_u + \dfrac{11}{4} {\bf H}^2_d
+ Y^{}_2 \Big( \dfrac{5}{4} {\bf H}^{}_d - \dfrac{9}{4} {\bf H}^{}_u \Big)
- \chi^{}_4 + \dfrac{3}{2} \lambda^2 \\ \nonumber
& & - 2 \, \lambda \big( 3 \, {\bf H}^{}_u + {\bf H}^{}_d \big)
+ \Big( \dfrac{223}{80}g_{\scr 1}^{2} 
+ \dfrac{135}{16}g_{\scr 2}^{2} + 16 \, g_{\scr 3}^{2} \Big) {\bf H}^{}_u
- \Big( \dfrac{43}{80}g_{\scr 1}^{2} 
- \dfrac{9}{16} g_{\scr 2}^{2} + 16 \, g_{\scr 3}^{2} \Big) {\bf H}^{}_d \\
& & + \dfrac{5}{2} Y^{}_4 + \dfrac{1187}{600} g_{\scr 1}^{4}
- \dfrac{9}{20} g_{\scr 1}^{2} g_{\scr 2}^{2}
+  \dfrac{19}{15} g_{\scr 1}^{2} g_{\scr 3}^{2}
- \dfrac{23}{4} g_{\scr 2}^{4} + 9 \, g_{\scr 2}^{2} g_{\scr 3}^{2}
- 108 \, g_{\scr 3}^{4}
\\ \nonumber & & \\ \nonumber
\beta^{\scr (2)}_d & = & 
\dfrac{3}{2} {\bf H}^2_d - {\bf H}^{}_d {\bf H}^{}_u
- \dfrac{1}{4} {\bf H}^{}_u {\bf H}^{}_d + \dfrac{11}{4} {\bf H}^2_u
+ Y^{}_2 \Big( \dfrac{5}{4} {\bf H}^{}_u - \dfrac{9}{4} {\bf H}^{}_d \Big)
- \chi^{}_4 + \dfrac{3}{2} \lambda^2 \\ \nonumber
& & - 2 \, \lambda \big( 3 \, {\bf H}^{}_d + {\bf H}^{}_u \big)
+ \Big( \dfrac{187}{80}g_{\scr 1}^{2} 
+ \dfrac{135}{16}g_{\scr 2}^{2} + 16 \, g_{\scr 3}^{2} \Big) {\bf H}^{}_d
- \Big( \dfrac{79}{80}g_{\scr 1}^{2} 
- \dfrac{9}{16} g_{\scr 2}^{2} + 16 \, g_{\scr 3}^{2} \Big) {\bf H}^{}_u \\
& & + \dfrac{5}{2} Y^{}_4 - \dfrac{127}{600} g_{\scr 1}^{4}
- \dfrac{27}{20} g_{\scr 1}^{2} g_{\scr 2}^{2}
+ \dfrac{31}{15} g_{\scr 1}^{2} g_{\scr 3}^{2}
- \dfrac{23}{4} g_{\scr 2}^{4} + 9 \, g_{\scr 2}^{2} g_{\scr 3}^{2}
- 108 \, g_{\scr 3}^{4}
\end{eqnarray}
\begin{eqnarray} \nonumber
\beta^{\scr (2)}_e & = & 
\dfrac{3}{2} {\bf H}^2_e - \dfrac{9}{4} Y^{}_2 {\bf H}^{}_e
- \chi^{}_4 + \dfrac{3}{2} \lambda^2 - 6 \, \lambda \, {\bf H}^{}_e
+ \Big( \dfrac{387}{80}g_{\scr 1}^{2} 
+ \dfrac{135}{15} g_{\scr 2}^{2} \Big) {\bf H}^{}_e + \dfrac{5}{2} Y^{}_4 
\\ & & + \dfrac{1371}{200} g_{\scr 1}^{4}
+ \dfrac{27}{20} g_{\scr 1}^{2} g_{\scr 2}^{2}
- \dfrac{23}{4} g_{\scr 2}^{4}
\end{eqnarray}
\subsection[Renormierungsgruppengleichung f"ur die Higgs-Selbstkopplung]
{Renormierungsgruppengleichung f"ur die Higgs-Selbstkopplung}
\begin{eqnarray} \label{rgehgs}
\dfrac{d}{dt} \, \lambda & = & \Big( \, \dfrac{1}{16 \pi^2}
\beta^{\scr (1)}_\lambda + \dfrac{1}{(16 \pi^2)^2}
\beta^{\scr (2)}_\lambda \, \Big) \, \lambda 
\\ \nonumber & & \\
\beta^{\scr (1)}_\lambda & = & 
12 \, \lambda^2 - \Big( \dfrac{9}{5}g_{\scr 1}^{2} + 9 \, g_{\scr 2}^{2} \Big)
\, \lambda + \dfrac{9}{4} \Big( \dfrac{3}{25} g_{\scr 1}^{4} 
+ \dfrac{2}{5} g_{\scr 1}^{2} g_{\scr 2}^{2} 
+  g_{\scr 2}^{4} \Big) + 4 \, Y^{}_2 \, \lambda - 4 \, H^{}_2
\\ \nonumber & & \\ \nonumber
\beta^{\scr (2)}_\lambda & = & 
- 78 \, \lambda^3 + 18 \, \Big( \dfrac{3}{5} g_{\scr 1}^{2} 
+ 3 \, g_{\scr 2}^{2} \Big) \, \lambda^2
- \Big( \dfrac{2661}{100} g_{\scr 1}^{4} 
- \dfrac{117}{20} g_{\scr 1}^{2} g_{\scr 2}^{2} 
+ \dfrac{73}{8} g_{\scr 2}^{4} \Big) \, \lambda \\ \nonumber
& & - \dfrac{3411}{1000} g_{\scr 1}^{6}
- \dfrac{1677}{200} g_{\scr 1}^{4} g_{\scr 2}^{2}
- \dfrac{289}{40} g_{\scr 1}^{2} g_{\scr 2}^{4}
+ \dfrac{305}{8} g_{\scr 2}^{6}
- 64 \, g_{\scr 3}^{2} \textrm{Tr} \big( {\bf H}^2_u + {\bf H}^2_d
\big) \\ \nonumber & & 
- \dfrac{8}{5} g_{\scr 1}^{2} \textrm{Tr} \big( 2 \, {\bf H}^2_u 
-  {\bf H}^2_d + 3 \, {\bf H}^2_e \big)
- \dfrac{3}{2} g_{\scr 2}^{4} Y^{}_2 + 10 \, \lambda Y^{}_4 \\ \nonumber & &
+ \dfrac{3}{5} g_{\scr 1}^{2} \Big[ 
\Big( -\dfrac{57}{10} g^{2}_{\scr 1}
+ 21 \, g^{2}_{\scr 2} \Big) \, \textrm{Tr} \big( {\bf H}^{}_u \big)
+ \Big( \dfrac{3}{2} g^{2}_{\scr 1}
+ 9 \, g^{2}_{\scr 2} \Big) \, \textrm{Tr} \big( {\bf H}^{}_d \big)
 \\ \nonumber & &
+ \Big( -\dfrac{15}{2} g^{2}_{\scr 1} + 11 \, g^{2}_{\scr 2} \Big) \, 
\textrm{Tr} \big( {\bf H}^{}_e \big) \Big]
- 24 \, \lambda^2 Y^{}_2 - \lambda H^{}_2 
+ 6 \, \lambda \, \textrm{Tr} \big( {\bf H}^{}_u {\bf H}^{}_d \big)
\big) \\ & &
+ 20 \, \textrm{Tr} \big( 3\,{\bf H}^3_u+3\,{\bf H}^3_d+{\bf H}^3_e
- 12 \, \textrm{Tr} \big( {\bf H}^2_u {\bf H}^{}_d \big)
- 12 \, \textrm{Tr} \big( {\bf H}^{}_u {\bf H}^2_d \big)
\end{eqnarray}
\subsection[Renormierungsgruppengleichung f"ur die See-Saw-Massenmatrix]
{Renormierungsgruppengleichung f"ur die See-Saw-Massenmatrix}
\begin{eqnarray} \label{rgessn}
16 \pi^2 \, \dfrac{d}{dt} \, {\bf M}_\nu & = & 
\big( 2 \, \lambda - 3 \, g_{\scr 2}^{2} + 2 \, Y^{}_2 \big) \, {\bf M}_\nu
- \dfrac{1}{2} \big( \,
{\bf M}_\nu \, {\bf H}^{}_e + {\bf H}^{T}_e \, {\bf M}_\nu
\, \big)
\end{eqnarray}
\section{$SU(4)_{\scr C} \otimes SU(2)_{\scr L} \otimes SU(2)_{\scr R} 
\big[ \otimes D \big]$-Modell}
Die Renormierungsgruppengleichungen f"ur die Eichkopplungen h"angen vom 
Higgs-Teil\-chen\-in\-halt des betrachteten Modells ab, das hei"st von
Anzahl und Art der Darstellungen, in welchen Higgs-Teilchen mit Massen
der Gr"o"senordnung $\lesssim M_{\scr I}$ liegen. Folgende Variablen
spezif\/izieren diesen Teilcheninhalt:
\begin{eqnarray*}
N_{1} \; \, & = & \textrm{Anzahl der} \; ({\bf 1,2,2})_{10/120} \\
N_{15} & = & \textrm{Anzahl der} \; ({\bf 15,2,2})_{120/126} \\
\Delta_L & = & \textrm{Anzahl der} \; ({\bf \overline{10},3,1})_{126} \\
\Delta_R & = & \textrm{Anzahl der} \; ({\bf 10,1,3})_{126}
\end{eqnarray*} 
Dann haben die Renormierungsgruppengleichungen f"ur die Eichkopplungen 
in Zweischleifenordnung die folgende Gestalt, wobei Beitr"age durch die 
Yukawakopplungen der Fermionen vernachl"assigt werden:
\begin{eqnarray} \nonumber
\dfrac{d}{dt} \, \alpha^{}_{\scr 4C} & = & 
-\dfrac{1}{2\pi} \Big( \dfrac{32}{3}-\dfrac{16}{3}
N_{15} -3 \, \Delta_L -3 \, \Delta_R \Big) \, 
\alpha^{2}_{\scr 4C} \\ \nonumber
& & -\dfrac{1}{8\pi^2}
\Big[
 \Big(-\dfrac{9}{2}-48 \, N_{15} -72 \, \Delta_L \Big)
 \, \alpha^{}_{\scr 2L} \\ \nonumber & & \hspace{0.8cm}
+\Big( -\dfrac{9}{2}-48 \, N_{15} -72 \, \Delta_R \Big)
 \, \alpha^{}_{\scr 2R} \\ \label{rgeps1} & & \hspace{0.8cm}
+\Big( \dfrac{473}{6} - \dfrac{896}{3} N_{15} 
-186 \, \Delta_L -186 \, \Delta_R \Big)
 \, \alpha^{}_{\scr 4C}
\Big] \, \alpha^{2}_{\scr 4C} 
\end{eqnarray}
\begin{eqnarray} \nonumber
%\\ \nonumber
\dfrac{d}{dt} \, \alpha^{}_{\scr 2L} & = & 
-\dfrac{1}{2\pi} \Big( \dfrac{10}{3}-\dfrac{1}{3} N_{1}
-5 \, N_{15} - \dfrac{20}{3} \Delta_L \Big) \, 
\alpha^{2}_{\scr 2L} 
\\ \nonumber
& & -\dfrac{1}{8\pi^2}
\Big[
 \Big(-\dfrac{11}{3}-\dfrac{13}{3} N_{1} - 65 \, N_{15} 
- \dfrac{560}{3} \Delta_L \Big)
 \, \alpha^{}_{\scr 2L} \\ \nonumber & & \hspace{0.8cm}
+ \Big( -3 \, N_{1} - 45 \, N_{15} 
 \Big)
 \, \alpha^{}_{\scr 2R} \\ \label{rgeps2} & & \hspace{0.8cm}
+\Big( -\dfrac{45}{2} -240 \, N_{15} -360 \, \Delta_L
\Big)
 \, \alpha^{}_{\scr 4C}
\Big] \, \alpha^{2}_{\scr 2L} 
\\ \nonumber
\dfrac{d}{dt} \, \alpha^{}_{\scr 2R} & = & 
-\dfrac{1}{2\pi} \Big( \ \dfrac{10}{3} - \dfrac{1}{3} N_{1}
-5 \, N_{15} - \dfrac{20}{3} \Delta_R \Big) \, 
\alpha^{2}_{\scr 2R} \\ \nonumber
& & -\dfrac{1}{8\pi^2}
\Big[ \Big(  -3 \, N_{1} - 45 \, N_{15} \Big)
 \, \alpha^{}_{\scr 2L} \\ \nonumber & & \hspace{0.8cm}
+ \Big( -\dfrac{11}{3}-\dfrac{13}{3} N_{1} - 65 \, N_{15} 
- \dfrac{560}{3} \Delta_R \Big)
 \, \alpha^{}_{\scr 2R} \\ \label{rgeps3} & & \hspace{0.8cm}
+\Big( -\dfrac{45}{2} -240 \, N_{15} -360 \, \Delta_R \Big)
 \, \alpha^{}_{\scr 4C}
\Big] \, \alpha^{2}_{\scr 2R} 
\end{eqnarray}
\chapter{Symmetriebrechungsskalen und Eichkopplungen}
Die verwendeten Startwerte f"ur die Integration der
Renormierungsgruppengleichungen der Eichkopplungen sind in Tabelle
\ref{vwsmk} zusammengefa"st.

\begin{table}[h]
\begin{center}
\begin{tabular}{|l||c|c|c|c|}
\hline
Gr"o"se & $M_{\scr Z}$ & $\alpha^{}_{\scr 1}(M_{\scr Z})$
& $\alpha^{}_{\scr 2}(M_{\scr Z})$ & $\alpha^{}_{\scr 3}(M_{\scr Z})$ \\
\hline
Wert & 91.187 GeV &  $(59.45)^{-1}$ & $(29.80 )^{-1}$ & $(8.40)^{-1}$ \\
\hline
\end{tabular}
\end{center}
\caption{\label{vwsmk} Verwendete Startwerte f"ur $M_{\scr Z}$ und die 
Eichkopplungen bei $\mu=M_{\scr Z}$}
\end{table}

\section{$SU(4)_{\scr C} \otimes SU(2)_{\scr L} \otimes SU(2)_{\scr R} 
\otimes D$-Modell} \label{sukfpsd}
Wie in Abschnitt \ref{sbspsd} erl"autert wurde, h"angen der Wert von 
$M_{\scr I}$ und somit die Eichkopplungen bei $\mu=M_{\scr I}$ nicht vom
Teilcheninhalt des Modells zwischen $M_{\scr U}$ und $M_{\scr I}$
ab; Tabelle \ref{emdp} gibt die Werte dieser Gr"o"sen an.

\begin{table}[h]
\begin{center}
\begin{tabular}{|l||c|c|c|c|c|c|}
\hline
Gr"o"se & $M_{\scr I}$ & $\alpha^{}_{\scr 1}(M_{\scr I})$ 
& $\alpha^{}_{\scr 2}(M_{\scr I})$ & $\alpha^{}_{\scr 3}(M_{\scr I})$ 
& $\alpha^{}_{\scr 2R,L}(M_{\scr I})$ & $\alpha^{}_{\scr 4C}(M_{\scr I})$ \\
\hline
Wert & $5.66 \cdot 10^{13}$ GeV & $(41.51)^{-1}$ & $(43.26)^{-1}$ 
& $(39.05)^{-1}$ & $(43.26)^{-1}$ & $(39.08)^{-1}$ \\
\hline
\end{tabular}
\end{center}
\caption{\label{emdp} Werte f"ur $M_{\scr I}$ und die Eichkopplungen bei 
$\mu=M_{\scr I}$ im $G_{\scr \textrm{PS}} \otimes D$-Modell}
\end{table}

\noindent Die vom Teilchenspektrum abh"angigen Werte f"ur $M_{\scr U}$ und 
$\alpha^{}_{\scr U}(M_{\scr U})$ sind f"ur die untersuchten F"alle in
Tabelle \ref{skpsd} aufgelistet.
\begin{table}[h]
\begin{center}
\begin{tabular}{|c|c|c|c|c|c|}
\hline
Gr"o"se & Wert & Wert & Wert & Wert & Wert \\
\hline \hline
$N_{1}$ & 1 & 2 & 3 & 4 & 5 \\
$N_{15}$ & 1 & 1 & 1 & 1 & 1 \\
\hline
$M_{\scr U}$ [GeV] & $1.46 \cdot 10^{15}$ & $1.28 \cdot 10^{15}$
& $1.13 \cdot 10^{15}$ & $1.01 \cdot 10^{15}$ & $9.12 \cdot 10^{14}$ \\
$\alpha^{}_{\scr U}(M_{\scr U})$ & $(37.95)^{-1}$ & $(38.01)^{-1}$ 
& $(38.05)^{-1}$ & $(38.10)^{-1}$ & $(38.14)^{-1}$ \\
\hline
\end{tabular}
\end{center}
%\end{table}
%
%\begin{table}[h]
\begin{center}
\begin{tabular}{|c|c|c|c|c|c|}
\hline
Gr"o"se & Wert & Wert & Wert & Wert & Wert \\
\hline \hline
$N_{1}$ & 1 & 2 & 3 & 4 & 5 \\
$N_{15}$ & 2 & 2 & 2 & 2 & 2 \\
\hline
$M_{\scr U}$ [GeV] & $1.75 \cdot 10^{15}$ & $1.51 \cdot 10^{15}$
& $1.32 \cdot 10^{15}$ & $1.16 \cdot 10^{15}$ & $1.04 \cdot 10^{15}$ \\
$\alpha^{}_{\scr U}(M_{\scr U})$ & $(34.47)^{-1}$ & $(34.67)^{-1}$ 
& $(34.87)^{-1}$ & $(35.04)^{-1}$ & $(35.20)^{-1}$ \\
\hline
\end{tabular}
\end{center}
\begin{center}
\begin{tabular}{|c|c|c|c|c|c|}
\hline
Gr"o"se & Wert & Wert & Wert & Wert & Wert \\
\hline \hline
$N_{1}$ & 1 & 2 & 3 & 4 & 5 \\
$N_{15}$ & 3 & 3 & 3 & 3 & 3 \\
\hline
$M_{\scr U}$ [GeV] & $2.14 \cdot 10^{15}$ & $1.81 \cdot 10^{15}$
& $1.56 \cdot 10^{15}$ & $1.36 \cdot 10^{15}$ & $1.20 \cdot 10^{15}$ \\
$\alpha^{}_{\scr U}(M_{\scr U})$ & $(30.50)^{-1}$ & $(30.91)^{-1}$ 
& $(31.28)^{-1}$ & $(31.62)^{-1}$ & $(31.93)^{-1}$ \\
\hline
\end{tabular}
\end{center}
\caption[$M_{\scr U}$ und $\alpha^{}_{\scr U}(M_{\scr U})$ im
$G_{\scr \textrm{PS}} \otimes D$-Modell]{$M_{\scr U}$ und 
$\alpha^{}_{\scr U}(M_{\scr U})$ im $G_{\scr \textrm{PS}} \otimes D$-Modell 
in Abh"angigkeit vom Higgs-Spektrum \label{skpsd}}
\end{table}
\section{$SU(4)_{\scr C} \otimes SU(2)_{\scr L} 
\otimes SU(2)_{\scr R}$-Modell} \label{sukfps}
Im Gegensatz zum $SU(4)_{\scr C} \otimes SU(2)_{\scr L} 
\otimes SU(2)_{\scr R} \otimes D$-Modell h"angen in Theorien mit 
$G_{\scr \textrm{PS}}$ als Symmetriegruppe auch $M_{\scr I}$ und die 
Eichkopplungen bei der Skala $\mu=M_{\scr I}$ vom Higgs-Spektrum
ab. In Modellen mit $N_{15}=3$ und $N_{1}=1,\dots,5$ sowie $N_{15}=2$ und 
$N_{1}=1,2,3$ f\/indet keine Vereinheitlichung statt, f"ur die "ubrigen
F"alle f\/inden sich in Tabelle \ref{skps} die Werte der relevanten Gr"o"sen:
%
%\begin{table}[h]
\begin{center}
\begin{tabular}{|c|c|c|c|c|}
\hline
Gr"o"se & Wert & Wert & Wert & Wert \\
\hline \hline
$N_{1}$ & 1 & 2 & 3 & 4 \\
$N_{15}$ & 1 & 1 & 1 & 1 \\
\hline
$M_{\scr I}$ & $1.70 \cdot 10^{10}$ GeV & $7.25 \cdot 10^{10}$ GeV 
& $1.94 \cdot 10^{11}$ GeV & $3.99 \cdot 10^{11}$ GeV \\
$\alpha^{}_{\scr 1}(M_{\scr I})$ & $(46.85)^{-1}$ & $(45.89)^{-1}$ 
& $(45.24)^{-1}$ & $(44.77)^{-1}$ \\
$\alpha^{}_{\scr 2}(M_{\scr I})$ & $(39.22)^{-1}$ & $(39.94)^{-1}$ 
& $(40.43)^{-1}$ & $(40.79)^{-1}$ \\
$\alpha^{}_{\scr 3}(M_{\scr I})$ & $(29.95)^{-1}$ & $(31.58)^{-1}$ 
& $(32.69)^{-1}$ & $(33.49)^{-1}$ \\
$\alpha^{}_{\scr 2R}(M_{\scr I})$ & $(58.22)^{-1}$ & $(55.54)^{-1}$ 
& $(53.72)^{-1}$ & $(52.40)^{-1}$ \\
$\alpha^{}_{\scr 2L}(M_{\scr I})$ & $(39.22)^{-1}$ & $(39.94)^{-1}$ & 
$(40.43)^{-1}$ & $(40.79)^{-1}$ \\
$\alpha^{}_{\scr 4C}(M_{\scr I})$ & $(29.98)^{-1}$ & $(31.61)^{-1}$ & 
$(32.71)^{-1}$ & $(33.52)^{-1}$ \\
$M_{\scr U}$ & $6.01 \cdot 10^{16}$ GeV & $1.83 \cdot 10^{16}$ GeV 
& $8.05 \cdot 10^{15}$ GeV & $4.37 \cdot 10^{15}$ GeV \\
$\alpha^{}_{\scr U}(M_{\scr U})$ & $(32.37)^{-1}$ & $(33.69)^{-1}$ 
& $(34.57)^{-1}$ & $(35.19)^{-1}$ \\
\hline
\end{tabular}
\end{center}
%\end{table}

\begin{table}[h]
\begin{center}
\begin{tabular}{|c|c|c|c|}
\hline
Gr"o"se & Wert & Wert & Wert \\
\hline \hline
$N_{1}$ & 5 & 4 & 5 \\
$N_{15}$ & 1 & 2 & 2 \\
\hline
$M_{\scr I}$ & $6.94 \cdot 10^{11}$ GeV & $6.14 \cdot 10^{10}$ GeV 
& $2.09 \cdot 10^{11}$ GeV \\
$\alpha^{}_{\scr 1}(M_{\scr I})$ & $(44.41)^{-1}$ & $(46.00)^{-1}$ 
& $(45.20)^{-1}$ \\
$\alpha^{}_{\scr 2}(M_{\scr I})$ & $(41.07)^{-1}$ & $(39.86)^{-1}$ 
& $(40.47)^{-1}$ \\
$\alpha^{}_{\scr 3}(M_{\scr I})$ & $(34.12)^{-1}$ & $(31.39)^{-1}$ 
& $(32.77)^{-1}$ \\
$\alpha^{}_{\scr 2R}(M_{\scr I})$ & $(51.37)^{-1}$ & $(55.85)^{-1}$ 
& $(53.59)^{-1}$ \\
$\alpha^{}_{\scr 2L}(M_{\scr I})$ & $(41.07)^{-1}$ & $(39.86)^{-1}$ 
& $(40.47)^{-1}$ \\
$\alpha^{}_{\scr 4C}(M_{\scr I})$ & $(34.14)^{-1}$ & $(31.42)^{-1}$ 
& $(32.80)^{-1}$ \\
$M_{\scr U}$ & $2.72 \cdot 10^{15}$ GeV & $1.31 \cdot 10^{16}$ GeV 
& $5.68 \cdot 10^{15}$ GeV \\
$\alpha^{}_{\scr U}(M_{\scr U})$ & $(35.65)^{-1}$ & $(20.08)^{-1}$ 
& $(23.76)^{-1}$ \\
\hline
\end{tabular}
\end{center}
\caption[Symmetriebrechungsskalen und Eichkopplungen im 
$G_{\scr \textrm{PS}}$-Modell]{Symmetriebrechungsskalen und Eichkopplungen im 
$G_{\scr \textrm{PS}}$-Modell in Abh"angigkeit vom Higgs-Spektrum \label{skps}}
\end{table}
\vspace*{5cm}
\newpage
\chapter{Eigenschaften der un\-ter\-such\-ten L"osungen} \label{appmm}
\section{Massenmatrizen, Massen und Mischungswinkel bei $M_{\scr I}$} 
\label{appmm1}
Tabelle \ref{parmiglf} gibt f"ur die drei Beispielmodelle 
aus Abschnitt \ref{numsolmm} die nichtverschwindenden Eintr"age der
Massenmatrizen der geladenen Fermionen sowie deren Massen und
Mischungswinkel bei $M_{\scr I}$ an, wobei f"ur die Mischungsmatrizen die
Parametrisierung (\ref{ckmpar}) benutzt wird. In Tabelle \ref{parmintr} 
sind die nichtverschwindenden Elemente der Dirac-, Majorana- und 
See-Saw-Massenmatrizen sowie Massen und Mischungen der leichten
Neutrinos zusammengefa"st.
\begin{table}[h]
\begin{center}
\begin{tabular}{|c|r|r|r|}
\hline
Parameter & Wert in Modell\,1 & Wert in Modell\,2a & Wert in Modell\,2b \\
\hline \hline
$({\bf M}_u)_{\scr 12}$ & $-198.37$ MeV & $-173.83$ MeV & $91.71$ MeV \\
$({\bf M}_u)_{\scr 21}$ & $1.98$ MeV & $2.26$ MeV & $-4.29$ MeV \\
$({\bf M}_u)_{\scr 23}$ & $-5539.4$ MeV & $-5617.2$ MeV & $-7532.3$ MeV \\
$({\bf M}_u)_{\scr 32}$ & $4993.8$ MeV & $5216.5$ MeV & $4365.4$ MeV \\
$({\bf M}_u)_{\scr 33}$ & $100940$ MeV & $100920$ MeV & $100840$ MeV \\
\hline
$m_u$ & $1.16$ MeV & $1.16$ MeV & $1.16$ MeV \\
$m_c$ & $337.6$ MeV & $337.6$ MeV & $337.6$ MeV \\
$m_t$ & $101210$ MeV & $101210$ MeV & $101210$ MeV \\
\hline
$\theta^{\scr (u)}_{\scr L12}$ & $-0.627$ & $-0.540$ & $0.275$ \\
$\theta^{\scr (u)}_{\scr L23}$ & $-0.055$ & $-0.055$ & $-0.074$ \\
$\theta^{\scr (u)}_{\scr L31}$ & $0.000$ & $0.000$ & $0.000$ \\
\hline
$\theta^{\scr (u)}_{\scr R12}$ & $0.005$ & $0.006$ & $-0.012$ \\
$\theta^{\scr (u)}_{\scr R23}$ & $0.049$ & $0.051$ & $0.043$ \\
$\theta^{\scr (u)}_{\scr R31}$ & $0.000$ & $0.000$ & $0.000$ \\
\hline 
\end{tabular}
\end{center}
\end{table}

\begin{table}[p]
\begin{center}
\begin{tabular}{|c|r|r|r|}
\hline
Parameter & Wert in Modell\,1 & Wert in Modell\,2a & Wert in Modell\,2b \\
\hline \hline
$({\bf M}_d)_{\scr 12}$ & $-33.66$ MeV & $-28.18$ MeV & $29.23$ MeV \\
$({\bf M}_d)_{\scr 21}$ & $6.03$ MeV & $7.57$ MeV & $-4.98$ MeV \\
$({\bf M}_d)_{\scr 23}$ & $-52.06$ MeV & $-52.43$ MeV & $-65.87$ MeV \\
$({\bf M}_d)_{\scr 32}$ & $1128.6$ MeV & $1152.5$ MeV & $854.63$ MeV \\
$({\bf M}_d)_{\scr 33}$ & $757.26$ MeV & $720.50$ MeV & $1056.1$ MeV \\
\hline
$m_d$ & $2.38$ MeV & $2.38$ MeV & $2.38$ MeV \\
$m_s$ & $47.4$ MeV & $47.4$ MeV & $47.4$ MeV \\
$m_b$ & $1360$ MeV & $1360$ MeV & $1360$ MeV \\
\hline
$\theta^{\scr (d)}_{\scr L12}$ & $-0.404$ & $-0.317$ & $0.498$ \\
$\theta^{\scr (d)}_{\scr L23}$ & $-0.021$ & $-0.020$ & $-0.038$ \\
$\theta^{\scr (d)}_{\scr L31}$ & $-0.021$ & $-0.018$ & $0.014$ \\
\hline
$\theta^{\scr (d)}_{\scr R12}$ & $0.117$ & $0.152$ & $-0.092$ \\
$\theta^{\scr (d)}_{\scr R23}$ & $0.979$ & $1.012$ & $0.679$ \\
$\theta^{\scr (d)}_{\scr R31}$ & $0.000$ & $0.000$ & $0.000$ \\
\hline \hline
$({\bf M}_e)_{\scr 12}$ & $-1.99$ MeV & $-2.27$ MeV & $2.05$ MeV \\
$({\bf M}_e)_{\scr 21}$ & $37.70$ MeV & $33.47$ MeV & $-32.16$ MeV \\
$({\bf M}_e)_{\scr 23}$ & $1232.3$ MeV & $-151.09$ MeV & $-193.32$ MeV \\
$({\bf M}_e)_{\scr 32}$ & $-155.77$ MeV & $1251.2$ MeV & $982.07$ MeV \\
$({\bf M}_e)_{\scr 33}$ & $1359.0$ MeV & $1342.2$ MeV & $1545.4$ MeV \\
\hline
$m_e$ & $0.513$ MeV & $0.513$ MeV & $0.513$ MeV \\
$m_\mu$ & $108.14$ MeV & $108.14$ MeV & $108.14$ MeV \\
$m_\tau$ & $1838.3$ MeV & $1838.3$ MeV & $1838.3$ MeV \\
\hline
$\theta^{\scr (e)}_{\scr L12}$ & $-0.018$ & $-0.015$ & $0.015$ \\
$\theta^{\scr (e)}_{\scr L23}$ & $0.733$ & $-0.060$ & $-0.089$ \\
$\theta^{\scr (e)}_{\scr L31}$ & $0.000$ & $-0.001$ & $0.001$ \\
\hline
$\theta^{\scr (e)}_{\scr R12}$ & $0.262$ & $0.314$ & $-0.301$ \\
$\theta^{\scr (e)}_{\scr R23}$ & $-0.063$ & $0.747$ & $0.561$ \\
$\theta^{\scr (e)}_{\scr R31}$ & $0.014$ & $-0.001$ & $0.002$ \\
\hline
\end{tabular}
\end{center}
\caption{Massenmatrixelemente, Massen und Mischungswinkel der
  geladenen Fermionen bei $M_{\scr I}$ \label{parmiglf}}
\end{table}

\begin{table}[p]
\begin{center}
\begin{tabular}{|c|r|r|r|}
\hline
Parameter & Wert in Modell\,1 & Wert in Modell\,2a & Wert in Modell\,2b \\
\hline \hline
$({\bf M}^{\scr (\textrm{Dir})}_\nu)_{\scr 12}$ & $-2184.0$ MeV 
& $-1459.5$ MeV & $1617.7$ MeV \\
$({\bf M}^{\scr (\textrm{Dir})}_\nu)_{\scr 21}$ & $-1983.7$ MeV 
& $-1283.5$ MeV & $1521.7$ MeV \\
$({\bf M}^{\scr (\textrm{Dir})}_\nu)_{\scr 23}$ & $8260.6$ MeV 
& $31082.8$ MeV & $32167.7$ MeV \\
$({\bf M}^{\scr (\textrm{Dir})}_\nu)_{\scr 32}$ & $-8806.2$ MeV 
& $-31483.5$ MeV & $-35334.6$ MeV \\
$({\bf M}^{\scr (\textrm{Dir})}_\nu)_{\scr 33}$ & $63207.4$ MeV 
& $70062.5$ MeV & $73367.5$ MeV \\
\hline
$({\bf M}^{\scr (\textrm{Maj})}_{\nu \scr{R}})_{\scr 12}$ 
& $y_1 M_{\scr R}$ & $y_1 M_{\scr R}$ & $y_1 M_{\scr R}$ \\
$({\bf M}^{\scr (\textrm{Maj})}_{\nu \scr{R}})_{\scr 21}$ 
& $y_1 M_{\scr R}$ & $y_1 M_{\scr R}$ & $y_1 M_{\scr R}$ \\
$({\bf M}^{\scr (\textrm{Maj})}_{\nu \scr{R}})_{\scr 33}$ 
& $19 \, y_1 M_{\scr R}$ & $24 \, y_1 M_{\scr R}$ & $-18 \, y_1 M_{\scr R}$ \\
\hline
$(y_1 M_{\scr R}/M_{\scr I}) \cdot ({\bf M}_\nu)_{\scr 12,21}$ 
& $-0.0706$ eV & $-0.0305$ eV & $-0.0401$ eV \\
$(y_1 M_{\scr R}/M_{\scr I}) \cdot ({\bf M}_\nu)_{\scr 22}$ 
& $-0.0585$ eV & $-0.656$ eV & $0.936$ eV \\
$(y_1 M_{\scr R}/M_{\scr I}) \cdot ({\bf M}_\nu)_{\scr 23,32}$ 
& $-0.732$ eV & $-2.136$ eV & $3.011$ eV \\
$(y_1 M_{\scr R}/M_{\scr I}) \cdot ({\bf M}_\nu)_{\scr 33}$ 
& $-3.425$ eV & $-3.331$ eV & $4.870$ eV \\
\hline
$(y_1 M_{\scr R}/M_{\scr I}) \cdot m_{\nu_1}$ & $-0.0365$ eV 
& $-0.0013$ eV & $0.0017$ eV \\
$(y_1 M_{\scr R}/M_{\scr I}) \cdot m_{\nu_2}$ & $0.130$ eV 
& $0.528 \,$ eV & $-0.695$ eV \\
$(y_1 M_{\scr R}/M_{\scr I}) \cdot m_{\nu_3}$ & $-3.577$ eV 
& $-4.514$ eV & $6.500$ eV \\
\hline
$\theta^{\scr (\nu)}_{\scr 12}$ & $-0.487$ & $-0.050$ & $0.051$ \\
$\theta^{\scr (\nu)}_{\scr 23}$ & $0.205$ & $0.506$ & $0.496$ \\
$\theta^{\scr (\nu)}_{\scr 31}$ & $0.004$ & $0.003$ & $-0.003$ \\
\hline
$m_{\nu_2}/m_{\nu_1}$ & $-3.57$ & $-406.2$ & $-401.2$ \\
$m_{\nu_3}/m_{\nu_2}$ & $-27.42$ & $-8.54$ & $-9.35$ \\
$\left( \dfrac{m_{\nu_3}^2-m_{\nu_2}^2}{m_{\nu_2}^2-m_{\nu_1}^2} \right)$ 
& $815.1$ & $72.0$ & $86.4$ \\
\hline
\end{tabular}
\end{center}
\caption{Dirac-, Majorana- und See-Saw-Massenmatrixelemente sowie Massen
  und Mischungswinkel der leichten Neutrinos bei $M_{\scr I}$ \label{parmintr}}
\end{table}

\section{$SO(10)$-Higgs-Parameter bei $M_{\scr I}$} \label{appmm2}
Tabelle \ref{hgspar} fa"st f"ur die betrachteten L"osungen die
numerischen Werte der $SO(10)$-Higgs-Parameter aus 
(\ref{parmrel1}-\ref{parmrel2}) zusammen.
\vspace*{1cm}
\begin{table}[h]
\begin{center}
\begin{tabular}{|c|r|r|r|}
\hline
Parameter & Wert in Modell\,1 & Wert in Modell\,2a & Wert in Modell\,2b \\
\hline \hline
$x_1 \, \upsilon^{\scr (1)}_d$ & $-5.90$ MeV 
& $-3.83$ MeV & $5.33$ MeV \\
$y_1 \, \omega^{\scr (1)}_d$ & $-7.92$ MeV 
& $-6.48$ MeV & $6.79$ MeV \\
$z_1 \, \tilde \upsilon^{\scr (1)}_d$ & $-19.84$ MeV 
& $-17.87$ MeV & $17.10$ MeV \\
$\tilde x_1 \, \upsilon^{\scr (1)}_d$ & $907.69$ MeV 
& $875.94$ MeV & $1178.45$ MeV \\
$\tilde y_1 \, \omega^{\scr (1)}_d$ & $-150.42$ MeV 
& $-155.43$ MeV & $-122.31$ MeV \\
$x_2 \, \upsilon^{\scr (2)}_d$ & $538.27$ MeV 
& $550.05$ MeV & $394.38$ MeV \\
$z_2 \, \tilde \upsilon^{\scr (2)}_d$ & $-269.23$ MeV 
& $-627.14$ MeV & $-492.11$ MeV \\
$z_2 \, \tilde \omega^{\scr (2)}_d$ & $-321.09$ MeV 
& $24.67$ MeV & $31.86$ MeV \\
$x_1 \, \upsilon^{\scr (1)}_u$ & $-594.61$ MeV 
& $-407.21$ MeV & $425.21$ MeV \\
$y_1 \, \omega^{\scr (1)}_u$ & $496.42$ MeV 
& $321.43$ MeV & $-381.50$ MeV \\
$z_1 \, \tilde \upsilon^{\scr (1)}_u$ & $-100.18$ MeV 
& $-88.05$ MeV & $48.00$ MeV \\
$\tilde x_1 \, \upsilon^{\scr (1)}_u$ & $91503.26$ MeV 
& $93205.41$ MeV & $93968.44$ MeV \\
$\tilde y_1 \, \omega^{\scr (1)}_u$ & $9431.94$ MeV 
& $7714.29$ MeV & $6866.97$ MeV \\
$x_2 \, \upsilon^{\scr (2)}_u$ & $-272.82$ MeV 
& $-200.35$ MeV & $-1583.48$ MeV \\
$z_2 \, \tilde \upsilon^{\scr (2)}_u$ & $-1816.59$ MeV 
& $3758.17$ MeV & $3976.16$ MeV \\
$z_2 \, \tilde \omega^{\scr (2)}_u$ & $-3450$ \;\;\,\, MeV 
& $-9175$ \;\;\,\, MeV & $-9925$ \;\;\,\, MeV \\
$\tilde x_1/x_1$ & $-153.89$ 
& $-228.89$ & $220.99$ \\
$\tilde y_1/y_1$ & $19$ \;\,\, \ 
& $24$ \;\,\, \ & $-18$ \;\,\, \ \\
\hline
\end{tabular}
\end{center}
\caption[$SO(10)$-Higgs-Parameter bei $M_{\scr I}$]{$SO(10)$-Higgs-Parameter 
bei $M_{\scr I}$ f"ur die untersuchten Modelle \label{hgspar}}
\end{table}

\section{Massenmatrizen, Massen und Mischungswinkel bei $M_{\scr Z}$}
\label{appmm3}
Tabelle \ref{parmz} gibt f"ur die drei untersuchten Modelle
die Eintr"age der Massenmatrizen der geladenen Fermionen und der 
leichten Neutrinos sowie deren Massen und Mischungswinkel bei $M_{\scr Z}$ an.

\begin{table}[p]
\begin{center}
\begin{tabular}{|c|r|r|r|}
\hline
Parameter & Wert in Modell\,1 & Wert in Modell\,2a & Wert in Modell\,2b \\
\hline \hline
$({\bf M}_u)_{\scr 11}$ & $0.00$ MeV & $0.00$ MeV & $0.00$ MeV \\
$({\bf M}_u)_{\scr 12}$ & $-397.86$ MeV & $-348.64$ MeV & $183.95$ MeV \\
$({\bf M}_u)_{\scr 13}$ & $1.94$ MeV & $1.78$ MeV & $-0.76$ MeV \\
$({\bf M}_u)_{\scr 21}$ & $3.98$ MeV & $4.54$ MeV & $-8.61$ MeV \\
$({\bf M}_u)_{\scr 22}$ & $56.66$ MeV & $60.01$ MeV & $67.28$ MeV \\
$({\bf M}_u)_{\scr 23}$ & $-9964.3$ MeV & $-10104$ MeV & $-13548$ MeV \\
$({\bf M}_u)_{\scr 31}$ & $0.02$ MeV & $0.03$ MeV & $-0.07$ MeV \\
$({\bf M}_u)_{\scr 32}$ & $8927.8$ MeV & $9326.1$ MeV & $7806.8$ MeV \\
$({\bf M}_u)_{\scr 33}$ & $180390$ MeV & $180360$ MeV & $180210$ MeV \\
\hline
$m_u$ & $2.33$ MeV & $2.33$ MeV & $2.33$ MeV \\
$m_c$ & $677.2$ MeV & $677.2$ MeV & $677.2$ MeV \\
$m_t$ & $180885$ MeV & $180882$ MeV & $180886$ MeV \\
\hline
$\theta^{\scr (u)}_{\scr L12}$ & $-0.627$ & $-0.540$ & $0.275$ \\
$\theta^{\scr (u)}_{\scr L23}$ & $-0.055$ & $-0.056$ & $-0.075$ \\
$\theta^{\scr (u)}_{\scr L31}$ & $0.000$ & $0.000$ & $0.000$ \\
\hline
$\theta^{\scr (u)}_{\scr R12}$ & $0.005$ & $0.006$ & $-0.012$ \\
$\theta^{\scr (u)}_{\scr R23}$ & $0.049$ & $0.051$ & $0.043$ \\
$\theta^{\scr (u)}_{\scr R31}$ & $0.000$ & $0.000$ & $0.000$ \\
\hline \hline
$({\bf M}_d)_{\scr 11}$ & $0.00$ MeV & $0.00$ MeV & $0.00$ MeV \\
$({\bf M}_d)_{\scr 12}$ & $-66.32$ MeV & $-55.51$ MeV & $57.58$ MeV \\
$({\bf M}_d)_{\scr 13}$ & $-0.02$ MeV & $-0.01$ MeV & $0.01$ MeV \\
$({\bf M}_d)_{\scr 21}$ & $11.88$ MeV & $14.91$ MeV & $-9.81$ MeV \\
$({\bf M}_d)_{\scr 22}$ & $-14.08$ MeV & $-14.58$ MeV & $-14.49$ MeV \\
$({\bf M}_d)_{\scr 23}$ & $-112.01$ MeV & $-112.41$ MeV & $-147.71$ MeV \\
$({\bf M}_d)_{\scr 31}$ & $-0.08$ MeV & $-0.10$ MeV & $0.08$ MeV \\
$({\bf M}_d)_{\scr 32}$ & $2489.1$ MeV & $2541.8$ MeV & $1884.4$ MeV \\
$({\bf M}_d)_{\scr 33}$ & $1670.8$ MeV & $1589.7$ MeV & $2329.8$ MeV \\
\hline
$m_d$ & $4.69$ MeV & $4.69$ MeV & $4.69$ MeV \\
$m_s$ & $93.4$ MeV & $93.4$ MeV & $93.4$ MeV \\
$m_b$ & $2999$ MeV & $2999$ MeV & $2999$ MeV \\
\hline
$\theta^{\scr (d)}_{\scr L12}$ & $-0.404$ & $-0.317$ & $0.498$ \\
$\theta^{\scr (d)}_{\scr L23}$ & $-0.025$ & $-0.024$ & $-0.041$ \\
$\theta^{\scr (d)}_{\scr L31}$ & $-0.018$ & $-0.016$ & $0.012$ \\
\hline
$\theta^{\scr (d)}_{\scr R12}$ & $0.117$ & $0.152$ & $-0.092$ \\
$\theta^{\scr (d)}_{\scr R23}$ & $0.979$ & $1.011$ & $0.679$ \\
$\theta^{\scr (d)}_{\scr R31}$ & $0.000$ & $0.000$ & $0.000$ \\
\hline
\end{tabular}
\end{center}
\end{table}

\begin{table}[p]
\begin{center}
\begin{tabular}{|c|r|r|r|}
\hline
Parameter & Wert in Modell\,1 & Wert in Modell\,2a & Wert in Modell\,2b \\
\hline \hline
$({\bf M}_e)_{\scr 11}$ & $0.00$ MeV & $0.00$ MeV & $0.00$ MeV \\
$({\bf M}_e)_{\scr 12}$ & $-1.89$ MeV & $-2.16$ MeV & $1.95$ MeV \\
$({\bf M}_e)_{\scr 13}$ & $0.00$ MeV & $0.00$ MeV & $0.00$ MeV \\
$({\bf M}_e)_{\scr 21}$ & $35.83$ MeV & $31.82$ MeV & $-30.57$ MeV \\
$({\bf M}_e)_{\scr 22}$ & $0.00$ MeV & $0.00$ MeV & $0.00$ MeV \\
$({\bf M}_e)_{\scr 23}$ & $1171.3$ MeV & $-143.61$ MeV & $-183.75$ MeV \\
$({\bf M}_e)_{\scr 31}$ & $0.00$ MeV & $0.00$ MeV & $0.00$ MeV \\
$({\bf M}_e)_{\scr 32}$ & $-148.06$ MeV & $1189.2$ MeV & $933.45$ MeV \\
$({\bf M}_e)_{\scr 33}$ & $1291.7$ MeV & $1275.8$ MeV & $1468.9$ MeV \\
\hline
$m_e$ & $0.488$ MeV & $0.488$ MeV & $0.488$ MeV \\
$m_\mu$ & $102.79$ MeV & $102.78$ MeV & $102.79$ MeV \\
$m_\tau$ & $1747.3$ MeV & $1747.3$ MeV & $1747.3$ MeV \\
\hline
$\theta^{\scr (e)}_{\scr L12}$ & $-0.018$ & $-0.015$ & $0.015$ \\
$\theta^{\scr (e)}_{\scr L23}$ & $0.733$ & $-0.060$ & $-0.089$ \\
$\theta^{\scr (e)}_{\scr L31}$ & $0.000$ & $-0.001$ & $0.001$ \\
\hline
$\theta^{\scr (e)}_{\scr R12}$ & $0.262$ & $0.314$ & $-0.301$ \\
$\theta^{\scr (e)}_{\scr R23}$ & $-0.063$ & $0.747$ & $0.561$ \\
$\theta^{\scr (e)}_{\scr R31}$ & $0.014$ & $-0.001$ & $0.002$ \\
\hline \hline
$(y_1 M_{\scr R}/M_{\scr I}) \cdot ({\bf M}_\nu)_{\scr 12,21}$ 
& $-0.0474$ eV & $-0.0205$ eV & $-0.0269$ eV \\
$(y_1 M_{\scr R}/M_{\scr I}) \cdot ({\bf M}_\nu)_{\scr 22}$ 
& $-0.0393$ eV & $-0.440$ eV & $0.629$ eV \\
$(y_1 M_{\scr R}/M_{\scr I}) \cdot ({\bf M}_\nu)_{\scr 23,32}$ 
& $-0.492$ eV & $-1.435$ eV & $2.022$ eV \\
$(y_1 M_{\scr R}/M_{\scr I}) \cdot ({\bf M}_\nu)_{\scr 33}$ 
& $-2.300$ eV & $-2.237$ eV & $3.271$ eV \\
$(y_1 M_{\scr R}/M_{\scr I}) \cdot ({\bf M}_\nu)_{\scr 11,13,31}$ 
& $0.000$ eV & $0.000$ eV & $0.000$ eV \\
\hline
$(y_1 M_{\scr R}/M_{\scr I}) \cdot m_{\nu_1}$ 
& $-0.0245$ eV & $-8.73 \cdot 10^{-4}$ eV & $1.16 \cdot 10^{-3}$ eV \\
$(y_1 M_{\scr R}/M_{\scr I}) \cdot m_{\nu_2}$ 
& $0.0876$ eV & $0.355 \,$ eV & $-0.467$ eV \\
$(y_1 M_{\scr R}/M_{\scr I}) \cdot m_{\nu_3}$ 
& $-2.402$ eV & $-3.031$ eV & $4.365$ eV \\
\hline
$\theta^{\scr (\nu)}_{\scr 12}$ & $-0.487$ & $-0.050$ & $0.051$ \\
$\theta^{\scr (\nu)}_{\scr 23}$ & $0.205$ & $0.506$ & $0.496$ \\
$\theta^{\scr (\nu)}_{\scr 31}$ & $0.004$ & $0.003$ & $-0.003$ \\
\hline
$m_{\nu_2}/m_{\nu_1}$ & $-3.57$ & $-406.3$ & $-401.2$ \\
$m_{\nu_3}/m_{\nu_2}$ & $-27.43$ & $-8.54$ & $-9.35$ \\
$\left( \dfrac{m_{\nu_3}^2-m_{\nu_2}^2}{m_{\nu_2}^2-m_{\nu_1}^2} \right)$ 
& $815.4$ & $71.9$ & $86.4$ \\
\hline
\end{tabular}
\end{center}
\caption{Massenmatrixelemente, Massen und Mischungswinkel der
  geladenen Fermionen und der leichten Neutrinos bei $M_{\scr Z}$ 
\label{parmz}}
\end{table}
\newpage
\section{CKM-Matrix und leptonische Mischungsmatrix bei $M_{\scr Z}$}
\label{appmm4}
Im folgenden werden die CKM-Matrix {\bf V} und die leptonische
Mischungsmatrix {\bf U} f"ur die in Abschnitt \ref{numsolmm}
analysierten Modelle angegeben.

\noindent Modell\,1:
\begin{equation}
{\bf V} \; = \; \left( \begin{array}{rrr}
0.9752 & 0.2212 & 0.0030 \\
- 0.2211 & 0.9746 & 0.0352 \\
0.0049 & -0.0350 & 0.9994
\end{array} \right) \; , \quad
{\bf U} \; = \; \left( \begin{array}{rrr}
0.891 & -0.455 & -0.005 \\
0.391 & 0.771 & -0.504 \\
0.233 & 0.447 & 0.864
\end{array} \right)
\end{equation}
Modell\,2a:
\begin{equation}
{\bf V} \; = \; \left( \begin{array}{rrr}
0.9752 & 0.2212 & 0.0030 \\
- 0.2211 & 0.9746 & 0.0353 \\
0.0049 & -0.0351 & 0.9994
\end{array} \right) \; , \quad
{\bf U} \; = \; \left( \begin{array}{rrr}
0.999 & -0.039 & 0.012 \\
0.026 & 0.844 & 0.536 \\
-0.031 & -0.535 & 0.844
\end{array} \right)
\end{equation}
Modell\,2b:
\begin{equation}
{\bf V} \; = \; \left( \begin{array}{rrr}
0.9752 & 0.2211 & 0.0025 \\
- 0.2211 & 0.9746 & 0.0355 \\
0.0054 & -0.0352 & 0.9994
\end{array} \right) \; , \quad
{\bf U} \; = \; \left( \begin{array}{rrr}
0.999 & 0.038 & -0.012 \\
-0.025 & 0.833 & 0.552 \\
0.031 & -0.552 & 0.834
\end{array} \right)
\end{equation}
\chapter{Berechnung der Nukleon\-zer\-falls\-ra\-ten}
\section{$A^{}_{i}$-Koef\/f\/izienten} \label{aicel}
In diesem Abschnitt wird die explizite Form der $A^{}_{i}$-Koef\/f\/izienten
dargestellt, welche in der ef\/fektiven Lagrangedichte (\ref{eldfnd})
f"ur die Nukleonenzerf"alle auftreten. Hierbei sind 
$\tilde G = g_{\scr{U}}^2/2M^2_{\scr X,Y}$ und 
$\tilde G' = g_{\scr{U}}^2/2M^2_{\scr X',Y'}$, wobei 
$M^2_{\scr X,Y} = M^2_{\scr X',Y'} \approx M^2_{\scr U}$ angenommen wird.
\begin{eqnarray*}
  A^{}_1 \; & = & \tilde G \; 
\big( ({\bf U}_{\scr R})_{11}^{} ({\bf U}_{\scr L})_{11}^{}
    + ({\bf U}_{\scr R})_{21}^{} ({\bf U}_{\scr L})_{21}^{} 
    + ({\bf U}_{\scr R})_{31}^{} ({\bf U}_{\scr L})_{31}^{} \big) \\
& & \hspace{0.4cm} \cdot
\big( ({\bf E}_{\scr R})_{11}^{} ({\bf D}_{\scr L})_{11}^{} 
    + ({\bf E}_{\scr R})_{21}^{} ({\bf D}_{\scr L})_{21}^{} 
    + ({\bf E}_{\scr R})_{31}^{} ({\bf D}_{\scr L})_{31}^{} \big)
\\
& + & \tilde G \; 
\big( ({\bf U}_{\scr R})_{11}^{} ({\bf D}_{\scr L})_{11}^{} 
+     ({\bf U}_{\scr R})_{21}^{} ({\bf D}_{\scr L})_{21}^{} 
+     ({\bf U}_{\scr R})_{31}^{} ({\bf D}_{\scr L})_{31}^{} \big) \\
& & \hspace{0.4cm} \cdot
\big( ({\bf E}_{\scr R})_{11}^{} ({\bf U}_{\scr L})_{11}^{} 
    + ({\bf E}_{\scr R})_{21}^{} ({\bf U}_{\scr L})_{21}^{} 
    + ({\bf E}_{\scr R})_{31}^{} ({\bf U}_{\scr L})_{31}^{} \big)
\\
A^{}_2 \; & = & \tilde G \; 
\big( ({\bf U}_{\scr R})_{11}^{} ({\bf U}_{\scr L})_{11}^{}
    + ({\bf U}_{\scr R})_{21}^{} ({\bf U}_{\scr L})_{21}^{} 
    + ({\bf U}_{\scr R})_{31}^{} ({\bf U}_{\scr L})_{31}^{} \big) \\
& & \hspace{0.4cm} \cdot
\big( ({\bf E}_{\scr L})_{11}^{} ({\bf D}_{\scr R})_{11}^{} 
    + ({\bf E}_{\scr L})_{21}^{} ({\bf D}_{\scr R})_{21}^{} 
    + ({\bf E}_{\scr L})_{31}^{} ({\bf D}_{\scr R})_{31}^{} \big)
\\
& + & \tilde G' \; 
\big( ({\bf D}_{\scr R})_{11}^{} ({\bf U}_{\scr L})_{11}^{} 
+     ({\bf D}_{\scr R})_{21}^{} ({\bf U}_{\scr L})_{21}^{} 
+     ({\bf D}_{\scr R})_{31}^{} ({\bf U}_{\scr L})_{31}^{} \big) \\
& & \hspace{0.5cm} \cdot
\big( ({\bf E}_{\scr L})_{11}^{} ({\bf U}_{\scr R})_{11}^{} 
    + ({\bf E}_{\scr L})_{21}^{} ({\bf U}_{\scr R})_{21}^{} 
    + ({\bf E}_{\scr L})_{31}^{} ({\bf U}_{\scr R})_{31}^{} \big)
\\
A^{}_3 \; & = & \tilde G \; 
\big( ({\bf U}_{\scr R})_{11}^{} ({\bf U}_{\scr L})_{11}^{}
    + ({\bf U}_{\scr R})_{21}^{} ({\bf U}_{\scr L})_{21}^{} 
    + ({\bf U}_{\scr R})_{31}^{} ({\bf U}_{\scr L})_{31}^{} \big) \\
& & \hspace{0.4cm} \cdot
\big( ({\bf E}_{\scr R})_{12}^{} ({\bf D}_{\scr L})_{11}^{}
    + ({\bf E}_{\scr R})_{22}^{} ({\bf D}_{\scr L})_{21}^{} 
    + ({\bf E}_{\scr R})_{32}^{} ({\bf D}_{\scr L})_{31}^{} \big)
\\
& + & \tilde G \; 
\big( ({\bf U}_{\scr R})_{11}^{} ({\bf D}_{\scr L})_{11}^{} 
+     ({\bf U}_{\scr R})_{21}^{} ({\bf D}_{\scr L})_{21}^{} 
+     ({\bf U}_{\scr R})_{31}^{} ({\bf D}_{\scr L})_{31}^{} \big) \\
& & \hspace{0.4cm} \cdot
\big( ({\bf E}_{\scr R})_{12}^{} ({\bf U}_{\scr L})_{11}^{} 
    + ({\bf E}_{\scr R})_{22}^{} ({\bf U}_{\scr L})_{21}^{} 
    + ({\bf E}_{\scr R})_{32}^{} ({\bf U}_{\scr L})_{31}^{} \big)
\\
A^{}_4 \; & = & \tilde G \; 
\big( ({\bf U}_{\scr R})_{11}^{} ({\bf U}_{\scr L})_{11}^{}
    + ({\bf U}_{\scr R})_{21}^{} ({\bf U}_{\scr L})_{21}^{} 
    + ({\bf U}_{\scr R})_{31}^{} ({\bf U}_{\scr L})_{31}^{} \big) \\
& & \hspace{0.4cm} \cdot
\big( ({\bf E}_{\scr L})_{12}^{} ({\bf D}_{\scr R})_{11}^{} 
    + ({\bf E}_{\scr L})_{22}^{} ({\bf D}_{\scr R})_{21}^{} 
    + ({\bf E}_{\scr L})_{32}^{} ({\bf D}_{\scr R})_{31}^{} \big)
\\
& + & \tilde G' \; 
\big( ({\bf D}_{\scr R})_{11}^{} ({\bf U}_{\scr L})_{11}^{} 
+     ({\bf D}_{\scr R})_{21}^{} ({\bf U}_{\scr L})_{21}^{} 
+     ({\bf D}_{\scr R})_{31}^{} ({\bf U}_{\scr L})_{31}^{} \big) \\
& & \hspace{0.5cm} \cdot
\big( ({\bf E}_{\scr L})_{12}^{} ({\bf U}_{\scr R})_{11}^{} 
    + ({\bf E}_{\scr L})_{22}^{} ({\bf U}_{\scr R})_{21}^{} 
    + ({\bf E}_{\scr L})_{32}^{} ({\bf U}_{\scr R})_{31}^{} \big)
\end{eqnarray*}
\pagebreak
\begin{eqnarray*}
A^{}_5 \; & = & \tilde G \; 
\big( ({\bf U}_{\scr R})_{11}^{} ({\bf U}_{\scr L})_{11}^{}
    + ({\bf U}_{\scr R})_{21}^{} ({\bf U}_{\scr L})_{21}^{} 
    + ({\bf U}_{\scr R})_{31}^{} ({\bf U}_{\scr L})_{31}^{} \big) \\
& & \hspace{0.4cm} \cdot
\big( ({\bf E}_{\scr R})_{11}^{} ({\bf D}_{\scr L})_{12}^{} 
    + ({\bf E}_{\scr R})_{21}^{} ({\bf D}_{\scr L})_{22}^{} 
    + ({\bf E}_{\scr R})_{31}^{} ({\bf D}_{\scr L})_{32}^{} \big)
\\
& + & \tilde G \; 
\big( ({\bf U}_{\scr R})_{11}^{} ({\bf D}_{\scr L})_{12}^{} 
+     ({\bf U}_{\scr R})_{21}^{} ({\bf D}_{\scr L})_{22}^{} 
+     ({\bf U}_{\scr R})_{31}^{} ({\bf D}_{\scr L})_{32}^{} \big) \\
& & \hspace{0.4cm} \cdot
\big( ({\bf E}_{\scr R})_{11}^{} ({\bf U}_{\scr L})_{11}^{} 
    + ({\bf E}_{\scr R})_{21}^{} ({\bf U}_{\scr L})_{21}^{} 
    + ({\bf E}_{\scr R})_{31}^{} ({\bf U}_{\scr L})_{31}^{} \big)
\\
A^{}_6 \; & = & \tilde G \; 
\big( ({\bf U}_{\scr R})_{11}^{} ({\bf U}_{\scr L})_{11}^{}
    + ({\bf U}_{\scr R})_{21}^{} ({\bf U}_{\scr L})_{21}^{} 
    + ({\bf U}_{\scr R})_{31}^{} ({\bf U}_{\scr L})_{31}^{} \big) \\
& & \hspace{0.4cm} \cdot
\big( ({\bf E}_{\scr L})_{11}^{} ({\bf D}_{\scr R})_{12}^{} 
    + ({\bf E}_{\scr L})_{21}^{} ({\bf D}_{\scr R})_{22}^{} 
    + ({\bf E}_{\scr L})_{31}^{} ({\bf D}_{\scr R})_{32}^{} \big)
\\
& + & \tilde G' \; 
\big( ({\bf D}_{\scr R})_{12}^{} ({\bf U}_{\scr L})_{11}^{} 
+     ({\bf D}_{\scr R})_{22}^{} ({\bf U}_{\scr L})_{21}^{} 
+     ({\bf D}_{\scr R})_{32}^{} ({\bf U}_{\scr L})_{31}^{} \big) \\
& & \hspace{0.5cm} \cdot
\big( ({\bf E}_{\scr L})_{11}^{} ({\bf U}_{\scr R})_{11}^{} 
    + ({\bf E}_{\scr L})_{21}^{} ({\bf U}_{\scr R})_{21}^{} 
    + ({\bf E}_{\scr L})_{31}^{} ({\bf U}_{\scr R})_{31}^{} \big)
\\
A^{}_7 \; & = & \tilde G \; 
\big( ({\bf U}_{\scr R})_{11}^{} ({\bf U}_{\scr L})_{11}^{}
    + ({\bf U}_{\scr R})_{21}^{} ({\bf U}_{\scr L})_{21}^{} 
    + ({\bf U}_{\scr R})_{31}^{} ({\bf U}_{\scr L})_{31}^{} \big) \\
& & \hspace{0.4cm} \cdot
\big( ({\bf E}_{\scr R})_{12}^{} ({\bf D}_{\scr L})_{12}^{} 
    + ({\bf E}_{\scr R})_{22}^{} ({\bf D}_{\scr L})_{22}^{} 
    + ({\bf E}_{\scr R})_{32}^{} ({\bf D}_{\scr L})_{32}^{} \big)
\\
& + & \tilde G \; 
\big( ({\bf U}_{\scr R})_{11}^{} ({\bf D}_{\scr L})_{12}^{} 
+     ({\bf U}_{\scr R})_{21}^{} ({\bf D}_{\scr L})_{22}^{} 
+     ({\bf U}_{\scr R})_{31}^{} ({\bf D}_{\scr L})_{32}^{} \big) \\
& & \hspace{0.4cm} \cdot
\big( ({\bf E}_{\scr R})_{12}^{} ({\bf U}_{\scr L})_{11}^{} 
    + ({\bf E}_{\scr R})_{22}^{} ({\bf U}_{\scr L})_{21}^{} 
    + ({\bf E}_{\scr R})_{32}^{} ({\bf U}_{\scr L})_{31}^{} \big)
\\
A^{}_8 \; & = & \tilde G \; 
\big( ({\bf U}_{\scr R})_{11}^{} ({\bf U}_{\scr L})_{11}^{}
    + ({\bf U}_{\scr R})_{21}^{} ({\bf U}_{\scr L})_{21}^{} 
    + ({\bf U}_{\scr R})_{31}^{} ({\bf U}_{\scr L})_{31}^{} \big) \\
& & \hspace{0.4cm} \cdot
\big( ({\bf E}_{\scr L})_{12}^{} ({\bf D}_{\scr R})_{12}^{} 
    + ({\bf E}_{\scr L})_{22}^{} ({\bf D}_{\scr R})_{22}^{} 
    + ({\bf E}_{\scr L})_{32}^{} ({\bf D}_{\scr R})_{32}^{} \big)
\\
& + & \tilde G' \; 
\big( ({\bf D}_{\scr R})_{12}^{} ({\bf U}_{\scr L})_{11}^{} 
+     ({\bf D}_{\scr R})_{22}^{} ({\bf U}_{\scr L})_{21}^{} 
+     ({\bf D}_{\scr R})_{32}^{} ({\bf U}_{\scr L})_{31}^{} \big) \\
& & \hspace{0.5cm} \cdot
\big( ({\bf E}_{\scr L})_{12}^{} ({\bf U}_{\scr R})_{11}^{} 
    + ({\bf E}_{\scr L})_{22}^{} ({\bf U}_{\scr R})_{21}^{} 
    + ({\bf E}_{\scr L})_{32}^{} ({\bf U}_{\scr R})_{31}^{} \big) 
\\
A^{}_9 \; & = & - \tilde G \; 
\big( ({\bf U}_{\scr R})_{11}^{} ({\bf D}_{\scr L})_{11}^{} 
+     ({\bf U}_{\scr R})_{21}^{} ({\bf D}_{\scr L})_{21}^{} 
+     ({\bf U}_{\scr R})_{31}^{} ({\bf D}_{\scr L})_{31}^{} \big) \\
& & \hspace{0.7cm} \cdot
\big( ({\bf N}_{\scr L})_{11}^{} ({\bf D}_{\scr R})_{11}^{} 
    + ({\bf N}_{\scr L})_{21}^{} ({\bf D}_{\scr R})_{21}^{} 
    + ({\bf N}_{\scr L})_{31}^{} ({\bf D}_{\scr R})_{31}^{} \big)
\\
& & - \tilde G' \; 
\big( ({\bf D}_{\scr R})_{11}^{} ({\bf D}_{\scr L})_{11}^{} 
+     ({\bf D}_{\scr R})_{21}^{} ({\bf D}_{\scr L})_{21}^{} 
+     ({\bf D}_{\scr R})_{31}^{} ({\bf D}_{\scr L})_{31}^{} \big) \\
& & \hspace{0.8cm} \cdot
\big( ({\bf N}_{\scr L})_{11}^{} ({\bf U}_{\scr R})_{11}^{} 
    + ({\bf N}_{\scr L})_{21}^{} ({\bf U}_{\scr R})_{21}^{} 
    + ({\bf N}_{\scr L})_{31}^{} ({\bf U}_{\scr R})_{31}^{} \big)
\\
A^{}_{10} & = & - \tilde G \; 
\big( ({\bf U}_{\scr R})_{11}^{} ({\bf D}_{\scr L})_{11}^{} 
+     ({\bf U}_{\scr R})_{21}^{} ({\bf D}_{\scr L})_{21}^{} 
+     ({\bf U}_{\scr R})_{31}^{} ({\bf D}_{\scr L})_{31}^{} \big) \\
& & \hspace{0.7cm} \cdot
\big( ({\bf N}_{\scr L})_{12}^{} ({\bf D}_{\scr R})_{11}^{} 
    + ({\bf N}_{\scr L})_{22}^{} ({\bf D}_{\scr R})_{21}^{} 
    + ({\bf N}_{\scr L})_{32}^{} ({\bf D}_{\scr R})_{31}^{} \big)
\\
& & - \tilde G' \; 
\big( ({\bf D}_{\scr R})_{11}^{} ({\bf D}_{\scr L})_{11}^{} 
+     ({\bf D}_{\scr R})_{21}^{} ({\bf D}_{\scr L})_{21}^{} 
+     ({\bf D}_{\scr R})_{31}^{} ({\bf D}_{\scr L})_{31}^{} \big) \\
& & \hspace{0.8cm} \cdot
\big( ({\bf N}_{\scr L})_{12}^{} ({\bf U}_{\scr R})_{11}^{} 
    + ({\bf N}_{\scr L})_{22}^{} ({\bf U}_{\scr R})_{21}^{} 
    + ({\bf N}_{\scr L})_{32}^{} ({\bf U}_{\scr R})_{31}^{} \big)
\\
A^{}_{11} & = & - \tilde G \; 
\big( ({\bf U}_{\scr R})_{11}^{} ({\bf D}_{\scr L})_{11}^{} 
+     ({\bf U}_{\scr R})_{21}^{} ({\bf D}_{\scr L})_{21}^{} 
+     ({\bf U}_{\scr R})_{31}^{} ({\bf D}_{\scr L})_{31}^{} \big) \\
& & \hspace{0.7cm} \cdot
\big( ({\bf N}_{\scr L})_{11}^{} ({\bf D}_{\scr R})_{12}^{} 
    + ({\bf N}_{\scr L})_{21}^{} ({\bf D}_{\scr R})_{22}^{} 
    + ({\bf N}_{\scr L})_{31}^{} ({\bf D}_{\scr R})_{32}^{} \big)
\\
& & - \tilde G' \; 
\big( ({\bf D}_{\scr R})_{12}^{} ({\bf D}_{\scr L})_{11}^{} 
+     ({\bf D}_{\scr R})_{22}^{} ({\bf D}_{\scr L})_{21}^{} 
+     ({\bf D}_{\scr R})_{32}^{} ({\bf D}_{\scr L})_{31}^{} \big) \\
& & \hspace{0.8cm} \cdot 
\big( ({\bf N}_{\scr L})_{11}^{} ({\bf U}_{\scr R})_{11}^{} 
    + ({\bf N}_{\scr L})_{21}^{} ({\bf U}_{\scr R})_{21}^{} 
    + ({\bf N}_{\scr L})_{31}^{} ({\bf U}_{\scr R})_{31}^{} \big)
\\
A^{}_{12} & = & - \tilde G \; 
\big( ({\bf U}_{\scr R})_{11}^{} ({\bf D}_{\scr L})_{11}^{} 
+     ({\bf U}_{\scr R})_{21}^{} ({\bf D}_{\scr L})_{21}^{} 
+     ({\bf U}_{\scr R})_{31}^{} ({\bf D}_{\scr L})_{31}^{} \big) \\
& & \hspace{0.7cm} \cdot
\big( ({\bf N}_{\scr L})_{12}^{} ({\bf D}_{\scr R})_{12}^{} 
    + ({\bf N}_{\scr L})_{22}^{} ({\bf D}_{\scr R})_{22}^{} 
    + ({\bf N}_{\scr L})_{32}^{} ({\bf D}_{\scr R})_{32}^{} \big)
\\
& & - \tilde G' \; 
\big( ({\bf D}_{\scr R})_{12}^{} ({\bf D}_{\scr L})_{11}^{} 
+     ({\bf D}_{\scr R})_{22}^{} ({\bf D}_{\scr L})_{21}^{} 
+     ({\bf D}_{\scr R})_{32}^{} ({\bf D}_{\scr L})_{31}^{} \big) \\
& & \hspace{0.8cm} \cdot 
\big( ({\bf N}_{\scr L})_{12}^{} ({\bf U}_{\scr R})_{11}^{} 
    + ({\bf N}_{\scr L})_{22}^{} ({\bf U}_{\scr R})_{21}^{} 
    + ({\bf N}_{\scr L})_{32}^{} ({\bf U}_{\scr R})_{31}^{} \big)
\end{eqnarray*}
\pagebreak
\begin{eqnarray*}
A^{}_{13} & = & - \tilde G \; 
\big( ({\bf U}_{\scr R})_{11}^{} ({\bf D}_{\scr L})_{12}^{} 
+     ({\bf U}_{\scr R})_{21}^{} ({\bf D}_{\scr L})_{22}^{} 
+     ({\bf U}_{\scr R})_{31}^{} ({\bf D}_{\scr L})_{32}^{} \big) \\
& & \hspace{0.7cm} \cdot
\big( ({\bf N}_{\scr L})_{11}^{} ({\bf D}_{\scr R})_{11}^{} 
    + ({\bf N}_{\scr L})_{21}^{} ({\bf D}_{\scr R})_{21}^{} 
    + ({\bf N}_{\scr L})_{31}^{} ({\bf D}_{\scr R})_{31}^{} \big)
\\
& & - \tilde G' \; 
\big( ({\bf D}_{\scr R})_{11}^{} ({\bf D}_{\scr L})_{12}^{} 
+     ({\bf D}_{\scr R})_{21}^{} ({\bf D}_{\scr L})_{22}^{} 
+     ({\bf D}_{\scr R})_{31}^{} ({\bf D}_{\scr L})_{32}^{} \big) \\
& & \hspace{0.8cm} \cdot
\big( ({\bf N}_{\scr L})_{11}^{} ({\bf U}_{\scr R})_{11}^{} 
    + ({\bf N}_{\scr L})_{21}^{} ({\bf U}_{\scr R})_{21}^{} 
    + ({\bf N}_{\scr L})_{31}^{} ({\bf U}_{\scr R})_{31}^{} \big)
\\
A^{}_{14} & = & - \tilde G \; 
\big( ({\bf U}_{\scr R})_{11}^{} ({\bf D}_{\scr L})_{12}^{} 
+     ({\bf U}_{\scr R})_{21}^{} ({\bf D}_{\scr L})_{22}^{} 
+     ({\bf U}_{\scr R})_{31}^{} ({\bf D}_{\scr L})_{32}^{} \big) \\
& & \hspace{0.7cm} \cdot
\big( ({\bf N}_{\scr L})_{12}^{} ({\bf D}_{\scr R})_{11}^{} 
    + ({\bf N}_{\scr L})_{22}^{} ({\bf D}_{\scr R})_{21}^{} 
    + ({\bf N}_{\scr L})_{32}^{} ({\bf D}_{\scr R})_{31}^{} \big)
\\
& & - \tilde G' \; 
\big( ({\bf D}_{\scr R})_{11}^{} ({\bf D}_{\scr L})_{12}^{} 
+     ({\bf D}_{\scr R})_{21}^{} ({\bf D}_{\scr L})_{22}^{} 
+     ({\bf D}_{\scr R})_{31}^{} ({\bf D}_{\scr L})_{32}^{} \big) \\
& & \hspace{0.8cm} \cdot
\big( ({\bf N}_{\scr L})_{12}^{} ({\bf U}_{\scr R})_{11}^{} 
    + ({\bf N}_{\scr L})_{22}^{} ({\bf U}_{\scr R})_{21}^{} 
    + ({\bf N}_{\scr L})_{32}^{} ({\bf U}_{\scr R})_{31}^{} \big)
\\
A^{}_{15} & = & - \tilde G \; 
\big( ({\bf U}_{\scr R})_{11}^{} ({\bf D}_{\scr L})_{11}^{} 
+     ({\bf U}_{\scr R})_{21}^{} ({\bf D}_{\scr L})_{21}^{} 
+     ({\bf U}_{\scr R})_{31}^{} ({\bf D}_{\scr L})_{31}^{} \big) \\
& & \hspace{0.7cm} \cdot
\big( ({\bf N}_{\scr L})_{13}^{} ({\bf D}_{\scr R})_{11}^{} 
    + ({\bf N}_{\scr L})_{23}^{} ({\bf D}_{\scr R})_{21}^{} 
    + ({\bf N}_{\scr L})_{33}^{} ({\bf D}_{\scr R})_{31}^{} \big)
\\
& & - \tilde G' \; 
\big( ({\bf D}_{\scr R})_{11}^{} ({\bf D}_{\scr L})_{11}^{} 
+     ({\bf D}_{\scr R})_{21}^{} ({\bf D}_{\scr L})_{21}^{} 
+     ({\bf D}_{\scr R})_{31}^{} ({\bf D}_{\scr L})_{31}^{} \big) \\
& & \hspace{0.8cm} \cdot
\big( ({\bf N}_{\scr L})_{13}^{} ({\bf U}_{\scr R})_{11}^{} 
    + ({\bf N}_{\scr L})_{23}^{} ({\bf U}_{\scr R})_{21}^{} 
    + ({\bf N}_{\scr L})_{33}^{} ({\bf U}_{\scr R})_{31}^{} \big)
\\
A^{}_{16} & = & - \tilde G \; 
\big( ({\bf U}_{\scr R})_{11}^{} ({\bf D}_{\scr L})_{11}^{} 
+     ({\bf U}_{\scr R})_{21}^{} ({\bf D}_{\scr L})_{21}^{} 
+     ({\bf U}_{\scr R})_{31}^{} ({\bf D}_{\scr L})_{31}^{} \big) \\
& & \hspace{0.7cm} \cdot
\big( ({\bf N}_{\scr L})_{13}^{} ({\bf D}_{\scr R})_{12}^{} 
    + ({\bf N}_{\scr L})_{23}^{} ({\bf D}_{\scr R})_{22}^{} 
    + ({\bf N}_{\scr L})_{33}^{} ({\bf D}_{\scr R})_{32}^{} \big)
\\
& & - \tilde G' \; 
\big( ({\bf D}_{\scr R})_{12}^{} ({\bf D}_{\scr L})_{11}^{} 
+     ({\bf D}_{\scr R})_{22}^{} ({\bf D}_{\scr L})_{21}^{} 
+     ({\bf D}_{\scr R})_{32}^{} ({\bf D}_{\scr L})_{31}^{} \big) \\
& & \hspace{0.8cm} \cdot 
\big( ({\bf N}_{\scr L})_{13}^{} ({\bf U}_{\scr R})_{11}^{} 
    + ({\bf N}_{\scr L})_{23}^{} ({\bf U}_{\scr R})_{21}^{} 
    + ({\bf N}_{\scr L})_{33}^{} ({\bf U}_{\scr R})_{31}^{} \big)
\\
A^{}_{17} & = & - \tilde G \; 
\big( ({\bf U}_{\scr R})_{11}^{} ({\bf D}_{\scr L})_{12}^{} 
+     ({\bf U}_{\scr R})_{21}^{} ({\bf D}_{\scr L})_{22}^{} 
+     ({\bf U}_{\scr R})_{31}^{} ({\bf D}_{\scr L})_{32}^{} \big) \\
& & \hspace{0.7cm} \cdot
\big( ({\bf N}_{\scr L})_{13}^{} ({\bf D}_{\scr R})_{11}^{} 
    + ({\bf N}_{\scr L})_{23}^{} ({\bf D}_{\scr R})_{21}^{} 
    + ({\bf N}_{\scr L})_{33}^{} ({\bf D}_{\scr R})_{31}^{} \big)
\\
& & - \tilde G' \;
\big( ({\bf D}_{\scr R})_{11}^{} ({\bf D}_{\scr L})_{12}^{} 
+     ({\bf D}_{\scr R})_{21}^{} ({\bf D}_{\scr L})_{22}^{} 
+     ({\bf D}_{\scr R})_{31}^{} ({\bf D}_{\scr L})_{32}^{} \big) \\
& & \hspace{0.8cm} \cdot
 \big( ({\bf N}_{\scr L})_{13}^{} ({\bf U}_{\scr R})_{11}^{} 
     + ({\bf N}_{\scr L})_{23}^{} ({\bf U}_{\scr R})_{21}^{} 
     + ({\bf N}_{\scr L})_{33}^{} ({\bf U}_{\scr R})_{31}^{} \big)
\end{eqnarray*}
\section{Meson-Wellenfunktionen} \label{wfm}
Die Tabellen \ref{wfpm} und \ref{wfvm} geben die Spin-Flavour-Anteile
der Wellenfunktionen f"ur die pseudoskalaren beziehungsweise
Vektormesonen an.
\begin{table}[h]
\begin{center}
\begin{tabular}{|c|c|c|c|}
\hline
Meson & Masse (MeV) & Spin $S$ & Wellenfunktion \\
\hline \hline
$\pi^+$ & 139.570 & 0 & $-1/\sqrt{2} \,
(d^{\scr C} \!\! \uparrow u \! \downarrow 
- d^{\scr C} \!\! \downarrow u \! \uparrow)$ \\
$\pi^0$ \ & 134.976 & 0 & $1/2 \,
(u^{\scr C} \!\! \uparrow u \! \downarrow 
- u^{\scr C} \!\! \downarrow u \! \uparrow
- d^{\scr C} \!\! \uparrow d \! \downarrow 
+ d^{\scr C} \!\! \downarrow d \! \uparrow)$ \\
$\pi^-$ & 139.570 & 0 & $1/\sqrt{2} \,
(u^{\scr C} \!\! \uparrow d \! \downarrow 
- u^{\scr C} \!\! \downarrow d \! \uparrow)$ \\
$K^+$ & 493.677 & 0 & $1/\sqrt{2} \,
(s^{\scr C} \!\! \uparrow u \! \downarrow 
- s^{\scr C} \!\! \downarrow u \! \uparrow)$ \\
$K^0$ \ & 497.672 & 0 & $1/\sqrt{2} \,
(s^{\scr C} \!\! \uparrow d \! \downarrow 
- s^{\scr C} \!\! \downarrow d \! \uparrow)$ \\
$\eta$ \; & 547.30 \ & 0 & $1/\sqrt{12} \,
(u^{\scr C} \!\! \uparrow u \! \downarrow 
- u^{\scr C} \!\! \downarrow u \! \uparrow
+ d^{\scr C} \!\! \uparrow d \! \downarrow 
- d^{\scr C} \!\! \downarrow d \! \uparrow$ \\ & & &
$- 2 \, s^{\scr C} \!\! \uparrow s \! \downarrow 
+ 2 \, s^{\scr C} \!\! \downarrow s \! \uparrow)$ \\
\hline
\end{tabular}
\end{center}
\caption[Wellenfunktionen der pseudoskalaren Mesonen]{Wellenfunktionen
  der pseudoskalaren Mesonen (Massenwerte aus \cite{pdg}) \label{wfpm}}
\end{table}
\begin{table}[h]
\begin{center}
\begin{tabular}{|c|c|c|c|c|}
\hline
Meson & Masse (MeV) & Spin $S$ & $S_z$ & Wellenfunktion \\
\hline \hline
$\rho_{\scr (+)}^+$ & 770.0 \, & 1 & $+1$ & 
$-d^{\scr C} \!\! \uparrow u \! \uparrow$ \\
$\rho_{\scr (0)}^+$ & 770.0 \, & 1 & $0$ & 
$-1/\sqrt{2} \, (d^{\scr C} \!\! \uparrow u \! \downarrow 
+ d^{\scr C} \!\! \downarrow u \! \uparrow)$ \\
$\rho_{\scr (-)}^+$ & 770.0 \, & 1 & $-1$ & 
$-d^{\scr C} \!\! \downarrow u \! \downarrow$ \\
\hline
$\rho_{\scr (+)}^0$ & 770.0 \, & 1 & $+1$ & $1/\sqrt{2} \, (
u^{\scr C} \!\! \uparrow u \! \uparrow
- d^{\scr C} \!\! \uparrow d \! \uparrow)$ \\
$\rho_{\scr (0)}^0$ & 770.0 \, & 1 & $0$ & 
$1/2 \, (u^{\scr C} \!\! \uparrow u \! \downarrow 
+ u^{\scr C} \!\! \downarrow u \! \uparrow
- d^{\scr C} \!\! \uparrow d \! \downarrow 
- d^{\scr C} \!\! \downarrow d \! \uparrow)$ \\
$\rho_{\scr (-)}^0$ & 770.0 \, & 1 & $-1$ & $1/\sqrt{2} \, (
u^{\scr C} \!\! \downarrow u \! \downarrow
- d^{\scr C} \!\! \downarrow d \! \downarrow)$ \\
\hline
$\rho_{\scr (+)}^-$ & 770.0 \, & 1 & $+1$ & 
$u^{\scr C} \!\! \uparrow d \! \uparrow$ \\
$\rho_{\scr (0)}^-$ \ & 770.0 \, & 1 & $0$ & 
$1/\sqrt{2} \, (u^{\scr C} \!\! \uparrow d \! \downarrow 
+ u^{\scr C} \!\! \downarrow d \! \uparrow)$ \\
$\rho_{\scr (-)}^-$ & 770.0 \, & 1 & $-1$ & 
$u^{\scr C} \!\! \downarrow d \! \downarrow$ \\
\hline
$\omega_{\scr (+)}$ & 781.94 & 1 & $+1$ & $1/\sqrt{2} \, (
u^{\scr C} \!\! \uparrow u \! \uparrow
+ d^{\scr C} \!\! \uparrow d \! \uparrow)$ \\
$\omega_{\scr (0)}$ \ & 781.94 & 1 & $0$ & 
$1/2 \, (u^{\scr C} \!\! \uparrow u \! \downarrow 
+ u^{\scr C} \!\! \downarrow u \! \uparrow
+ d^{\scr C} \!\! \uparrow d \! \downarrow 
+ d^{\scr C} \!\! \downarrow d \! \uparrow)$ \\
$\omega_{\scr (-)}$ & 781.94 & 1 & $-1$ & $1/\sqrt{2} \, (
u^{\scr C} \!\! \downarrow u \! \downarrow
+ d^{\scr C} \!\! \downarrow d \! \downarrow)$ \\
\hline
$K_{\scr (+)}^{*+}$ & 891.66 & 1 & $+1$ & 
$s^{\scr C} \!\! \uparrow u \! \uparrow$ \\
$K_{\scr (0)}^{*+}$ & 891.66 & 1 & $0$ & 
$1/\sqrt{2} \, (s^{\scr C} \!\! \uparrow u \! \downarrow 
+ s^{\scr C} \!\! \downarrow u \! \uparrow)$ \\
$K_{\scr (-)}^{*+}$ & 891.66 & 1 & $-1$ & 
$s^{\scr C} \!\! \downarrow u \! \downarrow$ \\
\hline
$K_{\scr (+)}^{*0}$ & 896.1 \, & 1 & $+1$ & 
$s^{\scr C} \!\! \uparrow d \! \uparrow$ \\
$K_{\scr (0)}^{*0}$ & 896.1 \, & 1 & $0$ & 
$1/\sqrt{2} \, (s^{\scr C} \!\! \uparrow d \! \downarrow 
+ s^{\scr C} \!\! \downarrow d \! \uparrow)$ \\
$K_{\scr (-)}^{*0}$ & 896.1 \, & 1 & $-1$ & 
$s^{\scr C} \!\! \downarrow d \! \downarrow$ \\
\hline
\end{tabular}
\end{center}
\caption[Wellenfunktionen der Vektormesonen]{Wellenfunktionen der 
Vektormesonen (Massenwerte aus \cite{pdg}) \label{wfvm}}
\end{table}
\section{Phasenraumfaktoren}
Die Phasenraumfaktoren, welche die $SU(6)$-Spin-Flavour-Symmetrie des
in Abschnitt \ref{hmtrxel} verwendeten nichtrelativistischen Quark-Modells
brechen, sind durch 
\begin{equation}
\rho_{\scr p,n} \; = \; (1-\chi_{\scr p,n}^2)(1-\chi_{\scr p,n}^4) 
\end{equation}
gegeben, wobei $\chi_{\scr p,n} = m_{\scr \textrm{Meson}}/m_{\scr p,n}$ 
der Quotient aus Mesonmasse und Masse des Protons beziehungsweise
Neutrons ist ($m_{\scr p}=938.27231$ MeV, $m_{\scr n}=939.56563$ MeV
\cite{pdg}). Tabelle \ref{phspf} gibt die entsprechenden Werte f"ur die
relevanten Mesonen an.
\begin{table}[h]
\begin{center}
\begin{tabular}{|c|c|c||c|c|c|}
\hline
Meson & $\rho_{\scr p}$ & $\rho_{\scr n}$ & 
Meson & $\rho_{\scr p}$ & $\rho_{\scr n}$ \\
\hline \hline
$\pi^0 $ & \quad 0.97889 \quad & \quad 0.97895 \quad &
$\rho^0 $ & \quad 0.17842 \quad & \quad 0.18025 \quad \\
$\pi^\pm $ & \quad 0.97739 \quad & \quad 0.97746 \quad &
$\rho^\pm $ & \quad 0.17842 \quad & \quad 0.18025 \quad \\
$K^\pm$ & \quad 0.66774 \quad & \quad 0.66875 \quad &
$\omega$ & \quad 0.15812 \quad & \quad 0.15993 \quad \\
$K^0$ & \quad 0.66178 \quad & \quad 0.66280 \quad &
$K^{*\pm}$ & \quad 0.01787 \quad & \quad 0.01877 \quad \\
$\eta$ & \quad 0.58338 \quad & \quad 0.58462 \quad &
$K^{*0}$ & \quad 0.01476 \quad & \quad 0.01560 \quad \\
\hline
\end{tabular}
\end{center}
\caption[Phasenraumfaktoren f"ur Proton- und Neutronzerf"alle]
{Phasenraumfaktoren f"ur Proton- und Neutronzerf"alle in Abh"angigkeit
  vom Meson im Endzustand \label{phspf}}
\end{table}
\section{"Ubergangsamplituden}
In diesem Abschnitt sind die Resultate f"ur die hadronischen 
"Ubergangsamplituden und -wahrscheinlichkeiten der
Nukleonenzerf"alle zusammengefa"st. Tabelle \ref{aepnd}
zeigt die Amplituden der elementaren Zefallsprozesse des
Protons und des Neutrons. Alle dort nicht aufgelisteten
Amplituden lassen sich durch Symmetrie"uberlegungen herleiten. So
unterscheiden sich die Amplituden mit Antileptonen der zweiten und dritten
Familie im Endzustand von denen mit $e^{\scr +}$ und $\nu_{\scr e}^{\scr C}$ 
lediglich durch die Vorfaktoren $A_i$; diese lassen sich (\ref{eldfnd}) 
entnehmen. Die "Ubergangsamplituden mit linksh"andigen Antileptonen
erh"alt man aus denjenigen mit rechtsh"andigen Antileptonen, indem man 
$\bar l^{\scr C}_{\scr R}$ durch $\bar l^{\scr C}_{\scr L}$ ersetzt
und die Spineinstellungen $\uparrow$ und $\downarrow$ der Quarks
vertauscht. Auch hier "andern sich lediglich die Vorfaktoren $A_i$
gem"a"s (\ref{eldfnd}), die numerischen Werte der Amplituden in der
dritten Spalte bleiben gleich.
Tabelle \ref{trprob} listet die "Ubergangswahrscheinlichkeiten f"ur
Zerfallsprozesse mit physikalischen Teilchen in den Endzust"anden auf.
\begin{table}[h]
\begin{center}
\begin{tabular}{|c|c|c|c|}
\hline
Zerfallsproze"s & Lagrangedichte-Term & Amplitude ($\cdot \sqrt{30}$)
& Vorfaktor \\
\hline \hline
$p \! \uparrow \; \rightarrow \;
e^{\scr +}_{\scr R} u^{\scr C} \!\! \uparrow u \! \downarrow$ 
& $\big( \bar u_{\scr L}^{\scr C}  \gamma^{\scr \mu} u_{\scr L}^{} \big)
\big( \bar e^{\scr +}_{\scr R} \gamma_{\scr \mu} d_{\scr R}^{} \big)$ 
& $-4$ & $A^{}_2$ \\
$p \! \uparrow \; \rightarrow \;
e^{\scr +}_{\scr R} u^{\scr C} \!\! \downarrow u \! \uparrow$ 
& $\big( \bar u_{\scr L}^{\scr C}  \gamma^{\scr \mu} u_{\scr L}^{} \big)
\big( \bar e^{\scr +}_{\scr R} \gamma_{\scr \mu} d_{\scr R}^{} \big)$ 
& $-8$ & $A^{}_2$ \\
$p \! \uparrow \; \rightarrow \;
e^{\scr +}_{\scr R} d^{\scr C} \!\! \uparrow d \! \downarrow$ 
& $\big( \bar u_{\scr L}^{\scr C}  \gamma^{\scr \mu} u_{\scr L}^{} \big)
\big( \bar e^{\scr +}_{\scr R} \gamma_{\scr \mu} d_{\scr R}^{} \big)$ 
& $-8$ & $A^{}_2$ \\
$p \! \uparrow \; \rightarrow \;
e^{\scr +}_{\scr R} d^{\scr C} \!\! \downarrow d \! \uparrow$ 
& $\big( \bar u_{\scr L}^{\scr C}  \gamma^{\scr \mu} u_{\scr L}^{} \big)
\big( \bar e^{\scr +}_{\scr R} \gamma_{\scr \mu} d_{\scr R}^{} \big)$ 
& $+2$ & $A^{}_2$ \\
\hline
$p \! \downarrow \; \rightarrow \;
e^{\scr +}_{\scr R} u^{\scr C} \!\! \downarrow u \! \downarrow$ 
& $\big( \bar u_{\scr L}^{\scr C}  \gamma^{\scr \mu} u_{\scr L}^{} \big)
\big( \bar e^{\scr +}_{\scr R} \gamma_{\scr \mu} d_{\scr R}^{} \big)$ 
& $-10$ & $A^{}_2$ \\
$p \! \downarrow \; \rightarrow \;
e^{\scr +}_{\scr R} d^{\scr C} \!\! \downarrow d \! \downarrow$ 
& $\big( \bar u_{\scr L}^{\scr C}  \gamma^{\scr \mu} u_{\scr L}^{} \big)
\big( \bar e^{\scr +}_{\scr R} \gamma_{\scr \mu} d_{\scr R}^{} \big)$ 
& $-2$ & $A^{}_2$ \\
\hline
$p \! \uparrow \; \rightarrow \;
e^{\scr +}_{\scr R} s^{\scr C} \!\! \uparrow d \! \downarrow$ 
& $\big( \bar u_{\scr L}^{\scr C}  \gamma^{\scr \mu} u_{\scr L}^{} \big)
\big( \bar e^{\scr +}_{\scr R} \gamma_{\scr \mu} s_{\scr R}^{} \big)$ 
& $-8$ & $A^{}_6$ \\
$p \! \uparrow \; \rightarrow \;
e^{\scr +}_{\scr R} s^{\scr C} \!\! \downarrow d \! \uparrow$ 
& $\big( \bar u_{\scr L}^{\scr C}  \gamma^{\scr \mu} u_{\scr L}^{} \big)
\big( \bar e^{\scr +}_{\scr R} \gamma_{\scr \mu} s_{\scr R}^{} \big)$ 
& $+2$ & $A^{}_6$ \\
\hline
$p \! \downarrow \; \rightarrow \;
e^{\scr +}_{\scr R} s^{\scr C} \!\! \downarrow d \! \downarrow$ 
& $\big( \bar u_{\scr L}^{\scr C}  \gamma^{\scr \mu} u_{\scr L}^{} \big)
\big( \bar e^{\scr +}_{\scr R} \gamma_{\scr \mu} s_{\scr R}^{} \big)$ 
& $-2$ & $A^{}_6$ \\
\hline
$p \! \uparrow \; \rightarrow \;
\nu_{\scr e R}^{\scr C} d^{\scr C} \!\! \uparrow u \! \downarrow$ 
& $\big( \bar u_{\scr L}^{\scr C}  \gamma^{\scr \mu} d_{\scr L}^{} \big)
\big( \bar \nu_{\scr e R}^{\scr C} \gamma_{\scr \mu} d_{\scr R}^{} \big)$ 
& $+4$ & $A^{}_9$ \\
$p \! \uparrow \; \rightarrow \;
\nu_{\scr e R}^{\scr C} d^{\scr C} \!\! \downarrow u \! \uparrow$ 
& $\big( \bar u_{\scr L}^{\scr C}  \gamma^{\scr \mu} d_{\scr L}^{} \big)
\big( \bar \nu_{\scr e R}^{\scr C} \gamma_{\scr \mu} d_{\scr R}^{} \big)$ 
& $-10$ & $A^{}_9$ \\
\hline
$p \! \downarrow \; \rightarrow \;
\nu_{\scr e R}^{\scr C} d^{\scr C} \!\! \downarrow u \! \downarrow$ 
& $\big( \bar u_{\scr L}^{\scr C}  \gamma^{\scr \mu} d_{\scr L}^{} \big)
\big( \bar \nu_{\scr e R}^{\scr C} \gamma_{\scr \mu} d_{\scr R}^{} \big)$ 
& $-8$ & $A^{}_9$ \\
\hline
$p \! \uparrow \; \rightarrow \;
\nu_{\scr e R}^{\scr C} s^{\scr C} \!\! \uparrow u \! \downarrow$ 
& $\big( \bar u_{\scr L}^{\scr C}  \gamma^{\scr \mu} d_{\scr L}^{} \big)
\big( \bar \nu_{\scr e R}^{\scr C} \gamma_{\scr \mu} s_{\scr R}^{} \big)$ 
& $+4$ & $A^{}_{11}$ \\
$p \! \uparrow \; \rightarrow \;
\nu_{\scr e R}^{\scr C} s^{\scr C} \!\! \downarrow u \! \uparrow$ 
& $\big( \bar u_{\scr L}^{\scr C}  \gamma^{\scr \mu} d_{\scr L}^{} \big)
\big( \bar \nu_{\scr e R}^{\scr C} \gamma_{\scr \mu} s_{\scr R}^{} \big)$ 
& $+2$ & $A^{}_{11}$ \\
$p \! \uparrow \; \rightarrow \;
\nu_{\scr e R}^{\scr C} s^{\scr C} \!\! \uparrow u \! \downarrow$ 
& $\big( \bar u_{\scr L}^{\scr C}  \gamma^{\scr \mu} s_{\scr L}^{} \big)
\big( \bar \nu_{\scr e R}^{\scr C} \gamma_{\scr \mu} d_{\scr R}^{} \big)$ 
& $0$ & $A^{}_{13}$ \\
$p \! \uparrow \; \rightarrow \;
\nu_{\scr e R}^{\scr C} s^{\scr C} \!\! \downarrow u \! \uparrow$ 
& $\big( \bar u_{\scr L}^{\scr C}  \gamma^{\scr \mu} s_{\scr L}^{} \big)
\big( \bar \nu_{\scr e R}^{\scr C} \gamma_{\scr \mu} d_{\scr R}^{} \big)$ 
& $-12$ & $A^{}_{13}$ \\
\hline
$p \! \downarrow \; \rightarrow \;
\nu_{\scr e R}^{\scr C} s^{\scr C} \!\! \downarrow u \! \downarrow$ 
& $\big( \bar u_{\scr L}^{\scr C}  \gamma^{\scr \mu} d_{\scr L}^{} \big)
\big( \bar \nu_{\scr e R}^{\scr C} \gamma_{\scr \mu} s_{\scr R}^{} \big)$ 
& $+4$ & $A^{}_{11}$ \\
$p \! \downarrow \; \rightarrow \;
\nu_{\scr e R}^{\scr C} s^{\scr C} \!\! \downarrow u \! \downarrow$ 
& $\big( \bar u_{\scr L}^{\scr C}  \gamma^{\scr \mu} s_{\scr L}^{} \big)
\big( \bar \nu_{\scr e R}^{\scr C} \gamma_{\scr \mu} d_{\scr R}^{} \big)$ 
& $-12$ & $A^{}_{13}$ \\
\hline
\end{tabular}
\end{center}
\end{table}
\begin{table}[h]
\begin{center}
\begin{tabular}{|c|c|c|c|}
\hline
Zerfallsproze"s & Lagrangedichte-Term & Amplitude ($\cdot \sqrt{30}$)
& Vorfaktor \\
\hline \hline
$n \! \uparrow \; \rightarrow \;
e^{\scr +}_{\scr R} u^{\scr C} \!\! \uparrow d \! \downarrow$ 
& $\big( \bar u_{\scr L}^{\scr C}  \gamma^{\scr \mu} u_{\scr L}^{} \big)
\big( \bar e^{\scr +}_{\scr R} \gamma_{\scr \mu} d_{\scr R}^{} \big)$ 
& $-4$ & $A^{}_2$ \\
$n \! \uparrow \; \rightarrow \;
e^{\scr +}_{\scr R} u^{\scr C} \!\! \downarrow d \! \uparrow$ 
& $\big( \bar u_{\scr L}^{\scr C}  \gamma^{\scr \mu} u_{\scr L}^{} \big)
\big( \bar e^{\scr +}_{\scr R} \gamma_{\scr \mu} d_{\scr R}^{} \big)$ 
& $+10$ & $A^{}_2$ \\
\hline
$n \! \downarrow \; \rightarrow \;
e^{\scr +}_{\scr R} u^{\scr C} \!\! \downarrow d \! \downarrow$ 
& $\big( \bar u_{\scr L}^{\scr C}  \gamma^{\scr \mu} u_{\scr L}^{} \big)
\big( \bar e^{\scr +}_{\scr R} \gamma_{\scr \mu} d_{\scr R}^{} \big)$ 
& $+8$ & $A^{}_2$ \\
\hline
$n \! \uparrow \; \rightarrow \;
\nu_{\scr e R}^{\scr C} u^{\scr C} \!\! \uparrow u \! \downarrow$ 
& $\big( \bar u_{\scr L}^{\scr C}  \gamma^{\scr \mu} d_{\scr L}^{} \big)
\big( \bar \nu_{\scr e R}^{\scr C} \gamma_{\scr \mu} d_{\scr R}^{} \big)$ 
& $+8$ & $A^{}_9$ \\
$n \! \uparrow \; \rightarrow \;
\nu_{\scr e R}^{\scr C} u^{\scr C} \!\! \downarrow u \! \uparrow$ 
& $\big( \bar u_{\scr L}^{\scr C}  \gamma^{\scr \mu} d_{\scr L}^{} \big)
\big( \bar \nu_{\scr e R}^{\scr C} \gamma_{\scr \mu} d_{\scr R}^{} \big)$ 
& $-2$ & $A^{}_9$ \\
$n \! \uparrow \; \rightarrow \;
\nu_{\scr e R}^{\scr C} d^{\scr C} \!\! \uparrow d \! \downarrow$ 
& $\big( \bar u_{\scr L}^{\scr C}  \gamma^{\scr \mu} d_{\scr L}^{} \big)
\big( \bar \nu_{\scr e R}^{\scr C} \gamma_{\scr \mu} d_{\scr R}^{} \big)$ 
& $+4$ & $A^{}_9$ \\
$n \! \uparrow \; \rightarrow \;
\nu_{\scr e R}^{\scr C} d^{\scr C} \!\! \downarrow d \! \uparrow$ 
& $\big( \bar u_{\scr L}^{\scr C}  \gamma^{\scr \mu} d_{\scr L}^{} \big)
\big( \bar \nu_{\scr e R}^{\scr C} \gamma_{\scr \mu} d_{\scr R}^{} \big)$ 
& $+8$ & $A^{}_9$ \\
\hline
$n \! \downarrow \; \rightarrow \;
\nu_{\scr e R}^{\scr C} u^{\scr C} \!\! \downarrow u \! \downarrow$ 
& $\big( \bar u_{\scr L}^{\scr C}  \gamma^{\scr \mu} d_{\scr L}^{} \big)
\big( \bar \nu_{\scr e R}^{\scr C} \gamma_{\scr \mu} d_{\scr R}^{} \big)$ 
& $+2$ & $A^{}_9$ \\
$n \! \downarrow \; \rightarrow \;
\nu_{\scr e R}^{\scr C} d^{\scr C} \!\! \downarrow d \! \downarrow$ 
& $\big( \bar u_{\scr L}^{\scr C}  \gamma^{\scr \mu} d_{\scr L}^{} \big)
\big( \bar \nu_{\scr e R}^{\scr C} \gamma_{\scr \mu} d_{\scr R}^{} \big)$ 
& $+10$ & $A^{}_9$ \\
\hline
$n \! \uparrow \; \rightarrow \;
\nu_{\scr e R}^{\scr C} s^{\scr C} \!\! \uparrow d \! \downarrow$ 
& $\big( \bar u_{\scr L}^{\scr C}  \gamma^{\scr \mu} d_{\scr L}^{} \big)
\big( \bar \nu_{\scr e R}^{\scr C} \gamma_{\scr \mu} s_{\scr R}^{} \big)$ 
& $+4$ & $A^{}_{11}$ \\
$n \! \uparrow \; \rightarrow \;
\nu_{\scr e R}^{\scr C} s^{\scr C} \!\! \downarrow d \! \uparrow$ 
& $\big( \bar u_{\scr L}^{\scr C}  \gamma^{\scr \mu} d_{\scr L}^{} \big)
\big( \bar \nu_{\scr e R}^{\scr C} \gamma_{\scr \mu} s_{\scr R}^{} \big)$ 
& $-4$ & $A^{}_{11}$ \\
$n \! \uparrow \; \rightarrow \;
\nu_{\scr e R}^{\scr C} s^{\scr C} \!\! \uparrow d \! \downarrow$ 
& $\big( \bar u_{\scr L}^{\scr C}  \gamma^{\scr \mu} s_{\scr L}^{} \big)
\big( \bar \nu_{\scr e R}^{\scr C} \gamma_{\scr \mu} d_{\scr R}^{} \big)$ 
& $0$ & $A^{}_{13}$ \\
$n \! \uparrow \; \rightarrow \;
\nu_{\scr e R}^{\scr C} s^{\scr C} \!\! \downarrow d \! \uparrow$ 
& $\big( \bar u_{\scr L}^{\scr C}  \gamma^{\scr \mu} s_{\scr L}^{} \big)
\big( \bar \nu_{\scr e R}^{\scr C} \gamma_{\scr \mu} d_{\scr R}^{} \big)$ 
& $+12$ & $A^{}_{13}$ \\
\hline
$n \! \downarrow \; \rightarrow \;
\nu_{\scr e R}^{\scr C} s^{\scr C} \!\! \downarrow d \! \downarrow$ 
& $\big( \bar u_{\scr L}^{\scr C}  \gamma^{\scr \mu} d_{\scr L}^{} \big)
\big( \bar \nu_{\scr e R}^{\scr C} \gamma_{\scr \mu} s_{\scr R}^{} \big)$ 
& $-2$ & $A^{}_{11}$ \\
$n \! \downarrow \; \rightarrow \;
\nu_{\scr e R}^{\scr C} s^{\scr C} \!\! \downarrow d \! \downarrow$ 
& $\big( \bar u_{\scr L}^{\scr C}  \gamma^{\scr \mu} s_{\scr L}^{} \big)
\big( \bar \nu_{\scr e R}^{\scr C} \gamma_{\scr \mu} d_{\scr R}^{} \big)$ 
& $+12$ & $A^{}_{13}$ \\
\hline
\end{tabular}
\end{center}
\caption{"Ubergangsamplituden f"ur die Elementarprozesse der
  Nukleonenzerf"alle \label{aepnd}}
\end{table}
\begin{table}[h]
\begin{center}
\begin{tabular}{|l|c||l|c|}
\hline
Zerfallsproze"s & $| \, \textrm{Amplitude} \cdot \textrm{Vorfaktor} \, |^2$ 
& Zerfallsproze"s & $| \, \textrm{Amplitude} \cdot \textrm{Vorfaktor} \, |^2$ 
\\
\hline \hline
$p \! \uparrow \; \rightarrow \; e^{\scr +}_{\scr R} \pi^0$ 
& $(49/30) \, A^2_2$ &
$n \! \uparrow \; \rightarrow \; e^{\scr +}_{\scr R} \pi^-$ 
& $(49/15) \, A^2_2$ \\
$p \! \uparrow \; \rightarrow \; e^{\scr +}_{\scr R} K^0$ 
& $(5/3) \, A^2_6$ &
$n \! \uparrow \; \rightarrow \; \nu_{\scr e R}^{\scr C} \pi^0$ 
& $(49/30) \, A^2_9$ \\
$p \! \uparrow \; \rightarrow \; e^{\scr +}_{\scr R} \eta$ 
& $(1/10) \, A^2_2$ &
$n \! \uparrow \; \rightarrow \; \nu_{\scr e R}^{\scr C} K^0$ 
& $(4/15)(2\,A_{11}-3\,A_{13})^2$ \\
$p \! \uparrow \; \rightarrow \; \nu_{\scr e R}^{\scr C} \pi^+$ 
& $(49/15) \, A^2_9$ &
$n \! \uparrow \; \rightarrow \; \nu_{\scr e R}^{\scr C} \eta$ 
& $(1/10) \, A^2_9$ \\
$p \! \uparrow \; \rightarrow \; \nu_{\scr e R}^{\scr C} K^+$ 
& $(1/15)(A_{11}+6\,A_{13})^2$ &
$n \! \uparrow \; \rightarrow \; e^{\scr +}_{\scr R} \rho_{\scr (0)}^-$ 
& $(3/5) \, A^2_2$ \\
$p \! \uparrow \; \rightarrow \; e^{\scr +}_{\scr R} \rho_{\scr (0)}^0$ 
& $(3/10) \, A^2_2$ &
$n \! \downarrow \; \rightarrow \; e^{\scr +}_{\scr R} \rho_{\scr (-)}^-$ 
& $(32/15) \, A^2_2$ \\
$p \! \downarrow \; \rightarrow \; e^{\scr +}_{\scr R} \rho_{\scr (-)}^0$ 
& $(16/15) \, A^2_2$ &
$n \! \uparrow \; \rightarrow \; \nu_{\scr e R}^{\scr C} \rho_{\scr (0)}^0$ 
& $(3/10) \, A^2_9$ \\
$p \! \uparrow \; \rightarrow \; e^{\scr +}_{\scr R} \omega_{\scr (0)}$ 
& $(27/10) \, A^2_2$ &
$n \! \downarrow \; \rightarrow \; \nu_{\scr e R}^{\scr C} \rho_{\scr (-)}^0$ 
& $(16/15) \, A^2_9$ \\
$p \! \downarrow \; \rightarrow \; e^{\scr +}_{\scr R} \omega_{\scr (-)}$ 
& $(12/5) \, A^2_2$ &
$n \! \uparrow \; \rightarrow \; \nu_{\scr e R}^{\scr C} \omega_{\scr (0)}^0$ 
& $(27/10) \, A^2_9$ \\
$p \! \uparrow \; \rightarrow \; e^{\scr +}_{\scr R} K_{\scr (0)}^{*0}$ 
& $(3/5) \, A^2_6$ &
$n \! \downarrow \; \rightarrow \; \nu_{\scr e R}^{\scr C} 
\omega_{\scr (-)}^0$ & $(12/5) \, A^2_9$ \\
$p \! \downarrow \; \rightarrow \; e^{\scr +}_{\scr R} K_{\scr (-)}^{*0}$ 
& $(2/15) \, A^2_6$ &
$n \! \uparrow \; \rightarrow \; \nu_{\scr e R}^{\scr C} K_{\scr (0)}^{*0}$ 
& $(12/5) \, A^2_{13}$ \\
$p \! \uparrow \; \rightarrow \; \nu_{\scr e R}^{\scr C} \rho_{\scr (0)}^+$ 
& $(3/5) \, A^2_9$ &
$n \! \downarrow \; \rightarrow \; \nu_{\scr e R}^{\scr C} K_{\scr (-)}^{*0}$ 
& $(2/15)(A_{11}-6\,A_{13})^2$ \\
$p \! \downarrow \; \rightarrow \; \nu_{\scr e R}^{\scr C} \rho_{\scr (-)}^+$ 
& $(32/15) \, A^2_9$ & & \\
$p \! \uparrow \; \rightarrow \; \nu_{\scr e R}^{\scr C} K_{\scr (0)}^{*+}$ 
& $(3/5)(A_{11}-2\,A_{13})^2$ & & \\
$p \! \downarrow \; \rightarrow \; \nu_{\scr e R}^{\scr C} K_{\scr (-)}^{*+}$ 
& $(8/15)(A_{11}-3\,A_{13})^2$ & & \\
\hline
\end{tabular}
\end{center}
\caption{"Ubergangswahrscheinlichkeiten f"ur Zerfallsprozesse der Nukleonen
  mit physikalischen Endzust"anden \label{trprob}}
\end{table}

\section{Experimentelle Grenzen f"ur die Nukleonzerfallsraten}
Tabelle \ref{exptab} zeigt die in \cite{pdg} angegebenen Untergrenzen
f"ur die inversen partiellen Zerfallsraten der Nukleonen aufgrund der 
experimentellen Nichtbeobachtung dieser Prozesse. Die aktuellsten
Resultate des Super-Kamiokande-Experiments Nukleonenzerf"alle betref\/fend
sind in Tabelle \ref{ndsumt} zu f\/inden.
\begin{table}[h]
\begin{center}
\begin{tabular}{|l|c||l|c|}
\hline
Zerfallskanal & $1/\Gamma_i \; (10_{}^{30} \textrm{yr})$
& Zerfallskanal & $1/\Gamma_i \; (10_{}^{30} \textrm{yr})$ \\
\hline
\hline
$p \; \rightarrow \; \epi$ & $>$ 550 \;
& $n \; \rightarrow \; \epin$ & $>$ 130 \;\\
$p \; \rightarrow \; \ek$  & $>$ 150 \;
& $n \; \rightarrow \; \mpin$ & $>$ 100 \; \\
$p \; \rightarrow \; \et$  & $>$ 140 \;
& $n \; \rightarrow \; \eron$ & $>$ 58 \\
$p \; \rightarrow \; \mpi$ & $>$ 270 \;
& $n \; \rightarrow \; \mron$ & $>$ 23 \\
$p \; \rightarrow \; \muk$ & $>$ 120 \;
& $n \; \rightarrow \; \nu^{\scr C}_{} \pi^0$ & $>$ 100 \; \\
$p \; \rightarrow \; \mt$  & $>$ 69
& $n \; \rightarrow \; \nu^{\scr C}_{} K^0$ & $>$ 86 \\
$p \; \rightarrow \; \ero$ & $>$ 75
& $n \; \rightarrow \; \nu^{\scr C}_{} \eta$ & $>$ 54 \\
$p \; \rightarrow \; \eo$  & $>$ 45
& $n \; \rightarrow \; \nu^{\scr C}_{} \rho^0$ & $>$ 19 \\
$p \; \rightarrow \; \eks$ & $>$ 52
& $n \; \rightarrow \; \nu^{\scr C}_{} \omega$ & $>$ 43 \\
$p \; \rightarrow \; \mro$ & $>$ 110 \;
& $n \; \rightarrow \; \nu^{\scr C}_{} K^{*0}$ & $>$ 22 \\
$p \; \rightarrow \; \mo$  & $>$ 57 & & \\
$p \; \rightarrow \; \nu^{\scr C}_{} \pi^+$ & $>$ 25 & & \\
$p \; \rightarrow \; \nu^{\scr C}_{} K^+$ & $>$ 100 \; & & \\
$p \; \rightarrow \; \nu^{\scr C}_{} \rho^+$ & $>$ 27 & & \\
$p \; \rightarrow \; \nu^{\scr C}_{} K^{*+}$ & $>$ 20 & & \\
\hline
\end{tabular}
\caption[Experimentelle Grenzen f"ur die Lebensdauern $1/\Gamma_i$
der Nukleonen]{Experimentelle Grenzen f"ur die Lebensdauern 
$1/\Gamma_i$ der Nukleonen \cite{pdg} \label{exptab}}
\end{center}
\end{table}
\section{Zerfallsraten der Nukleonen}
In diesem Abschnitt sind in den Tabellen \ref{pdrt}-\ref{tdrnt} die 
numerischen Resultate f"ur die partiellen und totalen
Nukleonzerfallsraten zusammengefa"st. Neben den Ergebnissen f"ur die
drei untersuchten Modelle ist zum Vergleich auch der Fall
verschwindender Fermionmischungen ber"ucksichtigt worden. W"ahrend die
Eintr"age 0.0 f"ur partielle Raten $< 0.05 \%$ stehen, sind Zerf"alle in
die mit "`---"' gekennzeichneten Kan"ale in f"uhrender Ordnung nicht
erlaubt. Es gilt $\Gamma = \sum_j \Gamma_j$ sowie $\tau = 1/\Gamma$ und
weiterhin in nat"urlichen Einheiten ($\hbar=c=1$) 
1 GeV$=4.77 \cdot 10^{31}$ yr$^{-1}$.
\begin{table}[h]
\begin{center}
\begin{tabular}{|l|r|r|r|r|}
\hline
Zerfallskanal & Raten in \% & Raten in \% & Raten in \% & Raten in \% \\
des Protons & (keine Mischungen) & in Modell\,1 & in Modell\,2a 
& in Modell\,2b \\
\hline
\hline
$p \; \rightarrow \; \epi$  
& 33.6 \; & 21.4 \; & 25.1 \; & 27.8 \; \\
$p \; \rightarrow \; \ek$   
&  --- \; &  3.1 \; &  2.6 \; &  4.5 \; \\
$p \; \rightarrow \; \et$   
&  1.2 \; &  0.8 \; &  0.9 \; &  1.0 \; \\
$p \; \rightarrow \; \mpi$  
&  --- \; &  8.5 \; &  5.7 \; &  5.6 \; \\
$p \; \rightarrow \; \muk$  
&  5.8 \; &  2.6 \; &  0.9 \; &  1.8 \; \\
$p \; \rightarrow \; \mt$   
&  --- \; &  0.3 \; &  0.2 \; &  0.2 \; \\
$p \; \rightarrow \; \ero$ 
&  5.1 \; &  3.3 \; &  3.8 \; &  4.2 \; \\
$p \; \rightarrow \; \eo$   
& 16.9 \; & 10.8 \; & 12.7 \; & 14.0 \; \\
$p \; \rightarrow \; \eks$  
&  --- \; &  0.0 \; &  0.0 \; &  0.0 \; \\
$p \; \rightarrow \; \mro$  
&  --- \; &  1.3 \; &  0.9 \; &  0.8 \; \\
$p \; \rightarrow \; \mo$   
&  --- \; &  4.3 \; &  2.9 \; &  2.8 \; \\
$p \; \rightarrow \; \nep$  
& 32.3 \; & 25.6 \; & 35.6 \; & 27.7 \; \\
$p \; \rightarrow \; \nek$  
&  --- \; &  2.0 \; &  2.0 \; &  4.1 \; \\
$p \; \rightarrow \; \nmpi$ 
&  --- \; &  8.9 \; &  0.5 \; &  0.3 \; \\
$p \; \rightarrow \; \nmk$  
&  0.1 \; &  1.3 \; &  0.2 \; &  0.3 \; \\
$p \; \rightarrow \; \nero$ 
&  4.9 \; &  3.9 \; &  5.4 \; &  4.2 \; \\
$p \; \rightarrow \; \neks$ 
&  --- \; &  0.4 \; &  0.3 \; &  0.6 \; \\
$p \; \rightarrow \; \nmro$ 
&  --- \; &  1.4 \; &  0.1 \; &  0.0 \; \\
$p \; \rightarrow \; \nmks$ 
&  0.1 \; &  0.0 \; &  0.0 \; &  0.0 \; \\
$p \; \rightarrow \; \ntp$  
&  --- \; &  0.1 \; &  0.1 \; &  0.0 \; \\
$p \; \rightarrow \; \ntk$  
&  --- \; &  0.1 \; &  0.1 \; &  0.0 \; \\
$p \; \rightarrow \; \ntro$ 
&  --- \; &  0.0 \; &  0.0 \; &  0.0 \; \\
$p \; \rightarrow \; \ntks$ 
&  --- \; &  0.0 \; &  0.0 \; &  0.0 \; \\
\hline
$p \; \rightarrow \; e^{\scr +} X^{\scr 0}$ 
& 56.8 \; & 39.4 \; & 45.1 \; & 51.5 \; \\
$p \; \rightarrow \; \mu^{\scr +} X^{\scr 0}$ 
&  5.8 \; & 17.0 \; & 10.6 \; & 11.2 \; \\
$p \; \rightarrow \; \nu^{\scr C} X^{\scr +}$ 
& 37.4 \; & 43.7 \; & 44.3 \; & 37.2 \; \\
\hline
\end{tabular}
\caption[Partielle Zerfallsraten $\Gamma_i/\Gamma$ des Protons]
{Partielle Zerfallsraten $\Gamma_i/\Gamma$ des Protons f"ur
  den Fall der Vernachl"assigung von Fermionmischungen und f"ur die 
  untersuchten L"osungen \label{pdrt}}
\end{center}
\end{table}
\begin{table}[h]
\begin{center}
\begin{tabular}{|l|c|c|c|}
\hline
Gr"o"se & Wert in Modell\,1 & Wert in Modell\,2a & Wert in Modell\,2b \\
\hline
\hline
$\Gamma_p$ & $2.54 \cdot 10_{}^{\scr -35} \; \textrm{yr}_{}^{\scr -1}$
& $2.57 \cdot 10_{}^{\scr -35} \; \textrm{yr}_{}^{\scr -1}$
& $2.83 \cdot 10_{}^{\scr -35} \; \textrm{yr}_{}^{\scr -1}$ \\
$\tau_p$ & $3.94 \cdot 10_{}^{\scr 34} \; \textrm{yr}$ 
& $3.89 \cdot 10_{}^{\scr 34} \; \textrm{yr}$ 
& $3.53 \cdot 10_{}^{\scr 34} \; \textrm{yr}$ \\
\hline
\end{tabular}
\caption{Totale Zerfallsraten und Lebensdauern des Protons \label{tdrpt}}
\end{center}
\end{table}
\begin{table}[h]
\begin{center}
\begin{tabular}{|l|r|r|r|r|}
\hline
Zerfallskanal & Raten in \% & Raten in \% & Raten in \% & Raten in \% \\
des Neutrons & (keine Mischungen) & in Modell\,1 & in Modell\,2a 
& in Modell\,2b \\
\hline
\hline
$n \; \rightarrow \; \epin$  
& 62.9 \; & 40.1 \; & 46.2 \; & 49.9 \; \\
$n \; \rightarrow \; \mpin$  
&  --- \; & 15.8 \; & 10.4 \; & 10.0 \; \\
$n \; \rightarrow \; \eron$  
&  9.7 \; &  6.2 \; &  7.1 \; &  7.7 \; \\
$n \; \rightarrow \; \mron$  
&  --- \; &  2.4 \; &  1.6 \; &  1.5 \; \\
$n \; \rightarrow \; \nepn$  
& 15.1 \; & 12.0 \; & 16.4 \; & 12.5 \; \\
$n \; \rightarrow \; \nekn$  
&  --- \; &  6.8 \; &  4.9 \; &  8.2 \; \\
$n \; \rightarrow \; \netn$  
&  0.6 \; &  0.4 \; &  0.6 \; &  0.5 \; \\
$n \; \rightarrow \; \nmpin$ 
&  --- \; &  4.2 \; &  0.2 \; &  0.1 \; \\
$n \; \rightarrow \; \nmkn$  
&  1.7 \; &  0.1 \; &  1.0 \; &  0.9 \; \\
$n \; \rightarrow \; \nmtn$  
&  --- \; &  0.2 \; &  0.0 \; &  0.0 \; \\
$n \; \rightarrow \; \neron$ 
&  2.3 \; &  1.8 \; &  2.5 \; &  1.9 \; \\
$n \; \rightarrow \; \neon$  
&  7.7 \; &  6.1 \; &  8.4 \; &  6.4 \; \\
$n \; \rightarrow \; \neksn$ 
&  --- \; &  0.2 \; &  0.2 \; &  0.4 \; \\
$n \; \rightarrow \; \nmron$ 
&  --- \; &  0.6 \; &  0.0 \; &  0.0 \; \\
$n \; \rightarrow \; \nmon$  
&  --- \; &  2.1 \; &  0.1 \; &  0.1 \; \\
$n \; \rightarrow \; \nmksn$ 
&  0.0 \; &  0.1 \; &  0.0 \; &  0.0 \; \\
$n \; \rightarrow \; \ntpn$  
&  --- \; &  0.0 \; &  0.0 \; &  0.0 \; \\
$n \; \rightarrow \; \ntkn$  
&  --- \; &  0.7 \; &  0.4 \; &  0.0 \; \\
$n \; \rightarrow \; \nttn$  
&  --- \; &  0.0 \; &  0.0 \; &  0.0 \; \\
$n \; \rightarrow \; \ntron$ 
&  --- \; &  0.0 \; &  0.0 \; &  0.0 \; \\
$n \; \rightarrow \; \nton$  
&  --- \; &  0.0 \; &  0.0 \; &  0.0 \; \\
$n \; \rightarrow \; \ntksn$ 
&  --- \; &  0.0 \; &  0.0 \; &  0.0 \; \\
\hline
$n \; \rightarrow \; e^{\scr +} X^{\scr -}$ 
& 72.6 \; & 46.3 \; & 53.3 \; & 57.6 \; \\
$n \; \rightarrow \; \mu^{\scr +} X^{\scr -}$ 
&  --- \; & 18.2 \; & 12.0 \; & 11.5 \; \\
$n \; \rightarrow \; \nu^{\scr C} X^{\scr 0}$ 
& 27.4 \; & 35.3 \; & 34.7 \; & 31.0 \; \\
\hline
\end{tabular}
\caption[Partielle Zerfallsraten $\Gamma_i/\Gamma$ des Neutrons]
{Partielle Zerfallsraten $\Gamma_i/\Gamma$ des Neutrons f"ur den Fall der 
Vernachl"assigung von Fermionmischungen und f"ur die untersuchten
L"osungen \label{ndrt}}
\end{center}
\end{table}
\begin{table}[h]
\begin{center}
\begin{tabular}{|l|c|c|c|}
\hline
Gr"o"se & Wert in Modell\,1 & Wert in Modell\,2a & Wert in Modell\,2b \\
\hline
\hline
$\Gamma_n$ & $2.72 \cdot 10_{}^{\scr -35} \; \textrm{yr}_{}^{\scr -1}$
& $2.80 \cdot 10_{}^{\scr -35} \; \textrm{yr}_{}^{\scr -1}$
& $3.14 \cdot 10_{}^{\scr -35} \; \textrm{yr}_{}^{\scr -1}$ \\
$\tau_n$ & $3.68 \cdot 10_{}^{\scr 34} \; \textrm{yr}$ 
& $3.57 \cdot 10_{}^{\scr 34} \; \textrm{yr}$ 
& $3.18 \cdot 10_{}^{\scr 34} \; \textrm{yr}$ \\
\hline
\end{tabular}
\caption{Totale Zerfallsraten und Lebensdauern des Neutrons \label{tdrnt}}
\end{center}
\end{table}

%%% Local Variables: 
%%% mode: latex
%%% TeX-master: t
%%% End: 

\end{appendix}

\newpage

\thispagestyle{empty}

\mbox{}

\newpage

\thispagestyle{empty}

\chapter*{Schlu"swort}
\addcontentsline{toc}{chapter}{Schlu"swort}

\noindent An dieser Stelle m"ochte ich mich bei all denen bedanken, die
mich w"ahrend der letzten drei Jahre auf die eine oder andere Weise 
unterst"utzt haben.

Zun"achst gilt mein Dank Prof. Dr. Yoav Achiman, der dieses Projekt
angeregt und betreut hat. Er hat durch sein st"andiges Interesse, die 
stete Bereitschaft zu ausgiebigen Diskussionen und zahlreiche 
physikalische Ideen wesentlich zum Gelingen dieser Arbeit beigetragen.
Dank auch den "ubrigen Mitgliedern des Pr"ufungsausschusses,
Prof. Dr. Peter Kroll als Zweitgutachter der Dissertation sowie
Prof. Dr. J"urgen Drees und Dr. Peer Ueberholz.

Dr. Mina Parida war mir w"ahrend seines Forschungsaufenthaltes in
Wuppertal mit seinem Fachwissen eine sehr gro"se Hilfe und stets bereit, 
sich mit meinem Fragen zu besch"aftigen. Ebenso danke ich Prof. Dr. Franco 
Buccella, Dr. Ofelia Pisanti und Dr. Pietro Santorelli f"ur die
aufschlu"sreichen Diskussionen w"ahren der Sommerschule auf Capri.

Dank auch allen Kollegen am Institut, insbesondere aber meinen Mitstreitern
aus F\,10.09. Marcus Richter, Dr. Armin Seyfried, Thorsten Struckmann und
Dr. Jochen Viehof\/f verdanke ich eine "uberaus angenehme Arbeitsatmosph"are 
w"ahrend der vergangenen drei Jahre. Auch die Hilfe von Norbert Eicker in
EDV-Dingen sowie von Sabine Hof\/fmann und Anita Wied aus dem Sekretariat 
in allen organisatorischen Fragen ist nicht zu untersch"atzen.

Meiner Familie und meinem Freundeskreis geb"uhrt ebenfalls gro"ser Dank. 
Meine Eltern haben mich w"ahrend meines gesamten Studiums in jeder Weise
tatkr"aftig unterst"utzt. Den Freunden, allen voran aber Christine
Rompel, sei f"ur ihre gro"se Geduld und Aufmunterung vor allem in der
Endphase meiner Promotion gedankt.

Der Deutschen Forschungsgemeinschaft schlie"slich bin ich f"ur die 
f\/inanzielle Unter\-st"ut\-zung in Form eines Stipendiums im Rahmen des
Graduiertenkollegs "`Feldtheoretische und numerische Methoden der 
Elementarteilchen- und Statistischen Physik"' "uberaus dankbar.

\vspace*{1.5cm}

\noindent Carsten Merten

\noindent Wuppertal, Dezember 1999

\newpage

\thispagestyle{empty}

\mbox{ }

\newpage

\thispagestyle{empty}

\mbox{ }

\newpage

\thispagestyle{empty}

\mbox{ }

\newpage

\thispagestyle{empty}

\mbox{ }

\newpage

\thispagestyle{empty}

\mbox{ }

\end{document}